\documentclass[pra,onecolumn, english,notitlepage,nofootinbib,aps,superscriptaddress,showkeys]{revtex4}

\usepackage{babel,calc,amsmath,amsthm,amssymb,graphicx,subfigure,xcolor,longtable,braket,bm,cases,makecell}

\usepackage[T1]{fontenc}
\usepackage[unicode=true]{hyperref}
\hypersetup{
colorlinks=true,       		
linkcolor=blue,          	
citecolor=red,            
urlcolor=magenta,           	
}
\newtheorem*{theorem*}{Theorem}
\newtheorem{theorem}{Theorem}

\newtheorem*{corollary*}{Corollary}

\newtheorem*{lemma*}{Lemma}

\newtheorem*{proposition*}{Proposition}

\newtheorem*{conjecture*}{Conjecture}
\theoremstyle{definition}

\newtheorem*{definition*}{Definition}
\theoremstyle{remark}

\newtheorem*{remark*}{Remark}

\begin{document}



\title{Exactly Solvable Quantum Model with Spin-Dependent Coulomb Interaction}

\author{Jiang-Lin Zhou}
\affiliation{School of Physics, Nankai University, Tianjin 300071, People's Republic of China}

\author{Yu-Xuan Zhang}
\affiliation{School of Physics, Nankai University, Tianjin 300071, People's Republic of China}

\author{Choo Hiap Oh}
\affiliation{Centre for Quantum Technologies and Department of Physics, National University of Singapore, 117543, Singapore}

\author{Jing-Ling Chen}
\email{chenjl@nankai.edu.cn}
\affiliation{Theoretical Physics Division, Chern Institute of Mathematics, Nankai University, Tianjin 300071, People's Republic of China}

\date{\today}

\begin{abstract}
In this work, we report an exactly solvable quantum model featuring a spin-dependent Coulomb interaction, described by the spin vector potential \(\vec{\mathcal{A}} = k (\vec{r} \times \vec{S}) / r^2\) together with a Coulomb-type scalar potential \(\varphi = \kappa / r\) . The model is governed by the Schr\"odinger-type Hamiltonian \(\mathcal{H}_{\rm S} = \vec{\Pi}^2 / (2M) + q \varphi\) in nonrelativistic quantum mechanics and by the Dirac-type Hamiltonian \(\mathcal{H}_{\rm D} = c \vec{\alpha} \cdot \vec{\Pi} + \beta M c^2 + q \varphi\) in relativistic quantum mechanics, where \(\vec{\Pi} = \vec{p} - (q/c)\vec{\mathcal{A}}\) is the canonical momentum. We demonstrate two main results: (i) Just as the Coulomb-type scalar potential \(\mathcal{S}_{\rm Maxwell} = \{\vec{\mathcal{A}} = 0,\ \varphi = \kappa / r\}\) is a local exact solution of Maxwell's equations on $r\neq0$, the gauge potential \(\mathcal{S}_{\rm YM} = \{\vec{\mathcal{A}} = k (\vec{r} \times \vec{S}) / r^2,\ \varphi = \kappa / r\}\) constitutes a local exact solution of the Yang--Mills equations on the punctured region $r\neq0$. (ii) Both Hamiltonians \(\mathcal{H}_{\rm S}\) and \(\mathcal{H}_{\rm D}\) can be solved exactly in the presence of this spin-dependent Coulomb interaction. The resulting energy spectra are derived, and they naturally reduce to those of the ordinary hydrogen atom when the spin-dependent terms are neglected. Finally, we clarify the quantization conditions and the fixed-background interpretation of the model.
\end{abstract}


\maketitle


\section{Introduction}

Yang--Mills (YM) theory, as a non-Abelian generalization of Maxwell's electrodynamics, stands as one of the cornerstones of modern theoretical physics~\cite{yang1954}. It provides the fundamental framework for describing gauge interactions in particle physics, successfully unifying the electromagnetic, weak, and strong forces within the Standard Model~\cite{Weinberg1995,Zee2010}. In its vacuum form, the YM equations extend Maxwell's equations by introducing non-commuting gauge potentials, leading to rich nonlinear structures and self-interactions that have profound implications in both high-energy physics and condensed matter systems~\cite{Wen2004,Sachdev2011}.

The physical role of gauge potentials was already clarified in the Abelian theory through the Dirac monopole and the Aharonov--Bohm effect. These examples show that the vector potential carries geometric and phase information beyond the local field strength~\cite{Dirac1931,AharonovBohm1959}. In the non-Abelian setting, monopole and dyon configurations such as the Wu--Yang monopole, the 't Hooft--Polyakov monopole, and the Julia--Zee dyon further illustrate how gauge potentials organize internal symmetry, topology, and dynamics~\cite{WuYang1975,tHooft1974,JuliaZee1975}.

In parallel, the concept of spin --- an intrinsic angular momentum carried by elementary particles --- has become indispensable in quantum theory. Beyond its role in defining particle statistics and magnetic moments, spin serves as a key degree of freedom in emerging fields such as spintronics~\cite{2001WolfS.A.,2004Zutic}, spin Hall effects~\cite{2008HostenOnur,2015Sinova}, and quantum information processing~\cite{Nielsen2000}. While spin-dependent interactions --- including the spin-orbit interaction, the Dzyaloshinskii--Moriya interaction~\cite{1958Dzyaloshinsky,1960Moriya}, and various exchange couplings~\cite{1999Flugge} --- are well established in quantum mechanics and condensed matter physics, their origin within a first-principle gauge-theoretical framework remains an active area of inquiry. From the viewpoint of symmetry, spin is also a carrier of an internal representation of the rotation algebra. This makes spin a natural bridge between spatial rotations and non-Abelian internal degrees of freedom, and it is precisely this bridge that will appear below in the diagonal rotational symmetry of the spin vector potential.

Recently, the notion of \emph{spin vector potential} was proposed as a gauge-field structure associated with the spin of a particle, analogous to the magnetic vector potential generated by a circulating current~\cite{Chenspin2025}. In that work, it was shown that such a potential can induce a spin-dependent Aharonov--Bohm phase, detectable in principle via interferometric experiments. This spin Aharonov--Bohm effect supports the possible physical relevance of the spin vector potential and suggests that it may play a role in mediating spin-spin and spin-orbit interactions, including the Dzyaloshinskii--Moriya interaction and the magnetic dipole-dipole coupling~\cite{1999Flugge}. However, a first-principles realization within YM theory was lacking.

Here we fill this gap and go beyond: we show that the very same vector potential, together with a Coulomb-type scalar potential $\varphi=\kappa/r$ defined away from the origin, forms a local exact solution of the YM equations on the punctured region $r\neq0$. More importantly, this gauge configuration defines an \emph{exactly solvable quantum model} for a charged particle moving in a spin-dependent Coulomb field, both non-relativistically and relativistically.

Our main results are twofold:
\begin{enumerate}
\item The set $\mathcal{S}_{\rm YM}=\{\vec{\mathcal{A}}=k(\vec{r}\times\vec{S})/r^2,\; \varphi=\kappa/r\}$ locally satisfies the sourceless YM equations on $\mathbb R^3\setminus\{0\}$, where both the spin vector potential and the Coulomb-type scalar potential are understood away from the origin, reducing to the standard Coulomb-type solution when spin is neglected. Two branches appear: $g\hbar k=1$ (Wu--Yang-type monopole) and $g\hbar k=2$ (flat-connection-like).
\item The corresponding Schr\"odinger and Dirac Hamiltonians are exactly solvable due to a diagonal rotational symmetry $\vec{J}=\vec{\ell}+\vec{S}$ (or $\vec{J}=\vec{\ell}+\vec{S}_1+\vec{S}_2$ in the Dirac case). The energy spectra are derived in closed form, reducing to the hydrogenic spectra as $k\to0$.
\end{enumerate}

The main text presents the physical construction, the essential Yang--Mills solution, the symmetry reduction, and the resulting exact spectra. Longer component-wise checks of the Yang--Mills equations and detailed algebraic reductions are provided in the Supplemental Material \cite{SM}, which is used as a technical companion to the derivations below.

The paper is organized as follows. In Section~\ref{sec:interaction}, we present the spin-dependent Coulomb interaction by studying Maxwell's equations and the Yang--Mills equations. In Section~\ref{sec:Schrodinger}, we exactly solve the quantum model with Schr\"odinger's equation. In Section~\ref{sec:Dirac}, we exactly solve the quantum model with Dirac's equation. Finally, in Section~\ref{sec:quantization} we summarize the quantization mechanism and the fixed-background interpretation, and in Section~\ref{sec:conclusion} we give the conclusion.

\section{The spin-dependent Coulomb interaction}\label{sec:interaction}

\subsection{Brief Review of The Yang--Mills Equations}
The Yang-Mills equations in vacuum without sources are given by \cite{yang1954}
\begin{subequations}
            \begin{align}
                  & D_\mu {F}^{\mu\nu} \equiv\partial_\mu {F}^{\mu\nu}
                        + {\rm i}\,g\Bigl[\mathbb{A}_\mu,\ {F}^{\mu\nu}\Bigr]=0,
                        \label{eq:YM1a} \\
                  & D_\mu {F}_{\nu\gamma} +D_\nu {F}_{\gamma\mu}
                        +D_\gamma {F}_{\mu\nu} =0, \label{eq:YM1b}
            \end{align}
\end{subequations}
where the field-strength tensor is
\begin{eqnarray}\label{eq:nonabelian}
            {F}_{\mu\nu} =\partial_\mu \mathbb{A}_{\nu}
                  -\partial_\nu \mathbb{A}_{\mu}
                  +\mathrm{i}\,g[\mathbb{A}_{\mu},\ \mathbb{A}_{\nu}], \;\;(\mu, \nu=0, 1, 2, 3),
\end{eqnarray}
$g$ is the coupling parameter, and $\mathbb{A}_\mu$'s are the potentials. Note that we have used the four-vectorial notations $x=(x_0, x_1, x_2, x_3)$, with $x_1 \equiv x$, $x_2 \equiv y$, $x_3 \equiv z$, $x_0 \equiv c\,t$, and
\begin{eqnarray}
&&\mathbb{A}_\mu=(\varphi,-\vec{\mathcal{A}})=(\varphi,-\mathcal{A}_x,-\mathcal{A}_y,-\mathcal{A}_z), \nonumber\\
&& {\partial}_{\mu}=\left(\frac{1}{c}\dfrac{\partial}{\partial t},\vec{\nabla}\right), \;\;\;\; \vec{\nabla}=(\dfrac{\partial}{\partial x}, \dfrac{\partial}{\partial y}, \dfrac{\partial}{\partial z}).
\end{eqnarray}
 Here $\varphi$ is called the \emph{scalar potential}, and $\vec{\mathcal{A}}$ is called the \emph{vector potential}.
 The YM equations are the natural generalization of Maxwell's equations from the Abelian potentials to the non-Abelian ones. When the commutator $[\mathbb{A}_{\mu}, \mathbb{A}_{\nu}]=0$, the definition of ${F}_{\mu\nu}$ reduces to the Abelian case, namely, $F_{\mu\nu}=\partial_\mu \mathbb{A}_\nu -\partial_\nu \mathbb{A}_\mu$ for the situation of Maxwell's equations.

 The matrix form of field tensor $F_{\mu\nu}$ is given by \cite{1999Jackson}
   \begin{eqnarray}
       F_{\mu\nu}=\begin{bmatrix}
            0 & \mathcal{E}_x & \mathcal{E}_y & \mathcal{E}_z \\
            -\mathcal{E}_x & 0 & -\mathcal{B}_z & \mathcal{B}_y \\
            -\mathcal{E}_y & \mathcal{B}_z & 0 & -\mathcal{B}_x \\
            -\mathcal{E}_z & -\mathcal{B}_y & \mathcal{B}_x & 0
      \end{bmatrix},
   \end{eqnarray}
   or
     \begin{eqnarray}
                        {F}^{\mu\nu} =\begin{bmatrix}
                              0 & -\mathcal{E}_x & -\mathcal{E}_y & -\mathcal{E}_z \\
                              \mathcal{E}_x & 0 & -\mathcal{B}_z & \mathcal{B}_y \\
                              \mathcal{E}_y & \mathcal{B}_z & 0 & -\mathcal{B}_x \\
                              \mathcal{E}_z & -\mathcal{B}_y & \mathcal{B}_x & 0
                        \end{bmatrix},
                  \end{eqnarray}
Based on which, one may write down the vector forms of ``magnetic'' field and ``electric'' field as
      \begin{subequations}
            \begin{eqnarray}
                  \vec{\mathcal{B}} &=& \vec{\nabla}\times\vec{\mathcal{A}}
                        -{\rm i}\,g\left(\vec{\mathcal{A}}\times\vec{\mathcal{A}}\right),
                      \label{eq:mag-2}  \\
                  \vec{\mathcal{E}} &=& -\dfrac{1}{c}
                              \dfrac{\partial\,\vec{\mathcal{A}}}{\partial\,t}
                        -\vec{\nabla}\varphi-{\rm i}\,g\left[\varphi,\ \vec{\mathcal{A}}
                        \right]. \label{eq:ele-2}
            \end{eqnarray}
      \end{subequations}
 Furthermore, in terms of the language of $\big\{\vec{\mathcal{E}}, \vec{\mathcal{B}}, \vec{\mathcal{A}}, \varphi\big\}$, the YM equations (\ref{eq:YM1a}) can be recast to the following easily understandable form
       \begin{subequations}{\small
            \begin{eqnarray}
                  && \vec{\nabla}\cdot\vec{\mathcal{E}} {-}\boxed{{\rm i}g\Bigl(
                        \vec{\mathcal{A}}\cdot\vec{\mathcal{E}}
                        -\vec{\mathcal{E}}\cdot\vec{\mathcal{A}}\Bigr)}=0, \label{eq:DivEYM} \\
                  && -\dfrac{1}{c} \dfrac{\partial}{\partial t} \vec{\mathcal{B}}
                        -\vec{\nabla}\times\vec{\mathcal{E}} {-}\boxed{{\rm i}g\biggl(\Bigl[\varphi,
                                    \vec{\mathcal{B}}\Bigr]
                              -\vec{\mathcal{A}}\times\vec{\mathcal{E}}
                              -\vec{\mathcal{E}}\times\vec{\mathcal{A}}\biggr)}=0, \label{eq:CurlEYM} \\
                  && \vec{\nabla}\cdot\vec{\mathcal{B}} {-}\boxed{{\rm i}g\Bigl(
                        \vec{\mathcal{A}}\cdot\vec{\mathcal{B}}
                        -\vec{\mathcal{B}}\cdot\vec{\mathcal{A}}\Bigr)}=0, \label{eq:DivBYM} \\
                  && -\dfrac{1}{c} \dfrac{\partial}{\partial t} \vec{\mathcal{E}}
                        +\vec{\nabla}\times\vec{\mathcal{B}}  {-}\boxed{{\rm i} g\biggl(\Bigl[\varphi,
                                    \vec{\mathcal{E}}\Bigr]
                              +\vec{\mathcal{A}}\times\vec{\mathcal{B}}
                              +\vec{\mathcal{B}}\times\vec{\mathcal{A}}\biggr)}=0. \label{eq:CurlBYM}
            \end{eqnarray} }
      \end{subequations}
  \noindent Here the ``boxed'' terms come from the nonlinear terms in YM equations, such as ${\rm i}\,g [\mathbb{A}_\mu,\ {F}^{\mu\nu}]$, which represent the self-interaction between the potentials $\big\{\vec{\mathcal{A}}, \varphi\big\}$ and the fields $\big\{\vec{\mathcal{E}}, \vec{\mathcal{B}}\big\}$. If one neglects the ``boxed'' terms (or just let $g=0$), then the above expressions reduce to the usual forms of Maxwell's equations in vacuum. Namely,
\begin{subequations}
            \begin{eqnarray}
                  && \vec{\nabla}\cdot\vec{\mathcal{E}}=0, \label{eq:DivEYM-M} \\
                  && -\dfrac{1}{c} \dfrac{\partial}{\partial\,t} \vec{\mathcal{B}}
                        -\vec{\nabla}\times\vec{\mathcal{E}}=0, \label{eq:CurlEYM-M} \\
                  && \vec{\nabla}\cdot\vec{\mathcal{B}}=0, \label{eq:DivBYM-M} \\
                  && -\dfrac{1}{c} \dfrac{\partial}{\partial\,t} \vec{\mathcal{E}}
                        +\vec{\nabla}\times\vec{\mathcal{B}}=0. \label{eq:CurlBYM-M}
            \end{eqnarray}
      \end{subequations}

\subsection{A Simple but Fundamental Local Solution of Maxwell's Equations}

As a comparison, let us first study a local exact solution of Maxwell's equations on $r\neq0$, i.e., \emph{the Coulomb-type potential}. Although such a simple but fundamental result is well-known, for the convenience of explanation, we would like to express it as the following theorem.
\begin{theorem}\label{Ther1}
Let the vector potential and the scalar potential be
\begin{eqnarray}\label{eq:Coul-1a}
    &&  \vec{\mathcal{A}}=0, \;\;\;\;\;      \varphi = \dfrac{\kappa}{r},
\end{eqnarray}
then the set of potentials $\{\vec{\mathcal{A}}, \varphi\}$ is a local exact solution of Maxwell's equations on the punctured region $r\neq0$.
Here $\kappa$ is a real constant number.
\end{theorem}

\begin{proof}
By substituting Eq. (\ref{eq:Coul-1a}) into Eq. (\ref{eq:mag-2}) and Eq. (\ref{eq:ele-2}), we have the magnetic field and the electric field as
\begin{eqnarray}\label{eq:Coul-2}
     && \vec{\mathcal{B}} =0, \;\;\;\; \vec{\mathcal{E}} =  \dfrac{\kappa}{r^3} \vec{r}.
\end{eqnarray}
Because $\{\vec{\mathcal{A}}, \varphi\}$ is time-independent, thus $\vec{\mathcal{B}}$  and $\vec{\mathcal{E}}$ are static fields. Let us come to check Maxwell's equations as shown in Eq. (\ref{eq:DivEYM-M})-Eq. (\ref{eq:CurlBYM-M}). Because $\vec{\mathcal{B}}=0$ and $\vec{\mathcal{E}}$ is time-independent, then Eq. (\ref{eq:DivBYM-M}) and Eq. (\ref{eq:CurlBYM-M}) are automatically satisfied. Furthermore, one may easily verify that $\vec{\nabla}\cdot\vec{\mathcal{E}}=0$ and $\vec{\nabla}\times\vec{\mathcal{E}}=0$ are satisfied for $r\neq0$. Thus
Eq. (\ref{eq:DivEYM-M}) and Eq. (\ref{eq:CurlEYM-M}) hold. This ends the proof.
\end{proof}
Let us denote $\varphi_{\rm cl} =\kappa/r$ as the Coulomb-type potential on $r\neq0$. Theorem \ref{Ther1} tells us that the Coulomb-type potential is a simple but fundamental local solution of Maxwell's equations. To be precise, the exact solution is the set $ \mathcal{S}_{\rm Maxwell}=\{\vec{\mathcal{A}}=0, \varphi=\kappa/r\}$ on $\mathbb R^3\setminus\{0\}$, although in this case the vector potential is equal to zero. Or inversely, in a certain sense, one may say that Maxwell's equations provide a field-theoretical origin for the Coulomb interaction.

\subsection{A Local Exact Solution of the Yang--Mills Equations}

As it is well-known that, the YM equations are the natural generalizations of Maxwell's equations from the Abelian potentials to the non-Abelian ones. Usually, the non-Abelian operators can be realized by the spin operators. Let $\vec{S}=(S_x, S_y, S_z)$ be the spin-$s$ angular momentum operator (with the spin value $s=0, 1/2, 1, ...$), which satisfies the following
commutation relations
\begin{eqnarray}\label{angularm1}
 [S_x, S_y]= \mathrm{i} \hbar S_z, \; [S_y, S_z]= \mathrm{i} \hbar S_x,\;  [S_z, S_x]= \mathrm{i} \hbar S_y,
\end{eqnarray}
or in the vector form  as $\vec{S} \times \vec{S} = \mathrm{i} \hbar\; \vec{S}$, where $\hbar$ is Planck's constant.
The square of $\vec{S}$ satisfies $\vec{S}^{\,2} = s(s+1) \hbar^2$. In particular, when $s=1/2$, the spin operator $\vec{S}$ can be realized by
\begin{eqnarray}
      \vec{S} = \frac{\hbar}{2} \vec{\sigma},
\end{eqnarray}
with $\vec{\sigma}=(\sigma_x, \sigma_y, \sigma_z)$ being the vector of Pauli matrices, $\vec{S}^2 = (3/4)\hbar^2 \openone$, $\openone$ is the $2\times 2$ identity matrix,  and
\begin{eqnarray}
 \label{S-3}
  && \sigma_x=
  \left(
    \begin{array}{cc}
      0 & 1 \\
      1 & 0 \\
    \end{array}
  \right),\sigma_y=
  \left(
    \begin{array}{cc}
      0 & -{\rm i} \\
      {\rm i} & 0 \\
    \end{array}
  \right), \sigma_z=
  \left(
    \begin{array}{cc}
      1 & 0 \\
      0 & -1 \\
    \end{array}
  \right).
   \end{eqnarray}
In this work, for simplicity, we restrict to the spin-$1/2$ case. The generalization to the high spin values can be done similarly.

\begin{figure}[h]
    \includegraphics[width=0.80\textwidth]{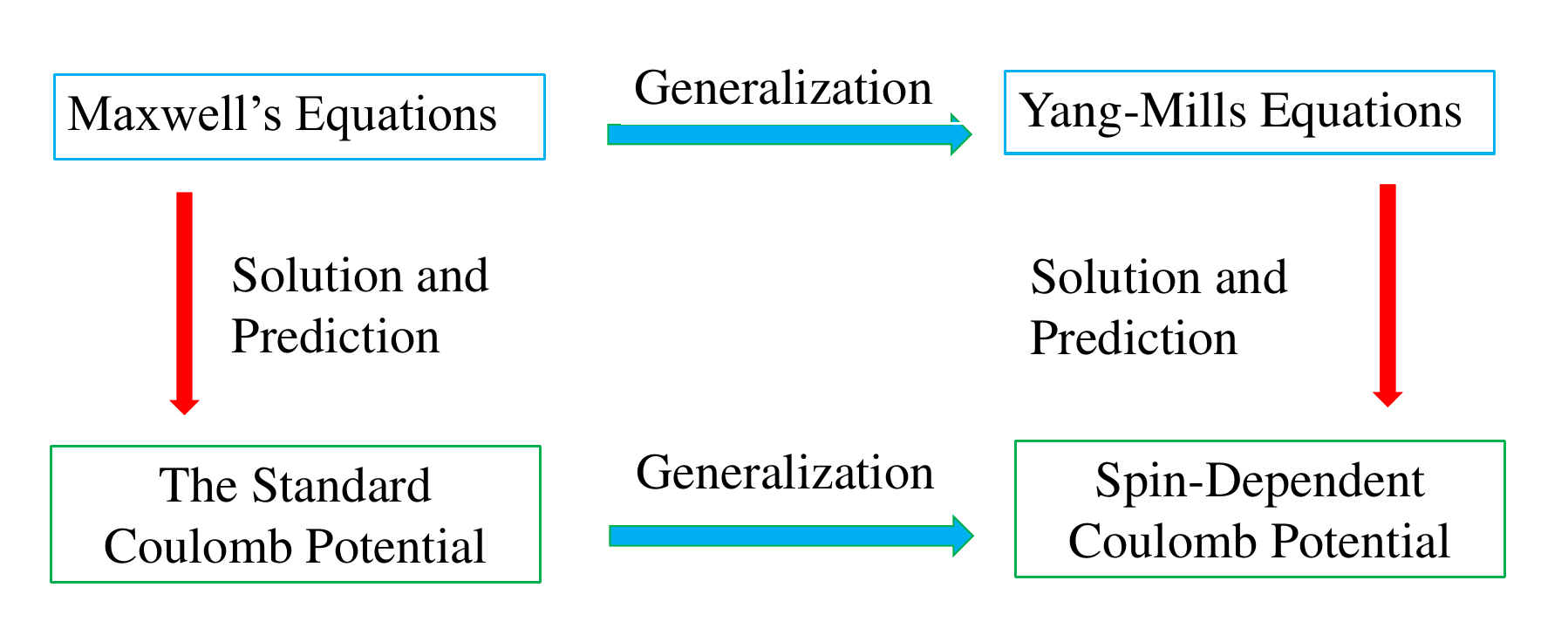}
    \caption{{\bf Illustration of the Spin-Dependent Coulomb Interaction}. The standard Coulomb potential is a simple but fundamental local solution (i.e., $ \mathcal{S}_{\rm Maxwell}$) of Maxwell's equations away from the origin, hence Maxwell's equations predict the existence of the standard Coulomb potential in Nature. The YM equations are the natural generalizations of Maxwell's equations from the Abelian potentials to the non-Abelian ones, and the non-Abelian operators can be realized by the spin operators, then one can expect that there will be a counterpart solution of potentials (i.e., $ \mathcal{S}_{\rm YM}$) that depends on spin. Thus based on the YM equations, one can predict a reasonable form of spin-dependent Coulomb potential, which naturally reduces the standard Coulomb potential if without considering the spin.}\label{fig:logic}
 \end{figure}

One may observe that the local exact solution $ \mathcal{S}_{\rm Maxwell}=\{\vec{\mathcal{A}}=0, \varphi=\kappa/r\}$ for Maxwell's equations on $r\neq0$ is spin-independent. This gives rise to a natural question: How will the standard Coulomb interaction be modified when it meets spin? Mathematically, let us denote the set $ \mathcal{S}_{\rm YM}=\{\vec{\mathcal{A}}, \varphi\}$ as a local exact solution of the YM equations on $r\neq0$. As one expects, the solution $ \mathcal{S}_{\rm YM}$ is spin-dependent, which naturally reduces to the solution $ \mathcal{S}_{\rm Maxwell}$ when the spin is neglected.
One of our purposes is to derive  a reasonable spin-dependent Coulomb interaction based on the YM equations. The logic of the prediction is illustrated in Fig. \ref{fig:logic}.

Now, let us come to study a local exact solution (i.e., $ \mathcal{S}_{\rm YM}=\{\vec{\mathcal{A}}, \varphi\}$) of the YM equations on the punctured region $r\neq0$. For the vector potential $\vec{\mathcal{A}}$, we just take it as the spin vector potential proposed in \cite{Chenspin2025}, i.e.,
\begin{eqnarray}\label{eq:svp-1}
    &&  \vec{\mathcal{A}} \propto \frac{\vec{r} \times \vec{S}}{r^2}.
\end{eqnarray}
By using an elegant method related to the angular momentum operators, Ref. \cite{Chenspin2025} has provided an intuitive derivation for the spin vector potential based on physical insights. Explicitly, let us denote $\vec{\ell}=\vec{r}\times \vec{p}$ as the orbital angular momentum operator, where $\vec{r}$ is the position and $\vec{p}$ is the linear momentum. Then the ``total'' angular momentum operator $\vec{J}=\vec{\ell}+\vec{S}$ can be recast to the form $ \vec{J}  = \vec{r} \times [\vec{p}-(q/c) \vec{A}]+ S_r \hat{e}_r$, with
$\vec{S}=S_r \hat{e}_r+ S_\theta \hat{e}_\theta+S_\phi\hat{e}_\phi$ expanded in the spherical coordinates, $\hat{e}_r=\vec{r}/r$, and $r=|\vec{r}|$.
Therefore, when $\vec{\Pi}=[\vec{p}-(q/c) \vec{A}]$ is viewed as the canonical momentum, then one immediately extracts the spin vector potentials as shown in Eq. (\ref{eq:svp-1}).

For the scalar potential $\varphi$, when we restrict to the spin-$1/2$ case, it is sufficient to consider the form $\varphi = f_1(r) (\vec{r} \cdot \vec{S}) +  f_2(r)$, where $f_1(r)$ and $f_2(r)$ are two functions that need to be determined by the YM equations. We find that a set of spin potentials $\mathcal{S}_{\rm YM}=\{\vec{\mathcal{A}}, \varphi\}$ can be a local exact solution of the YM equations on $r\neq0$. This result can be expressed in the following theorem.
\begin{theorem}\label{Ther2}
If the spin operator $\vec{S}$ is considered, then the Yang-Mills equations can have the following local exact solutions for spin vector and scalar potentials on $\mathbb R^3\setminus\{0\}$
\begin{subequations}
\begin{eqnarray}\label{eq:Coul-1}
    &&  \vec{\mathcal{A}}= k \frac{\vec{r} \times \vec{S}}{r^2}, \label{eq:A}\\
    &&       \varphi = f_1(r) \left(\vec{r} \cdot \vec{S}\right) +  f_2(r), \label{eq:phi}
\end{eqnarray}
\end{subequations}
where $k$ is a real number, $f_1(r)$ and $f_2(r)$ are two functions given in Table \ref{tab:gk}.
\end{theorem}

\begin{table}[h]
\caption{Solutions of $f_1(r)$ and $f_2(r)$ on the domain $r\neq0$. All Coulomb-type $1/r$ terms appearing in the table, in particular $f_2(r)=\kappa/r$, are understood on the same punctured domain. When $g=0$ and $k\neq 0$, there are no suitable solutions. We are interested in the nontrivial case of $g\neq 0$ and $k\neq 0$. For simplicity, we may choose $\kappa_2=0$ and $\kappa_3=0$, in this situation, the ``electric'' field is equal to the usual electric field with Coulomb potential, i.e., $\vec{\mathcal{E}}=(\kappa/r^3) \vec{r}$, the ``magnetic'' field $\mathcal{B}=0$ if $g\hbar k=2$ and $\mathcal{B}=-k [(\vec{r}\cdot\vec{S})/r^4] \; \vec{r}$ if $g\hbar k=1$.}\label{tab:gk}
                  \centering\begin{tabular}{|c|c|c|}
                  \hline
             (i) $g=0, k=0$  \;\; &  (ii) $g\neq 0, k=0$ \;\;& (iii)  $g\neq 0, k\neq 0$ \;\; \\
                  \hline
               $ \; \;f_1(r) =\dfrac{\kappa_2}{r^3}+\kappa_3 $   &$ f_1(r) =0$  & If $g\hbar k = 2$: $f_1(r) = 0$    \\
               $f_2(r)=\dfrac{\kappa}{r}\;(r\neq0)$      & $f_2(r)=\dfrac{\kappa}{r}\;(r\neq0)$  &  If $g\hbar k = 1$:  \\
               & & $f_1(r) = \dfrac{\kappa_2}{r} + \dfrac{\kappa_3}{r^2}$ \\
                  & &   $f_2(r) = \dfrac{\kappa}{r}\;(r\neq0)$\\
               \hline
               \end{tabular}
                 \end{table}

The entries in Table~\ref{tab:gk} should be understood as admissible solution branches of the radial Yang--Mills equations, rather than as time-evolution initial conditions. These branches determine the possible background potentials. Because the Coulomb term $\kappa/r$ and the spin vector potential are singular at the origin, all branches containing such terms are considered on $r\neq0$.

For the purpose of constructing a minimal spin-dependent Coulomb interaction, we focus on the case where the scalar potential reduces to the pure Coulomb form $\varphi=\kappa/r$ on $r\neq0$, i.e., we set $\kappa_2=\kappa_3=0$. This simple configuration, given by
\begin{eqnarray}\label{eq:YMsolu-simple}
        \mathcal S_{\rm YM}^{\rm(min)}=\left\{\vec{\mathcal A}=k\frac{\vec r\times\vec S}{r^2},\quad \varphi=\frac{\kappa}{r}\right\},\quad r\neq0,
\end{eqnarray}
will be used in the quantum models of Sections III and IV. For this minimal branch, the spin vector potential fixes the angular part of the Hamiltonian and Dirac operator, while the scalar potential remains of Coulomb type; this is the structural reason why the subsequent radial equations remain exactly solvable. The following proof establishes the general solution; the special case $\kappa_2=\kappa_3=0$ is a particular realization.

\begin{proof}Let us prove the theorem in the nontrivial case $g\neq0$ and $k\neq0$; the Abelian and $k=0$ limits are listed in Table~\ref{tab:gk}. The spin-$1/2$ identities
\begin{eqnarray}\label{eq:spin-id}
        (\vec r\cdot\vec S)^2=\frac{r^2\hbar^2}{4}\openone,\qquad
        (\vec r\times\vec S)\cdot\vec S={\rm i}\hbar(\vec r\cdot\vec S),
\end{eqnarray}
show that the scalar spin potential may be chosen in the form
\begin{eqnarray}
        \varphi=f_1(r)(\vec r\cdot\vec S)+f_2(r),
\end{eqnarray}
up to terms proportional to the identity. Thus the problem is reduced to determining the two radial functions $f_1(r)$ and $f_2(r)$ from the YM equations.

  For the vector potential
\begin{eqnarray}
        \vec{\mathcal A}=k\frac{\vec r\times\vec S}{r^2},
\end{eqnarray}
one has
\begin{eqnarray}\label{eq:curlA}
        \vec\nabla\times\vec{\mathcal A}=-2k\frac{\vec r\cdot\vec S}{r^4}\vec r.
\end{eqnarray}
The non-Abelian commutator part gives
\begin{eqnarray}\label{eq:AA}
        -{\rm i}g(\vec{\mathcal A}\times\vec{\mathcal A})
        =g\hbar k^2\frac{\vec r\cdot\vec S}{r^4}\vec r.
\end{eqnarray}
Substituting Eqs.~(\ref{eq:curlA}) and (\ref{eq:AA}) into Eq.~(\ref{eq:mag-2}) yields the ``magnetic'' field
  \begin{eqnarray}\label{eq:mag-3}
        \vec{\mathcal B}=-(2-g\hbar k)k\frac{\vec r\cdot\vec S}{r^4}\vec r.
\end{eqnarray}
  For the electric field, we first note that
\begin{eqnarray}\label{eq:gradphi}
        \vec\nabla\varphi=
        f_1'(r)\frac{\vec r\cdot\vec S}{r}\vec r+f_1(r)\vec S+f_2'(r)\frac{\vec r}{r},
\end{eqnarray}
and
\begin{eqnarray}\label{eq:comm}
        [\vec r\cdot\vec S,\vec r\times\vec S]
        ={\rm i}\hbar\left[r^2\vec S-\vec r(\vec r\cdot\vec S)\right].
\end{eqnarray}
Hence Eq.~(\ref{eq:ele-2}) gives
\begin{eqnarray}\label{eq:ele-3}
   \vec{\mathcal E}
        &=&-{\biggl\{}f_1'(r)\frac{\vec r\cdot\vec S}{r}+f_2'(r)\frac{1}{r}
          +g\hbar k f_1(r)\frac{\vec r\cdot\vec S}{r^2}{\biggr\}}\vec r \nonumber\\
        && +(g\hbar k-1)f_1(r)\vec S.
\end{eqnarray}

The remaining calculation consists of substituting Eqs.~(\ref{eq:mag-3}) and  (\ref{eq:ele-3})   into the YM equations (\ref{eq:DivEYM})--(\ref{eq:CurlBYM}). The Bianchi-type equations  (\ref{eq:CurlEYM}) and  (\ref{eq:DivBYM}) are   satisfied identically by the field-strength construction. The two dynamical equations reduce to the following radial conditions:
 \begin{subequations}  \label{eq:radial-conditions}
\begin{eqnarray}
&& f_2''(r)+\frac{2}{r}f_2'(r)=0, \label{eq:con-1}\\
&& f_1''(r)+\frac{4}{r}f_1'(r)+\frac{2g\hbar k(2-g\hbar k)}{r^2}f_1(r)=0, \label{eq:con-2}\\
&& (1-g\hbar k)\left[g\hbar f_1^2(r)+(2-g\hbar k)k\frac{1}{r^4}\right]=0. \label{eq:con-3}
\end{eqnarray}
\end{subequations}
  Equation~(\ref{eq:con-1}) gives $f_2(r)=\kappa/r+c_1$, and the additive constant $c_1$ may be set to zero. To solve the second equation, let
\begin{eqnarray}
        \alpha=g\hbar k.
\end{eqnarray}
  Then Eq.~(\ref{eq:con-2}) becomes
\begin{eqnarray}\label{eq:f1-alpha-v13}
        f_1''(r)+\frac{4}{r}f_1'(r)+\frac{2\alpha(2-\alpha)}{r^2}f_1(r)=0.
\end{eqnarray}
For a power-law trial function $f_1(r)=r^m$, one obtains
\begin{eqnarray}
        m^2+3m+2\alpha(2-\alpha)=0.
\end{eqnarray}
When $\alpha=1$, the roots are $m=-1$ and $m=-2$, giving
\begin{eqnarray}
        f_1(r)=\frac{\kappa_2}{r}+\frac{\kappa_3}{r^2}.
\end{eqnarray}
When $\alpha=2$, Eq.~(\ref{eq:f1-alpha-v13}) alone allows a constant term and a $1/r^3$ term, but Eq.~(\ref{eq:con-3}) imposes $f_1(r)=0$ in the nontrivial YM branch. If $(1-\alpha)(2-\alpha)\neq0$, Eq.~(\ref{eq:con-3}) would require $f_1(r)=C r^{-2}$. Substituting this into Eq.~(\ref{eq:f1-alpha-v13}) gives the residual $-2C(\alpha-1)^2 r^{-4}$, so no additional real branch is obtained except the already listed $\alpha=1$ case. Therefore the nontrivial solutions are precisely those shown in Table~\ref{tab:gk}. For the simple Coulomb scalar branch $\kappa_2=\kappa_3=0$, we obtain
  \begin{eqnarray}\label{eq:YMsolu}
        \mathcal S_{\rm YM}=\left\{\vec{\mathcal A}=k\frac{\vec r\times\vec S}{r^2},\quad \varphi=\frac{\kappa}{r}\right\},\quad r\neq0.
\end{eqnarray}
  This solution reduces to $\mathcal S_{\rm Maxwell}$ when the spin-dependent vector part is removed. This ends the proof.
\end{proof}

\subsection{Branch Structure and Relation to Known Yang--Mills Configurations}

The two nontrivial branches obtained above have different geometric meanings. For $g\hbar k=1$, Eq.~(\ref{eq:mag-3}) gives the nonzero radial field
\begin{eqnarray}\label{eq:WYbranch-v13}
        \vec{\mathcal B}=-k\frac{\vec r\cdot\vec S}{r^4}\vec r.
\end{eqnarray}
Writing $S_a=\hbar T_a$, the spatial potential can be written as
\begin{eqnarray}\label{eq:WYform-v13}
        \mathcal A_i = k\hbar\,\epsilon_{ija}\frac{x_j}{r^2}T_a.
\end{eqnarray}
On the branch $g\hbar k=1$, this becomes
\begin{eqnarray}\label{eq:WYform-branch-v13}
        \mathcal A_i = \frac{1}{g}\epsilon_{ija}\frac{x_j}{r^2}T_a,
\end{eqnarray}
up to the conventional sign choice for the Wu-Yang monopole \cite{WuYang1975}. This has the same tensorial structure as the Wu-Yang monopole connection in an $su(2)$ representation. Thus the branch $g\hbar k=1$ may be read as a Wu-Yang-type Yang-Mills branch written from the viewpoint of the spin vector potential.

The nontriviality of this branch can be seen directly from a simple quadratic field-strength measure. From Eq.~(\ref{eq:WYbranch-v13}),
  \begin{eqnarray}\label{eq:B2-v13}
        \operatorname{tr}_{s}\sum_i \mathcal B_i\mathcal B_i
        =k^2\frac{\operatorname{tr}_{s}\left[(\hat r\cdot\vec S)^2\right]}{r^4}.
\end{eqnarray}
  For spin $1/2$, $(\hat r\cdot\vec S)^2=\hbar^2\openone/4$, and hence the un-normalized spin trace gives $k^2\hbar^2/(2r^4)$. This confirms that the branch is not a trivial zero-field configuration and also displays the singular behavior at the origin. Throughout this paper the Yang--Mills solution is therefore understood locally on the punctured region $r\neq0$.

For $g\hbar k=2$, Eq.~(\ref{eq:con-3}) gives $f_1(r)=0$ in the nontrivial YM branch, and Eq.~(\ref{eq:mag-3}) gives $\vec{\mathcal B}=0$ in the simple Coulomb scalar sector, although the vector potential itself is not identically zero. This branch may therefore be viewed as a flat-connection-like branch on the punctured space.

In the general solution (Table~\ref{tab:gk}), the $g\hbar k=1$ branch admits nonzero $f_1(r)$ terms, which may be interpreted as non-Abelian temporal components with a hedgehog-like internal structure. This is reminiscent of the temporal component in dyon-type ansatzes, although the present construction does not introduce a Higgs triplet or a regular finite-energy core. However, for the minimal Coulombic realization adopted in this work (with $\kappa_2=\kappa_3=0$), the scalar potential reduces to $\varphi=\kappa/r$ and these additional terms vanish. The branch structure discussed above is presented for completeness.

\subsection{Diagonal Rotational Symmetry of the Spin Vector Potential}

The same spin vector potential also has an intrinsic symmetry that explains why the quantum models in the next sections are exactly solvable. The potential
\begin{eqnarray}
        \vec{\mathcal A}=k\frac{\vec r\times\vec S}{r^2}
\end{eqnarray}
is not an ordinary scalar central potential. Instead, it is covariant under simultaneous rotations of the coordinate vector $\vec r$ and the spin operator $\vec S$. For the nonrelativistic problem the relevant generator is
\begin{eqnarray}\label{eq:Jnonrel-v13}
        \vec J=\vec\ell+\vec S,\qquad \vec\ell=\vec r\times\vec p.
\end{eqnarray}
Indeed,
\begin{eqnarray}
        {}[J_i,r_j]&=&i\hbar\epsilon_{ijk}r_k,\nonumber\\
        {}[J_i,p_j]&=&i\hbar\epsilon_{ijk}p_k,\nonumber\\
        {}[J_i,S_j]&=&i\hbar\epsilon_{ijk}S_k.
\end{eqnarray}
Since $r^2$ is invariant and $\vec r\times\vec S$ transforms as a vector, one has
\begin{eqnarray}\label{eq:AJcomm-v13}
        [J_i,\mathcal A_j]=i\hbar\epsilon_{ijk}\mathcal A_k.
\end{eqnarray}
Consequently, the covariant momentum $\Pi_j=p_j-(q/c)\mathcal A_j$ also transforms as a vector under $\vec J$, and $\vec\Pi^{\,2}$ is a rotational scalar. For the Coulomb scalar component $\varphi=\kappa/r$ understood on $r\neq0$, the Hamiltonian commutes with the diagonal angular momentum,
\begin{eqnarray}\label{eq:HJcomm-v13}
        [\mathcal H,J_i]=0,\qquad [\mathcal H,\vec J^{\,2}]=0,\qquad [\mathcal H,J_z]=0.
\end{eqnarray}
This is the symmetry reason for the separation of variables used in the following sections.

\section{Exactly Solving Quantum Model with Schr{\"o}dinger's Equation}\label{sec:Schrodinger}

We consider the Schr\"odinger-type Hamiltonian
\begin{eqnarray}\label{eq:H-1a-1a0}
\mathcal{H}&=& \dfrac{\vec{\Pi}^2}{2M}  + q \varphi
= \dfrac{1}{2M}\left(\vec p-\frac{q}{c}\vec{\mathcal A}\right)^2+\frac{q\kappa}{r},
\end{eqnarray}
which describes a nonrelativistic hydrogen-like atom with a spin-dependent Coulomb interaction.
Here the canonical momentum is \(\vec{\Pi}=\vec{p}-(q/c)\vec{\mathcal A}\), the vector potential is
\(\vec{\mathcal A}=k(\vec r\times\vec S)/r^2\), and the scalar potential is the Coulomb form \(\varphi=\kappa/r\) on $r\neq0$.
For a hydrogen-like atom we take \(q=-e\) and \(\kappa=Ze\), so that \(q\kappa = -Ze^2<0\), i.e., an attractive Coulomb potential.
\(M\) is the mass of the particle, and \(\vec S\) is the spin-\(1/2\) operator, \(\vec S=(\hbar/2)\vec\sigma\).

\subsection{Diagonal rotational symmetry}

The spin vector potential is not an ordinary central potential, but it is covariant under simultaneous rotations of \(\vec r\) and \(\vec S\).
Define the total angular momentum
\begin{eqnarray}\label{eq:Jschr}
        \vec J=\vec\ell+\vec S,\qquad \vec\ell=\vec r\times\vec p .
\end{eqnarray}
Using the commutation relations
\begin{eqnarray}
[J_i,r_j]={\rm i}\hbar\epsilon_{ijk}r_k,\quad
[J_i,p_j]={\rm i}\hbar\epsilon_{ijk}p_k,\quad
[J_i,S_j]={\rm i}\hbar\epsilon_{ijk}S_k,
\end{eqnarray}
one finds
\begin{eqnarray}
[J_i,\mathcal A_j]={\rm i}\hbar\epsilon_{ijk}\mathcal A_k,\qquad
[J_i,\Pi_j]={\rm i}\hbar\epsilon_{ijk}\Pi_k .
\end{eqnarray}
Consequently, \(\vec\Pi^{\,2}\) is a scalar under rotations generated by \(\vec J\).
Since the Coulomb potential \(q\kappa/r\) is also rotationally invariant on the punctured domain, the Hamiltonian commutes with \(\vec J\):
\begin{eqnarray}\label{eq:HJcomm}
        [\mathcal H,J_i]=0,\qquad [\mathcal H,\vec J^{\,2}]=0,\qquad [\mathcal H,J_z]=0 .
\end{eqnarray}
This diagonal symmetry is the key to separating variables in the presence of the spin-dependent vector potential.

\subsection{Reduction to a radial equation}

For spin-\(1/2\), the vector potential satisfies \(\vec\nabla\cdot\vec{\mathcal A}=0\) (Coulomb gauge).
Using the identity
\begin{eqnarray}
\vec{\mathcal A}\cdot\vec p = k\frac{(\vec r\times\vec S)\cdot\vec p}{r^2}
= -k\frac{\vec\ell\cdot\vec S}{r^2},
\end{eqnarray}
and
\begin{eqnarray}
\vec{\mathcal A}^{\,2} = \frac{k^2}{r^4}(\vec r\times\vec S)\cdot(\vec r\times\vec S)
= \frac{k^2}{r^2}\left[\vec S^{\,2}-(\hat r\cdot\vec S)^2\right]
= \frac{k^2\hbar^2}{2r^2},
\end{eqnarray}
where we have used \((\hat r\cdot\vec S)^2 = \hbar^2/4\) for spin \(1/2\).
Substituting these expressions into \(\vec\Pi^{\,2}\) gives
\begin{eqnarray}\label{eq:HSForm1}
\frac{\vec\Pi^{\,2}}{2M}
&=&\frac{1}{2M}\left(\vec p^{\,2}-2\frac{q}{c}\vec{\mathcal A}\cdot\vec p+\frac{q^2}{c^2}\vec{\mathcal A}^{\,2}\right)\nonumber\\
&=&\frac{1}{2M}\left[\vec p^{\,2}+2\tilde k\frac{\vec\ell\cdot\vec S}{r^2}+\frac{\tilde k^2\hbar^2}{2r^2}\right],
\end{eqnarray}
where we introduced the dimensionless coupling
\begin{eqnarray}
\tilde k = \frac{qk}{c}.
\end{eqnarray}
Here \(k\) has dimensions such that \(\tilde k\) is dimensionless; in atomic units this is automatic.

The total Hamiltonian therefore reads
\begin{eqnarray}\label{eq:H-expanded-v13}
\mathcal H&=&\frac{1}{2M}\left[\vec p^{\,2}+2\tilde k\frac{\vec\ell\cdot\vec S}{r^2}+\frac{\tilde k^2\hbar^2}{2r^2}\right]+\frac{q\kappa}{r}.
\end{eqnarray}
We now work in the common eigenbasis of \(\vec J^{\,2}\), \(\vec\ell^{\,2}\) and \(J_z\).
For a spin-\(1/2\) particle, the eigenfunctions are spinor spherical harmonics \(\Phi_{ljm_j}(\theta,\phi)\).
The operator \(\vec\ell\cdot\vec S\) acts as
\begin{eqnarray}\label{eq:lS-v13}
\vec\ell\cdot\vec S\,\Phi_{ljm_j}=W\hbar^2\,\Phi_{ljm_j},\qquad
W=\frac{1}{2}\left[j(j+1)-l(l+1)-\frac{3}{4}\right].
\end{eqnarray}
Explicitly, \(W = l/2\) for \(j=l+1/2\) and \(W = -(l+1)/2\) for \(j=l-1/2\).

Writing the wave function as \(\Psi(r,\theta,\phi)=R(r)\,\Phi_{ljm_j}(\theta,\phi)\) and using
\begin{eqnarray}
\vec p^{\,2} = -\hbar^2\left(\frac{1}{r^2}\frac{\partial}{\partial r}r^2\frac{\partial}{\partial r} - \frac{\vec\ell^{\,2}}{\hbar^2 r^2}\right),
\end{eqnarray}
with \(\vec\ell^{\,2}\Phi_{ljm_j}=l(l+1)\hbar^2\Phi_{ljm_j}\), we obtain the radial equation
\begin{eqnarray}\label{eq:ReduceRadialEq}
\frac{{\rm d}^2R}{{\rm d}r^2}+\frac{2}{r}\frac{{\rm d}R}{{\rm d}r}
-\frac{\Omega}{r^2}R-\frac{2Mq\kappa}{\hbar^2 r}R+\epsilon R=0,
\end{eqnarray}
where
\begin{eqnarray}
\epsilon=\frac{2ME}{\hbar^2},\qquad
\Omega=l(l+1)+2\tilde k W+\frac{\tilde k^2}{2}.
\end{eqnarray}
Equation \eqref{eq:ReduceRadialEq} is a Coulomb-type radial equation with an effective centrifugal term \(\Omega/r^2\).
To identify the small-\(r\) behaviour, we note that as \(r\to0\) the dominant terms give
\begin{eqnarray}
\frac{{\rm d}^2R}{{\rm d}r^2}+\frac{2}{r}\frac{{\rm d}R}{{\rm d}r}-\frac{\Omega}{r^2}R=0,
\end{eqnarray}
whose solutions are \(R\sim r^{\lambda}\) and \(R\sim r^{-\lambda-1}\), where \(\lambda\) satisfies
\begin{eqnarray}
\lambda(\lambda+1)=\Omega.
\end{eqnarray}
Thus we define
\begin{eqnarray}\label{eq:lambda}
\lambda=\frac{-1+\sqrt{1+4\Omega}}{2},
\end{eqnarray}
taking the positive root to ensure normalizability at the origin.

\subsection{Bound-state solutions and energy spectrum}

For the attractive case \(q\kappa<0\) (hydrogen-like atom), the radial equation \eqref{eq:ReduceRadialEq} is identical in form to the standard Coulomb radial equation with angular momentum \(l\) replaced by \(\lambda\).
The normalizable solution that is regular at the origin is
\begin{eqnarray}\label{eq:R-sol}
R(r)=r^{\lambda}e^{-\sqrt{-\epsilon}r}\,
{}_1F_1\!\left(
\lambda+1+\frac{Mq\kappa}{\hbar^2\sqrt{-\epsilon}};\;
2\lambda+2;\;
2\sqrt{-\epsilon}r
\right).
\end{eqnarray}
The confluent hypergeometric series must terminate to avoid divergence at infinity; this requires
\begin{eqnarray}
\lambda+1+\frac{Mq\kappa}{\hbar^2\sqrt{-\epsilon}}=-N,\qquad N=0,1,2,\ldots,
\end{eqnarray}
where \(N\) is the radial quantum number (the number of nodes of the radial wave function).
Solving for the energy yields
\begin{eqnarray}\label{eq:energy-1}
\boxed{
E=-\dfrac{M(q\kappa)^2}{2\hbar^2}\dfrac{1}{\left(N+\lambda+1\right)^2},
\qquad (N=0,1,2,\ldots) } .
\end{eqnarray}
The principal quantum number is \(n = N+\lambda+1\), and the spectrum resembles that of a hydrogen atom but with the effective angular momentum \(\lambda\) given by Eq.~\eqref{eq:lambda}.

\subsection{Nonrelativistic limit}

When the spin coupling is turned off, \(k\to0\) (hence \(\tilde k\to0\)), we have
\begin{eqnarray}
\Omega \to l(l+1),\qquad \lambda \to l .
\end{eqnarray}
Then Eq.~\eqref{eq:energy-1} reduces to the well-known hydrogen-like spectrum
\begin{eqnarray}
E \to -\frac{M(q\kappa)^2}{2\hbar^2}\frac{1}{(N+l+1)^2},
\end{eqnarray}
which coincides with the standard Bohr--Sommerfeld formula.
Thus the spin-dependent Coulomb interaction introduces a shift in the effective angular momentum, but the overall structure remains exactly solvable.

The spectral formula makes explicit the Aharonov--Bohm-type effect associated with the spin vector potential. In the branch $g\hbar k=2$ listed in Table~\ref{tab:gk}, the Yang--Mills magnetic field satisfies $\vec{\mathcal B}=0$ for $r\neq0$, whereas the potential $\vec{\mathcal A}=k(\vec r\times\vec S)/r^2$ is still singular at the origin. Thus the quantum problem is not the same as the free Coulomb problem on the full space; it is defined on a punctured configuration space and the wave function can retain global information about the singular spin-gauge background. This information is visible directly in the Hamiltonian
\begin{eqnarray}
\mathcal H_{\rm S}
=\frac{\vec p^{\,2}}{2M}+\frac{1}{2Mr^2}
\left(2\tilde k\,\vec\ell\cdot\vec S+\frac{\tilde k^2\hbar^2}{2}\right)+\frac{q\kappa}{r},
\end{eqnarray}
and hence in the effective centrifugal parameter
\begin{eqnarray}
\Omega=l(l+1)+2\tilde k W+\frac{\tilde k^2}{2}.
\end{eqnarray}
For the field-strength-free branch $g\hbar k=2$, one may equivalently write $\tilde k=2q/(g\hbar c)$, so that
\begin{eqnarray}
\Omega_{g\hbar k=2}=l(l+1)+\frac{4q}{g\hbar c}W+\frac{2q^2}{g^2\hbar^2c^2}.
\end{eqnarray}
Therefore the $k$-dependent shift of $\Omega$, $\lambda$, and finally $E$ in Eq.~\eqref{eq:energy-1} is the bound-state spectral manifestation of the spin Aharonov--Bohm-type structure. The effect is not inferred from a local field strength in the region $r\neq0$, but from the singular gauge potential and the global wave-function problem it defines.

\section{Exactly Solving Quantum Model with Dirac's Equation}\label{sec:Dirac}

We now consider the relativistic hydrogen-like system with the spin-dependent Coulomb interaction. The Dirac-type Hamiltonian reads

\begin{eqnarray}
\mathcal{H}=c\,\vec{\alpha}_1\cdot\vec{\Pi}+\beta_1Mc^2+q\varphi,
\qquad
\vec{\Pi}=\vec p-\frac{q}{c}\vec{\mathcal{A}},
\end{eqnarray}
with the spin vector potential and scalar Coulomb potential both understood on $r\neq0$
\begin{eqnarray}
\vec{\mathcal{A}}=k\,\frac{\vec r\times\vec S_2}{r^2},\qquad
\varphi=\frac{\kappa}{r}.
\end{eqnarray}
Here \(\vec S_2=\frac{\hbar}{2}\vec\sigma_2\) acts on the spin degree of freedom of the gauge potential, while \(\vec\alpha_1,\beta_1\) are the Dirac matrices acting on the relativistic spinor space. The subscript distinguishes the two spin spaces.

\subsection{Diagonal rotational symmetry}

The Hamiltonian commutes with the {total angular momentum}
\begin{eqnarray}
\vec J=\vec\ell+\vec S_1+\vec S_2,
\end{eqnarray}
where \(\vec S_1=\frac{\hbar}{2}\vec\sigma_1\) is the Dirac spin operator.
Indeed, one verifies
\begin{eqnarray}
[\mathcal H,\vec J]=0,\qquad [\mathcal H,\vec J^{2}]=0,\qquad [\mathcal H,J_z]=0,
\end{eqnarray}
because \(\vec\Pi\) transforms as a vector under simultaneous rotations of orbital and both spin spaces, while \(1/r\) is rotationally invariant.
Therefore we can classify eigenstates by the quantum numbers \(j,m_j\) of \(\vec J^{2},J_z\).

\subsection{Choice of angular basis and reduction to radial equations}

To separate variables, we need a basis that diagonalizes not only \(\vec J^{2},J_z\) but also the operator
\begin{eqnarray}
\hat r\cdot(\vec\alpha_1\times\vec\sigma_2),
\end{eqnarray}
which appears in the spin--gauge coupling. This operator commutes with \(\vec J\) and has eigenvalues \(\eta = 2, -2, 0\) (each with degeneracy).
In each block of fixed \(\eta\), the angular part is completely fixed, and the Dirac equation reduces to a {system of four radial equations}.

Introducing the dimensionless parameters
\begin{eqnarray}
\widetilde E=\frac{E}{\hbar c},\quad
\widetilde M=\frac{Mc^2}{\hbar c},\quad
\widetilde k=\frac{qk}{2c},\quad
\widetilde\kappa=\frac{q\kappa}{\hbar c},
\end{eqnarray}
{For the Coulomb sector, we use Gaussian units. Since $\varphi(r)=\kappa/r$, the parameter $\kappa$ has the dimensions of electric charge, while $q\kappa$ has the dimensions of energy multiplied by length. For a hydrogen-like system, $q=-e$ and $\kappa=Ze$, so}
\begin{eqnarray}
{q\kappa=-Ze^2,\qquad
\widetilde\kappa=\frac{q\kappa}{\hbar c}=-Z\alpha,\qquad
\frac{q^2\kappa^2}{\hbar^2c^2}=\widetilde\kappa^{\,2}=Z^2\alpha^2,}
\end{eqnarray}
{where $\alpha=e^2/(\hbar c)$ is the fine-structure constant in Gaussian units. This identification makes the connection between the relativistic spectrum in Eq.~(\ref{eq:E}) and the standard Sommerfeld formula explicit.}

Here and throughout this subsection, tilded quantities denote dimensionless rescaled variables and should not be confused with the original dimensional parameters. The radial functions \(F_1,F_2,f_3,g_3\) (for a certain angular channel) satisfy \cite{SM}
\begin{eqnarray}\label{eq:four}
&&F_1'+\frac{2\widetilde k}{r}F_1+\frac{l+1}{r}g_3
   +\Bigl(\frac{\widetilde\kappa}{r}-\widetilde E+\widetilde M\Bigr)F_2=0, \nonumber\\
&&g_3'+\frac{1}{r}g_3+\frac{l+2}{r}F_1
   +\Bigl(\frac{\widetilde\kappa}{r}-\widetilde E+\widetilde M\Bigr)f_3=0, \nonumber\\
&&F_2'-\frac{2(\widetilde k-1)}{r}F_2-\frac{l+1}{r}f_3
   -\Bigl(\frac{\widetilde\kappa}{r}-\widetilde E-\widetilde M\Bigr)F_1=0, \nonumber\\
&& f_3'+\frac{1}{r}f_3-\frac{l+2}{r}F_2
   -\Bigl(\frac{\widetilde\kappa}{r}-\widetilde E-\widetilde M\Bigr)g_3=0 .
\end{eqnarray}
Here $F'_i=dF_i/dr$, $(i=1, 2)$, $f'_3=df_3/dr$, $g'_3=dg_3/dr$, \(l\) is the orbital angular momentum quantum number appearing in the chosen angular eigenstate. The terms proportional to \(\widetilde k\) originate directly from the spin vector potential.

\subsection{Asymptotic behavior}

\textbf{Large \(r\).} For bound states (\(|\widetilde E|<\widetilde M\)), the \(1/r\) terms become negligible. Setting

\begin{eqnarray}
F_1,F_2,f_3,g_3\sim e^{-\varepsilon r},\qquad
\varepsilon=\sqrt{\widetilde M^2-\widetilde E^2},
\end{eqnarray}
yields the exponential decay and the ratio between upper and lower components:
\begin{eqnarray}
f_3 = \mu\,g_3,\qquad F_2 = \mu\,F_1,\qquad
\mu=\sqrt{\frac{\widetilde M+\widetilde E}{\widetilde M-\widetilde E}} .
\end{eqnarray}

\textbf{Small \(r\).} Inserting the power-law ansatz
\begin{eqnarray}
F_1=d_1r^{\nu-1},\quad F_2=d_2r^{\nu-1},\quad
f_3=d_3r^{\nu-1},\quad g_3=d_4r^{\nu-1},
\end{eqnarray}
and keeping only the most singular terms in (\ref{eq:four}), we obtain a \(4\times4\) {indicial system}. Its nontrivial solution condition leads to the characteristic equation
\begin{eqnarray}\bigl[\nu^2+\widetilde\kappa^2-\Lambda_+^2\bigr]\,
\bigl[\nu^2+\widetilde\kappa^2-\Lambda_-^2\bigr]=0,
\end{eqnarray}
where
\begin{eqnarray}
\Lambda_{\pm}=
\frac{2\widetilde k-1\pm\sqrt{(2\widetilde k-1)^2+4(l+1)(l+2)}}{2}.
\end{eqnarray}
The branch that continuously connects to the ordinary Coulomb--Dirac problem (when \(\widetilde k\to0\)) is
\begin{eqnarray}\label{eq:nu}
\nu = \sqrt{\Lambda_+^2-\widetilde\kappa^2},\qquad
\Lambda_+ = \frac{2\widetilde k-1+\sqrt{(2\widetilde k-1)^2+4(l+1)(l+2)}}{2}.
\end{eqnarray}
For a relativistic bound state in this branch, the characteristic exponent must be real and the regular small-$r$ behavior must be selected. Hence the parameters must satisfy
\begin{eqnarray}
\Lambda_+^2-\widetilde\kappa^{\,2}\geq 0,
\end{eqnarray}
with the positive square-root branch in Eq.~(\ref{eq:nu}). This is the relativistic analogue of the non-negativity condition on the effective centrifugal parameter in the Schr\"odinger problem.

\subsection{Series solution and quantization}

We factor out the asymptotic behavior:
\begin{eqnarray}F_1=\frac{1}{\mu}h_1(r)e^{-\varepsilon r}r^{\nu-1},\;
F_2=h_2(r)e^{-\varepsilon r}r^{\nu-1},\;
f_3=h_3(r)e^{-\varepsilon r}r^{\nu-1},\;
g_3=\frac{1}{\mu}h_4(r)e^{-\varepsilon r}r^{\nu-1}.
\end{eqnarray}
Substituting into (\ref{eq:four}) and expanding \(h_i(r)\) in power series, one finds that the system can be reduced to a single confluent hypergeometric equation for a linear combination. For generic parameters the confluent hypergeometric series does not give a normalizable bound-state wave function at infinity. Therefore the bound-state boundary condition requires the series to terminate as a polynomial. This termination gives the quantization condition
\begin{eqnarray}
N+\nu+\widetilde q = 0,\qquad N=0,1,2,\dots,
\end{eqnarray}
where
\begin{eqnarray}
\widetilde q = \frac{\widetilde\kappa}{2}\Bigl(\mu-\frac{1}{\mu}\Bigr).
\end{eqnarray}
Using the definitions of \(\mu\) and \(\widetilde\kappa\), this condition determines the energy eigenvalues.

\subsection{Energy spectrum}

After straightforward algebra, the bound-state energy is obtained as
\begin{eqnarray}\label{eq:E}
\boxed{\;
E = \frac{Mc^{2}}{\sqrt{1+\dfrac{q^{2}\kappa^{2}}{\hbar^{2}c^{2}}\dfrac{1}{(N+\nu)^{2}}}}
\;},\qquad N=0,1,2,\dots,
\end{eqnarray}
with \(\nu\) given by (\ref{eq:nu}).
In the {nonrelativistic limit} \(E\approx Mc^{2}+E_{\text{nr}}\), expanding (\ref{eq:E}) to leading order yields
\begin{eqnarray}
E_{\text{nr}} = -\frac{M(q\kappa)^{2}}{2\hbar^{2}}\frac{1}{(N+\nu)^{2}},
\end{eqnarray}
which reproduces the Schr{\" o}dinger-type spectrum derived in Sec.~\ref{sec:Schrodinger}, with \(\nu\) here playing the role of the effective angular momentum \(\lambda\) from that section.

When the spin vector potential is turned off (\(k\to0\)), we have \(\widetilde k\to0\), \(\Lambda_+\to l+1\), and \(\nu\to\sqrt{(l+1)^2-\widetilde\kappa^2}\), which is exactly the standard exponent in the Dirac--Coulomb problem. Equation (\ref{eq:E}) then reduces to the well-known Sommerfeld fine-structure formula.
The relativistic spectrum contains the same information in a different form. Even in the field-strength-free branch away from the origin, the singular spin vector potential remains in the Dirac operator through the $1/r$ angular couplings in the radial equations, for example through the terms $2\widetilde k F_1/r$ and $-2(\widetilde k-1)F_2/r$. These terms determine the angular eigenvalue
\begin{eqnarray}
\Lambda_+(\widetilde k)=\frac{2\widetilde k-1+
\sqrt{(2\widetilde k-1)^2+4(l+1)(l+2)}}{2},
\end{eqnarray}
and hence the characteristic exponent
\begin{eqnarray}
\nu=\sqrt{\Lambda_+^2-\widetilde\kappa^2}.
\end{eqnarray}
Consequently, the relativistic energy in Eq.~(\ref{eq:E}) keeps a memory of the same singular spin-gauge background through \(\widetilde k\). This is the Dirac counterpart of the bound-state spin Aharonov--Bohm-type effect displayed above in the Schr\"odinger spectrum.
Thus the Dirac equation with spin-dependent Coulomb interaction is also exactly solvable, and its spectrum smoothly interpolates between the ordinary hydrogenic relativistic spectrum and a new family of spin--gauge modified energy levels.

\section{Quantization Conditions}\label{sec:quantization}

We now make explicit how quantization enters the two exact solutions. The quantization conditions used above are not additional assumptions; they are consequences of the same bound-state requirements that appear in the ordinary Coulomb problem, namely regularity at the singular point and normalizability at spatial infinity. In both the Schr\"odinger and Dirac cases, the spin vector potential modifies the angular part of the problem, while the radial equation remains of Coulomb--hypergeometric type. The allowed energies are therefore obtained by the usual polynomial-termination mechanism, but with spin-gauge-dependent effective angular parameters.

\subsection{Schr\"odinger equation}

For the nonrelativistic radial equation \eqref{eq:ReduceRadialEq}, the bound-state condition requires $\epsilon<0$. The behavior near $r=0$ is fixed by the indicial relation $\lambda(\lambda+1)=\Omega$, and the regular branch is chosen so that the radial wave function is square integrable at the origin. With this branch fixed, let
\begin{eqnarray}
\rho=2\sqrt{-\epsilon}\,r,
\qquad
R(r)=r^{\lambda}e^{-\rho/2}F(\rho).
\end{eqnarray}
Substituting this form into Eq.~\eqref{eq:ReduceRadialEq} gives the Kummer equation
\begin{eqnarray}
\rho F''+(2\lambda+2-\rho)F'
-\left(\lambda+1+\frac{Mq\kappa}{\hbar^2\sqrt{-\epsilon}}\right)F=0.
\end{eqnarray}
The solution regular at the origin is the confluent hypergeometric function
${}_1F_1(a;b;\rho)$ with
\begin{eqnarray}
a=\lambda+1+\frac{Mq\kappa}{\hbar^2\sqrt{-\epsilon}},
\qquad b=2\lambda+2.
\end{eqnarray}
For a generic value of $a$, the factor ${}_1F_1(a;b;\rho)$ grows exponentially for large $\rho$ and destroys the square-integrability of $R(r)$. Thus the series must terminate, i.e.
\begin{eqnarray}
a=-N,\qquad N=0,1,2,\ldots .
\end{eqnarray}
This termination condition is precisely the nonrelativistic quantization condition and gives Eq.~\eqref{eq:energy-1}. Equivalently, the effective principal quantum number is $N+\lambda+1$, so the spin vector potential affects the spectrum through $\lambda$, since $\lambda(\lambda+1)=\Omega$ and $\Omega$ contains the $k$-dependent spin-gauge terms. In this sense the quantization condition itself has the familiar Coulomb form, while the admissible angular channel has been deformed by the spin-dependent gauge background.

\subsection{Dirac equation}

The relativistic quantization condition is obtained in the same way, after the angular reduction of the Dirac equation. First, a bound state requires an exponentially decreasing radial factor, which gives
\begin{eqnarray}
|\widetilde E|<\widetilde M,
\qquad
\mu=\sqrt{\frac{\widetilde M-\widetilde E}{\widetilde M+\widetilde E}}>0.
\end{eqnarray}
Second, regular behavior near the origin requires the characteristic exponent to be real and the regular branch to be selected:
\begin{eqnarray}
\Lambda_+^2-\widetilde\kappa^2\ge0,
\qquad
\nu=\sqrt{\Lambda_+^2-\widetilde\kappa^2}.
\end{eqnarray}
Third, after extracting the leading behavior at $r=0$ and the exponential decay at infinity, the remaining radial function again satisfies a confluent-hypergeometric-type equation. Normalizability selects the polynomial solution and gives
\begin{eqnarray}
N+\nu+\widetilde q=0,
\qquad
N=0,1,2,\ldots,
\qquad
\widetilde q=\frac{\widetilde\kappa}{2}\left(\mu-\frac{1}{\mu}\right).
\end{eqnarray}
Solving this condition gives the relativistic spectrum Eq.~(\ref{eq:E}). Therefore the integer $N$, the regular exponent $\nu$, and the energy $E$ are fixed simultaneously by the bound-state boundary conditions. The difference from the ordinary Dirac--Coulomb problem is concentrated in the angular eigenvalue $\Lambda_+(\widetilde k)$ and hence in $\nu$; the radial quantization mechanism remains the same Coulomb-type termination condition.

\subsection{Fixed-background interpretation}

The quantization discussed in this section is the spectral quantization of a one-particle wave equation in a prescribed classical Yang--Mills background. It should be distinguished from the field quantization of QED or of a full Yang--Mills theory. In the present construction the background potentials are first obtained as local classical solutions of the sourceless Yang--Mills equations on the punctured region, and the Schr\"odinger and Dirac equations are then solved exactly in that fixed background. The integer $N$ and the allowed energies are therefore produced by the bound-state boundary conditions of the external-field problem.

More explicitly, after the classical background
\begin{eqnarray}
\mathcal A^{\rm cl}_i=k\frac{(\vec r\times\vec S)_i}{r^2},
\qquad
\varphi^{\rm cl}=\frac{\kappa}{r},
\qquad r\neq0,
\quad \text{with the origin excluded for both potentials},
\end{eqnarray}
is fixed, the two spectral problems solved in this paper are
\begin{eqnarray}
\mathcal H_{\rm S}[\mathcal A^{\rm cl},\varphi^{\rm cl}]\psi=E\psi,
\qquad
\mathcal H_{\rm D}[\mathcal A^{\rm cl},\varphi^{\rm cl}]\Psi=E\Psi,
\end{eqnarray}
where
\begin{eqnarray}
\mathcal H_{\rm S}=\frac{1}{2M}\left(\vec p-\frac{q}{c}\vec{\mathcal A}^{\rm cl}\right)^2+q\varphi^{\rm cl},
\qquad
\mathcal H_{\rm D}=c\vec\alpha\cdot\left(\vec p-\frac{q}{c}\vec{\mathcal A}^{\rm cl}\right)+\beta Mc^2+q\varphi^{\rm cl}.
\end{eqnarray}
Thus the background is not integrated over in the present analysis; it is a fixed classical input to the one-particle spectral problem. The punctured-domain condition is not only a property of the spin vector potential; it also applies to the Coulomb scalar potential $\kappa/r$.

This viewpoint also clarifies how the present exact model is related to possible field-theoretic treatments. If matter fields are subsequently quantized in the same prescribed background, the eigenfunctions and spectra obtained here may be used as the corresponding one-particle basis or as fixed-background input for Green-function and controlled perturbative analyses. If the gauge field itself is quantized, however, radiative corrections, renormalization, and ultraviolet issues such as Landau-pole or zero-charge behavior would in general modify the effective interaction. Such effects define a different, higher-level problem and are not part of the exact solvability claim made here. Thus the present model is best regarded as an analytically controllable low-energy benchmark that displays how the spin-dependent Coulomb background changes angular momentum channels and spectra.

\section{Conclusion}\label{sec:conclusion}

In this work, we have demonstrated that the spin vector potential
\(\vec{\mathcal{A}}=k(\vec{r}\times\vec{S})/r^2\), together with the Coulomb-type scalar potential
\(\varphi=\kappa/r\) defined away from the origin, constitutes a family of local exact solutions to the sourceless Yang--Mills equations
in the region \(r\neq0\). This solution, denoted as \(\mathcal{S}_{\rm YM}\), naturally reduces to the
standard Coulomb solution \(\mathcal{S}_{\rm Maxwell}\) when spin effects are neglected, thereby
establishing a direct link between spin-dependent interactions and non-Abelian gauge theory.
The branch structure places this solution in a recognizable Yang--Mills context: the branch
\(g\hbar k=1\) corresponds locally to a Wu--Yang-type monopole connection in the spin representation,
while the branch \(g\hbar k=2\) (with \(f_1(r)=0\) and \(\vec{\mathcal{B}}=0\)) gives a flat-connection-like
structure in the simple Coulomb scalar sector. The field invariant in Eq.~(\ref{eq:B2-v13}) confirms
the nontrivial character of the monopole-like branch and also indicates the singular behavior at
\(r=0\).

A second key result is the identification of the diagonal rotational symmetry of this potential.
In the Schr\"odinger problem the conserved generator is \(\vec{J}=\vec{\ell}+\vec{S}\), while in the
Dirac problem it is \(\vec{J}=\vec{\ell}+\vec{S}_1+\vec{S}_2\). This symmetry is the common reason
why both quantum equations can be reduced to radial Coulomb-type equations and solved exactly.
The resulting energy spectra are derived in closed form and reduce to the standard hydrogenic
spectra when the spin coupling is turned off (\(k\to0\)), while introducing spin-dependent
corrections that could be probed in precision atomic or condensed-matter experiments.

Our findings provide a Yang--Mills gauge-theoretical realization of the previously proposed spin
vector potential, which has been implicated in phenomena such as the spin Aharonov--Bohm effect
and various spin-mediated interactions. By embedding spin-dependent potentials within the YM
framework, we show that internal spin degrees of freedom can be naturally incorporated into
gauge fields, offering a unified perspective on how spin couples to fundamental interactions.
From an experimental standpoint, the spin Aharonov--Bohm effect offers a promising platform
for testing the physical relevance of the spin vector potential via interferometric setups,
while spectroscopic measurements in atoms or quantum dots may reveal the predicted
energy-level modifications.

Several open questions remain for future investigation. These include: (i) extending the
solution to higher-spin representations and more general gauge groups; (ii) investigating
the role of such spin-dependent potentials in emergent phenomena such as spin liquids,
skyrmions, and topological insulators; (iii) exploring possible cosmological or gravitational
implications of spin--gauge couplings, especially in contexts where spin-dependent and
spin--gravity interactions are considered~\cite{peres1978,Wilczek1984,Dobrescu2006,Fadeev2019,Du2023,Tarallo2014,Duan2016,Zhang2023,Cong2025}.
On the Yang--Mills side, further work could also address global patching, regularization
near the origin, finite-energy extensions, and possible couplings to Higgs-type fields.
We hope that these results will stimulate further theoretical and experimental explorations
at the intersection of gauge theory, spin physics, and quantum engineering.
\vspace{6pt}



\begin{acknowledgments}
This work is supported by the Quantum Science and Technology-National Science and Technology Major Project (Grant No. 2024ZD0301000), and the National Natural Science Foundation of China (Grant No. 12275136).
\end{acknowledgments}

J. L. Z. and Y. X. Z. contributed equally to this work.

\vspace{5mm}

\noindent \textbf{Competing Interests:} The authors declare no competing interests.

%
%
%
%

\end{document}



\title{Supplemental Material for\\``Exactly Solvable Quantum Model with Spin-Dependent Coulomb Interaction''}

\author{Jiang-Lin Zhou}
\affiliation{School of Physics, Nankai University, Tianjin 300071, People's Republic of China}

\author{Yu-Xuan Zhang}
\affiliation{School of Physics, Nankai University, Tianjin 300071, People's Republic of China}

\author{Choo Hiap Oh}
\affiliation{Centre for Quantum Technologies and Department of Physics, National University of Singapore, 117543, Singapore}

\author{Jing-Ling Chen}
\email{chenjl@nankai.edu.cn}
\affiliation{Theoretical Physics Division, Chern Institute of Mathematics, Nankai University, Tianjin 300071, People's Republic of China}

\date{\today}

\maketitle

\tableofcontents\newpage

\newpage

\section{The Yang-Mills Equations}

The Yang-Mills theory is one of cornerstones in modern physics. The Yang-Mills equations in vacuum without sources (i.e., the 4-vector current ${J^\nu}=0$) are given by \cite{Yang1954}
      \begin{subequations}
            \begin{align}
                  & D_\mu {F}^{\mu\nu} \equiv\partial_\mu {F}^{\mu\nu}
                        +{\rm i}\,g\Bigl[\mathbb{A}_\mu ,\ {F}^{\mu\nu}\Bigr]=0,
                        \label{eq:YM1a} \\
                  & D_\mu {F}_{\nu\gamma} +D_\nu {F}_{\gamma\mu}
                        +D_\gamma {F}_{\mu\nu} =0. \label{eq:YM1b}
            \end{align}
      \end{subequations}
Here the covariant derivative $D_{\mu}$ is defined as
\begin{eqnarray}
D_{\mu}=\partial_{\mu}+{\rm i} g \mathbb{A}_{\mu},  \quad\;\;\;\;  (\mu=0, 1, 2, 3),
\end{eqnarray}
and the field strength tensor is defined as
\begin{eqnarray}
{\rm i} g F_{\mu\nu}=[D_\mu,D_\nu], \quad \;\;\;  (\mu, \nu=0, 1, 2, 3),
\end{eqnarray}
i.e.,
      \begin{equation}\label{eq:nonabelian}
            {F}_{\mu\nu} =\partial_\mu \mathbb{A}_{\nu}
                  -\partial_\nu \mathbb{A}_{\mu}
                  +\mathrm{i}\,g[\mathbb{A}_{\mu},\ \mathbb{A}_{\nu}], \;\;\;\;\; (\mu, \nu=0, 1, 2, 3),
      \end{equation}
where $g$ is the coupling parameter, and $\mathbb{A}_\mu$'s are the potentials. Note that we have used the four-vectorial notations $x^\mu=(x_0, x_1, x_2, x_3)$, $x_\mu=(x_0, -x_1, -x_2, -x_3)$, with $x_1 \equiv x$, $x_2 \equiv y$, $x_3 \equiv z$, $x_0 \equiv c\,t$, and
      \begin{eqnarray}
      &&\mathbb{A}_\mu=(\varphi,-\vec{\mathcal{A}})=(\varphi,-\mathcal{A}_x ,-\mathcal{A}_y ,-\mathcal{A}_z), \;\;\;\; \; {\partial}_{\mu} =\frac{\partial}{\partial x^\mu}=\left(\frac{1}{c}\dfrac{\partial}{\partial t},\vec{\nabla}\right),\;\;\;\;\; \vec{\nabla}=\left(\dfrac{\partial}{\partial x}, \dfrac{\partial}{\partial y}, \dfrac{\partial}{\partial z}\right).
\end{eqnarray}
 Here $\varphi$ is called the \emph{scalar potential}, and $\vec{\mathcal{A}}$ is called the \emph{vector potential}.
 The Yang-Mills equations are the natural generalization of Maxwell's equations from the Abelian potentials to the non-Abelian ones. When the commutator $[\mathbb{A}_{\mu}, \mathbb{A}_{\nu}]=0$, the definition of ${F}_{\mu\nu}$ reduces to the Abelian case (i.e., $F_{\mu\nu}=\partial_\mu \mathbb{A}_\nu -\partial_\nu \mathbb{A}_\mu$ for the situation of Maxwell's equations). When some of the commutators $[\mathbb{A}_{\mu}, \mathbb{A}_{\nu}]$ are not zero, then one says that the potentials are non-Abelian.

 The matrix form of field tensor $F_{\mu\nu}$ is given by \cite{1999Jackson}
   \begin{equation}
       F_{\mu\nu}=\begin{bmatrix}
            0 & \mathcal{E}_x & \mathcal{E}_y & \mathcal{E}_z \\
            -\mathcal{E}_x & 0 & -\mathcal{B}_z & \mathcal{B}_y \\
            -\mathcal{E}_y & \mathcal{B}_z & 0 & -\mathcal{B}_x \\
            -\mathcal{E}_z & -\mathcal{B}_y & \mathcal{B}_x & 0
      \end{bmatrix},
   \end{equation}
   or
     \begin{equation}
                        {F}^{\mu\nu} =\begin{bmatrix}
                              0 & -\mathcal{E}_x & -\mathcal{E}_y & -\mathcal{E}_z \\
                              \mathcal{E}_x & 0 & -\mathcal{B}_z & \mathcal{B}_y \\
                              \mathcal{E}_y & \mathcal{B}_z & 0 & -\mathcal{B}_x \\
                              \mathcal{E}_z & -\mathcal{B}_y & \mathcal{B}_x & 0
                        \end{bmatrix}.
                  \end{equation}
Based on which, we have the three components of ``magnetic'' field $\vec{\mathcal{B}}=\left(\mathcal{B}_x, \mathcal{B}_y, \mathcal{B}_z\right)$ as
   \begin{equation}\label{eq:mag-1}
      \begin{split}
            \mathcal{B}_x =&\ F_{32} =\dfrac{\partial\,\mathbb{A}_2}{\partial\,x_3}
                  -\dfrac{\partial\,\mathbb{A}_3}{\partial\,x_2}
                  +\mathrm{i}\,g[\mathbb{A}_3, \mathbb{A}_2]
            =-\dfrac{\partial\,\mathcal{A}_y}{\partial\,z}
                  +\dfrac{\partial\,\mathcal{A}_z}{\partial\,y}
                  +\mathrm{i}\,g[\mathcal{A}_z, \mathcal{A}_y]= \left[\dfrac{\partial \mathcal{A}_z}{\partial y}- \dfrac{\partial \mathcal{A}_y}{\partial z}\right] -\mathrm{i}\,g\left[\mathcal{A}_y, \mathcal{A}_z\right], \\
            \mathcal{B}_y =&\ F_{13} =\dfrac{\partial\,\mathbb{A}_3}{\partial\,x_1}
                  -\dfrac{\partial\,\mathbb{A}_1}{\partial\,x_3}
                  +\mathrm{i}\,g[\mathbb{A}_1, \mathbb{A}_3]
            =-\dfrac{\partial\,\mathcal{A}_z}{\partial\,x}
                  +\dfrac{\partial\,\mathcal{A}_x}{\partial\,z}
                  +\mathrm{i}\,g[\mathcal{A}_x, \mathcal{A}_z]= \left[\dfrac{\partial \mathcal{A}_x}{\partial z}- \dfrac{\partial \mathcal{A}_z}{\partial x}\right] -\mathrm{i}\,g\left[\mathcal{A}_z, \mathcal{A}_x\right], \\
            \mathcal{B}_z =&\ F_{21} =\dfrac{\partial\,\mathbb{A}_1}{\partial\,x_2}
                  -\dfrac{\partial\,\mathbb{A}_2}{\partial\,x_1}
                  +\mathrm{i}\,g[\mathbb{A}_2, \mathbb{A}_1]
            =-\dfrac{\partial\,\mathcal{A}_x}{\partial\,y}
                  +\dfrac{\partial\,\mathcal{A}_y}{\partial\,x}
                  +\mathrm{i}\,g[\mathcal{A}_y, \mathcal{A}_x]= \left[\dfrac{\partial \mathcal{A}_y}{\partial x}- \dfrac{\partial \mathcal{A}_x}{\partial y}\right] -\mathrm{i}\,g\left[\mathcal{A}_x, \mathcal{A}_y\right],
      \end{split}
   \end{equation}
   and the three components of ``electric'' field $\vec{\mathcal{E}}=\left(\mathcal{E}_x, \mathcal{E}_y, \mathcal{E}_z\right)$ as
   \begin{equation}\label{eq:ele-1}
      \begin{split}
            \mathcal{E}_x =&\ F_{01} =\dfrac{\partial\,\mathbb{A}_1}{\partial\,x_0}
                  -\dfrac{\partial\,\mathbb{A}_0}{\partial\,x_1}
                  +\mathrm{i}\,g[\mathbb{A}_0, \mathbb{A}_1]
            =-\frac{1}{c}\dfrac{\partial\,\mathcal{A}_x}{\partial\,t}
                  -\dfrac{\partial\,\varphi}{\partial\,x}
                  -\mathrm{i}\,g [\varphi,\ \mathcal{A}_x], \\
            \mathcal{E}_y =&\ F_{02} =\dfrac{\partial\,\mathbb{A}_2}{\partial\,x_0}
                  -\dfrac{\partial\,\mathbb{A}_0}{\partial\,x_2}
                  +\mathrm{i}\,g[\mathbb{A}_0, \mathbb{A}_2]
            =-\frac{1}{c}\dfrac{\partial\,\mathcal{A}_y}{\partial\,t}
                  -\dfrac{\partial\,\varphi}{\partial\,y}
                  -\mathrm{i}\,g[\varphi,\ \mathcal{A}_y], \\
            \mathcal{E}_z =&\ F_{03} =\dfrac{\partial\,\mathbb{A}_3}{\partial\,x_0}
                  -\dfrac{\partial\,\mathbb{A}_0}{\partial\,x_3}
                  +\mathrm{i}\,g[\mathbb{A}_0, \mathbb{A}_3]
            =-\frac{1}{c}\dfrac{\partial\,\mathcal{A}_z}{\partial\,t}
                  -\dfrac{\partial\,\varphi}{\partial\,z}
                  -\mathrm{i}\,g [\varphi,\ \mathcal{A}_z].
      \end{split}
   \end{equation}
   According to Eq. (\ref{eq:mag-1}) and Eq. (\ref{eq:ele-1}), one may write down the vector forms of ``magnetic'' field and ``electric'' field as
      \begin{subequations}
            \begin{eqnarray}
                  \vec{\mathcal{B}} &=& \vec{\nabla}\times\vec{\mathcal{A}}
                        -{\rm i}\,g\left(\vec{\mathcal{A}}\times\vec{\mathcal{A}}\right),
                      \label{eq:mag-2}  \\
                  \vec{\mathcal{E}} &=& -\dfrac{1}{c}
                              \dfrac{\partial\,\vec{\mathcal{A}}}{\partial\,t}
                        -\vec{\nabla}\varphi-{\rm i}\,g\left[\varphi,\ \vec{\mathcal{A}}
                        \right]. \label{eq:ele-2}
            \end{eqnarray}
      \end{subequations}
Based on them, we can write down the Yang-Mills equations by using the ``magnetic'' field and the ``electric'' field. We show the calculation as follows.

\begin{remark}Let us study the first Yang-Mills equation (\ref{eq:YM1a}).

(i) Let us take $\nu=0$, then
\begin{eqnarray}
&& F^{i0}=\mathcal{E}_i,
\end{eqnarray}
with $i=1, 2, 3$ ( or $i=x, y, z$). From
\begin{eqnarray}
& D_\mu {F}^{\mu\nu} \equiv\partial_\mu {F}^{\mu\nu}
{+}{\rm i}\,g\Bigl[\mathbb{A}_\mu,\ {F}^{\mu\nu}\Bigr]=0,
\end{eqnarray}
we can have
\begin{eqnarray}
& D_i {F}^{i 0} \equiv\partial_i {F}^{i 0}
{+}{\rm i}\,g\Bigl[{-\mathcal{A}_i},\ {F}^{i 0}\Bigr]=0,
\end{eqnarray}
i.e.,
\begin{eqnarray}
& {\partial_i {F}^{i 0}
-{\rm i}\,g\Bigl[\mathcal{A}_i,\ {F}^{i 0}\Bigr]=0,}
\end{eqnarray}
i.e.,
\begin{eqnarray}
& { \partial_i \mathcal{E}_i
-{\rm i}\,g\Bigl[\mathcal{A}_i,\ \mathcal{E}_i\Bigr]=0,}
\end{eqnarray}
i.e.,
\begin{eqnarray}
& \left(\dfrac{\partial \mathcal{E}_x}{\partial x}+ \dfrac{\partial \mathcal{E}_y}{\partial y}+ \dfrac{\partial \mathcal{E}_z}{\partial z}\right)
-{\rm i}\,g \Bigl\{(\mathcal{A}_x  \mathcal{E}_x+\mathcal{A}_y  \mathcal{E}_y+\mathcal{A}_z  \mathcal{E}_z)-
(\mathcal{E}_x \mathcal{A}_x+\mathcal{E}_y \mathcal{A}_y+\mathcal{E}_z \mathcal{A}_z)\Bigr\}=0.
\end{eqnarray}
Then finally we have
\begin{eqnarray}
&\vec{\nabla}\cdot\vec{\mathcal{E}}\ \textcolor{blue}{-{\rm i}\,g}  \Bigl(
                        \vec{\mathcal{A}}\cdot\vec{\mathcal{E}}
                        -\vec{\mathcal{E}}\cdot\vec{\mathcal{A}}\Bigr)=0.
\end{eqnarray}

(ii-1) Let us take $\nu=1$, then
\begin{eqnarray}
&& F^{01}=-\mathcal{E}_x, \;\;\; F^{21}= \mathcal{B}_z,   \;\;\; F^{31}=-\mathcal{B}_y.
\end{eqnarray}
From
\begin{eqnarray}
& D_\mu {F}^{\mu\nu} \equiv\partial_\mu {F}^{\mu\nu}
{+}{\rm i}\,g\Bigl[\mathbb{A}_\mu,\ {F}^{\mu\nu}\Bigr]=0,
\end{eqnarray}
we can have
\begin{eqnarray}
&& \partial_\mu {F}^{\mu 1}
+{\rm i}\,g\Bigl[\mathbb{A}_\mu,\ {F}^{\mu 1}\Bigr]=0,
\end{eqnarray}
i.e.,
\begin{eqnarray}
&& \left(\partial_0 {F}^{0 1}+ \partial_2 {F}^{2 1}+ \partial_3 {F}^{3 1} \right)
+{\rm i}\,g \left\{ \Bigl[\mathbb{A}_0,\ {F}^{0 1}\Bigr]+\Bigl[\mathbb{A}_2,\ {F}^{2 1}\Bigr]+\Bigl[\mathbb{A}_3,\ {F}^{3 1}\Bigr]\right\}=0,
\end{eqnarray}
i.e.,
\begin{eqnarray}
&& \left(\partial_0 {F}^{0 1}+ \partial_2 {F}^{2 1}+ \partial_3 {F}^{3 1} \right)
+{\rm i}\,g \left\{ \Bigl[\varphi,\ {F}^{0 1}\Bigr]-\Bigl[\mathcal{A}_y,\ {F}^{2 1}\Bigr]-\Bigl[\mathcal{A}_z,\ {F}^{3 1}\Bigr]\right\}=0,
\end{eqnarray}
i.e.,
\begin{eqnarray}
&& \left[\partial_0 (-\mathcal{E}_x)+ \partial_2 \mathcal{B}_z+ \partial_3 (-\mathcal{B}_y) \right]
+{\rm i}\,g \left\{ \Bigl[\varphi,\ -\mathcal{E}_x\Bigr]-\Bigl[\mathcal{A}_y,\ \mathcal{B}_z\Bigr]-\Bigl[\mathcal{A}_z,\ -\mathcal{B}_y\Bigr]\right\}=0,
\end{eqnarray}
i.e.,
\begin{eqnarray}
&& \left[ -\frac{1}{c}\frac{\partial \mathcal{E}_x}{\partial t}+ \frac{\partial \mathcal{B}_z}{\partial y}
- \frac{\partial \mathcal{B}_y}{\partial z} \right]
-{\rm i}\,g \left\{ \Bigl[\varphi,\ \mathcal{E}_x\Bigr]+\Bigl[\mathcal{A}_y,\ \mathcal{B}_z\Bigr]+\Bigl[\mathcal{B}_y,\ \mathcal{A}_z\Bigr]\right\}=0,
\end{eqnarray}
i.e.,
\begin{eqnarray}
&& -\dfrac{1}{c} \dfrac{\partial \mathcal{E}_x}{\partial\,t}
                        + \left(\vec{\nabla}\times\vec{\mathcal{B}}\right)_x  {-}{\rm i}\,g\biggl(\Bigl[\varphi,\
                                    \mathcal{E}_x\Bigr]
                              +\left(\vec{\mathcal{A}}\times\vec{\mathcal{B}}\right)_x
                              +\left(\vec{\mathcal{B}}\times\vec{\mathcal{A}}\right)_x \biggr)=0.
\end{eqnarray}

(ii-2) Let us take $\nu=2$, similarly, we have
\begin{eqnarray}
&& -\dfrac{1}{c} \dfrac{\partial \mathcal{E}_y}{\partial\,t}
                        + \left(\vec{\nabla}\times\vec{\mathcal{B}}\right)_y  {-}{\rm i}\,g\biggl(\Bigl[\varphi,\
                                    \mathcal{E}_y \Bigr]
                              +\left(\vec{\mathcal{A}}\times\vec{\mathcal{B}}\right)_y
                              +\left(\vec{\mathcal{B}}\times\vec{\mathcal{A}}\right)_y \biggr)=0.
\end{eqnarray}

(ii-3) Let us take $\nu=3$, similarly, we have
\begin{eqnarray}
&& -\dfrac{1}{c} \dfrac{\partial \mathcal{E}_z}{\partial\,t}
                        + \left(\vec{\nabla}\times\vec{\mathcal{B}}\right)_z  {-}{\rm i}\,g\biggl(\Bigl[\varphi,\
                                    \mathcal{E}_z \Bigr]
                              +\left(\vec{\mathcal{A}}\times\vec{\mathcal{B}}\right)_z
                              +\left(\vec{\mathcal{B}}\times\vec{\mathcal{A}}\right)_z \biggr)=0.
\end{eqnarray}

Based on the results in (ii-1), (ii-2) and (ii-3), we finally have
\begin{eqnarray}
-\dfrac{1}{c} \dfrac{\partial}{\partial\,t} \vec{\mathcal{E}}
                        +\vec{\nabla}\times\vec{\mathcal{B}}  {-{\rm i}\,g }\biggl(\Bigl[\varphi,\
                                    \vec{\mathcal{E}}\Bigr]
                              +\vec{\mathcal{A}}\times\vec{\mathcal{B}}
                              +\vec{\mathcal{B}}\times\vec{\mathcal{A}}\biggr)=0.
\end{eqnarray}

$\blacksquare$
\end{remark}

\begin{remark}Let us study the second Yang-Mills equation (\ref{eq:YM1b}).

(iii) Let us take $\mu\nu\gamma=123$, then
\begin{eqnarray}
&& F_{12}= -\mathcal{B}_z, \;\;\;  F_{23}= -\mathcal{B}_x, \;\;\; F_{31}= -\mathcal{B}_y.
\end{eqnarray}
From
\begin{eqnarray}
&& D_\mu {F}_{\nu\gamma} +D_\nu {F}_{\gamma\mu} +D_\gamma {F}_{\mu\nu} =0,
\end{eqnarray}
we have
\begin{eqnarray}
&& D_1 {F}_{23} +D_2 {F}_{31} +D_3 {F}_{12} =0,
\end{eqnarray}
i.e.,
\begin{eqnarray}
&& -D_1 \mathcal{B}_x -D_2 \mathcal{B}_y -D_3 \mathcal{B}_z =0,
\end{eqnarray}
i.e.,
\begin{eqnarray}
&& D_1 \mathcal{B}_x +D_2 \mathcal{B}_y +D_3 \mathcal{B}_z =0,
\end{eqnarray}
i.e.,
\begin{eqnarray}
&& \partial_1 \mathcal{B}_x +{\rm i}\,g\Bigl[\mathbb{A}_1, \mathcal{B}_x\Bigr]+
\partial_2 \mathcal{B}_y +{\rm i}\,g\Bigl[\mathbb{A}_2, \mathcal{B}_y\Bigr]+
\partial_3 \mathcal{B}_z +{\rm i}\,g\Bigl[\mathbb{A}_3, \mathcal{B}_z\Bigr]=0,
\end{eqnarray}
i.e.,
\begin{eqnarray}
&& \frac{\partial \mathcal{B}_x}{\partial x}-{\rm i}\,g\Bigl[\mathcal{A}_x, \mathcal{B}_x\Bigr]+
\frac{\partial \mathcal{B}_y}{\partial y}-{\rm i}\,g\Bigl[\mathcal{A}_y, \mathcal{B}_y\Bigr]+
\frac{\partial \mathcal{B}_z}{\partial z}-{\rm i}\,g\Bigl[\mathcal{A}_z, \mathcal{B}_z\Bigr]=0,
\end{eqnarray}
i.e.,
\begin{eqnarray}
&& \left(\frac{\partial \mathcal{B}_x}{\partial x}+\frac{\partial \mathcal{B}_y}{\partial y}+\frac{\partial \mathcal{B}_z}{\partial z}\right)
-{\rm i}\,g \left\{\Bigl[\mathcal{A}_x, \mathcal{B}_x\Bigr]+\Bigl[\mathcal{A}_y, \mathcal{B}_y\Bigr]+\Bigl[\mathcal{A}_z, \mathcal{B}_z\Bigr]\right\}=0.
\end{eqnarray}
We then finally have
\begin{eqnarray}
 && \vec{\nabla}\cdot\vec{\mathcal{B}}\ {-{\rm i}\,g}\Bigl(
 \vec{\mathcal{A}}\cdot\vec{\mathcal{B}}
 -\vec{\mathcal{B}}\cdot\vec{\mathcal{A}}\Bigr)=0.
\end{eqnarray}

(iv-1) Let us take $\mu\nu\gamma=012$, then
\begin{eqnarray}
&& F_{01}= \mathcal{E}_x, \;\;\;  F_{12}= -\mathcal{B}_z, \;\;\; F_{20}= -\mathcal{E}_y.
\end{eqnarray}
From
\begin{eqnarray}
&& D_\mu {F}_{\nu\gamma} +D_\nu {F}_{\gamma\mu} +D_\gamma {F}_{\mu\nu} =0,
\end{eqnarray}
we have
\begin{eqnarray}
&& D_0 {F}_{12} +D_1 {F}_{20} +D_2 {F}_{01} =0,
\end{eqnarray}
i.e.,
\begin{eqnarray}
&& -D_0 \mathcal{B}_z -D_1 \mathcal{E}_y +D_2 \mathcal{E}_x =0,
\end{eqnarray}
i.e.,
\begin{eqnarray}
&& -\partial_0 \mathcal{B}_z -{\rm i}\,g\Bigl[\mathbb{A}_0, \mathcal{B}_z\Bigr]
-\partial_1 \mathcal{E}_y -{\rm i}\,g\Bigl[\mathbb{A}_1, \mathcal{E}_y\Bigr]+
\partial_2 \mathcal{E}_x +{\rm i}\,g\Bigl[\mathbb{A}_2, \mathcal{E}_x\Bigr]=0,
\end{eqnarray}
i.e.,
\begin{eqnarray}
&& -\partial_0 \mathcal{B}_z -{\rm i}\,g\Bigl[\varphi, \mathcal{B}_z\Bigr]
-\partial_1 \mathcal{E}_y -{\rm i}\,g\Bigl[-\mathcal{A}_x, \mathcal{E}_y\Bigr]+
\partial_2 \mathcal{E}_x +{\rm i}\,g\Bigl[-\mathcal{A}_y, \mathcal{E}_x\Bigr]=0,
\end{eqnarray}
i.e.,
\begin{eqnarray}
&& -\frac{1}{c} \frac{\partial \mathcal{B}_z}{\partial t}
-\left(\frac{\partial \mathcal{E}_y}{\partial x} - \frac{\partial \mathcal{E}_x}{\partial y}\right)
-{\rm i}\,g \left\{\Bigl[\varphi, \mathcal{B}_z\Bigr]
+\Bigl[-\mathcal{A}_x, \mathcal{E}_y\Bigr] -\Bigl[-\mathcal{A}_y, \mathcal{E}_x\Bigr]\right\}=0,
\end{eqnarray}
i.e.,
\begin{eqnarray}
&& -\frac{1}{c} \frac{\partial \mathcal{B}_z}{\partial t}
-\left(\frac{\partial \mathcal{E}_y}{\partial x} - \frac{\partial \mathcal{E}_x}{\partial y}\right)
-{\rm i}\,g \left\{\Bigl[\varphi, \mathcal{B}_z\Bigr]
-\Bigl[\mathcal{A}_x, \mathcal{E}_y\Bigr] -\Bigl[\mathcal{E}_x, \mathcal{A}_y\Bigr]\right\}=0,
\end{eqnarray}
i.e.,
\begin{eqnarray}
&& -\dfrac{1}{c} \dfrac{\partial \mathcal{B}_z}{\partial\,t}
-\left(\vec{\nabla}\times\vec{\mathcal{E}}\right)_z
 -{\rm i}\,g\biggl(\Bigl[\varphi,\ \mathcal{B}_z\Bigr]
 -\left(\vec{\mathcal{A}}\times\vec{\mathcal{E}}\right)_z
 -\left(\vec{\mathcal{E}}\times\vec{\mathcal{A}}\right)_z\biggr)=0.
\end{eqnarray}

(iv-2) Let us take $\mu\nu\gamma=013$, similarly we have

\begin{eqnarray}
&& -\dfrac{1}{c} \dfrac{\partial \mathcal{B}_y}{\partial\,t}
-\left(\vec{\nabla}\times\vec{\mathcal{E}}\right)_y
 -{\rm i}\,g\biggl(\Bigl[\varphi,\ \mathcal{B}_y\Bigr]
 -\left(\vec{\mathcal{A}}\times\vec{\mathcal{E}}\right)_y
 -\left(\vec{\mathcal{E}}\times\vec{\mathcal{A}}\right)_y\biggr)=0.
\end{eqnarray}

(iv-3) Let us take $\mu\nu\gamma=023$, similarly we have

\begin{eqnarray}
&& -\dfrac{1}{c} \dfrac{\partial \mathcal{B}_x}{\partial\,t}
-\left(\vec{\nabla}\times\vec{\mathcal{E}}\right)_x
 -{\rm i}\,g\biggl(\Bigl[\varphi,\ \mathcal{B}_x\Bigr]
 -\left(\vec{\mathcal{A}}\times\vec{\mathcal{E}}\right)_x
 -\left(\vec{\mathcal{E}}\times\vec{\mathcal{A}}\right)_x\biggr)=0.
\end{eqnarray}

We then finally have
\begin{eqnarray}
&& -\dfrac{1}{c} \dfrac{\partial \vec{\mathcal{B}}}{\partial\,t}
-\vec{\nabla}\times\vec{\mathcal{E}}
 {-{\rm i}\,g}\biggl(\Bigl[\varphi,\ \vec{\mathcal{B}}\Bigr]
 -\vec{\mathcal{A}}\times\vec{\mathcal{E}}
 -\vec{\mathcal{E}}\times\vec{\mathcal{A}}\biggr)=0.
\end{eqnarray}

$\blacksquare$
\end{remark}

In summary, in terms of $\big\{\vec{\mathcal{E}}, \vec{\mathcal{B}}, \vec{\mathcal{A}}, \varphi\big\}$, the Yang-Mills equations
(\ref{eq:YM1a}) and (\ref{eq:YM1b}) can be recast to the following form
      \begin{subequations}
            \begin{eqnarray}
                  && \vec{\nabla}\cdot\vec{\mathcal{E}}\ {-} \; \boxed{{\rm i}\,g\Bigl(
                        \vec{\mathcal{A}}\cdot\vec{\mathcal{E}}
                        -\vec{\mathcal{E}}\cdot\vec{\mathcal{A}}\Bigr)}=0, \label{eq:DivEYM} \\
                  && -\dfrac{1}{c} \dfrac{\partial \vec{\mathcal{B}}}{\partial\,t}
                        -\vec{\nabla}\times\vec{\mathcal{E}} {-}\; \boxed{{\rm i}\,g\biggl(\Bigl[\varphi,\
                                    \vec{\mathcal{B}}\Bigr]
                              -\vec{\mathcal{A}}\times\vec{\mathcal{E}}
                              -\vec{\mathcal{E}}\times\vec{\mathcal{A}}\biggr)}=0, \label{eq:CurlEYM} \\
                  && \vec{\nabla}\cdot\vec{\mathcal{B}}\ {-} \; \boxed{{\rm i}\,g\Bigl(
                        \vec{\mathcal{A}}\cdot\vec{\mathcal{B}}
                        -\vec{\mathcal{B}}\cdot\vec{\mathcal{A}}\Bigr)}=0, \label{eq:DivBYM} \\
                  && -\dfrac{1}{c} \dfrac{\partial \vec{\mathcal{E}}}{\partial\,t}
                        +\vec{\nabla}\times\vec{\mathcal{B}}  {-} \; \boxed{{\rm i}\,g\biggl(\Bigl[\varphi,\
                                    \vec{\mathcal{E}}\Bigr]
                              +\vec{\mathcal{A}}\times\vec{\mathcal{B}}
                              +\vec{\mathcal{B}}\times\vec{\mathcal{A}}\biggr)}=0, \label{eq:CurlBYM}
            \end{eqnarray}
      \end{subequations}
      where the ``boxed'' expressions imply the self-interaction terms regarding to potentials $\big\{\vec{\mathcal{A}}, \varphi\big\}$ and fields $\big\{\vec{\mathcal{E}}, \vec{\mathcal{B}}\big\}$ in the Yang-Mills equations. The ``boxed'' terms come from the nonlinear terms in Eqs. (\ref{eq:YM1a}) and (\ref{eq:YM1b}), such as ${\rm i}\,g [\mathbb{A}_\mu ,\ {F}^{\mu\nu}]$. Note that Yang-Mills equations are both Lorentz covariant and local gauge covariant. Generally, due to the self-interaction terms, it is hard to obtain the exact solutions for the Yang-Mills equations.

      If one neglects the ``boxed'' terms (or just let $g=0$), then the above expressions are just the usual forms of Maxwell's equations (in vacuum without sources). Namely,
\begin{subequations}
            \begin{eqnarray}
                  && \vec{\nabla}\cdot\vec{\mathcal{E}}=0, \label{eq:DivEYM-M} \\
                  && -\dfrac{1}{c} \dfrac{\partial \vec{\mathcal{B}}}{\partial\,t}
                        -\vec{\nabla}\times\vec{\mathcal{E}}=0, \label{eq:CurlEYM-M} \\
                  && \vec{\nabla}\cdot\vec{\mathcal{B}}=0, \label{eq:DivBYM-M} \\
                  && -\dfrac{1}{c} \dfrac{\partial \vec{\mathcal{E}}}{\partial\,t}
                        +\vec{\nabla}\times\vec{\mathcal{B}}=0. \label{eq:CurlBYM-M}
            \end{eqnarray}
\end{subequations}

\newpage

\section{Coulomb-Type Potential: A Simple but Fundamental Solution of Maxwell's Equations}

\subsection{Coulomb-Type Potential as an Exact Solution of Maxwell's Equations}
The Coulomb-type potential is given by
\begin{eqnarray}\label{eq:cl}
    &&       \varphi_{\rm cl} = \dfrac{\kappa}{r},
\end{eqnarray}
where $\kappa$ is a real constant number. A significant feature of Coulomb-type potential is that the function $\varphi_{\rm cl}(r)$ is proportional to $1/r$. Consider two charged particles, particle 1 has a charge $q$ and particle 2 has a charge $Q$. When $\kappa$ is taken as
\begin{eqnarray}
    &&    \kappa=Q,
\end{eqnarray}
then
\begin{eqnarray}
    && V(r)=q\varphi_{\rm cl}=\dfrac{qQ}{r}
\end{eqnarray}
is just the standard Coulomb potential in electromagnetics, and $V(r)$ represents the standard Coulomb interaction between two charged particles. Surprisingly, in our nature, both the standard Coulomb potential in electromagnetics and the gravitational potential in Newtonian gravity theory  share the same form of Coulomb-type potential shown in Eq. (\ref{eq:cl}).

It gives rise to a natural question: Can one derive the Coulomb-type potential based on some certain fundamental theories? The answer is positive.
Actually the Coulomb-type potential is a simple but fundamental solution of Maxwell's equations. The result can be expressed in the following theorem.
\begin{theorem}\label{Ther1}
Let the vector potential and the scalar potential be
\begin{eqnarray}\label{eq:Coul-1a}
    &&  \vec{\mathcal{A}}=0, \;\;\;\;\;      \varphi = \dfrac{\kappa}{r},
\end{eqnarray}
then the set of potentials $\{\vec{\mathcal{A}}, \varphi\}$ is a {local} exact solution of Maxwell's equations {on the punctured region $r\neq0$}.
Here $\kappa$ is a real constant number.
\end{theorem}

\begin{proof}
By substituting Eq. (\ref{eq:Coul-1a}) into Eq. (\ref{eq:mag-2}) and Eq. (\ref{eq:ele-2}), we have the magnetic filed and the electric field as
\begin{subequations}
\begin{eqnarray}\label{eq:Coul-2}
      \vec{\mathcal{B}} &=& \vec{\nabla}\times\vec{\mathcal{A}}
                        -{\rm i}\,g\bigl(\vec{\mathcal{A}}\times\vec{\mathcal{A}}\bigr)=0, \\
      \vec{\mathcal{E}} &=&  -\dfrac{1}{c} \dfrac{\partial\,\vec{\mathcal{A}}}{\partial\,t}
                        -\vec{\nabla}\varphi-{\rm i}\,g\left[\varphi,\ \vec{\mathcal{A}}
                        \right]=-\vec{\nabla}\varphi \nonumber\\
                        &=& - \dfrac{\partial }{\partial x}\left(\dfrac{\kappa}{r}\right) \hat{e}_x
                        -\dfrac{\partial }{\partial y}\left(\dfrac{\kappa}{r}\right) \hat{e}_y
                        -\dfrac{\partial }{\partial z}\left(\dfrac{\kappa}{r}\right) \hat{e}_z\nonumber\\
                        &=& - \dfrac{\partial }{\partial r}\left(\dfrac{\kappa}{r}\right) \dfrac{x}{r}\hat{e}_x
                        -\dfrac{\partial }{\partial r}\left(\dfrac{\kappa}{r}\right) \dfrac{y}{r} \hat{e}_y
                        -\dfrac{\partial }{\partial r}\left(\dfrac{\kappa}{r}\right) \dfrac{z}{r} \hat{e}_z=\dfrac{\kappa}{r^2} \hat{e}_r,
\end{eqnarray}
\end{subequations}
with
\begin{eqnarray}
\hat{e}_r=\dfrac{\vec{r}}{r}, \;\;\; r=|\vec{r}|.
\end{eqnarray}
Because $\{\vec{\mathcal{A}}, \varphi\}$ is time-independent, thus $\vec{\mathcal{B}}$  and $\vec{\mathcal{E}}$ are static fields. Let us come to check Maxwell's equations as shown in Eq. (\ref{eq:DivEYM-M})-Eq. (\ref{eq:CurlBYM-M}). Because $\vec{\mathcal{B}}=0$ and $\vec{\mathcal{E}}$ is time-independent, then Eq. (\ref{eq:DivBYM-M}) and Eq. (\ref{eq:CurlBYM-M}) are automatically satisfied. Accordingly, Eq. (\ref{eq:DivEYM-M}) and Eq. (\ref{eq:CurlEYM-M}) become
\begin{subequations}
            \begin{eqnarray}
                  && \vec{\nabla}\cdot\vec{\mathcal{E}}=0, \label{eq:DivEYM-M-2} \\
                  && \vec{\nabla}\times\vec{\mathcal{E}}=0. \label{eq:CurlEYM-M-2}
                \end{eqnarray}
      \end{subequations}
We can have
\begin{eqnarray}
 \vec{\nabla}\cdot\vec{\mathcal{E}}&=& \vec{\nabla}\cdot \left(\dfrac{\kappa}{r^2} \hat{e}_r\right)
 =\kappa\, \vec{\nabla}\cdot \left(\dfrac{x}{r^3} \hat{e}_x+ \dfrac{y}{r^3} \hat{e}_y+\dfrac{z}{r^3} \hat{e}_z\right)
 = \kappa\,\left[\dfrac{\partial }{\partial x}\left(\dfrac{x}{r^3} \right)+ \dfrac{\partial }{\partial y}\left(\dfrac{y}{r^3} \right)
 +\dfrac{\partial }{\partial z}\left(\dfrac{z}{r^3} \right) \right]\nonumber\\
 &=& \kappa \left[\left(\dfrac{1}{r^3}-3\dfrac{x^2}{r^5} \right)+ \left(\dfrac{1}{r^3}-3\dfrac{y^2}{r^5} \right)+\left(\dfrac{1}{r^3}-3\dfrac{z^2}{r^5} \right)\right]=\kappa \left[\dfrac{3}{r^3}-3\dfrac{x^2+y^2+z^2}{r^5} \right]\nonumber\\
 &=& \kappa \left[\dfrac{3}{r^3}-3\dfrac{1}{r^3} \right]=0,
\end{eqnarray}
thus Eq. (\ref{eq:DivEYM-M-2}) is satisfied. Similarly, we have
\begin{eqnarray}
 \vec{\nabla}\times\vec{\mathcal{E}}&=& \vec{\nabla}\times \left(\dfrac{\kappa}{r^2} \hat{e}_r\right)
 =\kappa\, \vec{\nabla}\times \left(\dfrac{x}{r^3} \hat{e}_x+ \dfrac{y}{r^3} \hat{e}_y+\dfrac{z}{r^3} \hat{e}_z\right)\nonumber\\
 &=& \kappa\,\left[\left(\dfrac{\partial \left(\dfrac{z}{r^3}\right)}{\partial y}-\dfrac{\partial \left(\dfrac{y}{r^3}\right)}{\partial z}\right)\hat{e}_x+ \left(\dfrac{\partial \left(\dfrac{x}{r^3}\right)}{\partial z}-\dfrac{\partial \left(\dfrac{z}{r^3}\right)}{\partial x}\right)\hat{e}_y+ \left(\dfrac{\partial \left(\dfrac{y}{r^3}\right)}{\partial x}-\dfrac{\partial \left(\dfrac{x}{r^3}\right)}{\partial y}\right)\hat{e}_z\right]\nonumber\\
 &=& \kappa\, \left[\left(-3 \dfrac{yz}{r^5}+3 \dfrac{zy}{r^5}\right)\hat{e}_x+\left(-3 \dfrac{zx}{r^5}+3 \dfrac{xz}{r^5}\right)\hat{e}_y+\left(-3 \dfrac{xy}{r^5}+3 \dfrac{yx}{r^5}\right)\hat{e}_z\right]=0,
\end{eqnarray}
thus Eq. (\ref{eq:CurlEYM-M-2}) is satisfied. This ends the proof.
\end{proof}

Theorem \ref{Ther1} tells us that the Coulomb-type potential $\varphi_{\rm cl} =\kappa/r$ is a simple but fundamental solution of Maxwell's equations (Note: To be precise, the exact solution is the set $ \mathcal{S}_{\rm Maxwell}=\{\vec{\mathcal{A}}=0, \varphi=\kappa/r\}$, although in this case the vector potential is equal to zero). Or inversely, in a certain sense, one may say that Maxwell's equations predict a fundamental interaction in Nature --- the Coulomb interaction. Undoubtly, the Coulomb-type potential represents a kind of fundamental interactions in Nature. In the following, let us study this fundamental interaction from the angle of view of classical picture as well as quantum picture.

\subsection{Classical Picture: Binding Two Particles with Coulomb Potential}

Suppose there are two particles, i.e., particle 1 and particle 2. The first particle (e.g, an electron) has a charge $q$ (e.g., $q=-e=-|e|$, and $e=|e|$ is the unit of electric charge) and a mass $M$, and the second particle (e.g., atomic nucleus) has a charge $Q$ (e.g., $Q=Z e$, $Z$ is the number of protons, and $Z=1$ for a single proton). The particle 2 induces a scalar potential $\varphi=\dfrac{\kappa}{r}=\dfrac{Q}{r}$ and hence a static electric field $\vec{\mathcal{E}}=\dfrac{\kappa}{r^2} \hat{e}_r$. Due to the electric filed, particle 2 exerts a force on particle 1, and the expression of the force is given by
\begin{eqnarray}
\vec{\mathbb{F}} &=&q \vec{\mathcal{E}} = \dfrac{q \kappa}{r^2} \hat{e}_r=\dfrac{q Q}{r^2} \hat{e}_r.
\end{eqnarray}
Here $\vec{\mathbb{F}}$ is the standard Coulomb force in electromagnetics. It is a kind of central forces. If $q \kappa < 0$, then $\vec{\mathbb{F}}$ is an attractive force (or the centripetal force); and if $q \kappa > 0$, then $\vec{\mathbb{F}}$ is a repulsive force.

Now, under the action of Coulomb force $\vec{\mathbb{F}}$, we come to determine the equation of motion of $\vec{r}$ for particle 1. From Newton's second law, we have
\begin{eqnarray}
&& \vec{\mathbb{F}} = M \vec{a}=M \dfrac{{\rm d}^2 \vec{r}}{{\rm d} t^2}.
\end{eqnarray}
For simplicity, we consider the polar coordinate system (i.e., a two-dimensional plan). The particle 2 is located on the origin of the polar coordinate system, and the position of particle 1 is denoted by $\vec{r}=(r \cos\theta, r\sin\theta)$.
With the help of
\begin{eqnarray}
&& {\rm d} \hat{e}_r = \theta \, \hat{e}_\theta, \;\;\;\;\; {\rm d} \hat{e}_\theta= -\theta \, \hat{e}_r, \nonumber\\
&&\dfrac{{\rm d} \hat{e}_r}{{\rm d} t}= \dfrac{{\rm d} \left(\theta \, \hat{e}_\theta\right)}{{\rm d} t}=\dfrac{{\rm d} \theta}{{\rm d} t}\, \hat{e}_\theta= \dot{\theta}\, \hat{e}_\theta, \;\;\;\;\;
\dfrac{{\rm d} \hat{e}_\theta}{{\rm d} t}=\dfrac{{\rm d} \left(-\theta \, \hat{e}_r\right)}{{\rm d} t}=-\dfrac{{\rm d} \theta}{{\rm d} t}\, \hat{e}_r= -\dot{\theta}\, \hat{e}_r,
\end{eqnarray}
we have
\begin{eqnarray}
  \dfrac{{\rm d} \vec{r}}{{\rm d} t} &=& \dfrac{{\rm d} \left(r \hat{e}_r\right)}{{\rm d} t} = \dfrac{{\rm d} r}{{\rm d} t}  \hat{e}_r +r \dfrac{{\rm d} \hat{e}_r}{{\rm d} t}=
  \dot{r}  \hat{e}_r +r \dot{\theta}\, \hat{e}_\theta,
\end{eqnarray}
\begin{eqnarray}
  \dfrac{{\rm d}^2 \vec{r}}{{\rm d} t^2} &=& \dfrac{{\rm d} \left(\dot{r}  \hat{e}_r +r \dot{\theta}\, \hat{e}_\theta\right)}{{\rm d} t} =\ddot{r} \hat{e}_r +\dot{r} \dfrac{{\rm d} \hat{e}_r}{{\rm d} t}+\dot{r}\dot{\theta} \hat{e}_\theta+ r \ddot{\theta}\hat{e}_\theta +r \dot{\theta}\dfrac{{\rm d} \hat{e}_\theta}{{\rm d} t}\nonumber\\
  &=& \ddot{r} \hat{e}_r +\dot{r} \dot{\theta}\, \hat{e}_\theta + \dot{r}\dot{\theta} \hat{e}_\theta+ r \ddot{\theta}\hat{e}_\theta +r \dot{\theta}\left( -\dot{\theta}\, \hat{e}_r\right)\nonumber\\
  &=& \left(\ddot{r} - r \dot{\theta}^2 \right)\hat{e}_r+\left(r \ddot{\theta}+2 \dot{r}\dot{\theta}\right)\hat{e}_\theta.
\end{eqnarray}
From
\begin{eqnarray}
&& \vec{\mathbb{F}} =M \dfrac{{\rm d}^2 \vec{r}}{{\rm d} t^2}=  \dfrac{q  Q}{r^2} \, \hat{e}_r,
\end{eqnarray}
we have
\begin{eqnarray}\label{eq:rt-1}
 M \ddot{r} - M r \dot{\theta}^2 = |\vec{\mathbb{F}}|=\dfrac{q  Q}{r^2},
\end{eqnarray}
and
\begin{eqnarray}\label{eq:rt-2}
 M \left( r \ddot{\theta}+2 \dot{r}\dot{\theta}\right)=0.
\end{eqnarray}
From Eq. (\ref{eq:rt-2}), one has
\begin{eqnarray}
 r\, M \left( r \ddot{\theta}+2 \dot{r}\dot{\theta}\right)= M \left( r^2 \ddot{\theta}+2 r\, \dot{r}\dot{\theta}\right)=\dfrac{{\rm d} \left(M r^2 \dot{\theta}\right)}{{\rm d} t}=0,
\end{eqnarray}
which means the angular momentum is a conserved quantity, i.e.,
\begin{eqnarray}
&&\ell=M r^2 \dot{\theta}={\rm constant}.
\end{eqnarray}
Based on which, we have
\begin{eqnarray}
&& \dot{\theta} = \dfrac{{\rm d} \theta}{{\rm d} t}= \dfrac{\ell}{M r^2},
\end{eqnarray}
thus
\begin{eqnarray}
&& \dfrac{{\rm d} }{{\rm d} t}= \dfrac{{\rm d} \theta}{{\rm d} t}\dfrac{{\rm d}}{{\rm d} \theta}=\dfrac{\ell}{ M r^2} \dfrac{{\rm d}}{{\rm d} \theta}.
\end{eqnarray}
Then Eq. (\ref{eq:rt-1}) becomes
\begin{eqnarray}
 M \dfrac{{\rm d}}{{\rm d} t}\left[\dfrac{{\rm d} r}{{\rm d} t}\right] -M  r \dot{\theta}^2 = \dfrac{q Q}{r^2},
\end{eqnarray}
i.e.,
\begin{eqnarray}
  M\dfrac{\ell}{ M r^2} \dfrac{{\rm d}}{{\rm d} \theta}\left[\dfrac{\ell}{ M r^2} \dfrac{{\rm d} r}{{\rm d} \theta}\right] -M r \dot{\theta}^2 = \dfrac{q  Q}{r^2},
\end{eqnarray}
i.e.,
\begin{eqnarray}\label{eq:rt-1a}
  \dfrac{\ell^2}{M r^2} \dfrac{{\rm d}}{{\rm d} \theta}\left[\dfrac{1}{ r^2} \dfrac{{\rm d} r}{{\rm d} \theta}\right] -M r \dot{\theta}^2 = \dfrac{q  Q}{r^2}.
\end{eqnarray}
We now let
\begin{eqnarray}
&&u=\dfrac{1}{r},
\end{eqnarray}
then we have
\begin{eqnarray}
&& \dfrac{{\rm d} r}{{\rm d} \theta}=\dfrac{{\rm d} r}{{\rm d} u}\dfrac{{\rm d} u}{{\rm d} \theta}=-\dfrac{1}{u^2}\dfrac{{\rm d} u}{{\rm d} \theta}.
\end{eqnarray}
Then Eq. (\ref{eq:rt-1a}) becomes
\begin{eqnarray}
  \dfrac{\ell^2}{M r^2} \dfrac{{\rm d}}{{\rm d} \theta}\left[\dfrac{1}{ r^2} \left(-\dfrac{1}{u^2}\dfrac{{\rm d} u}{{\rm d} \theta}\right)\right] -M r \dot{\theta}^2 =  \dfrac{q  Q}{r^2},
\end{eqnarray}
i.e.,
\begin{eqnarray}
  -\dfrac{\ell^2}{M r^2} \dfrac{{\rm d}}{{\rm d} \theta}\left[\dfrac{1}{ r^2} \left(\dfrac{1}{u^2}\dfrac{{\rm d} u}{{\rm d} \theta}\right)\right] -\dfrac{1}{M r^3} \left(M r^2 \dot{\theta}\right)^2 = \dfrac{q Q}{r^2},
\end{eqnarray}
i.e.,
\begin{eqnarray}
  -\dfrac{\ell^2 u^2}{M} \dfrac{{\rm d}}{{\rm d} \theta}\left[u^2 \left(\dfrac{1}{u^2}\dfrac{{\rm d} u}{{\rm d} \theta}\right)\right] -\dfrac{\ell^2 u^3 }{M}  =  q  Q u^2,
\end{eqnarray}
i.e.,
\begin{eqnarray}
  -\dfrac{\ell^2 u^2}{M } \dfrac{{\rm d}^2 u}{{\rm d} \theta^2} -\dfrac{\ell^2 u^3 }{M }  =  q  Q u^2= q \kappa u^2,
\end{eqnarray}
i.e.,
\begin{eqnarray}
  -\dfrac{\ell^2 }{M } \dfrac{{\rm d}^2 u}{{\rm d} \theta^2} -\dfrac{\ell^2 }{M } u  = q \kappa,
\end{eqnarray}
i.e.,
\begin{eqnarray}\label{eq:rt-1b}
  \dfrac{{\rm d}^2 u}{{\rm d} \theta^2} + u  = -\dfrac{M q \kappa}{\ell^2}.
\end{eqnarray}
The solution of Eq. (\ref{eq:rt-1b}) is
\begin{eqnarray}\label{eq:rt-1c}
  u= -\dfrac{M q \kappa}{\ell^2} + C_1 \cos(\theta-\theta_0)=\dfrac{M {|q \kappa|}}{\ell^2} \left[1+{\rm e} \cos(\theta-\theta_0)\right],
\end{eqnarray}
here we have considered $\vec{\mathbb{F}}$ as the attractive force, thus $q \kappa=-|q \kappa| <0$. Then the equation of motion of $\vec{r}$ is given by
\begin{eqnarray}\label{eq:rt-1d}
  r=  \dfrac{\ell^2}{M {|q \kappa|}} \dfrac{1}{\left[1+{\rm e} \cos(\theta-\theta_0)\right]},
\end{eqnarray}
which is an equation of the conic section, with ${\rm e}$ being the eccentricity ratio. When ${\rm e}>1$, the trajectory of particle 1 is a hyperbola; when ${\rm e}=1$, the trajectory of particle 1 is a parabola; when $0\leq {\rm e}<1$, the trajectory of particle 1 is an ellipse (specially ${\rm e}=0$ corresponds to a circle).

\begin{remark}\textcolor{blue}{The Relation between Eccentricity Ratio ${\rm e}$ and Energy}. Classically, the energy of the system can be written as
\begin{eqnarray}
      E= \frac{p^2}{2 M} +q \varphi = \frac{1}{2}M v^2 + \dfrac{q \kappa}{r}.
\end{eqnarray}
Because the square of velocity is
\begin{equation}
      v^2 = \dot{r}^2 + r^2 \dot{\theta}^2,
\end{equation}
and the angular momentum is
\begin{equation}\label{eq:l-1}
      \ell= M r^2 \dot{\theta},
\end{equation}
then the energy becomes
\begin{eqnarray}\label{eq:E-2a}
      E= \frac{1}{2}M v^2 +\dfrac{q \kappa}{r} = \frac{1}{2}M \left(\dot{r}^2 + \frac{\ell^2}{M^2 r^2}\right)+ \dfrac{q \kappa}{r},
\end{eqnarray}
i.e.,
\begin{eqnarray}
      \dot{r}^2=\dfrac{2}{M}\left(E - \dfrac{ q \kappa}{r}\right)-\frac{\ell^2}{M^2 r^2}=\dfrac{2}{M}\left(E - \dfrac{ q \kappa}{r}-\frac{\ell^2}{2 M r^2}\right),
\end{eqnarray}
i.e.,
\begin{eqnarray}
      \dot{r}=\sqrt{\dfrac{2}{M}\left(E - \dfrac{ q \kappa}{r}-\frac{\ell^2}{2 M r^2}\right)},
\end{eqnarray}
i.e.,
\begin{eqnarray}
      {\rm d} r=\sqrt{\dfrac{2}{M}\left(E - \dfrac{ q \kappa}{r}-\frac{\ell^2}{2 M r^2}\right)}\, {\rm d} t.
\end{eqnarray}

The energy $E$ and the angular momentum $\ell$ are conserved. From Eq. (\ref{eq:l-1}) we have
\begin{equation}
      {\rm d} \theta =\frac{\ell}{M r^2} {\rm d}t.
\end{equation}
Then
we attain
\begin{equation}
      \frac{{\rm d} r }{\sqrt{\dfrac{2}{M}\left(E - \dfrac{ q \kappa}{r}-\dfrac{\ell^2}{2 M r^2}\right)}} = \dfrac{M r^2}{\ell} {\rm d}\theta,
\end{equation}
i.e.,
\begin{equation}
     \frac{\ell}{M r^2}  \frac{{\rm d} r }{\sqrt{\dfrac{2}{M}\left(E - \dfrac{ q \kappa}{r}-\dfrac{\ell^2}{2 M r^2}\right)}} = {\rm d}\theta,
\end{equation}
We designate $r= \dfrac{1}{u}$, then
\begin{eqnarray}
&& \dfrac{{\rm d} r}{{\rm d} \theta}=\dfrac{{\rm d} r}{{\rm d} u}\dfrac{{\rm d} u}{{\rm d} \theta}=-\dfrac{1}{u^2}\dfrac{{\rm d} u}{{\rm d} \theta},
\end{eqnarray}
thus
\begin{equation}
     \frac{\ell}{M r^2} \left[-\dfrac{1}{u^2}\right] \frac{{\rm d} u }{\sqrt{\dfrac{2}{M}\left(E - \dfrac{ q \kappa}{r}-\dfrac{\ell^2}{2 M r^2}\right)}} = {\rm d}\theta,
\end{equation}
i.e.,
\begin{equation}
     \frac{\ell u^2}{M} \left[-\dfrac{1}{u^2}\right] \frac{{\rm d}u }{\sqrt{\dfrac{2}{M}\left(E - { q \kappa}{u}-\dfrac{\ell^2 u^2}{2 M}\right)}} = {\rm d}\theta,
\end{equation}
i.e.,
\begin{equation}
      -\frac{{\rm d}u }{\sqrt{\dfrac{2 M}{\ell^2} \left(E - { q \kappa}{u}-\dfrac{\ell^2 u^2}{2 M}\right)}} = {\rm d}\theta,
\end{equation}
i.e.,
\begin{equation}
      -\frac{{\rm d} u }{\sqrt{\dfrac{2 M}{\ell^2}E- \dfrac{2 M q \kappa}{\ell^2} u-u^2 }} = {\rm d}\theta,
\end{equation}
i.e.,
\begin{equation}
      -\frac{{\rm d}u }{\sqrt{\dfrac{2 M}{\ell^2}E + \left(\dfrac{ M q \kappa}{\ell^2}\right)^2- \left(u{+}\dfrac{ M q \kappa}{\ell^2}\right)^2 }} = {\rm d}\theta,
\end{equation}
i.e.,
\begin{equation}
      -\dfrac{1}{\sqrt{\dfrac{2 M}{\ell^2}E + \left(\dfrac{ M q \kappa}{\ell^2}\right)^2}}\frac{{\rm d}\left(u {+} \dfrac{ M q \kappa}{\ell^2}\right)}{\sqrt{1- \left(\dfrac{u {+}\dfrac{ M q \kappa}{\ell^2}}{\sqrt{\dfrac{2 M}{\ell^2}E + \left(\dfrac{ M q \kappa}{\ell^2}\right)^2}}\right)^2 }} = {\rm d}\theta.
\end{equation}
Because
\begin{equation}
\dfrac{{\rm d} \arccos{x}}{{\rm d} x}=-\dfrac{1}{\sqrt{1-x^2}},
\end{equation}
then we have
\begin{equation}
\arccos{\left(\frac{u{+}\dfrac{ M q \kappa}{\ell^2}}{\sqrt{\dfrac{2 M}{\ell^2}E + \left(\dfrac{ M q \kappa}{\ell^2}\right)^2}}\right)}=\theta-\theta_0, \end{equation}
i.e.,
\begin{equation}
u{+}\dfrac{ M q \kappa}{\ell^2}={\sqrt{\dfrac{2 M}{\ell^2}E + \left(\dfrac{ M q \kappa}{\ell^2}\right)^2}}\, \cos\left(\theta-\theta_0\right), \end{equation}
i.e.,
\begin{equation}
u={-}\dfrac{ M q \kappa}{\ell^2}+{\sqrt{\dfrac{2 M}{\ell^2}E + \left(\dfrac{ M q \kappa}{\ell^2}\right)^2}}\, \cos\left(\theta-\theta_0\right),
\end{equation}
i.e.,
\begin{equation}
r=\frac{1}{\dfrac{ M {|q \kappa|}}{\ell^2}+{\sqrt{\dfrac{2 M}{\ell^2}E + \left(\dfrac{ M q \kappa}{\ell^2}\right)^2}}\, \cos\left(\theta-\theta_0\right)},
\end{equation}
i.e.,
\begin{equation}\label{eq:rt-1e}
r=\dfrac{\ell^2}{ M {|q \kappa|}} \frac{1}{1+\dfrac{\sqrt{\dfrac{2 M}{\ell^2}E + \left(\dfrac{ M q \kappa}{\ell^2}\right)^2}}{\dfrac{ M {|q \kappa|}}{\ell^2}}\, \cos\left(\theta-\theta_0\right)}.
\end{equation}
By comparing Eq. (\ref{eq:rt-1d}) and Eq. (\ref{eq:rt-1e}), we have the eccentricity ratio as
\begin{eqnarray}\label{eq:rt-1f}
{\rm e} &=& \dfrac{\sqrt{\dfrac{2 M}{\ell^2}E + \left(\dfrac{ M q \kappa}{\ell^2}\right)^2}}{\dfrac{ M {|q \kappa|}}{\ell^2}}=\sqrt{1+\dfrac{\dfrac{2 M}{\ell^2}E}{\left( \dfrac{M q \kappa}{\ell^2}\right)^2}}=\sqrt{1+\dfrac{2\ell^2}{M (q\kappa)^2}E},
\end{eqnarray}
which implies that the value of ${\rm e}$ is determined by the energy $E$.
$\blacksquare$
\end{remark}

\begin{remark}
Here, we would like to study again the relation between the eccentricity ratio ${\rm e}$ and the energy $E$ based on the following formula
\begin{eqnarray}\label{eq:rt-1d-a}
  r=  \dfrac{\ell^2}{M |q \kappa|} \dfrac{1}{\left[1+{\rm e} \cos(\theta-\theta_0)\right]}.
\end{eqnarray}
Still, we restrict to the case of $q \kappa < 0$ (i.e., $\vec{\mathbb{F}}$ is an attractive force).

From Eq. (\ref{eq:E-2a}) we have
\begin{eqnarray}\label{eq:E-2b}
      E &=& \frac{1}{2}M \left(\dot{r}^2 + \frac{\ell^2}{M^2 r^2}\right)+ \dfrac{q \kappa}{r}.
\end{eqnarray}
Based on Eq. (\ref{eq:rt-1d-a}) one has
\begin{eqnarray}\label{eq:rt-1d-b}
  \dot{r}&=&\dfrac{{\rm d} r}{{\rm d} t}=  \dfrac{{\rm d}}{{\rm d} t}\left[\dfrac{\ell^2}{M |q \kappa|} \dfrac{1}{\left[1+{\rm e} \cos(\theta-\theta_0)\right]}\right]=\dfrac{\ell^2}{M |q \kappa|} \dfrac{{\rm e} \sin(\theta-\theta_0)}{\left[1+{\rm e} \cos(\theta-\theta_0)\right]^2} \dfrac{{\rm d} \theta}{{\rm d} t} \nonumber\\
  &=& \dfrac{\ell^2 }{M |q \kappa|} \dfrac{{\rm e} \sin(\theta-\theta_0)}{\left[1+{\rm e} \cos(\theta-\theta_0)\right]^2} \dot{\theta}
  =\dfrac{\ell^2 }{M |q \kappa|} \dfrac{{\rm e} \sin(\theta-\theta_0)}{\left[1+{\rm e} \cos(\theta-\theta_0)\right]^2} \dfrac{\ell}{m r^2} \nonumber\\
  &=&\dfrac{\ell^3 }{M^2 |q \kappa|} \dfrac{{\rm e} \sin(\theta-\theta_0)}{r^2\left[1+{\rm e} \cos(\theta-\theta_0)\right]^2}  = \dfrac{\ell^3 }{M^2 |q \kappa|} {\rm e} \sin(\theta-\theta_0) \left(\dfrac{M |q \kappa|}{\ell^2}\right)^2 \nonumber\\
  &=& \dfrac{|q \kappa|}{\ell} {\rm e} \sin(\theta-\theta_0).
\end{eqnarray}
Then the energy becomes
\begin{eqnarray}\label{eq:E-2b-a}
      E &=& \frac{1}{2}M \left(\dot{r}^2 + \frac{\ell^2}{M^2 r^2}\right)+ \dfrac{q \kappa}{r}\nonumber\\
      &=& \frac{1}{2}M \left[\dfrac{|q \kappa|}{\ell} {\rm e} \sin(\theta-\theta_0)\right]^2+\dfrac{\ell^2}{2M} \left[\dfrac{M |q \kappa|}{\ell^2} \left[1+{\rm e} \cos(\theta-\theta_0)\right]\right]^2+ q \kappa \left[\dfrac{M |q \kappa|}{\ell^2} \left[1+{\rm e} \cos(\theta-\theta_0)\right]\right]\nonumber\\
      &=& \frac{M |q \kappa|^2}{2 \ell^2} {\rm e}^2 \sin^2(\theta-\theta_0)+ \frac{M |q \kappa|^2}{2 \ell^2}\left[1+{\rm e} \cos(\theta-\theta_0)\right]^2+ \frac{M |q \kappa| q \kappa }{\ell^2}\left[1+{\rm e} \cos(\theta-\theta_0)\right]\nonumber\\
      &=&  \frac{M |q \kappa|^2}{2 \ell^2} {\rm e}^2 \sin^2(\theta-\theta_0)+ \frac{M |q \kappa|^2}{2 \ell^2}\left[1+{\rm e} \cos(\theta-\theta_0)\right]^2 - \frac{M |q \kappa|^2}{\ell^2}\left[1+{\rm e} \cos(\theta-\theta_0)\right]\nonumber\\
      &=&  \frac{M |q \kappa|^2}{2 \ell^2} \left\{{\rm e}^2 \sin^2(\theta-\theta_0)+ \left[1+{\rm e} \cos(\theta-\theta_0)\right]^2 - 2\left[1+{\rm e} \cos(\theta-\theta_0)\right]\right\}\nonumber\\
       &=&  \frac{M |q \kappa|^2}{2 \ell^2} \left\{{\rm e}^2 \sin^2(\theta-\theta_0)+ \left[1+{\rm e} \cos(\theta-\theta_0)-1\right]^2 -1 \right\}\nonumber\\
       &=&  \frac{M |q \kappa|^2}{2 \ell^2} \left\{{\rm e}^2 \sin^2(\theta-\theta_0)+{\rm e}^2 \cos^2(\theta-\theta_0)  -1 \right\}\nonumber\\
       &=&  \frac{M |q \kappa|^2}{2 \ell^2} ({\rm e}^2  -1),
\end{eqnarray}
thus
\begin{eqnarray}\label{eq:rt-1f-a}
{\rm e} &=& \sqrt{1+\dfrac{2\ell^2}{M |q\kappa|^2}E},
\end{eqnarray}
which recovers Eq. (\ref{eq:rt-1f}). $\blacksquare$
\end{remark}

\begin{remark}
Classically, for the attractive force (i.e., $q\kappa<0$), the kinetic energy $\frac{1}{2}m v^2 \geq 0$
and the potential energy $\dfrac{q \kappa}{r}<0$, thus the total energy $E$ can be less than, or equal to,
or greater than 0. However, for the case of repulsive force $\vec{\mathbb{F}}$, one has $q \kappa > 0$.
Because both kinetic energy and potential energy are greater than 0, namely,
\begin{eqnarray}
      \frac{1}{2}m v^2 \geq 0, \;\;\;\;\; \dfrac{q \kappa}{r} >0,
\end{eqnarray}
it is easy to prove the energy of the physical system
\begin{eqnarray}
      E= \frac{1}{2}m v^2 +\dfrac{q \kappa}{r} >0,
\end{eqnarray}
thus the eccentricity ratio ${\rm e}>1$. This means that the trajectory of particle 1 is a hyperbola
if $\vec{\mathbb{F}}$ is repulsive, thus particle 1 can escape the constraint of particle 2, and particles 1
and 2 cannot form a stable physical system. $\blacksquare$
\end{remark}

In conclusion, when $\vec{\mathbb{F}}$ is an attractive force and the energy $E<0$, one has the eccentricity
ratio $0\leq {\rm e}<1$. Then particle 1 moves around particle 2, and its trajectory is an ellipse (or a circle
when ${\rm e}=0$). Thus, the attractive Coulomb-type force $\vec{\mathbb{F}}$ serves as a fundamental force to
bind two charged particles. In this way, it is a good reason to explain why an electron and a proton can form a
stable hydrogen atom. This is the classical picture of binding two particles with Coulomb potential.

\subsection{Quantum Picture: Binding Two Particles with Coulomb Potential}

In quantum mechanics, the position operator $\vec{r}$ and the linear momentum operator $\vec{p}$ are not commutative, they satisfy
the following basic commutation relation
\begin{eqnarray}\label{linearp3}
&& [r_\alpha, p_\beta]= \mathrm{i}\hbar \delta_{\alpha\beta}, \;\;\; (\alpha, \beta=x, y, z),
\end{eqnarray}
with $\delta_{\alpha\beta}$ being the Kronecker delta function, and $\hbar$ being Planck's constant. Explicitly, one has
\begin{eqnarray}\label{linearp3-a}
&& [x, p_x]= \mathrm{i}\hbar, \;\; [y, p_y]= \mathrm{i}\hbar, \;\; [z, p_z]= \mathrm{i}\hbar, \nonumber\\
&& [x, p_y]= [x, p_z]= 0, \;\;\;  [y, p_x]= [y, p_z]= 0, \;\;\;  [z, p_x]= [z, p_y]= 0.
\end{eqnarray}
Physically, it means that the position and the momentum of a particle cannot be measured simultaneously, thus the
concept of ``force'' becomes a bit out of place. To avoid such an embarrassment, scientists have used the term ``interaction'' to
replace the term ``force'', i.e., ``fundamental force'' turns to ``fundamental interaction''.

In classical picture, scientists adopted ``force'' (such as the Coulomb force) to describe the interaction between two particles.
However, in quantum picture, scientists adopted ``potential'' (such as the Coulomb potential) to describe the interaction between
two particles. In quantum mechanics, when a two-body system can have bounded states, then these two particles can form a stable physical
system, such as the well-known hydrogen atom.

In the following, we provide an example of a hydrogen atom, where particle 1 is the electron, and particle 2 is the proton.
The Schr{\" o}dinger-type Hamiltonian of the physical system is given by
\begin{equation}\label{H1}
      \mathcal{H}=\frac{\vec{p}^{\,2}}{2 M} +V(r)=\frac{\vec{p}^{\,2}}{2 M} +\frac{q \kappa}{r},
\end{equation}
with $q=-e$, $\kappa=Q=Ze$, and $Z=1$. The eigen-problem reads
\begin{equation}
\mathcal{H}\,\Psi(\vec{r})=E\,\Psi(\vec{r}),
\end{equation}
where $E$ is the energy and $\Psi(\vec{r})$ is the wavefunction.
Because
 \begin{eqnarray}\label{eq:E-18}
            \vec{p}^{\;2} =-\hbar^2 \dfrac{\partial^2}{\partial\,r^2}
                  -2\hbar^2 \dfrac{1}{r} \dfrac{\partial}{\partial\,r} +\dfrac{\vec{\ell}^{\;2}}{r^2},
      \end{eqnarray}
where
 \begin{eqnarray}
 \vec{\ell}=\vec{r}\times \vec{p}
 \end{eqnarray}
is the orbital angular momentum operator, then we have
      \begin{eqnarray}\label{eq:J-4}
            \mathcal{H} &=& \dfrac{1}{2 M} \left(-\hbar^2 \dfrac{\partial^2}{\partial\,r^2}
                  -2 \hbar^2 \dfrac{1}{r} \dfrac{\partial }{\partial\,r}
                  + \dfrac{\vec{\ell}^{\;2}}{ r^2}\right)+\frac{q \kappa}{r}.
      \end{eqnarray}
It can be proved that
\begin{eqnarray}
 [\mathcal{H}, \vec{\ell}^{\, 2}]=0, \;\;\;\;\; [\mathcal{H}, \ell_z]=0, \;\;\;\;\; [ \vec{\ell}^{\, 2}, \ell_z]=0,
 \end{eqnarray}
thus $\{H,\vec{\ell}^{\,2}, \ell_z\}$ forms a common set to solve the eigen-problem. Here $\ell_z$ is the $z$-component of $\vec{\ell}$.

Let us separate the wavefunction into the radial part and the angular part as follows
\begin{equation}
\Psi(\vec{r})=R(r)\, Y_{lm}(\theta,\phi),
\end{equation}
which satisfies
\begin{eqnarray}
&& \mathcal{H}\,R(r)\,Y_{lm}(\theta,\phi)=E\,R(r)\,Y_{lm}(\theta,\phi), \nonumber \\
&& \vec{\ell}^{\, 2}\,R(r)\,Y_{lm}(\theta,\phi)=l(l+1)\hbar^2\, R(r)\,Y_{lm}(\theta,\phi), \nonumber\\
&& {\ell}_z\,R(r)\,Y_{lm}(\theta,\phi)=m \hbar\, R(r)\,Y_{lm}(\theta,\phi),
\end{eqnarray}
or
\begin{eqnarray}
&& \mathcal{H}\,R(r)=E\,R(r), \nonumber \\
&& \vec{\ell}^{\, 2}\,Y_{lm}(\theta,\phi)=l(l+1)\hbar^2\, Y_{lm}(\theta,\phi), \nonumber\\
&& {\ell}_z\,Y_{lm}(\theta,\phi)=m \hbar\,Y_{lm}(\theta,\phi),
\end{eqnarray}
where the angular-part wavefunction $Y_{lm}(\theta,\phi)$ is the well-known \emph{spherical harmonics function}.

\begin{remark} \textcolor{blue}{The Eigen-Problem of $\{\vec{\ell}^{\; 2},\ell_z\}$.} In the spherical coordinate $(r,\theta,\phi)$, the
orbit angular momentum operator
            $\vec{\ell}=(\ell_x ,\ell_y ,\ell_z)$ is given by
            \begin{eqnarray}\label{eq:E-21}
                  \ell_x &=& {\rm i}\hbar\left(\sin\phi \dfrac{\partial}{\partial\,\theta}
                        +\cos\phi \cot\theta \dfrac{\partial}{\partial\,\phi}\right), \nonumber \\
                  \ell_y &=& {\rm i}\hbar\left(-\cos\phi \dfrac{\partial}{\partial\,\theta}
                        +\sin\phi \cot\theta \dfrac{\partial}{\partial\,\phi}\right), \nonumber \\
                  \ell_z &=& -{\rm i}\hbar\dfrac{\partial}{\partial\,\phi},
            \end{eqnarray}
            and its squared operator is given by
            \begin{eqnarray}\label{eq:E-22}
                  \vec{\ell}^{\;2} &=& -\hbar^2 \Biggl[\dfrac{1}{\sin\theta} \dfrac{\partial}{\partial\,\theta}
                              \left(\sin\theta \dfrac{\partial}{\partial\,\theta}\right)
                        +\dfrac{1}{{\sin^2}\theta} \dfrac{\partial^2}{\partial\,\phi^2}\Biggr],
            \end{eqnarray}
            or
            \begin{eqnarray}\label{eq:E-23}
                  \vec{\ell}^{\;2} &=& -\hbar^2 \left[ \dfrac{\partial^2}{\partial \theta^2}
                  +\cot\theta  \dfrac{\partial}{\partial \theta}
                  + (1+\cot^2\theta) \dfrac{\partial^2 }{\partial \phi^2} \right].
            \end{eqnarray}
            Since $[\vec{\ell}^{\;2}, \ell_z]=0$, one usually solves the eigen-problem of
            $\vec{\ell}^{\; 2}$ in the common set $\{\vec{\ell}^{\; 2}, \ell_z\}$, viz.
            \begin{eqnarray}\label{eq:E-24}
                  \vec{\ell}^{\;2} \; Y_{lm}(\theta, \phi)&=& l(l+1) \hbar^2 \;Y_{lm}(\theta, \phi), \nonumber\\
                  \ell_z\; Y_{lm}(\theta, \phi)&=& m \hbar \;Y_{lm}(\theta, \phi),
            \end{eqnarray}
            with quantum numbers $l=0, 1, 2, \cdots$, and $m=0, \pm 1, \pm 2, \cdots, \pm l$. When $m>0$, the spherical harmonics function reads

            \begin{eqnarray}
               \label{Funda}   Y_{lm} &=& \sqrt{\frac{2l+1}{4\pi} \frac{(l-|m|)!}{(l+|m|)!}}P^{|m|}_l (\cos \theta ) e^{{\rm i} m \phi} ,
              \end{eqnarray}
      and when $m<0$, it reads
      \begin{eqnarray}
            \label{Funda'}   Y_{lm} &=& (-1)^m \sqrt{\frac{2l+1}{4\pi} \frac{(l-|m|)!}{(l+|m|)!}}P^{|m|}_l (\cos \theta ) e^{{\rm i} m \phi} .
           \end{eqnarray}
           The function $P_{l}^m(\cos\theta)$ satisfies
            \begin{eqnarray}\label{eq:E-26}
                  \dfrac{1}{\sin\theta} \dfrac{{\rm d}}{{\rm d}\,\theta}\left[
                        \sin\theta \dfrac{{\rm d}\,P_{l}^m(\cos\theta)}{{\rm d}\,\theta}\right]
                  +\left[l(l+1)-\dfrac{m^2}{{\sin^2}\theta}\right]P_{l}^m(\cos\theta)=0,
            \end{eqnarray}
and
\begin{eqnarray}\label{eq:H-24-d}
P_l^m(z)=(-1)^m (1-z^2)^{m/2} \frac{{\rm d}^m P_l(z)}{{\rm d} z^m}.
		\end{eqnarray}
$\blacksquare$
\end{remark}

Next, we need to determine the radial-part wavefunction $R(r)$, which satisfies
\begin{eqnarray}
&& \left[\dfrac{1}{2 M} \left(-\hbar^2 \dfrac{\partial^2}{\partial\,r^2}
-2 \hbar^2 \dfrac{1}{r} \dfrac{\partial }{\partial\,r}
+ \dfrac{l(l+1)\hbar^2}{ r^2}\right)+\frac{q \kappa}{r}\right] R(r)=E\, R(r),
\end{eqnarray}
or
\begin{eqnarray}\label{eq:eigen-1}
&& \left[\dfrac{1}{2 M} \left(-\hbar^2 \dfrac{{\rm d}^2}{{\rm d} r^2}
-2 \hbar^2 \dfrac{1}{r} \dfrac{{\rm d} }{{\rm d}r}
+ \dfrac{l(l+1)\hbar^2}{ r^2}\right)-\frac{|q \kappa|}{r}\right] R(r)=E\, R(r),
\end{eqnarray}
for $q\kappa<0$ (which corresponds to the attractive potential).

Let us solve Eq. (\ref{eq:eigen-1}) by considering three different situations: $E< 0$, $E=0$, and $E >0$. The conclusion
is: (i) For $E<0$, the two-particle system can have physically admissible wavefunctions; (ii) For $E = 0$ and $E>0$, the
two-particle system cannot have physically admissible wavefunctions (i.e, the wavefuncions $\Psi(\vec{r})$ are divergent
and cannot be normalized), thus these cases are excluded. In the following, we focus on the eigen-problem with $E<0$.


Let us analyze the asymptotic behaviors of $R(r)$.

\emph{Case (i).---}When $r\rightarrow 0$, Eq. (\ref{eq:eigen-1})
                  becomes
                  \begin{eqnarray}\label{eq:F-8}
                        &&\dfrac{{\rm d}^2 R(r)}{{\rm d} r^2}
                              +\dfrac{2 }{r} \dfrac{{\rm d} R(r)}{{\rm d} r}
                              -\dfrac{l(l+1)}{r^2} R(r)=0.
                  \end{eqnarray}
                  In the region near to $r=0$, let $R(r) \propto r^\nu$, after substituting it into above equation, we have
                  \begin{eqnarray}\label{eq:F-9}
                        && \nu(\nu-1)r^{\nu-2}+2\nu r^{\nu-2} -l(l+1) r^{\nu-2}=0,
                  \end{eqnarray}
                  which leads to
                  \begin{eqnarray}\label{eq:F-10}
                        && \nu(\nu+1) -l(l+1)=0.
                  \end{eqnarray}
                  Based on which, one obtains
                  \begin{eqnarray}\label{eq:F-11}
                        && \nu_1=l, \;\;\; {\rm or }\;\;\; \nu_2= -(l+1).
                  \end{eqnarray}
                  To avoid $\lim\limits_{r\rightarrow 0} R(r)\rightarrow \infty$, from the
                  viewpoint of physics, we choose $\nu=\nu_1=l \geq 0$. When $l=0$, we specially have $\nu=0$.

            \emph{Case (ii).---} When $r \rightarrow \infty$, Eq.
                  (\ref{eq:eigen-1}) becomes
                  \begin{eqnarray}\label{eq:F-12}
                        &&\dfrac{{\rm d}^2 R(r)}{{\rm d}\,r^2} + \epsilon\,R(r)=0,
                  \end{eqnarray}
                  with
                   \begin{eqnarray}
                        &&\epsilon= \dfrac{2M E}{\hbar^2}<0.
                  \end{eqnarray}
                  Then we attain
                  \begin{eqnarray}\label{eq:F-13-01}
                        R(r)\propto {\rm e}^{\sqrt{-\epsilon}\,r}, \;\;\; {\rm or }\;\;\;
                        R(r)\propto {\rm e}^{-\sqrt{-\epsilon}\,r}.
                  \end{eqnarray}
                  To avoid $\lim\limits_{r\rightarrow \infty}R(r)\rightarrow \infty$, we
                  choose the solution $R(r)\propto {\rm e}^{-\sqrt{-\epsilon}\,r}$.

                  Based on the above analysis, we may set $R(r)$ as
                  \begin{eqnarray}\label{eq:F-13-t}
                        && R(r)=r^{\nu} {\rm e}^{-\sqrt{-\epsilon}\,r} F(r),
                  \end{eqnarray}
                  with $\nu=l$. After that, we can calculate
                  \begin{eqnarray}
                        \dfrac{{\rm d}\,R(r)}{{\rm d}\,r} &=&
                              \nu\,r^{\nu-1} {\rm e}^{-\sqrt{-\epsilon}\,r} F(r)
                              -\sqrt{-\epsilon}\,r^{\nu} {\rm e}^{-\sqrt{-\epsilon}\,r}\,F(r)
                              +r^{\nu} {\rm e}^{-\sqrt{-\epsilon}\,r}\,\dfrac{{\rm d}\,F(r)}{{\rm d}\,r} \nonumber \\
                        &=& r^{\nu} {\rm e}^{-\sqrt{-\epsilon}\,r} \Biggr[
                              \biggr(\dfrac{\nu}{r} -\sqrt{-\epsilon}\biggr)F(r)
                              +\dfrac{{\rm d} F(r)}{{\rm d} r}\Biggr] = r^{\nu} {\rm e}^{-\sqrt{-\epsilon}\,r}\,C(r),
                  \end{eqnarray}
                  thus
                  \begin{eqnarray}
                         \dfrac{{\rm d}^2\,R(r)}{{\rm d}\,r^2} &=&
                              \dfrac{{\rm d}^2\,\Bigl[
                                          r^{\nu} {\rm e}^{-\sqrt{-\epsilon}\,r}\,C(r)\Bigr]}
                                    {{\rm d}\,r^2} \nonumber \\
                        &=& r^{\nu} {\rm e}^{-\sqrt{-\epsilon}\,r} \Biggr[
                              \biggr(\dfrac{s}{r} -\sqrt{-\epsilon}\biggr)C(r)
                              +\dfrac{{\rm d}\,C(r)}{{\rm d}\,r}\Biggr] \nonumber \\
                        &=& r^{\nu} {\rm e}^{-\sqrt{-\epsilon}\,r} \Biggl\{
                              \biggr(\dfrac{\nu}{r} -\sqrt{-\epsilon}\biggr)\biggr[
                                    \biggr(\dfrac{\nu}{r} -\sqrt{-\epsilon}\biggr)F(r)
                                    +\dfrac{{\rm d} F(r)}{{\rm d} r}\biggr]
                              +\dfrac{{\rm d}}{{\rm d}\,r}\,\biggr[
                                    \biggr(\dfrac{\nu}{r} -\sqrt{-\epsilon}\biggr)F(r)
                                    +\dfrac{{\rm d} F(r)}{{\rm d} r}\biggr]\Biggr\} \nonumber \\
                        &=& r^{\nu} {\rm e}^{-\sqrt{-\epsilon}\,r} \Biggl\{
                              \biggr(\dfrac{\nu}{r} -\sqrt{-\epsilon}\biggr)\biggr[
                                    \biggr(\dfrac{\nu}{r} -\sqrt{-\epsilon}\biggr)F(r)
                                    +\dfrac{{\rm d} F(r)}{{\rm d} r}\biggr]
                              +\biggr[-\dfrac{\nu}{r^2} F(r)+\biggr(\dfrac{\nu}{r} -\sqrt{-\epsilon}\biggr)\dfrac{{\rm d} F(r)}{{\rm d} r}
                                    +\dfrac{{\rm d}^2 F(r)}{{\rm d}\,r^2}\biggr]\Biggr\} \nonumber \\
                        &=& r^{\nu} {\rm e}^{-\sqrt{-\epsilon}\,r} \Biggl\{
                              \dfrac{{\rm d}^2 F(r)}{{\rm d}\,r^2}
                              +2\biggr(\dfrac{\nu}{r} -\sqrt{-\epsilon}\biggr)
                                    \dfrac{{\rm d} F(r)}{{\rm d} r}
                              +\biggl[\Bigl(\dfrac{\nu}{r} -\sqrt{-\epsilon}\Bigr)^2
                                    -\dfrac{\nu}{r^2}\biggr]F(r)\Biggr\}.
                  \end{eqnarray}
                  Then,
                  \begin{eqnarray}
                        && \dfrac{{\rm d}^2\,R(r)}{{\rm d}\,r^2}
                                    +\dfrac{2}{r} \dfrac{{\rm d}\,R(r)}{{\rm d}\,r}
                                    -\dfrac{l(l+1)}{ r^2}R(r) - \dfrac{2 M q\kappa}{\hbar^2 r}R(r) +\epsilon R(r) \nonumber \\
                        &=& r^{\nu} {\rm e}^{-\sqrt{-\epsilon}\,r} \Biggl\{
                                    \dfrac{{\rm d}^2 F(r)}{{\rm d}\,r^2}
                                    +2\biggr(\dfrac{\nu}{r} -\sqrt{-\epsilon}\biggr)
                                          \dfrac{{\rm d} F(r)}{{\rm d} r}
                                    +\biggl[\Bigl(\dfrac{\nu}{r} -\sqrt{-\epsilon}\Bigr)^2
                                          -\dfrac{\nu}{r^2}\biggr]F(r)\Biggr\} \nonumber \\
                              && +\dfrac{2}{r} r^{\nu} {\rm e}^{-\sqrt{-\epsilon}\,r} \Biggr[
                                    \biggr(\dfrac{\nu}{r} -\sqrt{-\epsilon}\biggr)F(r)
                                    +\dfrac{{\rm d} F(r)}{{\rm d} r}\Biggr]
                              +\left(\epsilon- \dfrac{2M q\kappa}{\hbar^2 r}-\dfrac{l(l+1)}{ r^2}\right)r^{\nu}
                                    {\rm e}^{-\sqrt{-\epsilon}\,r} F(r)=0,
                  \end{eqnarray}
                  i.e.,
                   \begin{eqnarray}
                        &&\Biggl\{
                                    \dfrac{{\rm d}^2 F(r)}{{\rm d}\,r^2}
                                    +2\biggr(\dfrac{\nu}{r} -\sqrt{-\epsilon}\biggr)
                                          \dfrac{{\rm d} F(r)}{{\rm d} r}
                                    +\biggl[\Bigl(\dfrac{\nu}{r} -\sqrt{-\epsilon}\Bigr)^2
                                          -\dfrac{\nu}{r^2}\biggr]F(r)\Biggr\} \nonumber \\
                              && +\dfrac{2}{r}  \Biggr[
                                    \biggr(\dfrac{\nu}{r} -\sqrt{-\epsilon}\biggr)F(r)
                                    +\dfrac{{\rm d} F(r)}{{\rm d} r}\Biggr]
                              +\left(\epsilon- \dfrac{2M q\kappa}{\hbar^2 r}-\dfrac{l(l+1)}{ r^2}\right)  F(r)=0,
                  \end{eqnarray}
                  i.e.,
                   \begin{eqnarray}
                        &&     \dfrac{{\rm d}^2 F(r)}{{\rm d}\,r^2}
                                    + \left[2\biggr(\dfrac{\nu}{r} -\sqrt{-\epsilon}\biggr)+\dfrac{2}{r}  \right]
                                          \dfrac{{\rm d} F(r)}{{\rm d} r} \nonumber\\
                        &&            + \left\{\biggl[\Bigl(\dfrac{\nu}{r} -\sqrt{-\epsilon}\Bigr)^2
                                          -\dfrac{\nu}{r^2}\biggr]  +\dfrac{2}{r}  \Biggr[
                                    \biggr(\dfrac{\nu}{r} -\sqrt{-\epsilon}\biggr)\Biggr]
                              +\left(\epsilon- \dfrac{2M q\kappa}{\hbar^2 r}-\dfrac{l(l+1)}{ r^2}\right) \right\} F(r)=0,
                  \end{eqnarray}
                  i.e.,
                  \begin{eqnarray}
                        \dfrac{{\rm d}^2 F(r)}{{\rm d}\,r^2}
                              +\biggr[\dfrac{2(\nu+1)}{r} -2\sqrt{-\epsilon}\biggr]
                                    \dfrac{{\rm d}\,F(r)}{{\rm d}\,r}
                              +\biggr[\dfrac{\nu(\nu+1)-l(l+1)}{r^2}-\dfrac{2(\nu+1)\sqrt{-\epsilon}+\dfrac{2M q\kappa}{\hbar^2} }{r}\biggr]F(r)=0,
                  \end{eqnarray}
                  i.e.,
                  \begin{eqnarray}
                        \dfrac{{\rm d}^2 F(r)}{{\rm d}\,r^2}
                              +\biggr[\dfrac{2(\nu+1)}{r} -2\sqrt{-\epsilon}\biggr]
                                    \dfrac{{\rm d}\,F(r)}{{\rm d}\,r}
                              +\biggr[-\dfrac{2(\nu+1)\sqrt{-\epsilon}+\dfrac{2M q\kappa}{\hbar^2} }{r}\biggr]F(r)=0,
                  \end{eqnarray}
                 i.e.,
                  \begin{eqnarray}\label{eq:F-22}
                        r\dfrac{{\rm d}^2 F(r)}{{\rm d}\,r^2}
                              +\biggr[2(\nu+1) -2\sqrt{-\epsilon}\;r\biggr]
                                    \dfrac{{\rm d}\,F(r)}{{\rm d}\,r}
                              -\biggr[2(\nu+1)\sqrt{-\epsilon}+\dfrac{2M q\kappa}{\hbar^2}\biggr]F(r)=0,
                  \end{eqnarray}
                  i.e.,
                  \begin{eqnarray}\label{eq:F-28-t}
                        (\sqrt{-\epsilon})^2 r\dfrac{{\rm d}^2 F(z)}{{\rm d} (\sqrt{-\epsilon} \;r)^2}
                              +\sqrt{-\epsilon} \;\biggr[2(\nu+1) -2\sqrt{-\epsilon}\;r\biggr]
                                    \dfrac{{\rm d}\,F(z)}{{\rm d}\,(\sqrt{-\epsilon} \;r)}
                        -\biggr[2(\nu+1)\sqrt{-\epsilon}+\dfrac{2M q\kappa}{\hbar^2}\biggr]F(z)=0.
                  \end{eqnarray}
                  Let
                  \begin{eqnarray}\label{eq:F-27-01}
                        z=\sqrt{-\epsilon} \; r,
                  \end{eqnarray}\\
                  then Eq. (\ref{eq:F-28-t}) becomes
                  \begin{eqnarray}\label{eq:F-29-01}
                        \sqrt{-\epsilon}\,z\dfrac{{\rm d}^2 F(z)}{{\rm d}\,z^2}
                                    +\sqrt{-\epsilon} \;\biggr[2(\nu+1) -2z\biggr]
                                          \dfrac{{\rm d}\,F(z)}{{\rm d}\,z}
                              -\biggr[2(\nu+1)\sqrt{-\epsilon}+\dfrac{2M q\kappa}{\hbar^2}\biggr]F(z)=0,
                  \end{eqnarray}
                  i.e.,
                  \begin{eqnarray}
                        z\dfrac{{\rm d}^2 F(z)}{{\rm d}\,z^2}
                                    +\biggr[2(\nu+1) -2z\biggr]
                                          \dfrac{{\rm d}\,F(z)}{{\rm d}\,z}
                              -\biggr[2(\nu+1)+\dfrac{2M q\kappa}{\hbar^2 \sqrt{-\epsilon}}\biggr]F(z)=0,
                  \end{eqnarray}
                  i.e.,
                  \begin{eqnarray}\label{eq:F-30-d-t}
                        \tau\dfrac{{\rm d}^2 F(\tau)}{{\rm d}\,\tau^2}
                        +\biggr[2(\nu+1) -\tau\biggr]
                              \dfrac{{\rm d}\,F(\tau)}{{\rm d}\,\tau}
                        -\left[(\nu+1)+\dfrac{M q\kappa}{\hbar^2 \sqrt{-\epsilon}}\right]F(\tau)=0,
                  \end{eqnarray}
                  with $\tau=2\,z$.

                  Remarkably, let us set
                  \begin{eqnarray}\label{eq:F-31-01}
                        \alpha &=& (\nu+1)+\dfrac{M q\kappa}{\hbar^2 \sqrt{-\epsilon}}=(l+1)+\dfrac{M q\kappa}{\hbar^2 \sqrt{-\epsilon}}, \;\;\;\;\;\;
                        \gamma = 2(\nu+1)=2(l+1),
                  \end{eqnarray}
                  then Eq.({\ref{eq:F-30-d-t}}) can be rewritten as
                  \begin{eqnarray}\label{eq:F-23-01}
                        \tau\dfrac{{\rm d}^2\,F(\tau)}{{\rm d}\,\tau^2}
                        +(\gamma-\tau)\dfrac{{\rm d}\,F(\tau)}{{\rm d}\,\tau}-\alpha\,F(\tau)=0,
                  \end{eqnarray}
                  which is a confluent hypergeometric equation with general solution
                  \begin{eqnarray}\label{eq:F-24-01}
                        F(\tau)= C_1\,F(\alpha;\,\gamma;\,\tau)
                              +C_2\,\tau^{1-\gamma}\,F(\alpha+1-\gamma;\,2-\gamma;\,\tau),
                  \end{eqnarray}
                  where $C_1$, $C_2$ are two constants, and
                  \begin{eqnarray}\label{eq:F-25}
                        F(\alpha;\gamma;\tau) &=& 1+\sum_{k=1}^{\infty}
                              \dfrac{(\alpha)_k}{k!\,(\gamma)_k}\tau^k = 1+\dfrac{\alpha}{\gamma}\tau
                              +\dfrac{\alpha(\alpha+1)}{2!\,\gamma(\gamma+1)}\tau^2
                              +\dfrac{\alpha(\alpha+1)(\alpha+2)}{3!\,\gamma(\gamma+1)(\gamma+2)}\tau^3+\cdots
                  \end{eqnarray}\\
                  is the confluent hypergeometric function, including
                  \begin{eqnarray}\label{eq:F-26}
                        (\alpha)_k &=& \alpha(\alpha+1)\cdots(\alpha+k-1), \;\;\;\;\;
                        (\gamma)_k = \gamma(\gamma+1)\cdots(\gamma+k-1).
                  \end{eqnarray}\\
                  For convenience, one can denote the two hypergeometric functions
                  as
                  \begin{eqnarray}\label{eq:F-32-t}
                        w_1 &=& F(\alpha;\gamma;\tau) = F\left((l+1)+\dfrac{M q\kappa}{\hbar^2 \sqrt{-\epsilon}};\,
                              2(l+1);\,2 \sqrt{-\epsilon}\,r\right),
                  \end{eqnarray}
                  \begin{eqnarray}\label{eq:F-33-y}
                        w_2 &=& \tau^{1-\gamma}F(\alpha-\gamma+1;\,2-\gamma;\,\tau) = \left(2\sqrt{-\epsilon}\,r\right)^{1-\gamma}
                              F(\alpha-\gamma+1;\,2-\gamma;\,2 \sqrt{-\epsilon}\,r),
                  \end{eqnarray}
                  namely
                  \begin{eqnarray}
                        F(r)=C_1\,w_1 +C_2\,w_2.
                  \end{eqnarray}
                  By substituting Eq. (\ref{eq:F-32-t}) and Eq. (\ref{eq:F-33-y}) into Eq. (\ref{eq:F-13-t}), one then
                  obtains the radial function as
                  \begin{eqnarray}\label{eq:F-34-01}
                        R(r) &=& r^{\nu} {\rm e}^{-\sqrt{-\epsilon}\,r} F(r) = r^{\nu} {\rm e}^{-\sqrt{-\epsilon}\,r} (C_1\,w_1 +C_2\,w_2)=C_1R_1(r)+C_2R_2(r).
                  \end{eqnarray}

                 Let us study $R_2(r)$. We need to check the behavior of $R_2(r)$ when $r\rightarrow 0$, which is
                  \begin{eqnarray}
                       \lim_{r\rightarrow 0} R_2(r) &=& \lim_{r\rightarrow 0} r^{\nu} {\rm e}^{-\sqrt{-\epsilon}\,r}\left[\left(2\sqrt{-\epsilon}\,r\right)^{1-\gamma}
                              F(\alpha-\gamma+1;\,2-\gamma;\,2 \sqrt{-\epsilon}\,r)\right]\nonumber\\
                               &\propto & \lim_{r\rightarrow 0} r^{\nu} {\rm e}^{-\sqrt{-\epsilon}\,r}\left[\left(2\sqrt{-\epsilon}\,r\right)^{1-\gamma}
                             \right]\nonumber\\
                  &\propto& \lim_{r\rightarrow 0} \left[r^{l} r^{1-\gamma}\right]=\lim_{r\rightarrow 0} \left[ r^{l+1-\gamma}\right]=\lim_{r\rightarrow 0}  r^{-(l+1)} \rightarrow \infty,
                  \end{eqnarray}
                 thus $R_2(r)$ is not a physically admissible solution.

                  Let us study $R_1(r)$. On one hand, to avoid $\lim\limits_{r\rightarrow \infty}R_1(r)\rightarrow \infty$, the function $w_1$ has to be a truncated polynomial, i.e.,
                  \begin{eqnarray}\label{eq:n-1}
                       \alpha=(l+1)+\dfrac{M q\kappa}{\hbar^2 \sqrt{-\epsilon}}=-N, \;\;\;\ (N=0, 1, 2, ...).
                  \end{eqnarray}
                  [Note: One may easily observe that for the repulsive potential (i.e., $q\kappa>0$), Eq. (\ref{eq:n-1}) is not valid.]

                  On the other hand, we need to check the behavior of $R_1(r)$ when $r\rightarrow 0$, which is
                  \begin{eqnarray}
                       \lim_{r\rightarrow 0} R_1(r) &=& \lim_{r\rightarrow 0} r^{\nu} {\rm e}^{-\sqrt{-\epsilon}\,r}\left[F(\alpha;\gamma;\tau) \right]
                  = \lim_{r\rightarrow 0} r^{\nu} {\rm e}^{-\sqrt{-\epsilon}\,r}= \lim_{r\rightarrow 0} r^{s} = \lim_{r\rightarrow 0} r^{l}=0, \;\; {\rm or} \;\; 1,
                  \end{eqnarray}
                  thus $R_1(r)$ is the physical solution.

                  Based on Eq. (\ref{eq:n-1}). we have
                  \begin{eqnarray}
                      \dfrac{M q\kappa}{\hbar^2 \sqrt{-\epsilon}}=-N-(l+1),
                  \end{eqnarray}
                  i.e.,
                  \begin{eqnarray}
                     - \dfrac{M |q\kappa|}{\hbar^2 \sqrt{-\epsilon}}=-N-(l+1),
                  \end{eqnarray}
                  i.e.,
                  \begin{eqnarray}
                     \dfrac{M |q\kappa|}{\hbar^2 \sqrt{-\epsilon}}=N+(l+1).
                  \end{eqnarray}

                   Let us denote the principle quantum number as
                       \begin{eqnarray}
                     n=N+(l+1),
                  \end{eqnarray}
                  then we have
                  \begin{eqnarray}
                      \dfrac{M |q\kappa|}{\hbar^2 \sqrt{-\epsilon}}=n, \;\; (n=1, 2, 3,...)
                  \end{eqnarray}
                  i.e.,
                  \begin{eqnarray}
                      \dfrac{\hbar^2 \sqrt{-\epsilon}}{M |q\kappa|}=\dfrac{1}{n},
                  \end{eqnarray}
                  i.e.,
                  \begin{eqnarray}
                      -\dfrac{\hbar^4 \epsilon}{M^2 (q\kappa)^2}=\dfrac{1}{n^2},
                  \end{eqnarray}
                   i.e.,
                  \begin{eqnarray}
                     \dfrac{\hbar^4 \dfrac{2M E}{\hbar^2}}{M^2 (q\kappa)^2}=-\dfrac{1}{n^2},
                  \end{eqnarray}
                 i.e.,
                 \begin{eqnarray}\label{eq:hyd-1a}
                     E_n=-\dfrac{M (q\kappa)^2}{2\hbar^2}\dfrac{1}{n^2},  \;\; (n=1, 2, 3,...).
                  \end{eqnarray}
Let $q=-e$ and $\kappa=Q=Ze$, we have
                 \begin{eqnarray}\label{eq:hyd-1b}
                     E_n=-\dfrac{Z^2 M e^4}{2\hbar^2}\dfrac{1}{n^2},  \;\; (n=1, 2, 3,...),
                  \end{eqnarray}
which is just the energy of a hydrogen-like atom (or a hydrogen atom when $Z=1$).

\newpage

\section{Spin-Dependent Coulomb-Type Potential: An Exact Solution of Yang-Mills Equations}

Now, we face two facts in front of us: (i) The Coulomb-type potential is a simple but fundamental solution (i.e., denoted as $ \mathcal{S}_{\rm Maxwell}=\{\vec{\mathcal{A}}=0, \varphi=\kappa/r\}$) of Maxwell's equations;
(ii) The Yang-Mills equations are the natural generalizations of Maxwell's equations from the Abelian potentials to the non-Abelian ones.
Then it naturally gives rise to a scientific question: Can one derive some generalized Coulomb-type potentials based on Yang-Mills equations
(just like deriving the Coulomb-type potential from Maxwell's equations)?
The answer is positive. We find that the spin-dependent Coulomb-type potential is an exact and fundamental solution (i.e., denoted as
$\mathcal{S}_{\rm YM}=\{\vec{\mathcal{A}}, \varphi \}$) of Yang-Mills equations.

\subsection{ A Set of Spin Potentials $\mathcal{S}_{\rm YM}= \{\vec{\mathcal{A}}, \varphi\}$: An Exact Solution of Yang-Mills Equations}

As mentioned before, there are two kinds of potentials: the vector potential (denoted by $\vec{\mathcal{A}}$)
and the scalar potential (denoted by $\varphi$).  When $\vec{\mathcal{A}}$ and $\varphi$ contain the spin
operator $\vec{S}$, then we call $\vec{\mathcal{A}}$ and $\varphi$ as the spin vector potential and
the spin scalar potential, respectively. Or succinctly, we call $\mathcal{S}_{\rm YM}= \{\vec{\mathcal{A}}, \varphi\}$ as a set of spin potentials.

As it is well-known that, the Yang-Mills equations are the natural generalizations of Maxwell's equations from the Abelian potentials to the non-Abelian ones. Usually, the non-Abelian operators can be realized by the spin operators.
Let $\vec{S}=(S_x, S_y, S_z)$ be the spin-$s$ angular momentum operator (with the spin value $s=0, 1/2, 1, ...$), which satisfies the following
commutation relations
\begin{eqnarray}\label{angularm1}
&& [S_x, S_y]= \mathrm{i} \hbar\; S_z, \;\;\; [S_y, S_z]= \mathrm{i} \hbar\; S_x,\;\;\;  [S_z, S_x]= \mathrm{i} \hbar\; S_y.
\end{eqnarray}
or in the vector form  as
\begin{eqnarray}
&& \vec{S} \times \vec{S} = \mathrm{i} \hbar\; \vec{S}.
\end{eqnarray}
The square of $\vec{S}$ satisfies
\begin{eqnarray}
&& \vec{S}^{\,2} = s(s+1) \hbar^2.
\end{eqnarray}
Specially, when $s=1/2$, the spin operator $\vec{S}$ can be realized by
\begin{equation}
      \vec{S} = \frac{\hbar}{2} \vec{\sigma},
\end{equation}
where $\vec{\sigma}=(\sigma_x, \sigma_y, \sigma_z)$ is the vector of Pauli matrices, $\vec{S}^2 = \dfrac{3}{4} \hbar^2 \openone$, $\openone$ is the $2\times 2$ identity matrix, and
\begin{eqnarray}
 \label{S-3}
  && \sigma_x=
  \left(
    \begin{array}{cc}
      0 & 1 \\
      1 & 0 \\
    \end{array}
  \right),\;\;\;\sigma_y=
  \left(
    \begin{array}{cc}
      0 & -{\rm i} \\
      {\rm i} & 0 \\
    \end{array}
  \right),\;\;\;  \sigma_z=
  \left(
    \begin{array}{cc}
      1 & 0 \\
      0 & -1 \\
    \end{array}
  \right).
   \end{eqnarray}
We find that the set of spin potentials $\mathcal{S}_{\rm YM}=\{\vec{\mathcal{A}}, \varphi\}$ can be an exact solution of the Yang-Mills equation.
The result can be expressed in the following theorem.
\begin{theorem}\label{Ther2}
If the spin operator $\vec{S}$ is considered, then the Yang-Mills equations can have the following {local} exact solutions for spin vector and scalar potentials {on $\mathbb R^3\setminus\{0\}$}
\begin{subequations}
\begin{eqnarray}\label{eq:Coul-1}
    &&  \vec{\mathcal{A}}= k \frac{\vec{r} \times \vec{S}}{r^2}, \label{eq:A}\\
    &&       \varphi = f_1(r) \left(\vec{r} \cdot \vec{S}\right) +  f_2(r), \label{eq:phi}
\end{eqnarray}
\end{subequations}
where two functions $f_1(r)$ and $f_2(r)$ are given in Table \ref{tab:gk}.

\begin{table}[H]
\caption{Summary of the consistent solutions for $f_1(r)$ and $f_2(r)$. Note that for the case $g\neq 0$, $k\neq 0$, solutions exist only when $g\hbar k = 1$ or $2$.}\label{tab:gk}
\centering
\begin{tabular}{|c|c|c|c|}
\hline
(i) $g=0$, $k=0$ & (ii) $g=0$, $k\neq 0$ & (iii) $g\neq 0$, $k=0$ & (iv) $g\neq 0$, $k\neq 0$ \\
\hline
\makecell[l]{$f_1(r) = \dfrac{\kappa_2}{r^3} + \kappa_3$ \\ $f_2(r) = \dfrac{\kappa_1}{r}$} &
\makecell[l]{No solutions} &
\makecell[l]{$f_1(r) = 0$ \\ $f_2(r) = \dfrac{\kappa_1}{r}$} &
\makecell[l]{If $g\hbar k = 2$: $f_1(r) = 0$ \\ If $g\hbar k = 1$: $f_1(r) = \dfrac{\kappa_2}{r} + \dfrac{\kappa_3}{r^2}$ \\ $f_2(r) = \dfrac{\kappa_1}{r}$} \\
\hline
\end{tabular}
\end{table}

\end{theorem}
\begin{proof}
In Ref. \cite{Chenspin2025}, the researchers have presented a kind of spin-dependent vector potentials to investigate the
spin Aharonov-Bohm effect. The form of the spin vector potential is given by
\begin{equation}\label{eq:A1}
      \vec{\mathcal{A}} = k \frac{\vec{r} \times \vec{S}}{r^2},
\end{equation}
where $k$ is a proportional coefficient, and $\vec{S}$ is the spin-$s$ angular momentum operator.

Now the question becomes: For the fixed $\vec{\mathcal{A}}$ in Eq. (\ref{eq:A1}), then for what $\varphi$, the set of spin potentials
$\mathcal{S}_{\rm YM}=\{\vec{\mathcal{A}}, \varphi\}$ is an exact solution of Yang-Mills equations? To do so, let us determine the form of the
scalar potential $\varphi$ first. For simplicity, let us consider the case of spin value $s=1/2$.

Because $\varphi$ is (i) a scalar quantity; (ii) a function of $\vec{r}$ and $\vec{S}$, then it is a possible combination of the following quantities
\begin{eqnarray}
   &&   \vec{r}\cdot \vec{S}, \;\;\; \left(\vec{r}\cdot \vec{S}\right)^2, \;\;\; \left(\vec{r}\cdot \vec{S}\right)^n, \;\;\; \left( \vec{r} \times \vec{S}\right) \cdot \vec{S}, \;\;\; \vec{r} \cdot \left(\vec{S} \times \vec{S}\right), \;\;\; \left(\vec{S} \times \vec{S}\right) \cdot \vec{S}.
\end{eqnarray}
However, because
\begin{align}
   \left( \vec{r} \times \vec{S}\right) \cdot \vec{S} = \vec{r} \cdot \left(\vec{S} \times \vec{S}\right) ={\rm i} \hbar \left(\vec{r} \cdot \vec{S}\right),
\end{align}
\begin{align}
  \vec{S}\cdot \left( \vec{r} \times \vec{S}\right) =  - \vec{S}\cdot \left( \vec{S} \times \vec{r}\right)=-\left(\vec{S} \times \vec{S}\right)\cdot \vec{r}=-{\rm i} \hbar \left(\vec{r} \cdot \vec{S}\right),
\end{align}

\begin{align}
    \left(\vec{S} \times \vec{S}\right) \cdot \vec{S} = {\rm i}\hbar S^2 = \frac{{\rm i} 3\hbar^3 }{4}\openone,
\end{align}
\begin{align}
   \left(\vec{r} \cdot \vec{S}\right)^2  = \frac{r^2 \hbar^2}{4}\openone,
\end{align}
thus it is sufficient that $\varphi$ is of the following form
\begin{equation}
      \varphi = f_1(r) \left(\vec{r} \cdot \vec{S}\right) +  f_2(r),
\end{equation}
where $f_1(r)$ and $f_2(r)$ are two functions of $r$.

Here we have chosen $\vec{\mathcal{A}}$ and $\varphi$ as the time-independent spin potentials, therefore
\begin{eqnarray}
    \dfrac{\partial \vec{\mathcal{A}}}{\partial t}=0, \;\;\; \dfrac{\partial \varphi}{\partial t}=0,
\end{eqnarray}
which yields
\begin{eqnarray}
    \dfrac{\partial \vec{\mathcal{E}}}{\partial t}=0, \;\;\; \dfrac{\partial \vec{\mathcal{B}}}{\partial t}=0,
\end{eqnarray}
i.e., $\vec{\mathcal{E}}$ and $\vec{\mathcal{B}}$ are time-independent fields.

In the next, our task is to use Eq. (\ref{eq:DivEYM})-Eq. (\ref{eq:CurlBYM}) to determine two functions $f_1(r)$ and $f_2(r)$. Although  Eq. (\ref{eq:CurlEYM}) and Eq. (\ref{eq:DivBYM}) are  automatically satisfied because of the definition of the field strength tensor, we would like to verify it by calculation.

\emph{Analysis 1: The study of Eq. (\ref{eq:DivEYM})---}Let's calculate the ``electric''  field at first. We can have
\begin{eqnarray}
    \vec{\mathcal{E}}=  -\dfrac{1}{c} \dfrac{\partial\,\vec{\mathcal{A}}}{\partial\,t}-\nabla\varphi-{\rm i}\,g\left[\varphi,\ \vec{\mathcal{A}}\right]= -\nabla\varphi-{\rm i}\,g\left[\varphi,\ \vec{\mathcal{A}}\right]
\end{eqnarray}
Because
\begin{eqnarray}
\nabla\varphi=\nabla \left[  f_1(r) \left(\vec{r} \cdot \vec{S}\right) +  f_2(r)\right],
\end{eqnarray}

\begin{eqnarray}
\nabla \left[f_1(r) \left(\vec{r} \cdot \vec{S}\right)\right] = \left[\nabla f_1(r)\right] \left(\vec{r} \cdot \vec{S}\right) + f_1(r) \left[\nabla\left(\vec{r} \cdot \vec{S}\right)\right],
\end{eqnarray}
\begin{eqnarray}
 \nabla \left[ f_1(r) \right] &=& \dfrac{\partial f_1(r)}{\partial r}\dfrac{\partial r}{\partial x} \; \hat{e}_x +\dfrac{\partial f_1(r)}{\partial r}\dfrac{\partial r}{\partial y} \; \hat{e}_y+\dfrac{\partial f_1(r)}{\partial r}\dfrac{\partial r}{\partial z} \; \hat{e}_z\nonumber\\
&=& \dfrac{\partial f_1(r)}{\partial r} \dfrac{x}{r} \; \hat{e}_x +\dfrac{\partial f_1(r)}{\partial r} \dfrac{y}{r} \; \hat{e}_y+\dfrac{\partial f_1(r)}{\partial r} \dfrac{z}{r} \; \hat{e}_z\nonumber\\
&=& {\dfrac{\partial f_1(r)}{\partial r} \dfrac{\vec{r}}{r}},
\end{eqnarray}
\begin{eqnarray}
\nabla \left[ f_2(r) \right] &=&  {\dfrac{\partial f_2(r)}{\partial r} \dfrac{\vec{r}}{r}},
\end{eqnarray}
\begin{eqnarray}
\nabla \left[  \left(\vec{r} \cdot \vec{S}\right)\right]&=& \nabla \left[x S_x+ y S_y +z S_z  \right]=S_x\; \hat{e}_x + S_y\; \hat{e}_y + S_z\; \hat{e}_z= \vec{S},
\end{eqnarray}
we can have
\begin{eqnarray}
\nabla\varphi&=&\nabla \left[  f_1(r) \left(\vec{r} \cdot \vec{S}\right) +  f_2(r)\right]\nonumber\\
&=& {\dfrac{\partial f_1(r)}{\partial r} \left(\vec{r} \cdot \vec{S}\right)\dfrac{\vec{r}}{r}}+ f_1(r)\vec{S}+{\dfrac{\partial f_2(r)}{\partial r} \dfrac{\vec{r}}{r}}.
\end{eqnarray}
Because
\begin{eqnarray}
\left[\varphi, \vec{\mathcal{A}}\right] &=& \left[ f_1(r) \left(\vec{r} \cdot \vec{S}\right) + f_2(r), k\frac{\left(\vec{r} \times \vec{S}\right)}{r^2} \right] \nonumber\\
&=& k\frac{f_1(r)}{r^2} \left[ \left(\vec{r} \cdot \vec{S}\right) , \left(\vec{r} \times \vec{S}\right) \right],
\end{eqnarray}
\begin{eqnarray}
       \left[ \left(\vec{r} \cdot \vec{S}\right), \left(\vec{r} \times \vec{S}\right)_z \right] &=& \left[ x S_x+ y S_y +z S_z, x S_y- y S_x \right]= \left[ x S_x+ z S_z, x S_y \right]-\left[ y S_y +z S_z, y S_x \right] \nonumber\\
       &=& x^2 \left[S_x, S_y \right]+ z x \left[S_z, S_y \right]- y^2\left[S_y, S_x \right]-zy\left[S_z, S_x \right]\nonumber\\
       &=& x^2 ({\rm i} \hbar S_z) + z x (-{\rm i} \hbar S_x) - y^2 (-{\rm i} \hbar S_z)-zy ({\rm i} \hbar S_y)\nonumber\\
       &=& {\rm i} \hbar (x^2+ y^2)S_z -  {\rm i} \hbar (x S_x+y S_y) z\nonumber\\
       &=& {\rm i} \hbar r^2 S_z -  {\rm i} \hbar \left(\vec{r}\cdot \vec{S}\right) z,
\end{eqnarray}
\begin{eqnarray}
       \left[ \left(\vec{r} \cdot \vec{S}\right), \left(\vec{r} \times \vec{S}\right) \right]
       &=& {\rm i} \hbar r^2 \vec{S} -  {\rm i} \hbar \left(\vec{r}\cdot \vec{S}\right) \vec{r},
\end{eqnarray}
then we can have
\begin{eqnarray}
\left[\varphi, \vec{\mathcal{A}}\right] &=& \left[ f_1(r) \left(\vec{r} \cdot \vec{S}\right) + f_2(r), k\frac{\left(\vec{r} \times \vec{S}\right)}{r^2} \right] \nonumber\\
&=& k\frac{f_1(r)}{r^2} \left[ \left(\vec{r} \cdot \vec{S}\right) , \left(\vec{r} \times \vec{S}\right) \right]\nonumber\\
&=&{\rm i} \hbar k\frac{f_1(r)}{r^2} \left[ r^2 \vec{S} -   \left(\vec{r}\cdot \vec{S}\right) \vec{r}\right].
\end{eqnarray}
Then we have the ``electric'' field as
\begin{eqnarray}
    \vec{\mathcal{E}}&=&  -\dfrac{1}{c} \dfrac{\partial\,\vec{\mathcal{A}}}{\partial\,t}-\nabla\varphi-{\rm i}\,g\left[\varphi,\ \vec{\mathcal{A}}\right]= -\nabla\varphi-{\rm i}\,g\left[\varphi,\ \vec{\mathcal{A}}\right]\nonumber\\
&=&-{\dfrac{\partial f_1(r)}{\partial r} \left(\vec{r} \cdot \vec{S}\right)\dfrac{\vec{r}}{r}}- f_1(r)\vec{S}-{\dfrac{\partial f_2(r)}{\partial r} \dfrac{\vec{r}}{r}}+ g\hbar k\frac{f_1(r)}{r^2} \left[ r^2 \vec{S} -  \left(\vec{r}\cdot \vec{S}\right) \vec{r}\right]\nonumber\\
&=&{-}\left\{\dfrac{\partial f_1(r)}{\partial r} \dfrac{1}{r}  \left(\vec{r} \cdot \vec{S}\right)+\dfrac{\partial f_2(r)}{\partial r} \dfrac{1}{r} + f_1(r)\dfrac{g \hbar k}{r^2} \left(\vec{r} \cdot \vec{S}\right) \right\}\vec{r}+\left[-1+  g \hbar k \right] f_1(r) \vec{S}.
\end{eqnarray}
Then we can have
\begin{eqnarray}
    \vec{\nabla}\cdot \vec{\mathcal{E}} &=& {-} \vec{\nabla}\cdot\left\{  \left[\dfrac{\partial f_1(r)}{\partial r} \dfrac{1}{r}  \left(\vec{r} \cdot \vec{S}\right)+\dfrac{\partial f_2(r)}{\partial r} \dfrac{1}{r} +  g \hbar f_1(r)\dfrac{k}{r^2} \left(\vec{r} \cdot \vec{S}\right) \right]\vec{r}\right\}
      +\vec{\nabla}\cdot\left\{ \left[-1+  g \hbar k \right] f_1(r) \vec{S} \right\}.
\end{eqnarray}
Due to
\begin{eqnarray}\label{G-3}
\vec{G}=G_r \hat{e}_r+ G_\theta \hat{e}_\theta+G_\phi \hat{e}_\phi,
\end{eqnarray}
\begin{eqnarray}\label{N-1b}
\vec{\nabla}=\hat{e}_r \frac{\partial }{\partial r}+\hat{e}_\theta \frac{1}{r}\frac{\partial }{\partial \theta}+\hat{e}_\phi \frac{1}{r\sin\theta}\frac{\partial }{\partial \phi},
\end{eqnarray}
\begin{eqnarray}
 \label{G-12}
&&  \vec{\nabla}\cdot\vec{G}=\frac{1}{r^2}\biggr[ \frac{\partial (r^2 G_r)}{\partial r}\biggr]+\frac{1}{r\sin\theta}\biggr[ \frac{\partial (\sin\theta G_\theta)}{\partial \theta}\biggr]+\frac{1}{r\sin\theta}\biggr[ \frac{\partial G_\phi}{\partial \phi}\biggr].
 \end{eqnarray}
we can have
\begin{eqnarray}
&&\vec{\nabla}\cdot\left\{  \left[\dfrac{\partial f_1(r)}{\partial r} \dfrac{1}{r}  \left(\vec{r} \cdot \vec{S}\right)+\dfrac{\partial f_2(r)}{\partial r} \dfrac{1}{r} +  g \hbar f_1(r)\dfrac{k}{r^2} \left(\vec{r} \cdot \vec{S}\right) \right]\vec{r}\right\}\nonumber\\
&=& {\vec{\nabla}\cdot\left\{  \left[\dfrac{\partial f_1(r)}{\partial r} \dfrac{1}{r}  \left(\vec{r} \cdot \vec{S}\right)+\dfrac{\partial f_2(r)}{\partial r} \dfrac{1}{r} +  g \hbar f_1(r)\dfrac{k}{r^2} \left(\vec{r} \cdot \vec{S}\right) \right]r \hat{e}_r\right\} }\nonumber\\
&=&\frac{1}{r^2}\frac{\partial}{\partial r}\left\{  \left[\dfrac{\partial f_1(r)}{\partial r} \dfrac{1}{r}  \left(\vec{r} \cdot \vec{S}\right)+\dfrac{\partial f_2(r)}{\partial r} \dfrac{1}{r} +  g \hbar f_1(r)\dfrac{k}{r^2} \left(\vec{r} \cdot \vec{S}\right) \right]r^3\right\}\nonumber\\
&=&3\left[\dfrac{\partial f_1(r)}{\partial r} \dfrac{1}{r}  \left(\vec{r} \cdot \vec{S}\right)+\dfrac{\partial f_2(r)}{\partial r} \dfrac{1}{r} +  g \hbar f_1(r)\dfrac{k}{r^2} \left(\vec{r} \cdot \vec{S}\right) \right]+r\frac{\partial}{\partial r}\left[\dfrac{\partial f_1(r)}{\partial r} \dfrac{1}{r}  \left(\vec{r} \cdot \vec{S}\right)+\dfrac{\partial f_2(r)}{\partial r} \dfrac{1}{r} +  g \hbar f_1(r)\dfrac{k}{r^2} \left(\vec{r} \cdot \vec{S}\right) \right]\nonumber\\
&=&3\left[\dfrac{\partial f_1(r)}{\partial r} \dfrac{1}{r}  \left(\vec{r} \cdot \vec{S}\right)+\dfrac{\partial f_2(r)}{\partial r} \dfrac{1}{r} +  g \hbar f_1(r)\dfrac{k}{r^2} \left(\vec{r} \cdot \vec{S}\right) \right]+r\frac{\partial}{\partial r}\left[\dfrac{\partial f_1(r)}{\partial r}   \left(\hat{r} \cdot \vec{S}\right)+\dfrac{\partial f_2(r)}{\partial r} \dfrac{1}{r} +  g \hbar f_1(r)\dfrac{k}{r} \left(\hat{r} \cdot \vec{S}\right) \right]\nonumber\\
&=&3\left[\dfrac{\partial f_1(r)}{\partial r} \dfrac{1}{r}  \left(\vec{r} \cdot \vec{S}\right)+\dfrac{\partial f_2(r)}{\partial r} \dfrac{1}{r} +  g \hbar f_1(r)\dfrac{k}{r^2} \left(\vec{r} \cdot \vec{S}\right) \right]+r\left(\hat{r} \cdot \vec{S}\right)\frac{\partial}{\partial r}\left[\dfrac{\partial f_1(r)}{\partial r}   +  g \hbar k f_1(r)\dfrac{1}{r} \right]+ r\frac{\partial}{\partial r}\left[\dfrac{\partial f_2(r)}{\partial r} \dfrac{1}{r}\right] \nonumber\\
&=&3\left[\dfrac{\partial f_1(r)}{\partial r} \dfrac{1}{r}  \left(\vec{r} \cdot \vec{S}\right)+\dfrac{\partial f_2(r)}{\partial r} \dfrac{1}{r} +  g \hbar f_1(r)\dfrac{k}{r^2} \left(\vec{r} \cdot \vec{S}\right) \right]+\left(\vec{r} \cdot \vec{S}\right)\left[\dfrac{\partial^2 f_1(r)}{\partial r^2}   +  g \hbar k \dfrac{1}{r}\dfrac{\partial f_1(r)}{\partial r}- g \hbar k f_1(r)\dfrac{1}{r^2}  \right]\nonumber\\
&&+\dfrac{\partial^2 f_2(r)}{\partial r^2}-\frac{1}{r}\dfrac{\partial f_2(r)}{\partial r}  \nonumber\\
&=&\left[\dfrac{\partial^2 f_1(r)}{\partial r^2} +(3+g \hbar k) \dfrac{1}{r}\dfrac{\partial f_1(r)}{\partial r}+ 2 g \hbar k f_1(r)\dfrac{1}{r^2}\right] \left(\vec{r} \cdot \vec{S}\right)  +\left[\dfrac{\partial^2 f_2(r)}{\partial r^2}+2\frac{1}{r}\dfrac{\partial f_2(r)}{\partial r} \right].
\end{eqnarray}
Furthermore, we have
\begin{eqnarray}
    \vec{\nabla}\cdot\left\{ \left[-1+  g \hbar k \right] f_1(r) \vec{S} \right\}&=& \dfrac{\partial }{\partial x}\left\{ \left[-1+  g \hbar k \right] f_1(r) S_x \right\}+\dfrac{\partial }{\partial y}\left\{ \left[-1+  g \hbar k \right] f_1(r) S_y \right\}+\dfrac{\partial }{\partial z}\left\{ \left[-1+  g \hbar k \right] f_1(r) S_z \right\}\nonumber\\
    &=& \left(-1+  g \hbar k \right) \dfrac{\partial f_1(r)}{\partial x} S_x +\left(-1+  g \hbar k \right) \dfrac{\partial f_1(r)}{\partial y} S_y +\left(-1+  g \hbar k \right) \dfrac{\partial f_1(r)}{\partial z} S_z\nonumber\\
    &=& \left(-1+  g \hbar k \right) \dfrac{\partial f_1(r)}{\partial r} \dfrac{x}{r}S_x +\left(-1+  g \hbar k \right) \dfrac{\partial f_1(r)}{\partial r} \dfrac{y}{r} S_y +\left(-1+  g \hbar k \right) \dfrac{\partial f_1(r)}{\partial r} \dfrac{z}{r} S_z\nonumber\\
    &=& \left(-1+  g \hbar k \right) \dfrac{\partial f_1(r)}{\partial r} \dfrac{1}{r}\left(\vec{r} \cdot \vec{S}\right) \nonumber\\
    &=& -\left(1-  g \hbar k \right) \dfrac{\partial f_1(r)}{\partial r} \dfrac{1}{r}\left(\vec{r} \cdot \vec{S}\right).
\end{eqnarray}
Then,
\begin{eqnarray}
    \vec{\nabla}\cdot \vec{\mathcal{E}} &=& {-} \vec{\nabla}\cdot\left\{  \left[\dfrac{\partial f_1(r)}{\partial r} \dfrac{1}{r}  \left(\vec{r} \cdot \vec{S}\right)+\dfrac{\partial f_2(r)}{\partial r} \dfrac{1}{r} +  g \hbar f_1(r)\dfrac{k}{r^2} \left(\vec{r} \cdot \vec{S}\right) \right]\vec{r}\right\}
      +\vec{\nabla}\cdot\left\{ \left[-1+  g \hbar k \right] f_1(r) \vec{S} \right\}\nonumber\\
&=&-\left[\dfrac{\partial^2 f_1(r)}{\partial r^2} + \dfrac{4}{r}\dfrac{\partial f_1(r)}{\partial r}+ 2 g \hbar k f_1(r)\dfrac{1}{r^2}\right] \left(\vec{r} \cdot \vec{S}\right)  -\left[\dfrac{\partial^2 f_2(r)}{\partial r^2}+\frac{2}{r}\dfrac{\partial f_2(r)}{\partial r} \right].
\end{eqnarray}
Due to
 \begin{eqnarray}
 &&   \vec{r}\cdot \left(\vec{r} \times \vec{S}\right)=\left(\vec{r} \times \vec{S}\right)\cdot\vec{r}=0,
 \end{eqnarray}
one easily has
 \begin{eqnarray}
   {\rm i}\,g\Bigl( \vec{\mathcal{A}}\cdot\vec{\mathcal{E}}  -\vec{\mathcal{E}}\cdot\vec{\mathcal{A}}\Bigr) &=&
  {\rm i}\,g\left[-1+  g \hbar k \right] f_1(r) \left[ \vec{\mathcal{A}}\cdot\vec{S} -\vec{S}\cdot\vec{\mathcal{A}}\right]\nonumber\\
 &=& {\rm i}\,g \left[-1+  g \hbar k \right] f_1(r) \dfrac{k}{r^2}\left[ \left(\vec{r} \times \vec{S}\right)\cdot\vec{S} -\vec{S}\cdot\left(\vec{r} \times \vec{S}\right)\right]\nonumber\\
  &=&  {\rm i}\,g \left[-1+  g \hbar k \right] f_1(r) \dfrac{k}{r^2}\left[{\rm i} \hbar \vec{r} \cdot \vec{S} - (-{\rm i} \hbar \vec{r} \cdot \vec{S})\right]\nonumber\\
  &=& 2  \left[1-  g \hbar k \right] f_1(r)\dfrac{g \hbar k}{r^2} \left(\vec{r} \cdot \vec{S}\right).
 \end{eqnarray}
Therefore, from Eq. (\ref{eq:DivEYM}) we have
\begin{eqnarray}
&&\vec{\nabla}\cdot\vec{\mathcal{E}}\ {-} {{\rm i}\,g\Bigl(
                        \vec{\mathcal{A}}\cdot\vec{\mathcal{E}}
                        -\vec{\mathcal{E}}\cdot\vec{\mathcal{A}}\Bigr)}\nonumber\\
&=& -\left[ \left(\dfrac{\partial^2 f_1(r)}{\partial r^2}  +  \dfrac{\partial f_1(r)}{\partial r}\dfrac{4}{r}+ 2f_1(r)\dfrac{g \hbar k}{r^2} \right)\right] \left(\vec{r} \cdot \vec{S}\right) -\left[\dfrac{\partial^2 f_2(r)}{\partial r^2}+ 2 \dfrac{\partial f_2(r)}{\partial r}\dfrac{1}{r}\right]\nonumber\\
&& {-}2 \left[1-  g \hbar k \right] f_1(r)\dfrac{g \hbar k}{r^2} \left(\vec{r} \cdot \vec{S}\right)=0,
   \end{eqnarray}
i.e.,
\begin{eqnarray}
&& \left[ \left(\dfrac{\partial^2 f_1(r)}{\partial r^2}  +  \dfrac{\partial f_1(r)}{\partial r}\dfrac{4}{r}+ 2f_1(r)\dfrac{g \hbar k}{r^2} \right)\right] \left(\vec{r} \cdot \vec{S}\right) +\left[\dfrac{\partial^2 f_2(r)}{\partial r^2}+ 2 \dfrac{\partial f_2(r)}{\partial r}\dfrac{1}{r}\right]\nonumber\\
&& + 2 \left[1-  g \hbar k \right] f_1(r)\dfrac{g \hbar k}{r^2} \left(\vec{r} \cdot \vec{S}\right)=0,
   \end{eqnarray}
i.e.,
\begin{eqnarray}
&& \left[ \dfrac{\partial^2 f_1(r)}{\partial r^2}  +  \dfrac{\partial f_1(r)}{\partial r}\dfrac{4}{r}+ 2  f_1(r)\dfrac{g \hbar k(2-g \hbar k) }{r^2} \right] \left(\vec{r} \cdot \vec{S}\right) +\left[\dfrac{\partial^2 f_2(r)}{\partial r^2}+ 2 \dfrac{\partial f_2(r)}{\partial r}\dfrac{1}{r}\right]=0,
   \end{eqnarray}
i.e.,
\begin{eqnarray}
&& \left[ \dfrac{\partial^2 f_1(r)}{\partial r^2}  +  \dfrac{4}{r} \dfrac{\partial f_1(r)}{\partial r}+  \dfrac{2 g \hbar k(2-g \hbar k)}{r^2} f_1(r) \right] \left(\vec{r} \cdot \vec{S}\right) +\left[\dfrac{\partial^2 f_2(r)}{\partial r^2}+ \dfrac{2}{r} \dfrac{\partial f_2(r)}{\partial r}\right]=0.
   \end{eqnarray}
Based on above, Eq. (\ref{eq:DivEYM}) yields the following conditions for two functions $f_1(r)$ and $f_2(r)$:
\begin{subequations}
\begin{eqnarray}
&& \dfrac{\partial^2 f_2(r)}{\partial r^2}+ \dfrac{2}{r} \dfrac{\partial f_2(r)}{\partial r}=0, \label{eq:con-1}\\
&&  \dfrac{\partial^2 f_1(r)}{\partial r^2}  +  \dfrac{4}{r} \dfrac{\partial f_1(r)}{\partial r}+   \dfrac{2 g \hbar k(2-g \hbar k)}{r^2} f_1(r) =0. \label{eq:con-2}
\end{eqnarray}
\end{subequations}
\emph{Analysis 2: The study of Eq. (\ref{eq:CurlBYM})---}Let's calculate the ``magnetic'' field. Because
\begin{eqnarray}
(\vec{\nabla} \times \vec{\mathcal{A}})_z &=&k \left(\vec{\nabla}\times\frac{\vec{r}\times\vec{S}}{r^2}\right)_z
= k\left[\partial_x\left(\frac{(\vec{r}\times\vec{S})_y}{r^2}\right)-\partial_y\left(\frac{(\vec{r}\times\vec{S})_x}{r^2}\right)\right]\nonumber\\
&=&k\left[\partial_x\left(\frac{z S_x-xS_z}{r^2}\right)-\partial_y\left(\frac{yS_z-zS_y}{r^2}\right)\right]\nonumber\\
&=&k\left[ \frac{x^2 S_z - y^2 S_z - z^2 S_z - 2xz S_x}{r^4}-\frac{x^2 S_z - y^2 S_z + z^2 S_z + 2yz S_y}{r^4}\right]\nonumber\\
&=&k\frac{ -2z^2 S_z - 2xz S_x - 2yz S_y}{r^4}\nonumber\\
&=&-2k\frac{ x S_x +y S_y+ z S_z}{r^4}z\nonumber\\
&=&-2k \frac{\vec{r}\cdot \vec{S}}{r^4}z,
\end{eqnarray}
\begin{eqnarray}
(\vec{\nabla} \times \vec{\mathcal{A}}) &=&-2k \frac{\vec{r}\cdot \vec{S}}{r^4} {\vec{r}},
\end{eqnarray}
\begin{eqnarray}
( \vec{\mathcal{A}} \times \vec{\mathcal{A}})_z&=& k^2 (\frac{\vec{r}\times\vec{S}}{r^2}\times\frac{\vec{r}\times\vec{S}}{r^2})_z\nonumber\\
&=&k^2\frac{1}{r^4}[yS_z-zS_y,z S_x-xS_z]\nonumber\\
&=& {\rm i} \hbar k^2\frac{1}{r^4}(yzS_y+z^2S_z+zxS_x)\nonumber\\
&=& {\rm i} \hbar k^2\frac{1}{r^4}(xS_x+yS_y+zS_z)z\nonumber\\
&=& {\rm i} \hbar \; k^2 \; \frac{\vec{r}\cdot\vec{S}}{r^4} z,
\end{eqnarray}
\begin{eqnarray}
&& \vec{\mathcal{A}} \times \vec{\mathcal{A}}= {\rm i} \hbar \; k^2 \; \frac{\vec{r}\cdot\vec{S}}{r^4} \; \vec{r},
\end{eqnarray}
we have
\begin{eqnarray}
\vec{\mathcal{B}} &=& \vec{\nabla}\times\vec{\mathcal{A}} -{\rm i}\,g\bigl(\vec{\mathcal{A}}\times\vec{\mathcal{A}}\bigr) \nonumber\\
&=& -2k \frac{\vec{r}\cdot \vec{S}}{r^4}\vec{r} - {\rm i}\,g \left[{\rm i} \hbar \; k^2 \; \frac{\vec{r}\cdot\vec{S}}{r^4} \; \vec{r}\right]\nonumber\\
&=& -2k \frac{\vec{r}\cdot \vec{S}}{r^4}\vec{r} + g  \hbar \; k^2 \; \frac{\vec{r}\cdot\vec{S}}{r^4} \; \vec{r}\nonumber\\
&=&(-2+g \hbar k)k \frac{\vec{r}\cdot \vec{S}}{r^4}\vec{r}\nonumber\\
&=&-(2-g \hbar k)k \frac{\vec{r}\cdot \vec{S}}{r^4}\vec{r}.
\end{eqnarray}
Since $\vec{\mathcal{E}}$ is time-independent, then Eq. (\ref{eq:CurlBYM}) becomes
\begin{eqnarray}
 && \vec{\nabla}\times\vec{\mathcal{B}} {-} {\rm i}\,g\biggl(\Bigl[\varphi,\
                                    \vec{\mathcal{E}}\Bigr]
                              +\vec{\mathcal{A}}\times\vec{\mathcal{B}}
                              +\vec{\mathcal{B}}\times\vec{\mathcal{A}}\biggr)=0, \label{eq:CurlBYM-1}
            \end{eqnarray}
 Let us study the $z$-component of Eq. (\ref{eq:CurlBYM-1}), which is
\begin{eqnarray}
 && \left(\vec{\nabla}\times\vec{\mathcal{B}}\right)_z {-} {\rm i}\,g\biggl(\Bigl[\varphi,\
                                    {\mathcal{E}}_z\Bigr]
                              +\left(\vec{\mathcal{A}}\times\vec{\mathcal{B}}\right)_z
                              +\left(\vec{\mathcal{B}}\times\vec{\mathcal{A}}\right)_z\biggr)=0. \label{eq:CurlBYM-2}
 \end{eqnarray}
One can have
\begin{eqnarray}
   \left[\varphi, \vec{\mathcal{E}}_z\right] & =& \left[\varphi, -\left\{\dfrac{\partial f_1(r)}{\partial r} \dfrac{1}{r}  \left(\vec{r} \cdot \vec{S}\right)+\dfrac{\partial f_2(r)}{\partial r} \dfrac{1}{r} + f_1(r)\dfrac{g \hbar k}{r^2} \left(\vec{r} \cdot \vec{S}\right) \right\}z
      +\left[-1+  g \hbar k \right] f_1(r) {S}_z\right]\nonumber\\
   &=& \left(-1+  g \hbar k \right) f_1(r)\left[\varphi,  {S}_z\right]\nonumber\\
   &=& \left(-1+  g \hbar k \right) f_1(r)\left[f_1(r) \left(\vec{r} \cdot \vec{S}\right) +  f_2(r),  {S}_z\right]\nonumber\\
   &=& \left(-1+  g \hbar k \right) f^2_1(r)\left[\vec{r} \cdot \vec{S},  {S}_z\right]\nonumber\\
   &=& \left(-1+  g \hbar k \right) f^2_1(r)\left[x S_x+ y S_y+ z S_z,  {S}_z\right]\nonumber\\
   &=& \left(-1+  g \hbar k \right) f^2_1(r)\left[ (-{\rm i} \hbar) x S_y+ ({\rm i} \hbar) y S_x\right]\nonumber\\
      &=& \left(-1+  g \hbar k \right) f^2_1(r) {\left[-({\rm i} \hbar)\left(\vec{r}\times \vec{S}\right)_z\right]}\nonumber\\
   &=& ({\rm i} \hbar)\left(1-  g \hbar k \right) f^2_1(r) \left(\vec{r}\times \vec{S}\right)_z.
   \end{eqnarray}
Because
\begin{eqnarray}
   \vec{\mathcal{A}}\times\vec{\mathcal{B}} &=& \left(k \frac{\vec{r} \times \vec{S}}{r^2}\right)\times \left[-(2-g \hbar k)k \frac{\vec{r}\cdot \vec{S}}{r^4}\vec{r}\right]\nonumber\\
   &=& -(2-g \hbar k)k^2 \dfrac{1}{r^6}\left[\left( \vec{r} \times \vec{S}\right)\times\vec{r}\right] \left(\vec{r}\cdot \vec{S}\right)\nonumber\\
  &=& -(2-g \hbar k)k^2 \dfrac{1}{r^6} \left[(\vec{r}\cdot\vec{r})\vec{S}-\vec{r}(\vec{S}\cdot\vec{r})\right] \left(\vec{r}\cdot \vec{S}\right)\nonumber\\
  &=& -(2-g \hbar k)k^2  \left[\dfrac{1}{r^4}\vec{S}\left(\vec{r}\cdot \vec{S}\right)-\dfrac{1}{r^6}\vec{r}(\vec{r}\cdot\vec{S})^2\right],
\end{eqnarray}
\begin{eqnarray}
   \vec{\mathcal{B}}\times\vec{\mathcal{A}} &=& - \left(\vec{\mathcal{A}}\times\vec{\mathcal{B}}\right)^\dag = (2-g \hbar k)k^2  \left[\dfrac{1}{r^4}\left(\vec{r}\cdot \vec{S}\right) \vec{S}-\dfrac{1}{r^6}\vec{r}(\vec{r}\cdot\vec{S})^2\right],
\end{eqnarray}
 thus
\begin{eqnarray}
 && \left(\vec{\mathcal{A}}\times\vec{\mathcal{B}}\right)_z
  +\left(\vec{\mathcal{B}}\times\vec{\mathcal{A}}\right)_z
  =(2-g \hbar k)k^2  \dfrac{1}{r^4} \left[\vec{r}\cdot \vec{S}, S_z\right]={-}({\rm i} \hbar)(2-g \hbar k)k^2  \dfrac{1}{r^4}\left(\vec{r}\times \vec{S}\right)_z.
 \end{eqnarray}
 One can have
 \begin{eqnarray}
  &&\left(\vec{\nabla}\times\vec{\mathcal{B}}\right)_z =  \dfrac{\partial \mathcal{B}_y}{\partial x} - \dfrac{\partial \mathcal{B}_x}{\partial y}\nonumber\\
  &=&  \dfrac{\partial }{\partial x}\left[-(2-g \hbar k)k \frac{\vec{r}\cdot \vec{S}}{r^4}y\right] - \dfrac{\partial }{\partial y}\left[-(2-g \hbar k)k \frac{\vec{r}\cdot \vec{S}}{r^4}x\right]\nonumber\\
    &=&  -(2-g \hbar k)k\dfrac{\partial }{\partial x}\left[ \frac{\vec{r}\cdot \vec{S}}{r^4}\right]y +(2-g \hbar k)k \dfrac{\partial }{\partial y}\left[\frac{\vec{r}\cdot \vec{S}}{r^4}\right]x\nonumber\\
   &=&  -(2-g \hbar k)k \left\{\dfrac{\partial }{\partial x}\left[ \frac{1}{r^4}\right] \left(\vec{r}\cdot \vec{S}\right)y
   +\dfrac{\partial \left(\vec{r}\cdot \vec{S}\right)}{\partial x}\frac{1}{r^4} y \right\}+(2-g \hbar k)k \left\{\dfrac{\partial }{\partial y}\left[ \frac{1}{r^4}\right] \left(\vec{r}\cdot \vec{S}\right)x
   +\dfrac{\partial \left(\vec{r}\cdot \vec{S}\right)}{\partial y}\frac{1}{r^4} x \right\}\nonumber\\
   &=&  -(2-g \hbar k)k \left\{\dfrac{\partial }{\partial r}\left[ \frac{1}{r^4}\right] \dfrac{x}{r}\left(\vec{r}\cdot \vec{S}\right)y
   +S_x\frac{1}{r^4} y \right\}+(2-g \hbar k)k \left\{\dfrac{\partial }{\partial r}\left[ \frac{1}{r^4}\right] \dfrac{y}{r}\left(\vec{r}\cdot \vec{S}\right)x
   +S_y\frac{1}{r^4} x \right\}\nonumber\\
     &=& (2-g \hbar k)k \frac{1}{r^4} \left( x S_y-y S_x\right)\nonumber\\
   &=& (2-g \hbar k)k \frac{1}{r^4} \left(\vec{r}\times \vec{S}\right)_z.
 \end{eqnarray}
Consequently, one has
\begin{eqnarray}
  && \left(\vec{\nabla}\times\vec{\mathcal{B}}\right)_z {-} {\rm i}\,g\biggl(\Bigl[\varphi,\
                                    {\mathcal{E}}_z\Bigr]
                              +\left(\vec{\mathcal{A}}\times\vec{\mathcal{B}}\right)_z
                              +\left(\vec{\mathcal{B}}\times\vec{\mathcal{A}}\right)_z\biggr)\nonumber\\
 &=& (2-g \hbar k)k \frac{1}{r^4} \left(\vec{r}\times \vec{S}\right)_z {-} {\rm i}\,g \left\{ ({\rm i} \hbar)\left(1-  g \hbar k \right) f^2_1(r) \left(\vec{r}\times \vec{S}\right)_z {-} ({\rm i} \hbar)(2-g \hbar k)k^2  \dfrac{1}{r^4}\left(\vec{r}\times \vec{S}\right)_z\right\}\nonumber\\
 &=& {(2-g \hbar k)k \frac{1}{r^4} \left(\vec{r}\times \vec{S}\right)_z + g \left\{ \hbar\left(1-  g \hbar k \right) f^2_1(r) \left(\vec{r}\times \vec{S}\right)_z {-} \hbar (2-g \hbar k)k^2  \dfrac{1}{r^4}\left(\vec{r}\times \vec{S}\right)_z\right\}}\nonumber\\
 &=&\left[g\hbar \left(1-  g \hbar k \right) f^2_1(r) + (2-g \hbar k)k \frac{1}{r^4}{-}g\hbar k (2-g \hbar k)k \dfrac{1}{r^4}\right] \left(\vec{r}\times \vec{S}\right)_z\nonumber\\
 &=&\left[g\hbar \left(1-  g \hbar k \right) f^2_1(r) + \left(1 {-}  g \hbar k \right) (2-g \hbar k)k \frac{1}{r^4}\right] \left(\vec{r}\times \vec{S}\right)_z\nonumber\\
&=&\left(1-  g \hbar k \right)\left[g\hbar  f^2_1(r) + (2-g \hbar k)k \frac{1}{r^4}\right] \left(\vec{r}\times \vec{S}\right)_z,
 \end{eqnarray}
and generally
\begin{eqnarray}
   \left(\vec{\nabla}\times\vec{\mathcal{B}}\right) {-} {\rm i}\,g\biggl(\Bigl[\varphi,\
                                    \vec{\mathcal{E}}\Bigr]
                              +\left(\vec{\mathcal{A}}\times\vec{\mathcal{B}}\right)
                              +\left(\vec{\mathcal{B}}\times\vec{\mathcal{A}}\right)\biggr)=\left(1-  g \hbar k \right)\left[g\hbar  f^2_1(r) + (2-g \hbar k)k \frac{1}{r^4}\right] \left(\vec{r}\times \vec{S}\right).
 \end{eqnarray}
Therefore,  Eq. (\ref{eq:CurlBYM}) leads to the following condition for the function $f_1(r)$ as
\begin{eqnarray}\label{eq:con-3}
\left(1-  g \hbar k \right)\left[g\hbar  f^2_1(r) + (2-g \hbar k)k \frac{1}{r^4}\right]=0.
 \end{eqnarray}
\emph{Analysis 3: The study of Eq. (\ref{eq:DivBYM})---}Because
\begin{equation}
      \vec{\mathcal{A}} = k \frac{\vec{r} \times \vec{S}}{r^2},
\end{equation}
\begin{eqnarray}
\vec{\mathcal{B}} &=&-(2-g \hbar k)k \frac{\vec{r}\cdot \vec{S}}{r^4}\vec{r},
\end{eqnarray}
then we have
\begin{eqnarray}
 && {\rm i}\,g\Bigl( \vec{\mathcal{A}}\cdot\vec{\mathcal{B}}
                        -\vec{\mathcal{B}}\cdot\vec{\mathcal{A}}\Bigr)=0,
\end{eqnarray}
and
\begin{eqnarray}
      \vec{\nabla} \cdot \vec{\mathcal{B}} &=& -(2-g \hbar k)k \vec{\nabla} \cdot \left[\frac{\vec{r}\cdot \vec{S}}{r^4}\vec{r}\right]\nonumber\\
      &=& -(2-g \hbar k)k \left\{ \dfrac{\partial }{\partial x}\left(\frac{\vec{r}\cdot \vec{S}}{r^4}x\right)
      + \dfrac{\partial }{\partial y}\left(\frac{\vec{r}\cdot \vec{S}}{r^4}y\right) + \dfrac{\partial }{\partial z}\left(\frac{\vec{r}\cdot \vec{S}}{r^4}z\right) \right\}\nonumber\\
 &=& -(2-g \hbar k)k \left\{\left[ \dfrac{\partial }{\partial x}\left(\frac{\vec{r}\cdot \vec{S}}{r^4}\right)\right]x
      + \left[ \dfrac{\partial }{\partial y}\left(\frac{\vec{r}\cdot \vec{S}}{r^4}\right)\right]y + \left[\dfrac{\partial }{\partial z}\left(\frac{\vec{r}\cdot \vec{S}}{r^4}\right)\right]z + 3 \left(\frac{\vec{r}\cdot \vec{S}}{r^4}\right)\right\}\nonumber\\
 &=& -(2-g \hbar k)k \left\{\left[ \dfrac{\partial }{\partial r}\left(\frac{\vec{r}\cdot \vec{S}}{r^4}\right)\right]\dfrac{x^2}{r}
      +\left[ \dfrac{\partial }{\partial r}\left(\frac{\vec{r}\cdot \vec{S}}{r^4}\right)\right]\dfrac{y^2}{r} +\left[ \dfrac{\partial }{\partial r}\left(\frac{\vec{r}\cdot \vec{S}}{r^4}\right)\right]\dfrac{z^2}{r}+ 3 \left(\frac{\vec{r}\cdot \vec{S}}{r^4}\right)\right\}\nonumber\\
 &=& -(2-g \hbar k)k \left\{\left[ \dfrac{\partial }{\partial r}\left(\frac{\vec{r}\cdot \vec{S}}{r^4}\right)\right]r+ 3 \left(\frac{\vec{r}\cdot \vec{S}}{r^4}\right)\right\}\nonumber\\
  &=& -(2-g \hbar k)k \left\{\left[ \dfrac{\partial }{\partial r}\left(\frac{1}{r^3} \left(\hat{r}\cdot \vec{S}\right)\right)\right]r+ 3 \left(\frac{\vec{r}\cdot \vec{S}}{r^4}\right)\right\} \nonumber\\
  &=& -(2-g \hbar k)k \left\{\left[ \dfrac{\partial }{\partial r}\left(\frac{1}{r^3} \right)\right]r \left(\hat{r}\cdot \vec{S}\right)+ 3 \left(\frac{\vec{r}\cdot \vec{S}}{r^4}\right)\right\}\nonumber\\
    &=& -(2-g \hbar k)k \left\{\left[ \dfrac{\partial }{\partial r}\left(\frac{1}{r^3} \right)\right] \left(\vec{r}\cdot \vec{S}\right)+ 3 \left(\frac{\vec{r}\cdot \vec{S}}{r^4}\right)\right\}\nonumber\\
        &=& -(2-g \hbar k)k \left\{-3\dfrac{1}{r^4} \left(\vec{r}\cdot \vec{S}\right)+ 3 \left(\frac{\vec{r}\cdot \vec{S}}{r^4}\right)\right\}\nonumber\\
        &=&0.
\end{eqnarray}
therefore
\begin{eqnarray}
&& \vec{\nabla}\cdot\vec{\mathcal{B}}\ {-}{\rm i}\,g\Bigl(
                        \vec{\mathcal{A}}\cdot\vec{\mathcal{B}}
                        -\vec{\mathcal{B}}\cdot\vec{\mathcal{A}}\Bigr)=0
\end{eqnarray}
is automatically satisfied. Namely, Eq. (\ref{eq:DivBYM}) has no conditions for two functions $f_1(r)$ and $f_2(r)$.

\emph{Analysis 4: The study of Eq. (\ref{eq:CurlEYM})---} Since $\vec{\mathcal{B}}$ is time-independent, then Eq. (\ref{eq:CurlEYM}) becomes
\begin{eqnarray}
&&   -\vec{\nabla}\times\vec{\mathcal{E}} {-}{\rm i}\,g\biggl(\Bigl[\varphi,\
                                    \vec{\mathcal{B}}\Bigr]
                              -\vec{\mathcal{A}}\times\vec{\mathcal{E}}
                              -\vec{\mathcal{E}}\times\vec{\mathcal{A}}\biggr)=0. \label{eq:CurlEYM-1}
\end{eqnarray}
 Let us study the $z$-component of Eq. (\ref{eq:CurlEYM-1}), which is
\begin{eqnarray}
&&   -\left(\vec{\nabla}\times\vec{\mathcal{E}}\right)_z {-}{\rm i}\,g\biggl(\Bigl[\varphi,\
                                    {\mathcal{B}}_z\Bigr]
                              -\left(\vec{\mathcal{A}}\times\vec{\mathcal{E}}\right)_z
                              -\left(\vec{\mathcal{E}}\times\vec{\mathcal{A}}\right)_z\biggr)=0. \label{eq:CurlEYM-2}
\end{eqnarray}
One can have
\begin{eqnarray}
&&   \left[\varphi,\ {\mathcal{B}}_z\right]=  \left[f_1(r) \left(\vec{r} \cdot \vec{S}\right) +  f_2(r),\ -(2-g \hbar k)k \frac{\vec{r}\cdot \vec{S}}{r^4} z\right]=0.
\end{eqnarray}
and
\begin{eqnarray}
 && \left(\vec{\nabla}\times\vec{\mathcal{E}}\right)_z = \dfrac{\partial \mathcal{E}_y}{\partial x} - \dfrac{\partial \mathcal{E}_x}{\partial y}\nonumber\\
  &=& \dfrac{\partial }{\partial x}\left[-\left\{\dfrac{\partial f_1(r)}{\partial r} \dfrac{1}{r}  \left(\vec{r} \cdot \vec{S}\right)+\dfrac{\partial f_2(r)}{\partial r} \dfrac{1}{r} + f_1(r)\dfrac{g \hbar k}{r^2} \left(\vec{r} \cdot \vec{S}\right) \right\}y
      +\left[-1+  g \hbar k \right] f_1(r) S_y\right]\nonumber\\
  &&-\dfrac{\partial }{\partial y}\left[-\left\{\dfrac{\partial f_1(r)}{\partial r} \dfrac{1}{r}  \left(\vec{r} \cdot \vec{S}\right)+\dfrac{\partial f_2(r)}{\partial r} \dfrac{1}{r} + f_1(r)\dfrac{g \hbar k}{r^2} \left(\vec{r} \cdot \vec{S}\right) \right\}x
      +\left[-1+  g \hbar k \right] f_1(r) S_x\right]\nonumber\\
  &=& -\dfrac{\partial }{\partial x}\left\{\dfrac{\partial f_1(r)}{\partial r} \dfrac{1}{r}  \left(\vec{r} \cdot \vec{S}\right)+\dfrac{\partial f_2(r)}{\partial r} \dfrac{1}{r} + f_1(r)\dfrac{g \hbar k}{r^2} \left(\vec{r} \cdot \vec{S}\right) \right\} y
      +\dfrac{\partial }{\partial x}\left\{\left[-1+  g \hbar k \right] f_1(r) \right\}S_y\nonumber\\
  && +\dfrac{\partial }{\partial y}\left\{\dfrac{\partial f_1(r)}{\partial r} \dfrac{1}{r}  \left(\vec{r} \cdot \vec{S}\right)+\dfrac{\partial f_2(r)}{\partial r} \dfrac{1}{r} + f_1(r)\dfrac{g \hbar k}{r^2} \left(\vec{r} \cdot \vec{S}\right) \right\} x
      -\dfrac{\partial }{\partial y}\left\{\left[-1+  g \hbar k \right] f_1(r) \right\}S_x\nonumber\\
 &=& -\left\{ \left[ \dfrac{\partial }{\partial x }\left(\dfrac{\partial f_1(r)}{\partial r} \dfrac{1}{r} +  f_1(r)\dfrac{g \hbar k}{r^2} \right)\right]\left(\vec{r} \cdot \vec{S}\right)+ \left(\dfrac{\partial f_1(r)}{\partial r} \dfrac{1}{r} +  f_1(r)\dfrac{g \hbar k}{r^2} \right)
  \dfrac{\partial \left(\vec{r} \cdot \vec{S}\right) }{\partial x }
  +\dfrac{\partial }{\partial x}\left(\dfrac{\partial f_2(r)}{\partial r} \dfrac{1}{r}\right)  \right\}y \nonumber\\
  &&+ \left[-1+  g \hbar k \right] \dfrac{\partial f_1(r) }{\partial r} \dfrac{x}{r}S_y\nonumber\\
  &&+\left\{ \left[ \dfrac{\partial }{\partial y }\left(\dfrac{\partial f_1(r)}{\partial r} \dfrac{1}{r} +  f_1(r)\dfrac{g \hbar k}{r^2} \right)\right]\left(\vec{r} \cdot \vec{S}\right)+ \left(\dfrac{\partial f_1(r)}{\partial r} \dfrac{1}{r} +  f_1(r)\dfrac{g \hbar k}{r^2} \right)
  \dfrac{\partial \left(\vec{r} \cdot \vec{S}\right) }{\partial y }
  +\dfrac{\partial }{\partial y}\left(\dfrac{\partial f_2(r)}{\partial r} \dfrac{1}{r}\right)  \right\}x \nonumber\\
   &&- \left[-1+  g \hbar k \right] \dfrac{\partial f_1(r) }{\partial r} \dfrac{y}{r}S_x\nonumber\\
   &=& -\left\{ \left[ \dfrac{\partial }{\partial r }\left(\dfrac{\partial f_1(r)}{\partial r} \dfrac{1}{r} +  f_1(r)\dfrac{g \hbar k}{r^2} \right)\right]\dfrac{x}{r} \left(\vec{r} \cdot \vec{S}\right)+ \left(\dfrac{\partial f_1(r)}{\partial r} \dfrac{1}{r} +  f_1(r)\dfrac{g \hbar k}{r^2} \right)S_x
  +\dfrac{\partial }{\partial r}\left(\dfrac{\partial f_2(r)}{\partial r} \dfrac{1}{r}\right) \dfrac{x}{r} \right\}y \nonumber\\
  &&+ \left[-1+  g \hbar k \right] \dfrac{\partial f_1(r) }{\partial r} \dfrac{x}{r}S_y\nonumber\\
  &&+\left\{ \left[ \dfrac{\partial }{\partial r }\left(\dfrac{\partial f_1(r)}{\partial r} \dfrac{1}{r} +  f_1(r)\dfrac{g \hbar k}{r^2} \right)\right] \dfrac{y}{r}\left(\vec{r} \cdot \vec{S}\right)+ \left(\dfrac{\partial f_1(r)}{\partial r} \dfrac{1}{r} +  f_1(r)\dfrac{g \hbar k}{r^2} \right)S_y
  +\dfrac{\partial }{\partial r}\left(\dfrac{\partial f_2(r)}{\partial r} \dfrac{1}{r}\right)\dfrac{y}{r}  \right\}x \nonumber\\
   &&- \left[-1+  g \hbar k \right] \dfrac{\partial f_1(r) }{\partial r} \dfrac{y}{r}S_x\nonumber\\
  &=& \left(\dfrac{\partial f_1(r)}{\partial r} \dfrac{1}{r} +  f_1(r)\dfrac{g \hbar k}{r^2} \right)\left( x S_y- y S_x\right)
  + \left[-1+  g \hbar k \right] \dfrac{\partial f_1(r) }{\partial r} \dfrac{1}{r}\left( x S_y- y S_x\right)\nonumber\\
  &=& g \hbar k \left\{\dfrac{\partial f_1(r) }{\partial r} \dfrac{1}{r} +\dfrac{1}{r^2}f_1(r) \right\} \left( \vec{r}\times \vec{S}\right)_z.
\end{eqnarray}

\begin{remark}
Here we have used the following relations
\begin{eqnarray}
&&(\vec{r} \times \vec{S})\times \vec{r}=(\vec{r}\cdot\vec{r})\vec{S}-\vec{r}(\vec{S}\cdot\vec{r})=r^2\vec{S}-\vec{r}(\vec{r}\cdot\vec{S}), \end{eqnarray}
\begin{eqnarray}
&&(\vec{S} \times \vec{r})\times \vec{S}=(\vec{S}\cdot\vec{S})\vec{r}-\vec{S}(\vec{r}\cdot\vec{S})=\vec{S}^2 \vec{r}-\vec{S}(\vec{r}\cdot\vec{S}). \end{eqnarray}
In general, if $\vec{a}$ and $\vec{b}$ are commutative, then one has
\begin{eqnarray}
&&(\vec{a} \times \vec{b})\times \vec{a}=(\vec{a}\cdot\vec{a})\vec{b}-\vec{a}(\vec{b}\cdot\vec{a}).
\end{eqnarray}
Or, more generally, if $\vec{b}$ and $\vec{c}$ are commutative, then one has
\begin{eqnarray}
&&\vec{a} \times (\vec{b}\times \vec{c})=(\vec{a}\cdot\vec{c})\vec{b}-(\vec{a}\cdot \vec{b})\vec{c},\nonumber\\
&&(\vec{a} \times \vec{b})\times \vec{c}=(\vec{a}\cdot\vec{c})\vec{b}-\vec{a}(\vec{b}\cdot\vec{c}).
\end{eqnarray}
$\blacksquare$
\end{remark}
Furthermore, due to the above relations, one has
\begin{eqnarray}
  \vec{\mathcal{A}}\times\vec{\mathcal{E}}
&=&\left(k \frac{\vec{r} \times \vec{S}}{r^2}\right)\times \left[{-}\left\{\dfrac{\partial f_1(r)}{\partial r} \dfrac{1}{r}  \left(\vec{r} \cdot \vec{S}\right)+\dfrac{\partial f_2(r)}{\partial r} \dfrac{1}{r} + f_1(r)\dfrac{g \hbar k}{r^2} \left(\vec{r} \cdot \vec{S}\right) \right\}\vec{r}
      +\left[-1+  g \hbar k \right] f_1(r) \vec{S}\right]\nonumber\\
&=& -\dfrac{k}{r^2} \left[(\vec{r} \times \vec{S})\times \vec{r}\right]\left\{\dfrac{\partial f_1(r)}{\partial r} \dfrac{1}{r}  \left(\vec{r} \cdot \vec{S}\right)+\dfrac{\partial f_2(r)}{\partial r} \dfrac{1}{r} + f_1(r)\dfrac{g \hbar k}{r^2} \left(\vec{r} \cdot \vec{S}\right) \right\}-\dfrac{k}{r^2} \left[(\vec{r} \times \vec{S})\times \vec{S}\right]\left[1- g \hbar k \right] f_1(r)\nonumber\\
&=& -\dfrac{k}{r^2} \left[(\vec{r}\cdot\vec{r})\vec{S}-\vec{r}(\vec{S}\cdot\vec{r})\right]\left\{\dfrac{\partial f_1(r)}{\partial r} \dfrac{1}{r}  \left(\vec{r} \cdot \vec{S}\right)+\dfrac{\partial f_2(r)}{\partial r} \dfrac{1}{r} + f_1(r)\dfrac{g \hbar k}{r^2} \left(\vec{r} \cdot \vec{S}\right) \right\}\nonumber\\
&&+\dfrac{k}{r^2} \left[(\vec{S} \times \vec{r})\times \vec{S}\right]\left[1- g \hbar k \right] f_1(r)\nonumber\\
&=& -\dfrac{k}{r^2} \left[r^2\vec{S}-\vec{r}(\vec{r}\cdot\vec{S})\right]\left\{\dfrac{\partial f_1(r)}{\partial r} \dfrac{1}{r}  \left(\vec{r} \cdot \vec{S}\right)+\dfrac{\partial f_2(r)}{\partial r} \dfrac{1}{r} + f_1(r)\dfrac{g \hbar k}{r^2} \left(\vec{r} \cdot \vec{S}\right) \right\}\nonumber\\
&&+\dfrac{k}{r^2} \left[(\vec{S}\cdot\vec{S})\vec{r}-\vec{S}(\vec{r}\cdot\vec{S})\right]\left[1- g \hbar k \right] f_1(r),
\end{eqnarray}
thus
\begin{eqnarray}
  \left(\vec{\mathcal{A}}\times\vec{\mathcal{E}}\right)_z
&=& -\dfrac{k}{r^2} \left[r^2 {S}_z-z(\vec{r}\cdot\vec{S})\right]\left\{\dfrac{\partial f_1(r)}{\partial r} \dfrac{1}{r}  \left(\vec{r} \cdot \vec{S}\right)+\dfrac{\partial f_2(r)}{\partial r} \dfrac{1}{r} + f_1(r)\dfrac{g \hbar k}{r^2} \left(\vec{r} \cdot \vec{S}\right) \right\}\nonumber\\
&&+\dfrac{k}{r^2} \left[(\vec{S}\cdot\vec{S})z-S_z(\vec{r}\cdot\vec{S})\right]\left[1- g \hbar k \right] f_1(r).
\end{eqnarray}
Since $\vec{\mathcal{A}}$ and $\vec{\mathcal{E}}$ are hermitian, so
\begin{eqnarray}
 \left(\vec{\mathcal{E}}\times\vec{\mathcal{A}}\right)_z &=&-\left\{ \left(\vec{\mathcal{A}}\times\vec{\mathcal{E}}\right)_z\right\}^\dag= \left\{\dfrac{\partial f_1(r)}{\partial r} \dfrac{1}{r}  \left(\vec{r} \cdot \vec{S}\right)+\dfrac{\partial f_2(r)}{\partial r} \dfrac{1}{r} + f_1(r)\dfrac{g \hbar k}{r^2} \left(\vec{r} \cdot \vec{S}\right) \right\} \dfrac{k}{r^2} \left[r^2 {S}_z-z(\vec{r}\cdot\vec{S})\right]\nonumber\\
 &&-\dfrac{k}{r^2} \left[(\vec{S}\cdot\vec{S})z-(\vec{r}\cdot\vec{S})S_z\right]\left[1- g \hbar k \right] f_1(r).
\end{eqnarray}
Therefore
\begin{eqnarray}
&&  \left(\vec{\mathcal{A}}\times\vec{\mathcal{E}}\right)_z + \left(\vec{\mathcal{E}}\times\vec{\mathcal{A}}\right)_z\nonumber\\
&=& -\dfrac{k}{r^2} \left[r^2 {S}_z\right]\left\{\dfrac{\partial f_1(r)}{\partial r} \dfrac{1}{r}  \left(\vec{r} \cdot \vec{S}\right)+\dfrac{\partial f_2(r)}{\partial r} \dfrac{1}{r} + f_1(r)\dfrac{g \hbar k}{r^2} \left(\vec{r} \cdot \vec{S}\right) \right\}+\dfrac{k}{r^2} \left[-S_z(\vec{r}\cdot\vec{S})\right]\left[1- g \hbar k \right] f_1(r)\nonumber\\
&& +\left\{\dfrac{\partial f_1(r)}{\partial r} \dfrac{1}{r}  \left(\vec{r} \cdot \vec{S}\right)+\dfrac{\partial f_2(r)}{\partial r} \dfrac{1}{r} + f_1(r)\dfrac{g \hbar k}{r^2} \left(\vec{r} \cdot \vec{S}\right) \right\}\dfrac{k}{r^2} \left[r^2 {S}_z\right]+\dfrac{k}{r^2} \left[(\vec{r}\cdot\vec{S})S_z\right]\left[1- g \hbar k \right] f_1(r)\nonumber\\
&=& -\dfrac{k}{r^2} \left[r^2 {S}_z\right]\left\{\dfrac{\partial f_1(r)}{\partial r} \dfrac{1}{r}  \left(\vec{r} \cdot \vec{S}\right) + f_1(r)\dfrac{g \hbar k}{r^2} \left(\vec{r} \cdot \vec{S}\right) \right\}+\dfrac{k}{r^2} \left[-S_z(\vec{r}\cdot\vec{S})\right]\left[1- g \hbar k \right] f_1(r)\nonumber\\
&& +\left\{\dfrac{\partial f_1(r)}{\partial r} \dfrac{1}{r}  \left(\vec{r} \cdot \vec{S}\right)+ f_1(r)\dfrac{g \hbar k}{r^2} \left(\vec{r} \cdot \vec{S}\right) \right\}\dfrac{k}{r^2} \left[r^2 {S}_z\right]+\dfrac{k}{r^2} \left[(\vec{r}\cdot\vec{S})S_z\right]\left[1- g \hbar k \right] f_1(r)\nonumber\\
&=& -k \left\{\dfrac{\partial f_1(r)}{\partial r} \dfrac{1}{r}   + f_1(r)\dfrac{g \hbar k}{r^2}  \right\}\left[{S}_z \left(\vec{r} \cdot \vec{S}\right)\right]-k\dfrac{1- g \hbar k }{r^2}  f_1(r) \left[S_z(\vec{r}\cdot\vec{S})\right]\nonumber\\
&& + k \left\{\dfrac{\partial f_1(r)}{\partial r} \dfrac{1}{r}   + f_1(r)\dfrac{g \hbar k}{r^2}  \right\}\left[ \left(\vec{r} \cdot \vec{S}\right){S}_z\right]+k\dfrac{1- g \hbar k }{r^2}  f_1(r) \left[(\vec{r}\cdot\vec{S}){S}_z\right]\nonumber\\
&=& -k \left\{\dfrac{\partial f_1(r)}{\partial r} \dfrac{1}{r}   + \dfrac{1}{r^2}f_1(r)  \right\}\left[{S}_z \left(\vec{r} \cdot \vec{S}\right)\right]+ k \left\{\dfrac{\partial f_1(r)}{\partial r} \dfrac{1}{r}   + \dfrac{1}{r^2} f_1(r) \right\}\left[ \left(\vec{r} \cdot \vec{S}\right){S}_z\right]\nonumber\\
&=& -k \left\{\dfrac{\partial f_1(r)}{\partial r} \dfrac{1}{r}   + \dfrac{1}{r^2}f_1(r)  \right\}\left[{S}_z, \left(\vec{r} \cdot \vec{S}\right)\right] \nonumber\\
&=& -k \left\{\dfrac{\partial f_1(r)}{\partial r} \dfrac{1}{r}   + \dfrac{1}{r^2}f_1(r)  \right\}\left[{S}_z, x S_x+y S_y+z S_z\right]\nonumber\\
&=& -k \left\{\dfrac{\partial f_1(r)}{\partial r} \dfrac{1}{r}   + \dfrac{1}{r^2}f_1(r)  \right\} \left({\rm i}\hbar x S_y -{\rm i}\hbar y S_x\right)\nonumber\\
&=& - {\rm i}\hbar k \left\{\dfrac{\partial f_1(r)}{\partial r} \dfrac{1}{r}   + \dfrac{1}{r^2}f_1(r)  \right\} \left( \vec{r}\times \vec{S}\right)_z.
\end{eqnarray}
Based on which we have
\begin{eqnarray}
&&   -\left(\vec{\nabla}\times\vec{\mathcal{E}}\right)_z {-}{\rm i}\,g\biggl(\Bigl[\varphi,\
                                    {\mathcal{B}}_z\Bigr]
                              -\left(\vec{\mathcal{A}}\times\vec{\mathcal{E}}\right)_z
                              -\left(\vec{\mathcal{E}}\times\vec{\mathcal{A}}\right)_z\biggr)\nonumber\\
&=&   -\left(\vec{\nabla}\times\vec{\mathcal{E}}\right)_z +{\rm i}\,g\biggl(
                              \left(\vec{\mathcal{A}}\times\vec{\mathcal{E}}\right)_z
                              +\left(\vec{\mathcal{E}}\times\vec{\mathcal{A}}\right)_z\biggr)\nonumber\\
&=& -g \hbar k \left\{\dfrac{\partial f_1(r) }{\partial r} \dfrac{1}{r} +\dfrac{1}{r^2}f_1(r) \right\} \left( \vec{r}\times \vec{S}\right)_z
{{+}{\rm i}\,g \left[- {\rm i}\hbar k \left\{\dfrac{\partial f_1(r)}{\partial r} \dfrac{1}{r}   + \dfrac{1}{r^2}f_1(r)  \right\} \left( \vec{r}\times \vec{S}\right)_z\right]}\nonumber\\
&=&0.
\end{eqnarray}
Thus the $z$-component of Eq. (\ref{eq:CurlEYM}) is satisfied directly. Similarly, the $x$-component and $z$-component of Eq. (\ref{eq:CurlEYM}) are also satisfied. Namely, Eq. (\ref{eq:CurlEYM}) has no conditions for two functions $f_1(r)$ and $f_2(r)$.

Based on Eq. (\ref{eq:con-1}), Eq. (\ref{eq:con-2}),  and Eq. (\ref{eq:con-3}), we have the total conditions for $f_1(r)$ and $f_2(r)$ as follows:
\begin{subequations}
\begin{eqnarray}
&& \dfrac{\partial^2 f_2(r)}{\partial r^2}+ \dfrac{2}{r} \dfrac{\partial f_2(r)}{\partial r}=0, \label{eq:con-1a}\\
&&  \dfrac{\partial^2 f_1(r)}{\partial r^2}  +  \dfrac{4}{r} \dfrac{\partial f_1(r)}{\partial r}+ \dfrac{2 g \hbar k(2-g \hbar k)}{r^2} f_1(r) =0,  \label{eq:con-2a}\\
&&\left(1-  g \hbar k \right)\left[g\hbar  f^2_1(r) + (2-g \hbar k)k \frac{1}{r^4}\right]=0.\label{eq:con-3a}
\end{eqnarray}
\end{subequations}
From Eq. (\ref{eq:con-1a}), we have the solution for $f_2(r)$ as
\begin{eqnarray}
&& f_2(r)=\dfrac{\kappa_1}{r}+c_1,
\end{eqnarray}
where one can set the constant $c_1=0$ for it is trivial. In the following, let us study Eq. (\ref{eq:con-2a}) and Eq. (\ref{eq:con-3a}) and work out the consistent solutions of $f_1(r)$. We divide the study by two cases: $k=0$ and $k\neq 0$.

\subsubsection{The Case of $k=0$}

In this case, the spin vector potential is $\vec{A}=0$. From  Eq. (\ref{eq:con-2a}) and Eq. (\ref{eq:con-3a}) we have
\begin{subequations}
\begin{eqnarray}
&&  \dfrac{\partial^2 f_1(r)}{\partial r^2}  +  \dfrac{4}{r} \dfrac{\partial f_1(r)}{\partial r}=0,  \label{eq:con-2b}\\
&& g\hbar f^2_1(r)=0.\label{eq:con-3b}
\end{eqnarray}
\end{subequations}
Then we have two possible cases:

\emph{Case (i).} If $g \neq 0$, then $ f_1(r)=0$. In this situation, we have the scalar potential as
\begin{eqnarray}
    &&       \varphi = f_1(r) \left(\vec{r} \cdot \vec{S}\right) +  f_2(r)=\dfrac{\kappa_1}{r}, \label{eq:coul-1}
\end{eqnarray}
which is the usual Coulomb-type potential.

\emph{Case (ii).} If $g = 0$, then
\begin{eqnarray}
&& f_1(r)=\dfrac{\kappa_2}{r^3}+\kappa_3,
\end{eqnarray}
where $\kappa_2$ and $\kappa_3$ are some real constant numbers. In this situation, we have the scalar potential as
\begin{eqnarray}
    &&       \varphi = f_1(r) \left(\vec{r} \cdot \vec{S}\right) +  f_2(r)
    =\left(\dfrac{\kappa_2}{r^3}+\kappa_3\right) \left(\vec{r} \cdot \vec{S}\right)+\dfrac{\kappa_1}{r}. \label{eq:coul-2}
\end{eqnarray}

\subsubsection{The Case of $k\neq 0$}

In this case, the spin vector potential $\vec{A} \neq 0$. From  Eq. (\ref{eq:con-2a}) and Eq. (\ref{eq:con-3a}) we have
\begin{subequations}
\begin{eqnarray}
&&  \dfrac{\partial^2 f_1(r)}{\partial r^2}  +  \dfrac{4}{r} \dfrac{\partial f_1(r)}{\partial r}+   \dfrac{2 g \hbar k(2-g \hbar k)}{r^2} f_1(r) =0,  \label{eq:con-2c}\\
&& \left(1-  g \hbar k \right)\left[g\hbar  f^2_1(r) + (2-g \hbar k)k \frac{1}{r^4}\right]=0.\label{eq:con-3c}
\end{eqnarray}
\end{subequations}
If $g=0$, then from Eq. (\ref{eq:con-3c}) one must have $k=0$. Thus we exclude this case. In the following, we would like to consider $g\neq 0$ and $k\neq 0$.

There are three possible cases.

\emph{Case (i).} $(2-g \hbar k)=0$. Namely, one has $ g \hbar k=2$. In this case, we have
\begin{subequations}
\begin{eqnarray}
&&  \dfrac{\partial^2 f_1(r)}{\partial r^2}  +  \dfrac{4}{r} \dfrac{\partial f_1(r)}{\partial r}=0,  \label{eq:con-2d}\\
&& g\hbar \left(1-  g \hbar k \right) f^2_1(r) =0,\label{eq:con-3d}
\end{eqnarray}
\end{subequations}
whose solution is $f_1(r)=0$. In this situation, we have the scalar potential as
\begin{eqnarray}
    &&       \varphi = f_1(r) \left(\vec{r} \cdot \vec{S}\right) +  f_2(r)=\dfrac{\kappa_1}{r},
\end{eqnarray}
which is the usual Coulomb-type potential. From $(2-g \hbar k)=0$ we can obtain
\begin{eqnarray}
    &&      k=\frac{2}{g\hbar}.
\end{eqnarray}

\emph{Case (ii).} $(1 -  g \hbar k)=0$. Namely, one has $ g \hbar k=1$. In this case, we have
\begin{eqnarray}
&&  \dfrac{\partial^2 f_1(r)}{\partial r^2}  +  \dfrac{4}{r} \dfrac{\partial f_1(r)}{\partial r}+\dfrac{2}{r^2} f_1(r) =0,
\end{eqnarray}
whose solution is $f_1(r)= \dfrac{\kappa_2}{r} + \dfrac{\kappa_3}{r^2}$. In this situation, we have the scalar potential as
\begin{eqnarray}
    &&       \varphi = f_1(r) \left(\vec{r} \cdot \vec{S}\right) +  f_2(r)=\left(\frac{\kappa_2}{r} + \frac{\kappa_3}{r^2}\right)\left(\vec{r} \cdot \vec{S}\right)+\frac{\kappa_1}{r}
\end{eqnarray}
which is a general spin potential.

\emph{Case (iii).} $\left(1 -  g \hbar k \right) (2-g \hbar k)\neq 0$. From Eq. (\ref{eq:con-3c}) one has
\begin{eqnarray}
&&  f^2_1(r) =-  \frac{ (2-g \hbar k)k }
{g\hbar }\frac{1}{r^4},
\end{eqnarray}
when
\begin{eqnarray}
&& \lambda=-  \frac{ (2-g \hbar k)k }
{g\hbar }\geq 0,
\end{eqnarray}
we have
\begin{eqnarray}\label{eq:f1}
&&  f_1(r) = \frac{\sqrt{\lambda}}{r^2}.
\end{eqnarray}
After substituting Eq. (\ref{eq:f1}) into  Eq. (\ref{eq:con-2c}), we have
\begin{eqnarray}
 &&\dfrac{\partial^2 f_1(r)}{\partial r^2}  +  \dfrac{4}{r} \dfrac{\partial f_1(r)}{\partial r}+   \dfrac{ 2 g \hbar k(2-g \hbar k)}{r^2} f_1(r) \nonumber\\
 &=& -\frac{2\sqrt{\lambda}}{r^4}+\dfrac{ 2 g \hbar k(2-g \hbar k)}{r^2}f_1(r)\nonumber\\
 &=& -\frac{2\sqrt{\lambda}}{r^4}+\dfrac{ 2 g \hbar k(2-g \hbar k)}{r^2} \frac{\sqrt{\lambda}}{r^2} \nonumber\\
 &=& -\frac{2\sqrt{\lambda}}{r^4} [1- g \hbar k(2-g \hbar k)]\nonumber\\
 &=&-\frac{2\sqrt{\lambda}}{r^4}  (1-g \hbar k)^2 \neq 0.
   \end{eqnarray}
Thus we exclude this case. This ends the proof.
\end{proof}

\begin{remark}
From \emph{``Analysis 1''} to \emph{``Analysis 4''}, we have not used the relation $ (\vec{r} \cdot \vec{S})^2 = (r^2/4) \hbar^2 \openone$, which is satisfied by spin-$1/2$ particle, instead, we have merely used the definition of angular momentum operator, i.e., $[S_{a},S_{b}]={\rm i} \hbar \, \epsilon_{abc} S_c$. Thus the exact solution  is also valid for any spin-$s$ angular momentum $\vec{S}$.
For the spin operator $\vec{S}$ with higher-spin value, if the vector potential is fixed as $\vec{\mathcal{A}} = k \dfrac{\vec{r} \times \vec{S}}{r^2}$, then the more general form of the scalar potential could be
\begin{eqnarray}
    &&       \varphi = \sum_{j=0}^{2s} f_j(r) \left(\vec{r} \cdot \vec{S}\right)^j.
\end{eqnarray}
For example, for the spin-1 case we have
\begin{eqnarray}
    &&       \varphi = \sum_{j=0}^{2} f_j(r) \left(\vec{r} \cdot \vec{S}\right)^j= f_2(r) \left(\vec{r} \cdot \vec{S}\right)^2+
    f_1(r) \left(\vec{r} \cdot \vec{S}\right)+f_0(r),
\end{eqnarray}
one can use the relation $\left(\vec{r} \cdot \vec{S}\right)^3=\left(\vec{r} \cdot \vec{S}\right)$ to reduce any high-order term,
such as $\left(\vec{r} \cdot \vec{S}\right)^n$, to the lower-order terms. However, in this work, for simplicity we restrict our study to the
case of spin-$1/2$. $\blacksquare$
\end{remark}

\subsection{Discussion on the Solutions of Yang-Mills Equations}

Based on Theorem \ref{Ther2} and Table \ref{tab:gk}, we have obtained four exact solutions of Yang-Mills equations. Let us denote the solution of Yang-Mills equations
\begin{eqnarray}
    && \mathcal{S}_{\rm YM} =\left\{ \vec{\mathcal{A}}= k \frac{\vec{r} \times \vec{S}}{r^2}, \;
           \varphi = f_1(r) \left(\vec{r} \cdot \vec{S}\right) +  f_2(r)\right\},
\end{eqnarray}
explicitly, they are

\textcolor{blue}{\emph{Solution 1.--}} For $k=0$ and $g=0$, the solution is
\begin{eqnarray}
    && \mathcal{S}^{(1)}_{\rm YM} =\left\{ \vec{\mathcal{A}}= 0, \;
           \varphi = \left(\dfrac{\kappa_2}{r^3} + \kappa_3\right) \left(\vec{r} \cdot \vec{S}\right) +  \dfrac{\kappa_1}{r} \right\}.
\end{eqnarray}
Let us analyze its physical meaning. In this situation, $k=0$ implies that there is no vector potential, and $g=0$ implies that we are studying the
Maxwell-type equations (in vacuum without sources), namely,
\begin{subequations}
            \begin{eqnarray}
                  && \vec{\nabla}\cdot\vec{\mathcal{E}}=0, \label{eq:DivEYM-Mt} \\
                  && -\dfrac{1}{c} \dfrac{\partial \vec{\mathcal{B}}}{\partial\,t}
                        -\vec{\nabla}\times\vec{\mathcal{E}}=0, \label{eq:CurlEYM-Mt} \\
                  && \vec{\nabla}\cdot\vec{\mathcal{B}}=0, \label{eq:DivBYM-Mt} \\
                  && -\dfrac{1}{c} \dfrac{\partial \vec{\mathcal{E}}}{\partial\,t}
                        +\vec{\nabla}\times\vec{\mathcal{B}}=0. \label{eq:CurlBYM-Mt}
            \end{eqnarray}
\end{subequations}
The Maxwell-type equations and Maxwell's equations look the same. The difference between them is that the former can allow the operator solution (i.e., the ``magnetic'' field  and the ``electric'' field  can be operators). In other words, the operator solution of the Maxwell-type equations is just
$\mathcal{S}^{(1)}_{\rm YM}$. If one neglects the operator part of $\mathcal{S}^{(1)}_{\rm YM}$ (i.e., $\left(\dfrac{\kappa_2}{r^3} + \kappa_3\right) \left(\vec{r} \cdot \vec{S}\right)$), then the solution $\mathcal{S}^{(1)}_{\rm YM}$ naturally reduces to the exact solution of Maxwell's equations
$\mathcal{S}_{\rm Maxwell}=\{\vec{\mathcal{A}}=0, \varphi={\kappa}/{r}\}$. For this solution, the corresponding ``magnetic'' field and the ``electric'' field are given by
\begin{eqnarray}
\vec{\mathcal{B}} &=& \vec{\nabla}\times\vec{\mathcal{A}}
                        -{\rm i}\,g\left(\vec{\mathcal{A}}\times\vec{\mathcal{A}}\right)=0,
\end{eqnarray}
\begin{eqnarray}
\vec{\mathcal{E}} &=& -\dfrac{1}{c}
                              \dfrac{\partial\,\vec{\mathcal{A}}}{\partial\,t}
                        -\vec{\nabla}\varphi-{\rm i}\,g\left[\varphi,\ \vec{\mathcal{A}}
                        \right]=-\vec{\nabla}\varphi \nonumber\\
&=& -\vec{\nabla} \left[\left(\dfrac{\kappa_2}{r^3} + \kappa_3\right) \left(\vec{r} \cdot \vec{S}\right) +  \dfrac{\kappa_1}{r}\right] \nonumber\\
&=& \vec{\mathcal{E}}_1 +\vec{\mathcal{E}}_2, \label{eq:ele-2t}
\end{eqnarray}
where $\vec{\mathcal{E}}$ contains two parts. The second part is the usual electric field in the Coulomb-type potential, i.e.,
\begin{eqnarray}
\vec{\mathcal{E}}_2 &=& -\vec{\nabla}\left(\dfrac{\kappa_1}{r}\right)=\dfrac{\kappa}{r^2} \hat{e}_r.
\end{eqnarray}
The first part is calculated as
\begin{eqnarray}
\vec{\mathcal{E}}_1
&=& -\vec{\nabla} \left[\left(\dfrac{\kappa_2}{r^3} + \kappa_3\right) \left(\vec{r} \cdot \vec{S}\right)\right] \nonumber\\
&=& - \left[\frac{\partial \left(\dfrac{\kappa_2}{r^3} + \kappa_3\right) }{\partial x} \hat{e}_x +
\frac{\partial \left(\dfrac{\kappa_2}{r^3} + \kappa_3\right) }{\partial y} \hat{e}_y+
\frac{\partial \left(\dfrac{\kappa_2}{r^3} + \kappa_3\right) }{\partial z} \hat{e}_z \right]\left(\vec{r} \cdot \vec{S}\right)\nonumber\\
&& -\left(\dfrac{\kappa_2}{r^3} + \kappa_3\right) \left[\frac{\partial (\vec{r} \cdot \vec{S}) }{\partial x} \hat{e}_x+
\frac{\partial (\vec{r} \cdot \vec{S}) }{\partial y} \hat{e}_y+\frac{\partial (\vec{r} \cdot \vec{S}) }{\partial z} \hat{e}_z\right] \nonumber\\
&=& - \kappa_2 \left[\frac{\partial \left(\dfrac{1}{r^3}\right) }{\partial x} \hat{e}_x +
\frac{\partial \left(\dfrac{1}{r^3} \right) }{\partial y} \hat{e}_y+
\frac{\partial \left(\dfrac{1}{r^3} \right) }{\partial z} \hat{e}_z \right]\left(\vec{r} \cdot \vec{S}\right)-\left(\dfrac{\kappa_2}{r^3} + \kappa_3\right) \left(S_x \hat{e}_x+
S_y \hat{e}_y+ S_z \hat{e}_z\right) \nonumber\\
&=& - \kappa_2 \left[\frac{\partial \left(\dfrac{1}{r^3}\right) }{\partial r} \frac{x}{r}\hat{e}_x +
\frac{\partial \left(\dfrac{1}{r^3} \right) }{\partial r} \frac{y}{r} \hat{e}_y+
\frac{\partial \left(\dfrac{1}{r^3} \right) }{\partial r} \frac{z}{r} \hat{e}_z \right]\left(\vec{r} \cdot \vec{S}\right)-\left(\dfrac{\kappa_2}{r^3} + \kappa_3\right) \vec{S} \nonumber\\
&=& - \kappa_2 \left[ \frac{-3}{r^4} \left( \frac{x}{r}\hat{e}_x + \frac{y}{r} \hat{e}_y+\frac{z}{r} \hat{e}_z \right) \right]\left(\vec{r} \cdot \vec{S}\right)-\left(\dfrac{\kappa_2}{r^3} + \kappa_3\right) \vec{S} \nonumber\\
&=& \frac{3 \kappa_2}{r^4} \left(\vec{r} \cdot \vec{S}\right)\hat{e}_r -\left(\dfrac{\kappa_2}{r^3} + \kappa_3\right) \vec{S},
\end{eqnarray}
which is spin-dependent. The direction of $\vec{\mathcal{E}}_1$ is determined by $\hat{e}_r $ and $\vec{S}$.

\textcolor{blue}{\emph{Solution 2.--}} For $k=0$ and $g\neq 0$, the solution is
\begin{eqnarray}
    && \mathcal{S}^{(2)}_{\rm YM} =\left\{ \vec{\mathcal{A}}= 0, \;
           \varphi =  \dfrac{\kappa_1}{r} \right\},
\end{eqnarray}
which looks the same as the exact solution of Maxwell's equations
$\mathcal{S}_{\rm Maxwell}=\{\vec{\mathcal{A}}=0, \varphi={\kappa}/{r}\}$.

In this work, we are more interested in the case of $k\neq 0$, i.e., the vector potential and the scalar potential co-exist. From Table \ref{tab:gk}, we have known that for $k\neq 0$ and $g=0$, there is no solution. For $k\neq 0$ and $g\neq 0$, there are two different solutions, depending on $g\hbar k=2$ or $g\hbar k=1$.

\textcolor{blue}{\emph{Solution 3.--}} For $k\neq 0$, $g\neq 0$, and $g\hbar k=2$, the solution is
\begin{eqnarray}
    && \mathcal{S}^{(3)}_{\rm YM} =\left\{ \vec{\mathcal{A}}= k \frac{\vec{r} \times \vec{S}}{r^2}, \;
           \varphi = \dfrac{\kappa_1}{r} \right\}.
\end{eqnarray}
For this solution, the corresponding ``magnetic'' field and the ``electric'' field are given by
\begin{eqnarray}
\vec{\mathcal{B}} &=& \vec{\nabla}\times\vec{\mathcal{A}}
                        -{\rm i}\,g\left(\vec{\mathcal{A}}\times\vec{\mathcal{A}}\right)= -(2-g \hbar k)k \frac{\vec{r}\cdot \vec{S}}{r^4}\vec{r}=0,
\end{eqnarray}
\begin{eqnarray}
\vec{\mathcal{E}} &=& -\dfrac{1}{c}
                              \dfrac{\partial\,\vec{\mathcal{A}}}{\partial\,t}
                        -\vec{\nabla}\varphi-{\rm i}\,g\left[\varphi,\ \vec{\mathcal{A}}
                        \right]=-\vec{\nabla}\varphi =-\vec{\nabla}\left(\dfrac{\kappa_1}{r}\right)=\dfrac{\kappa}{r^2} \hat{e}_r.
\end{eqnarray}
Namely, the magnetic field is zero and the electric field is the usual electric field in the Coulomb-type potential.

\textcolor{blue}{\emph{Solution 4.--}} For $k\neq 0$, $g\neq 0$, and $g\hbar k=1$, the solution is
\begin{eqnarray}
    && \mathcal{S}^{(4)}_{\rm YM} =\left\{ \vec{\mathcal{A}}= k \frac{\vec{r} \times \vec{S}}{r^2}, \;
           \varphi =\left(\dfrac{\kappa_2}{r} + \dfrac{\kappa_3}{r^2}\right) \left(\vec{r} \cdot \vec{S}\right)+\dfrac{\kappa_1}{r} \right\}.
\end{eqnarray}
For this solution, the corresponding ``magnetic'' field and the ``electric'' field are given by
\begin{eqnarray}
\vec{\mathcal{B}} &=& \vec{\nabla}\times\vec{\mathcal{A}}
                        -{\rm i}\,g\left(\vec{\mathcal{A}}\times\vec{\mathcal{A}}\right)= -(2-g \hbar k)k \frac{\vec{r}\cdot \vec{S}}{r^4}\vec{r}
                        =-k \frac{\vec{r}\cdot \vec{S}}{r^4}\vec{r}=-k \frac{\vec{r}\cdot \vec{S}}{r^3}\hat{e}_r,
\end{eqnarray}
\begin{eqnarray}
\vec{\mathcal{E}} &=& -\dfrac{1}{c}\dfrac{\partial\,\vec{\mathcal{A}}}{\partial\,t}
                        -\vec{\nabla}\varphi-{\rm i}\,g\left[\varphi,\ \vec{\mathcal{A}}\right]=-\vec{\nabla}\varphi-{\rm i}\,g\left[\varphi,\ \vec{\mathcal{A}}\right] \nonumber\\
&=& {-}\left\{\dfrac{\partial f_1(r)}{\partial r} \dfrac{1}{r}  \left(\vec{r} \cdot \vec{S}\right)+\dfrac{\partial f_2(r)}{\partial r} \dfrac{1}{r} + f_1(r)\dfrac{g \hbar k}{r^2} \left(\vec{r} \cdot \vec{S}\right) \right\}\vec{r}+\left[-1+  g \hbar k \right] f_1(r) \vec{S}\nonumber\\
&=& {-}\left\{\dfrac{\partial f_1(r)}{\partial r} \dfrac{1}{r}  \left(\vec{r} \cdot \vec{S}\right)+\dfrac{\partial f_2(r)}{\partial r} \dfrac{1}{r} + f_1(r)\dfrac{1}{r^2} \left(\vec{r} \cdot \vec{S}\right) \right\}\vec{r}\nonumber\\
&=& {-}\left\{\dfrac{\partial \left(\dfrac{\kappa_2}{r} + \dfrac{\kappa_3}{r^2}\right)}{\partial r} \dfrac{1}{r}  \left(\vec{r} \cdot \vec{S}\right)
+\dfrac{\partial \left(\dfrac{\kappa_1}{r}\right)}{\partial r} \dfrac{1}{r} + \left(\dfrac{\kappa_2}{r} + \dfrac{\kappa_3}{r^2}\right)\dfrac{1}{r^2} \left(\vec{r} \cdot \vec{S}\right) \right\}\vec{r}\nonumber\\
&=& {-}\left\{ \left(\dfrac{-\kappa_2}{r^2} + \dfrac{-2\kappa_3}{r^3}\right) \dfrac{1}{r}  \left(\vec{r} \cdot \vec{S}\right)
+\dfrac{-\kappa_1}{r^2} \dfrac{1}{r} + \left(\dfrac{\kappa_2}{r} + \dfrac{\kappa_3}{r^2}\right)\dfrac{1}{r^2} \left(\vec{r} \cdot \vec{S}\right) \right\}\vec{r}\nonumber\\
&=& \left\{ \left(\dfrac{\kappa_2}{r^2} + \dfrac{2\kappa_3}{r^3}\right) \dfrac{1}{r}  \left(\vec{r} \cdot \vec{S}\right)
+\dfrac{\kappa_1}{r^2} \dfrac{1}{r} - \left(\dfrac{\kappa_2}{r} + \dfrac{\kappa_3}{r^2}\right)\dfrac{1}{r^2} \left(\vec{r} \cdot \vec{S}\right) \right\}\vec{r}\nonumber\\
&=& \left[ \kappa_3 \dfrac{\vec{r} \cdot \vec{S}}{r^3}
+\dfrac{\kappa_1}{r^2} \right] \hat{e}_r.
\end{eqnarray}
The directions of $\vec{\mathcal{B}}$  and $\vec{\mathcal{E}}$ are both $\hat{e}_r$.

\begin{remark}When one takes
\begin{eqnarray}
\kappa_2=0, \;\;\; \kappa_3=0,
\end{eqnarray}
then for solution 3 and solution 4, we both have $f_1(r)=0$. In this situation, we have the exact solution of Yang-Mills as
\begin{eqnarray}
    && \mathcal{S}_{\rm YM} \equiv\mathcal{S}^{(3)}_{\rm YM} =\mathcal{S}^{(4)}_{\rm YM} =\left\{ \vec{\mathcal{A}}= k \frac{\vec{r} \times \vec{S}}{r^2}, \;
           \varphi = \dfrac{\kappa_1}{r} \right\}.
\end{eqnarray}
In next sections, we find that the Schr{\" o}dinger-type and the Dirac-type Hamiltonians with spin-dependent Coulomb interaction (i.e., with the spin potentials $\mathcal{S}_{\rm YM}=\left\{\vec{\mathcal{A}}= k \dfrac{\vec{r} \times \vec{S}}{r^2},\; \varphi=\dfrac{\kappa}{r}\right\}$) can be exactly solved.
$\blacksquare$
\end{remark}

\newpage

\section{Exactly Solving Schr{\" o}dinger's Equation with Spin-Dependent Coulomb Interaction}

 The Schr{\" o}dinger-type Hamiltonian with spin-dependent Coulomb interaction (i.e., with the spin potentials $\mathcal{S}_{\rm YM}=\left\{\vec{\mathcal{A}}= k \dfrac{\vec{r} \times \vec{S}}{r^2},\; \varphi=\dfrac{\kappa}{r}\right\}$) is given by
\begin{eqnarray}\label{eq:H-1a-1a0}
\mathcal{H}&=& \dfrac{\vec{\Pi}^2}{2M}  +q \varphi= \dfrac{\left(\vec{p}-\dfrac{q}{c}\vec{\mathcal{A}}\right)^2}{2M}
 + \dfrac{q \kappa}{r},
\end{eqnarray}
where
\begin{eqnarray}
    &&  \vec{\mathcal{A}} = k \frac{\vec{r} \times \vec{S}}{r^2},
\end{eqnarray}
and the constraint {$g\hbar k=1, 2$}. Here $\kappa=Q$ (with $Q=Ze$, $Z$ is an integer, $Z=1$ for the hydrogen atom and $Z \geq 2$ for the hydrogen-like atom). For simplicity, we consider merely the spin-$1/2$ case, i.e., $\vec{S}=\dfrac{\hbar}{2}\vec{\sigma}$.
Because
\begin{eqnarray}
\vec{\nabla}\cdot \vec{\mathcal{A}} & =& \vec{\nabla}\cdot \left(k \frac{\vec{r} \times \vec{S}}{r^2}\right)\nonumber\\
&=& k \dfrac{\partial}{\partial x}\dfrac{\left(\vec{r}\times\vec{S}\right)_x}{r^2}+k \dfrac{\partial}{\partial y}\dfrac{\left(\vec{r}\times\vec{S}\right)_y}{r^2}+k \dfrac{\partial}{\partial z}\dfrac{\left(\vec{r}\times\vec{S}\right)_z}{r^2}\nonumber\\
&=& k \dfrac{\partial}{\partial x}\dfrac{\left(y S_z-z S_y \right)}{r^2}+k \dfrac{\partial}{\partial y}\dfrac{\left(z S_x-x S_z \right)}{r^2}
+k \dfrac{\partial}{\partial z}\dfrac{\left(x S_y-y S_x \right)}{r^2}\nonumber\\
&=& k \left(y S_z-z S_y \right) \dfrac{\partial}{\partial x}\dfrac{1}{r^2}
+k \left(z S_x-x S_z \right)\dfrac{\partial}{\partial y}\dfrac{1}{r^2}
+k \left(x S_y-y S_x \right) \dfrac{\partial}{\partial z}\dfrac{1}{r^2}\nonumber\\
&=& k \left(y S_z-z S_y \right) \left[\dfrac{\partial}{\partial r}\dfrac{1}{r^2}\right]\dfrac{\partial r}{\partial x}
+k \left(z S_x-x S_z \right) \left[\dfrac{\partial}{\partial r}\dfrac{1}{r^2}\right]\dfrac{\partial r}{\partial y}
+k \left(x S_y-y S_x \right) \left[\dfrac{\partial}{\partial r}\dfrac{1}{r^2}\right]\dfrac{\partial r}{\partial z}\nonumber\\
&=& k \left(y S_z-z S_y \right) \dfrac{-2}{r^3}\dfrac{x}{r}+k \left(z S_x-x S_z \right)\dfrac{-2}{r^3}\dfrac{y}{r}
+k \left(x S_y-y S_x \right) \dfrac{-2}{r^3}\dfrac{z}{r}\nonumber\\
&=& \dfrac{-2k}{r^4} \left[\vec{r}\cdot \left(\vec{r} \times \vec{S}\right)\right]=0,
\end{eqnarray}
\begin{eqnarray}
\vec{\mathcal{A}}\cdot\vec{p} &=&\ \dfrac{k\,\hbar}{2}
                  \dfrac{(\vec{r}\times\vec{\sigma})\cdot\vec{p}}{r^2} = -\dfrac{g\,\hbar}{2} \dfrac{(\vec{\sigma}\times\vec{r})\cdot\vec{p}}{r^2} = -\dfrac{k\,\hbar}{2} \dfrac{\vec{\sigma} \cdot(\vec{r}\times\vec{p})}{r^2}
                  = -\dfrac{k\,\hbar}{2} \dfrac{\vec{\ell}\cdot\vec{\sigma}}{r^2}=-k \dfrac{\vec{\ell}\cdot\vec{S}}{r^2},
\end{eqnarray}
      \begin{eqnarray}\label{eq:ASqure}
      \vec{\mathcal{A}}^{\, 2} &=& \left(\dfrac{k\,\hbar}{2} \dfrac{\vec{r}\times\vec{\sigma}}{r^2}\right)\cdot
            \left(\dfrac{k\,\hbar}{2} \dfrac{\vec{r}\times\vec{\sigma}}{r^2}\right) =\dfrac{k^2\,\hbar^2}{4\,r^4} (\vec{r}\times\vec{\sigma})\cdot(\vec{r}\times\vec{\sigma}) =\dfrac{k^2\,\hbar^2}{4\,r^4} \bigl[(\vec{r}\times\vec{\sigma})\times\vec{r}\bigr]\cdot\vec{\sigma} \nonumber\\
            &=& \dfrac{k^2\,\hbar^2}{4\,r^4} \bigl[(r^2 \vec{\sigma})-(\vec{\sigma} \cdot\vec{r})\vec{r}\bigr]\cdot\vec{\sigma}=\dfrac{k^2\,\hbar^2}{4\,r^4} \bigl(3r^2-r^2\bigr)= \dfrac{k^2\,\hbar^2}{2} \dfrac{1}{r^2},
            \end{eqnarray}
we then have
      \begin{eqnarray}\label{eq:HSForm1}
           \dfrac{1}{2M}\vec{\Pi}^2 & =&\ \dfrac{1}{2M} \left(
                  \vec{p}^{\,2} -{\dfrac{q}{c}}\vec{\mathcal{A}}\cdot\vec{p}-{\dfrac{q}{c}}\vec{p}\cdot\vec{\mathcal{A}}
                  +{\dfrac{q^2}{c^2}}\vec{\mathcal{A}}^{\,2}\right) \nonumber\\
           & =&\ \dfrac{1}{2M} \biggl[\vec{p}^{\,2} -{\dfrac{q}{c}}\vec{\mathcal{A}}\cdot\vec{p}
                  -{\dfrac{q}{c}}\left(\vec{\mathcal{A}}\cdot\vec{p}-{\rm i}\hbar\vec{\nabla}\cdot\vec{\mathcal{A}}\right)
                  +{\dfrac{q^2}{c^2}}\vec{\mathcal{A}}^2\biggr] \nonumber\\
           & =&\ \dfrac{1}{2M} \left(\vec{p}^{\, 2} -2{\dfrac{q}{c}}\vec{\mathcal{A}}\cdot\vec{p}+{\dfrac{q^2}{c^2}}\vec{\mathcal{A}}^{\,2}\right) \nonumber\\
           & =&\ \dfrac{1}{2M} \left(\vec{p}^{\,2} +2{\dfrac{q}{c}} k \,\dfrac{\vec{\ell}\cdot\vec{S}}{r^2}
                  +{\dfrac{q^2}{c^2}}\dfrac{k^2\,\hbar^2}{2} \dfrac{1}{r^2}\right),
      \end{eqnarray}
i.e.,
\begin{eqnarray}
           \dfrac{1}{2M}\vec{\Pi}^2
           & =&\ \dfrac{1}{2M} \left(\vec{p}^{\,2} +2 \tilde{k} \,\dfrac{\vec{\ell}\cdot\vec{S}}{r^2}
                  +\dfrac{\tilde{k}^2\,\hbar^2}{2} \dfrac{1}{r^2}\right),
\end{eqnarray}
with
\begin{eqnarray}
 \tilde{k}= {\dfrac{q}{c}} k.
\end{eqnarray}
{Note that $ \tilde{k}$ is a dimensionless number, or $k$ has a dimension of $[c/q]$.}

Then the Hamiltonian becomes
\begin{eqnarray}\label{eq:H-2a}
\mathcal{H}&=& \dfrac{1}{2M} \left(\vec{p}^{\,2} +2\tilde{k}\,\dfrac{\vec{\ell}\cdot\vec{S}}{r^2}
                  +\dfrac{\tilde{k}^2\,\hbar^2}{2} \dfrac{1}{r^2}\right)+  \dfrac{q \kappa}{r}.
\end{eqnarray}
Due to Eq. (\ref{eq:E-18}), we have the Hamiltonian as
      \begin{eqnarray}\label{eq:H-2b-b}
            \mathcal{H} &=& \dfrac{1}{2M} \left(-\hbar^2 \dfrac{\partial^2}{\partial\,r^2}
                  -2 \hbar^2 \dfrac{1}{r} \dfrac{\partial }{\partial\,r}
                  + \dfrac{\vec{\ell}^{\;2}}{ r^2}+2\tilde{k} \,\dfrac{\vec{\ell}\cdot\vec{S}}{r^2}
                  +\dfrac{\tilde{k}^2\,\hbar^2}{2} \dfrac{1}{r^2}\right) +  \dfrac{q \kappa}{r}\nonumber\\
      &=& \dfrac{1}{2M} \left\{-\hbar^2 \dfrac{\partial^2}{\partial\,r^2}
                  -2 \hbar^2 \dfrac{1}{r} \dfrac{\partial }{\partial\,r}
                  + \dfrac{\left[\vec{\ell}^{\;2}+2\tilde{k} \left(\vec{\ell}\cdot\vec{S}\right)
                  +\dfrac{\tilde{k}^2\,\hbar^2}{2}\right]}{r^2}\right\}+  \dfrac{q \kappa}{r}.
      \end{eqnarray}

 \begin{remark} \textcolor{blue}{The Eigen-Problem of $\vec{J}^{\; 2}$}. Let the ``total'' angular momentum operator is
 \begin{equation}\label{eq:DefL}
      \vec{J}=\vec{\ell}+\vec{S},
      \end{equation}
where $\vec{S}$ is the spin-$1/2$ angular momentum operator. The common eigenstates of $\{\vec{J}^{\; 2},\vec{\ell}^{\;2},J_z\}$ are given as
            \begin{eqnarray}\label{eq:E-28}
                  \vec{J}^{\; 2}\; \Phi_{ljm_j}(\theta, \phi)&=& j(j+1) \hbar^2 \; \Phi_{ljm_j}(\theta, \phi), \nonumber\\
                  \vec{\ell}^{\;2}\; \Phi_{ljm_j}(\theta, \phi)&=& l(l+1) \hbar^2 \; \Phi_{ljm_j}(\theta, \phi), \nonumber\\
                  J_z\; \Phi_{ljm_j}(\theta, \phi)&=& m_j \hbar \; \Phi_{ljm_j}(\theta, \phi).
            \end{eqnarray}
            More precisely, one has

            \emph{Case A.---}For the quantum number $j=l+1/2$, $(m_j=m+1/2)$, the eigenstate is given by
                  \begin{eqnarray}\label{eq:E-29}
                        \Phi^A_{ljm_j}(\theta, \phi) &=& \dfrac{1}{\sqrt{2l+1}} \begin{bmatrix}
                              \sqrt{l+m+1}\:Y_{lm}(\theta, \phi) \\
                              \sqrt{l-m}\:Y_{l,m+1}(\theta, \phi) \\
                        \end{bmatrix}.
                  \end{eqnarray}

            \emph{Case B.---}For the quantum number $j=l-1/2$, $l\neq 0$, $(m_j=m+1/2)$, the eigenstate
                  is given by
                  \begin{eqnarray}\label{eq:E-30}
                        \Phi^B_{ljm_j}(\theta, \phi)&=& \dfrac{1}{\sqrt{2l+1}} \begin{bmatrix}
                              -\sqrt{l-m}\:Y_{lm}(\theta, \phi) \\
                              \sqrt{l+m+1}\:Y_{l,m+1}(\theta, \phi) \\
                        \end{bmatrix}.
                  \end{eqnarray}
Since the right-hand side of Eq. (\ref{eq:E-29}) and Eq. (\ref{eq:E-30}) appear only quantum numbers $l$ and $m$, for simplicity, without any confusion we may denote $\Phi^A_{ljm_j}(\theta, \phi)\equiv\Phi^A_{lm}(\theta, \phi)$ and $\Phi^B_{ljm_j}(\theta, \phi)\equiv\Phi^B_{lm}(\theta, \phi)$. $\blacksquare$
\end{remark}

     \begin{remark}  \textcolor{blue}{The Eigen-Problem of $\vec{\ell}\cdot\vec{S}$.}
     Let the ``total'' angular momentum be
      \begin{eqnarray}
             \vec{J}= \vec{\ell}+\vec{S}.
            \end{eqnarray}
     Since
       \begin{eqnarray}
       \left[\vec{J}^{\,2},\ \vec{\ell}^{\, 2}\right]=0, \;\;\;\left[\vec{J}^{\,2},\ \vec{\ell}\cdot\vec{S}\right]=0, \;\;\;
       \left[\vec{\ell}^{\,2},\ \vec{\ell}\cdot\vec{S}\right]=0,
     \end{eqnarray}
     thus three operators $\{\vec{J}^{\,2}, \vec{\ell}^{\,2}, \vec{\ell}\cdot \vec{S}\}$ share the same eigenstates. From
            \begin{eqnarray}\label{eq:E-32}
             \vec{J}^{\, 2}= \left(\vec{\ell}+\vec{S}\right)^2,
            \end{eqnarray}
     one can have
            \begin{eqnarray}
                  \vec{\ell}\cdot \vec{S}  &=& \dfrac{1}{2}\left(\vec{J}^{\, 2} -\vec{\ell}^{\;2}-\dfrac{3}{4}\hbar^2\right).
            \end{eqnarray}
            We then have the eigenvalues and eigenstates for the operator $\vec{\ell}\cdot \vec{S}$ as follows:
            \begin{eqnarray}\label{eq:E-33}
                  \vec{\ell}\cdot \vec{S}\; \Phi_{ljm_j}(\theta, \phi) &&= \dfrac{\hbar^2}{2} \left[
                              j(j+1)-l(l+1)-\dfrac{3}{4}\right]\,\Phi_{ljm_j}(\theta, \phi) \nonumber\\
                    &&= \begin{cases}			
                        &\dfrac{l}{2} \,\hbar^2 \; \Phi^A_{ljm_j}(\theta, \phi), \;\;\;\;\;\;\;\;\;\;\; {\rm for}\;\; j=l+\dfrac{1}{2},\\
                        & -\dfrac{l+1}{2}\, \hbar^2 \; \Phi^B_{ljm_j}(\theta, \phi), \;\;\; {\rm for}\;\; j=l-\dfrac{1}{2},\; (l\neq 0),
                  \end{cases}
            \end{eqnarray}
            with $m_j=m+1/2$. $\blacksquare$
\end{remark}

We now separate the wavefunction as
            \begin{eqnarray}\label{eq:F-2}
                  \Psi(\vec{r})=\Psi(r, \theta, \phi)=R(r)\;\Phi_{ljm_j}(\theta, \phi),
            \end{eqnarray}
            where the two-component quantum state $\Phi_{ljm_j}(\theta, \phi)=\Phi^A_{ljm_j}(\theta, \phi)$,
            or $\Phi^B_{ljm_j}(\theta, \phi)$.
            We then have
            \begin{eqnarray}\label{eq:F-3}
                  \mathcal{H}\;R(r) \Phi_{ljm_j}(\theta, \phi) &=&E  \;R(r) \Phi_{ljm_j}(\theta, \phi),\nonumber\\
                  \vec{\ell}^{\; 2} \;R(r) \Phi_{ljm_j}(\theta, \phi) &=& l(l+1) \hbar^2  \;R(r) \Phi_{ljm_j}(\theta, \phi),\nonumber\\
                  \vec{\ell}\cdot\vec{S} \;R(r) \Phi_{ljm_j}(\theta, \phi) &=& W \hbar^2  \;R(r) \Phi_{ljm_j}(\theta, \phi),
            \end{eqnarray}
where $W=l/2$ or $-(l+1)/2$ ($l\neq 0$), depending on $\Phi_{ljm_j}(\theta, \phi)$ takes
            $\Phi^A_{ljm_j}(\theta, \phi)$ or $\Phi^B_{ljm_j}(\theta, \phi)$. By substituting the last two
            equations of (\ref{eq:F-3}) into the first equation of (\ref{eq:F-3}), we then have the following
            radial eigen-equation:
            \begin{eqnarray}\label{eq:F-4}
                  \dfrac{1}{2M} \left\{-\hbar^2 \dfrac{\partial^2}{\partial\,r^2}
                        -2\hbar^2 \dfrac{1}{r} \dfrac{\partial}{\partial\,r} +\left[l(l+1)\hbar^2+2\tilde{k} W\hbar^2
                  +\dfrac{\tilde{k}^2\,\hbar^2}{2}\right]\dfrac{1}{r^2}\right\}R(r)+ \dfrac{q \kappa}{r} R(r)
                  =E\,R(r),
            \end{eqnarray}
            i.e.,
            \begin{eqnarray}\label{eq:F-5a}
                  \dfrac{\hbar^2}{2M} \left\{-\dfrac{\partial^2}{\partial\,r^2}
                        -2 \dfrac{1}{r} \dfrac{\partial}{\partial\,r} +\left[l(l+1)+2\tilde{k} W
                  +\dfrac{\tilde{k}^2}{2}\right]\dfrac{1}{r^2}\right\}R(r)+ \dfrac{q \kappa}{r} R(r)
                  =E\,R(r),
            \end{eqnarray}
            i.e.,
             \begin{eqnarray}\label{eq:F-5}
                  \dfrac{\hbar^2}{2M} \left\{-\dfrac{{\rm d}^2}{{\rm d}\, r^2}
                        -2 \dfrac{1}{r} \dfrac{{\rm d}}{{\rm d}\,r} +\left[l(l+1)+2\tilde{k} W
                  +\dfrac{\tilde{k}k^2}{2}\right]\dfrac{1}{r^2}\right\}R(r)+ \dfrac{q \kappa}{r} R(r)
                  =E\,R(r).
            \end{eqnarray}
            Denote
            \begin{eqnarray}\label{eq:DefEpsilonKappa}
                  \epsilon &=& \dfrac{2M}{\hbar^2} E,\;\;\;\;\;\;                  \Omega = l(l+1)+2\tilde{k}W+\dfrac{\tilde{k}^2}{2},
            \end{eqnarray}
            then Eq. (\ref{eq:F-5}) becomes
            \begin{eqnarray}\label{eq:ReduceRadialEq}
                  \dfrac{{\rm d}^2\,R(r)}{{\rm d}\,r^2}
                        +\dfrac{2}{r} \dfrac{{\rm d}\,R(r)}{{\rm d}\,r}
                        -\dfrac{\Omega}{r^2} R(r)- \dfrac{2M q \kappa}{\hbar^2 r} R(r)+\epsilon R(r)=0.
            \end{eqnarray}

\emph{Observation 1.---}The parameter $\Omega$ in \Eq{eq:DefEpsilonKappa} is non-negative.

            \begin{proof}
                  \begin{itemize}
                        \item [(i)] For $W=\dfrac{l}{2}$, because $\tilde{k}\neq 0$ and $l\ge 0$, we then have
                              \begin{equation}\label{eq:KappaK1}
                                    \Omega= l(l+1)+2\tilde{k}W+\dfrac{\tilde{k}^2}{2}= l(l+1)+\tilde{k}l+\dfrac{\tilde{k}^2}{2}
                                    =\left(\dfrac{\tilde{k}}{2} +l\right)^2 +\dfrac{\tilde{k}^2}{4} +l > 0.
                              \end{equation}

                        \item [(ii)] For $ W =-\dfrac{l+1}{2}$, ($l\neq 0$), because $l\ge 1$, we have
                              \begin{equation}\label{eq:KappaK2}
                                    \begin{split}
                                          \Omega =&\ l(l+1)+2\tilde{k}W+\dfrac{\tilde{k}^2}{2} = l(l+1)-\tilde{k}(l+1)+\dfrac{\tilde{k}^2}{2} \\
                                          =&\ \dfrac{1}{2} \Bigl[\tilde{k}^2 -2\tilde{k}(l+1)+2l(l+1)\Bigr] = \dfrac{1}{2} \Bigl\{\bigl[\tilde{k}-(l+1)\bigr]^2 +2l(l+1)-(l+1)^2\Bigr\} \\
                                          =&\ \dfrac{1}{2} \Bigl\{\bigl[\tilde{k}-(l+1)\bigr]^2 +(l+1)(l-1)\Bigr\} \\
                                          =&\ \dfrac{1}{2} \Bigl\{\bigl[\tilde{k}-(l+1)\bigr]^2 +l^2 -1\Bigr\}\geq 0,
                                    \end{split}
                              \end{equation}
                  \end{itemize}
            This ends the proof.
            \end{proof}

%

Let us denote
\begin{eqnarray}
\Omega=\lambda(\lambda+1),
\end{eqnarray}
then Eq. (\ref{eq:F-5}) becomes
 \begin{eqnarray}\label{eq:F-5-1a}
 \dfrac{\hbar^2}{2M } \left\{-\dfrac{{\rm d}^2}{{\rm d}\, r^2}
 -2 \dfrac{1}{r} \dfrac{{\rm d}}{{\rm d}\,r} +\dfrac{\lambda(\lambda+1)}{r^2}\right\}R(r)+ \dfrac{q \kappa}{r} R(r) =E\,R(r).
 \end{eqnarray}
Let us compare Eq. (\ref{eq:F-5-1a}) and Eq. (\ref{eq:eigen-1}), i.e.,
\begin{eqnarray}\label{eq:eigen-1-1}
&& \left[\dfrac{1}{2M} \left(-\hbar^2 \dfrac{{\rm d}^2}{{\rm d} r^2}
-2 \hbar^2 \dfrac{1}{r} \dfrac{{\rm d} }{{\rm d}r}
+ \dfrac{l(l+1)\hbar^2}{ r^2}\right)+\frac{q \kappa}{r}\right] R(r)=E\, R(r).
\end{eqnarray}
One finds that they are almost the same except the following replacement:
\begin{eqnarray}
l \rightarrow \lambda.
\end{eqnarray}
The difference is that $l$ is an integer, while $\lambda$ is generally not an integer because $\tilde{k}\neq 0$. Therefore, for the case of attractive potential, i.e., $q \kappa<0$, the eigenfunctions and the energy spectrum are as follows:
\begin{eqnarray}
R(r) &=& r^{\lambda} {\rm e}^{-\sqrt{-\epsilon}\,r} F\left((\lambda+1)+\dfrac{M q\kappa}{\hbar^2 \sqrt{-\epsilon}};\,
                              2(\lambda+1);\,2 \sqrt{-\epsilon}\,r\right),
\end{eqnarray}
  \begin{eqnarray}
  &&\epsilon= \dfrac{2M E}{\hbar^2}<0,
\end{eqnarray}
\begin{eqnarray}\label{eq:energy-1}
E=-\dfrac{M (q\kappa)^2}{2\hbar^2}\dfrac{1}{\left(N+\lambda+1\right)^2} <0,  \;\; (N=0, 1, 2, 3,...).
\end{eqnarray}
where
\begin{eqnarray}
&& \lambda= \lambda_+=\dfrac{-1+ \sqrt{1+4\Omega}}{2}>0.
\end{eqnarray}

\begin{remark}Here we have chosen $\lambda=\lambda_+$ (instead of $\lambda_-$), the reason is: when $k \rightarrow0$, we expect $\lambda\rightarrow l$, $\Omega \rightarrow l(l+1)$, and accordingly all the above results reduce to those of the usual hydrogen-like atom. Explicitly, let $q=-e$, $\kappa=Q=Ze$, when $k \rightarrow0$ we let $\lambda=l$ and $n=N+l+1$, then Eq. (\ref{eq:energy-1}) becomes
\begin{eqnarray}
E_n=-\dfrac{Z^2 M e^4}{2\hbar^2}\dfrac{1}{n^2},  \;\; (n=1, 2, 3,...),
\end{eqnarray}
which is just Eq. (\ref{eq:hyd-1b}), i.e., the energy of a hydrogen-like atom. $\blacksquare$
\end{remark}

\begin{remark} Let us study the ground state energy with $Z=1$. Let $l=0$ and $N=0$, we can have $\Omega=\tilde{k}^2/2$ and
\begin{eqnarray}
&& \lambda= \dfrac{-1+ \sqrt{1+4\Omega}}{2}=\dfrac{-1+ \sqrt{1+2\tilde{k}^2}}{2}.
\end{eqnarray}
For $q=-e$, $\kappa=Q=Ze$, and $Z=1$, from Eq. (\ref{eq:energy-1}) we have the ground state energy with spin potentials as
\begin{eqnarray}\label{eq:energy-2}
E_{\rm ground}^{\rm spin}=-\dfrac{M e^4}{2\hbar^2}\dfrac{1}{\left(N+\lambda+1\right)^2} =
                     -\dfrac{M e^4}{2\hbar^2}\dfrac{1}{\left(1+\dfrac{-1+ \sqrt{1+2\tilde{k}^2}}{2}\right)^2}.
\end{eqnarray}
The ground state energy of the usual hydrogen atom (HA) is given by
\begin{eqnarray}
E_{\rm ground}^{\rm HA}=-\dfrac{M e^4}{2\hbar^2}.
\end{eqnarray}
In general, for $\tilde{k}\neq 0$, one finds that
\begin{eqnarray}
E_{\rm ground}^{\rm spin} >E_{\rm ground}^{\rm HA}=-\dfrac{M e^4}{2\hbar^2}.
\end{eqnarray}
If $\tilde{k}$ is small, then we can have
\begin{eqnarray}
E_{\rm ground}^{\rm spin}
& =&-\dfrac{M e^4}{2\hbar^2}\dfrac{1}{\left(1+\dfrac{-1+ \sqrt{1+2\tilde{k}^2}}{2}\right)^2} \approx -\dfrac{M e^4}{2\hbar^2}\dfrac{1}{\left(1+\dfrac{-1+ (1+\tilde{k}^2)}{2}\right)^2}\nonumber\\
&=&-\dfrac{M e^4}{2\hbar^2}\dfrac{1}{\left(1+\dfrac{\tilde{k}^2}{2}\right)^2} \approx -\dfrac{M e^4}{2\hbar^2} (1-\tilde{k}^2).
\end{eqnarray}
Thus
\begin{eqnarray}
\Delta E= E_{\rm ground}^{\rm spin}-E_{\rm ground}^{\rm HA}=
\dfrac{M e^4}{2\hbar^2} \tilde{k}^2.
\end{eqnarray}
$\blacksquare$
\end{remark}

\begin{remark}From {$g\hbar k=1, 2$}, one has
\begin{eqnarray}\label{eq:energy-4}
g=\frac{1}{\hbar k}, \;\; {\rm or} \;\; \frac{2}{\hbar k}.
\end{eqnarray}
Because the dimension of $k$ is $[k]=[c/q]$, then the dimension of the parameter $g$ is $[g]=1/[\hbar k]=[q/\hbar c]$. $\blacksquare$
\end{remark}

\newpage

\section{Exact Solution of Dirac's Equation with the Usual Coulomb Interaction}

In this section, for the convenience of comparison, let us review how to exactly solve Dirac's equation with the usual Coulomb interaction.
The Dirac-type Hamiltonian with the usual Coulomb interaction is given by
\begin{eqnarray}\label{eq:H-1a-2}
\mathcal{H}&=&c\, \vec{\alpha}\cdot \vec{p}+\beta M c^2+ q \varphi=c\,\vec{\alpha}\cdot \vec{p}+\beta M c^2+  \dfrac{q\kappa}{r}.
\end{eqnarray}
Here $\vec{\alpha}$ and $\beta$ are the well-known Dirac's matrices, i.e.,
      \begin{equation}
            \vec{\alpha}=
            \left(\begin{array}{cc}
                    0 & \vec{\sigma} \\
                  \vec{\sigma} & 0 \\
                  \end{array}
                \right)=\sigma_x\otimes\vec{\sigma},\;\;\;\;\;
             \beta=
             \left(\begin{array}{cc}
                  \openone & 0 \\
                  0 & -\openone \\
                  \end{array}
                \right)
           =\sigma_z\otimes\openone,
      \end{equation}
which satisfy the following relations
\begin{subequations}
\begin{eqnarray}
  && {\alpha}_x^2={\alpha}_y^2={\alpha}_z^2=\beta^2 =\mathbb{I}, \\
       && \{{\alpha}_j, \beta\}={\alpha}_j \beta+\beta{\alpha}_j=0, \;\;\;\; (j=x, y, z),
\end{eqnarray}
\end{subequations}
with $\openone$ being the $2\times 2$ identity matrix, and $\mathbb{I}=\openone\otimes\openone$ being the $4\times 4$ identity matrix.

One can let $\kappa=Q=Ze$ ($Z$ is an integer, $Z=1$ for the hydrogen atom and $Z \geq 2$ for the hydrogen-like atom). The eigen-problem of the relativistic hydrogen-like atom can be exactly solved, which can be found in textbook of quantum mechanics, such as \cite{1999Flugge}.

\subsection{The Discussion about $\vec{\sigma}\cdot \vec{p}$}

First, we consider the operator $\vec{\sigma}\cdot \vec{p}$, which is
\begin{eqnarray}
      \vec{\sigma}\cdot \vec{p} = -{\rm i}\hbar  \begin{pmatrix}
        \partial_z & \partial_x -{\rm i} \partial_y\\
        \partial_x + {\rm i} \partial_y& -\partial_z
    \end{pmatrix},
\end{eqnarray}
where $\partial_x =\dfrac{\partial }{\partial x }$ , $\partial_y =\dfrac{\partial }{\partial y }$ , $\partial_z =\dfrac{\partial }{\partial z }$.
For $\partial_x ,\partial_y,\partial_z$, we can represent them with $\partial_r , \partial_{\theta}, \partial_{\phi}$, namely
\begin{subequations}
    \begin{eqnarray}
        \partial_x =  \sin \theta \cos\phi \,\partial_r+\frac{\cos \theta \cos\phi}{r}\,\partial_{\theta } -\frac{\sin \phi}{r \sin\theta }\,\partial_{\phi},
    \end{eqnarray}
    \begin{eqnarray}
        \partial_y=  \sin \theta \sin\phi \, \partial_r+\frac{\cos \theta \sin\phi}{r}\, \partial_{\theta } +\frac{\cos \phi}{r \sin\theta }\, \partial_{\phi},
    \end{eqnarray}
    \begin{eqnarray}
        \partial_z =  \cos \theta \,\partial_r-\frac{\sin\theta }{r}\, \partial_{\theta }.
    \end{eqnarray}
\end{subequations}
According to some formulas about $Y_{lm}(\theta ,\phi)$, we can attain
\begin{subequations}
    \begin{eqnarray}
    \label{bas1}    (\partial_x -{\rm i} \partial_y )\,[F(r) Y_{lm}] &=&-\sqrt{\frac{(l+m)(l+m-1)}{(2l+1)(2l-1)}} \left(\frac{\partial F(r)}{\partial r}+\frac{l+1}{r} F(r)\right) Y_{l-1,m-1} \notag\\
        &&+\sqrt{\frac{(l-m+1)(l-m+2)}{(2l+1)(2l+3)}} \left(\frac{\partial F(r)}{\partial r} -\frac{l}{r} F(r)\right) Y_{l+1,m-1},
    \end{eqnarray}
    \begin{eqnarray}
      \label{bas2}  (\partial_x +{\rm i} \partial_y) \,[F(r) Y_{lm}]& =&\sqrt{\frac{(l-m)(l-m-1)}{(2l+1)(2l-1)}} \left(\frac{\partial F(r)}{\partial r} +\frac{l+1}{r} F(r)\right) Y_{l-1,m+1}\notag\\
        &&-\sqrt{\frac{(l+m+2)(l+m+1)}{(2l+1)(2l+3)}} \left(\frac{\partial F(r)}{\partial r} -\frac{l}{r} F(r)\right) Y_{l+1,m+1},
    \end{eqnarray}
    \begin{eqnarray}
        \label{bas3}  \partial_z \, [F(r) Y_{lm}] &=&\sqrt{\frac{(l+m)(l-m)}{(2l+1)(2l-1)}} \left(\frac{\partial F(r)}{\partial r}+\frac{l+1}{r} F(r)\right) Y_{l-1,m} \notag\\
        &&+\sqrt{\frac{(l+m+1)(l-m+1)}{(2l+1)(2l+3)}} \left(\frac{\partial F(r)}{\partial r} -\frac{l}{r} F(r)\right) Y_{l+1,m}.
    \end{eqnarray}
\end{subequations}
Then for $\Phi^A_{ljm_j}$ and $\Phi^B_{ljm_j}$, we can attain
\begin{subequations}
    \begin{eqnarray}
       ( \vec{\sigma}\cdot \vec{p}) \,[F(r) \Phi^A_{ljm_j}(\theta, \phi)] = {\rm i } \hbar  \left(\frac{\partial F(r)}{\partial r}-\frac{l}{r} F(r)\right) \Phi^B_{ljm_j}(\theta, \phi),
    \end{eqnarray}
    \begin{eqnarray}
        ( \vec{\sigma}\cdot \vec{p}) \,[F(r) \Phi^B_{ljm_j}(\theta, \phi)] = {\rm i } \hbar  \left(\frac{\partial F(r)}{\partial r}+\frac{l+2}{r} F(r)\right) \Phi^A_{ljm_j}(\theta, \phi).
     \end{eqnarray}
\end{subequations}
It means that based on  \Eq{bas1}, \Eq{bas2}, and \Eq{bas3}, one can eventually have
    \begin{eqnarray}\label{eq:alpha-2a}
       ( \vec{\alpha}\cdot \vec{p}) \begin{pmatrix}
        F(r)\Phi^A_{ljm_j}(\theta, \phi) \\
        G(r)\Phi^B_{ljm_j}(\theta, \phi)
       \end{pmatrix}={\rm i }\hbar
       \begin{pmatrix}
        \left(\frac{\partial G(r)}{\partial r}+\frac{l+2}{r} G(r)\right)\Phi^A_{ljm_j}(\theta, \phi) \\
        \left(\frac{\partial F(r)}{\partial r}-\frac{l}{r} F(r)\right)\Phi^B_{ljm_j}(\theta, \phi)
       \end{pmatrix}.
    \end{eqnarray}

\begin{proof} Here, we would like to prove  \Eq{bas1}, \Eq{bas2}, and \Eq{bas3}. In the beginning, let us prepare some formulas needed for the proof.

      There are some useful recursive formulas about $P_l(x)$, i.e.,
      \begin{subequations}
            \begin{eqnarray}
             \label{prin1}  (2l+1) x P_l(x) = (l+1) P_{l+1}(x) +l P_{l-1}(x) ,
            \end{eqnarray}
            \begin{eqnarray}
                \label{prin2}     P_l (x) = P_{l+1}' (x ) -2x P_l' (x ) + P_{l-1}' (x),
            \end{eqnarray}
            \begin{eqnarray}
                \label{prin3}    P'_{l+1}(x) = x P_l'(x) +(l+1)P_l(x),
            \end{eqnarray}
            \begin{eqnarray}
            \label{prin4}    P_{l-1}'(x) = x P'_l (x) -l P_l (x),
            \end{eqnarray}
    \end{subequations}
    where $P_l'(x) = \dfrac{d}{d x } P_l (x) $. According to \Eq{prin3} and  \Eq{prin4}, we have
\begin{eqnarray}
    (1-x^2 ) P_{l+1}' &=& P_{l+1}' - x(x P_{l+1}'  )= \left[ xP_l' +(l+1)P_l \right]- x \left[P_{l}'+(l+1)P_{l+1}\right]\nonumber\\
    &=& (l+1)P_l -(l+1)x P_{l+1}=(l+1)[P_l-x P_{l+1}],
\end{eqnarray}
or
\begin{eqnarray}
    (1-x^2 ) P_{l}' &=& l[P_{l-1}-x P_{l}].
\end{eqnarray}
In addition,
%
%
%
\begin{eqnarray}
    [(1-x^2)P_l']^{(|m|)} =  (1-x^2)P_l^{(|m|+1)} -2|m|xP_l^{(|m|)} -(|m|-1)|m|P_l^{(|m|-1)},
\end{eqnarray}
then
\begin{eqnarray}
   \label{P-1} (1-x^2)P_l^{(|m|+1)} = l \left[P_{l-1}^{(|m|)} -  (x P_l)^{(|m|)}\right] + 2|m| x P_{l}^{(|m|)}+ (|m|-1)|m| P_{l}^{(|m|-1)}.
\end{eqnarray}
Based on \Eq{prin4}, we can have
\begin{eqnarray}
    \label{P0} P_{l-1}^{(|m|)} -  (x P_l)^{(|m|)} =  [P_{l-1}' - x P_{l}' -P_l]^{(|m|-1) }= [-l P_l -P_l]^{(|m|-1) }= -(l+1) P_l^{(|m|-1) },
\end{eqnarray}
and
\begin{eqnarray}
  \label{P1}  &&|m| (\cos^2 \theta +1 ) P_l^{(|m|)} -\cos \theta \sin^2 \theta  P_l^{(|m|+1)} \nonumber\\
    &=&\left\{|m| (\cos^2 \theta +1 ) P_l^{(|m|)} -\cos \theta \left[ l P_{l-1}^{(|m|)} - l (\cos \theta P_l)^{(|m|)} + 2|m| \cos \theta P_{l}^{(|m|)}+ (|m|-1)|m| P_{l}^{(|m|-1)}\right] \right\}\notag\\
    &=&\left\{|m| (-\cos^2 \theta +1 ) P_l^{(|m|)} + {\cos \theta [(l+1)l-(|m|-1)|m| ][ P_l^{(|m|-1)} ]} \right\}.
\end{eqnarray}
Similarly, based on \Eq{prin1} and \Eq{prin2}, we have
\begin{eqnarray}
  \label{P2}  (2l+1)P_l = P_{l+1}' -P_{l-1}'.
\end{eqnarray}

(i) Firstly, we consider the case of $m>0$.

   \emph{Proof of \Eq{bas1}.---}For $ \,\partial_x -{\rm i} \partial_y $, we have
    \begin{eqnarray}
        \partial_x -{\rm i} \partial_y  = \sin \theta e^{-{\rm i} \phi }\frac{\partial}{\partial r} + \cos \theta e^{-{\rm i}  \phi} \dfrac{1}{r}\frac{\partial}{\partial \theta} -e^{-\rm i \phi } \dfrac{{\rm i}}{r \sin \theta }\frac{ \partial }{ \partial \phi }.
    \end{eqnarray}
    One can have
    \begin{eqnarray}
        \sin \theta e^{-{\rm i} \phi }\frac{\partial (F(r) Y_{lm})}{\partial r} &=&  \sin \theta e^{-{\rm i} \phi }\frac{\partial F(r) }{\partial r}Y_{lm}\notag\\
        & =& \frac{\partial F(r) }{\partial r} \left\{ -\sqrt{\frac{(l+m-1)(l+m)}{(2l+1)(2l-1)}}Y_{l-1,m-1} + \sqrt{\frac{(l-m+1)(l-m+2)}{(2l+1)(2l+3)}} Y_{l+1,m-1}\right\}.
    \end{eqnarray}
Because
\begin{eqnarray}
   \label{P4} \frac{\partial}{\partial \theta } \left\{ (\sin \theta )^{|m|} \frac{d^{|m|}}{d(\cos \theta )^{|m|}} P_l(\cos \theta)\right\} &= &\sin \theta \left\{|m| (\sin^2 \theta)^{\frac{|m|}{2}-1}\cos \theta P_l^{(|m|)} -(\sin^2 \theta)^{\frac{|m|}{2}}P_l^{(|m|+1)}  \right\}\notag\\
&&=\left\{|m| (\sin^2 \theta)^{\frac{|m|-1}{2}} \cos \theta P_l^{(|m|)} -(\sin^2 \theta)^{\frac{|m|+1}{2}}P_l^{(|m|+1)}  \right\},
\end{eqnarray}
where $P_l^{(|m|)} =\dfrac{\partial^{|m|}}{\partial (\cos\theta)^{|m|} } P_l (\cos \theta) $,
then we have
\begin{eqnarray}
    &&\cos \theta e^{-{\rm i}  \phi}\frac{\partial Y_{lm}}{r\partial \theta} -e^{-\rm i \phi }\frac{{\rm i} \partial Y_{lm} }{ r \sin \theta \partial \phi }\notag\\
    &=& (-1)^m \sqrt{\frac{2l+1}{4\pi} \frac{(l-|m|)!}{(l+|m|)!}}e^{{\rm i} (m-1) \phi} \left\{ |m| (\sin^2 \theta)^{\frac{|m|-1}{2}}(\cos^2 \theta +1 ) P_l^{(|m|)} -(\sin^2 \theta)^{\frac{|m|+1}{2}} \cos \theta P_l^{(|m|+1)}  \right\}\notag\\
    &=&(-1)^m \sqrt{\frac{2l+1}{4\pi} \frac{(l-|m|)!}{(l+|m|)!}}e^{{\rm i} (m-1) \phi} ((\sin^2 \theta)^{\frac{|m|-1}{2}}) \left\{|m| (\cos^2 \theta +1 ) P_l^{(|m|)} -\cos \theta \sin^2 \theta  P_l^{(|m|+1)}  \right\}.
\end{eqnarray}

Based on \Eq{P1}, we have
\begin{eqnarray}
  \label{Equ1'}  &&\cos \theta e^{-{\rm i}  \phi} \dfrac{1}{r}\frac{\partial Y_{lm}}{\partial \theta} -e^{-\rm i \phi }
  \dfrac{{\rm i} }{r \sin \theta}\frac{\partial Y_{lm} }{  \partial \phi }\notag\\
    &=&(-1)^m \sqrt{\frac{2l+1}{4\pi} \frac{(l-|m|)!}{(l+|m|)!}}e^{{\rm i} (m-1) \phi} ((\sin^2 \theta)^{\frac{|m|-1}{2}}) \left\{|m| (\cos^2 \theta +1 ) P_l^{(|m|)} -\cos \theta sin^2 \theta  P_l^{(|m|+1)}  \right\}\notag\\
    &=&(-1)^m \sqrt{\frac{2l+1}{4\pi} \frac{(l-|m|)!}{(l+|m|)!}}e^{{\rm i} (m-1) \phi} ((\sin^2 \theta)^{\frac{|m|-1}{2}}) \left\{|m| (-\cos^2 \theta +1 ) P_l^{(|m|)} +\cos \theta [(l+1)l-(|m|-1)|m| ][ P_l^{(|m|-1)} ] \right\}\notag\\
    &=& |m| \sin \theta e^{-{\rm i} \phi} Y_{lm} - \cos \theta [(l+1)l-(|m|-1)|m| ] \sqrt{\frac{1}{(l+|m|)(l-|m|+1)}} Y_{l,m-1}\notag\\
    &=& |m| \sin \theta e^{-{\rm i} \phi} Y_{lm} - \cos \theta \sqrt{{(l+|m|)(l-|m|+1)}} Y_{l,m-1},
\end{eqnarray}
which means
\begin{eqnarray}
    \cos \theta e^{-{\rm i}  \phi} \dfrac{1}{r}\frac{\partial Y_{lm}}{\partial \theta} -e^{-\rm i \phi }\dfrac{{\rm i} }{ r \sin \theta}\frac{\partial Y_{lm} }{ \partial \phi } =  -\sqrt{\frac{(l+m)(l+m-1)}{(2l+1)(2l-1)}} \frac{l+1}{r}  Y_{l-1,m-1}
    -\sqrt{\frac{(l-m+1)(l-m+2)}{(2l+1)(2l+3)}} \frac{l}{r} Y_{l+1,m-1}.\nonumber\\
\end{eqnarray}
Then we have

\begin{eqnarray}
    (\partial_x -{\rm i} \partial_y )[F(r) Y_{lm}] &=&-\sqrt{\frac{(l+m)(l+m-1)}{(2l+1)(2l-1)}} \left(\frac{\partial F(r)}{\partial r}+\frac{l+1}{r} F(r)\right) Y_{l-1,m-1} \notag\\
    &&+\sqrt{\frac{(l-m+1)(l-m+2)}{(2l+1)(2l+3)}} \left(\frac{\partial F(r)}{\partial r} -\frac{l}{r} F(r)\right) Y_{l+1,m-1}.
\end{eqnarray}
Thus \Eq{bas1} is proved.

 \emph{Proof of \Eq{bas2}.---}For $ \,\partial_x +{\rm i} \partial_y $, we have
    \begin{eqnarray}
        \partial_x +{\rm i} \partial_y  = \sin \theta e^{{\rm i} \phi } \frac{\partial}{\partial r} +\cos \theta e^{{\rm i}  \phi} \dfrac{1}{r}\frac{\partial}{\partial \theta} +e^{\rm i \phi } \dfrac{{\rm i} }{ r \sin \theta}\frac{\partial }{ \partial \phi }.
    \end{eqnarray}
 One can have
    \begin{eqnarray}
        &&\sin \theta e^{{\rm i} \phi }\frac{\partial \left[F(r) Y_{lm}\right]}{\partial r} =  \sin \theta e^{{\rm i} \phi }\frac{\partial F(r) }{\partial r}Y_{lm} \notag\\
        &=& \frac{\partial F(r) }{\partial r} \left\{ \sqrt{\frac{(l-m-1)(l-m)}{(2l+1)(2l-1)}}Y_{l-1,m+1} - \sqrt{\frac{(l+m+1)(l+m+2)}{(2l+1)(2l+3)}} Y_{l+1,m+1}\right\},
    \end{eqnarray}
and
\begin{eqnarray}
        &&\cos \theta e^{{\rm i}  \phi} \dfrac{1}{r}\frac{\partial Y_{lm}}{\partial \theta} +e^{\rm i \phi }
        \dfrac{{\rm i}}{r \sin \theta}\frac{ \partial Y_{lm} }{ \partial \phi }\notag\\
        &=& (-1)^m \sqrt{\frac{2l+1}{4\pi} \frac{(l-|m|)!}{(l+|m|)!}}e^{{\rm i} (m+1) \phi} \left\{|m| (\sin^2 \theta)^{\frac{|m|-1}{2}}(\cos^2 \theta -1 ) P_l^{(|m|)} -(\sin^2 \theta)^{\frac{|m|+1}{2}} \cos \theta P_l^{(|m|+1)}  \right\}\notag\\
        &=&(-1)^m \sqrt{\frac{2l+1}{4\pi} \frac{(l-|m|)!}{(l+|m|)!}}e^{{\rm i} (m+1) \phi} ((\sin^2 \theta)^{\frac{|m|+1}{2}}) \left\{-|m|  P_l^{(|m|)} -\cos \theta  P_l^{(|m|+1)}  \right\}.
    \end{eqnarray}
So based on \Eq{P2}, we have
\begin{eqnarray}
  \label{Equ1}  &&\cos \theta e^{{\rm i}  \phi} \dfrac{1}{r}\frac{\partial Y_{lm}}{\partial \theta} +e^{\rm i \phi }
  \dfrac{{\rm i} }{ r \sin \theta}\frac{\partial Y_{lm} }{ \partial \phi }\notag\\
    &=& |m|\sqrt{\frac{(l+|m|+1)(l+|m|+2)}{(2l+3)(2l+1)}} Y_{l+1,m+1} - |m|\sqrt{\frac{(l-|m|-1)(l-|m|)}{(2l-1)(2l+1)}} Y_{l-1,m+1} \nonumber\\
    &&+{\sqrt{(l-|m|)(l+|m|+1)}} \cos \theta Y_{l,m+1}\notag\\  \notag\\
    &=&|m|\sqrt{\frac{(l+|m|+1)(l+|m|+2)}{(2l+3)(2l+1)}} Y_{l+1,m+1} - |m|\sqrt{\frac{(l-|m|-1)(l-|m|)}{(2l-1)(2l+1)}} Y_{l-1,m+1} \notag\\
    &&+\sqrt{(l-m)(l+m+1)} \left\{\sqrt{\frac{(l+|m|+2)(l-|m|)}{(2l+3)(2l+1)}} Y_{l+1,m+1} +\sqrt{\frac{(l-|m|-1)(l+|m|+1)}{(2l-1)(2l+1)}} Y_{l-1,m+1} \right\}\notag\\
    &=&l\sqrt{\frac{(l+|m|+1)(l+|m|+2)}{(2l+3)(2l+1)}} Y_{l+1,m+1}+ (l+1)\sqrt{\frac{(l-|m|-1)(l-|m|)}{(2l-1)(2l+1)}} Y_{l-1,m+1},
\end{eqnarray}
then
\begin{eqnarray}
      (\partial_x +{\rm i} \partial_y) [F(r) Y_{lm}]& =&\sqrt{\frac{(l-m)(l-m-1)}{(2l+1)(2l-1)}} \left(\frac{\partial F(r)}{\partial r} +\frac{l+1}{r} F(r)\right) Y_{l-1,m+1}\notag\\
      &&-\sqrt{\frac{(l+m+2)(l+m+1)}{(2l+1)(2l+3)}} \left(\frac{\partial F(r)}{\partial r} -\frac{l}{r} F(r)\right) Y_{l+1,m+1}.
  \end{eqnarray}
Thus \Eq{bas2} is proved.

 \emph{Proof of \Eq{bas3}.---}For $\partial_z$, we have
     \begin{eqnarray}
        \partial_z =  \cos \theta \, \frac{\partial}{\partial r}-\sin\theta \frac{1}{r}\, \frac{\partial}{\partial \theta}.
    \end{eqnarray}
One can have
\begin{eqnarray}
    \cos \theta \frac{\partial [F(r) Y_{lm}]}{\partial r} =  \cos \theta \frac{\partial F(r) }{\partial r}Y_{lm}
    = \frac{\partial F(r) }{\partial r} \left[ \sqrt{\frac{(l-m)(l+m)}{(2l+1)(2l-1)}}Y_{l-1,m} + \sqrt{\frac{(l+m+1)(l-m+1)}{(2l+1)(2l+3)}} Y_{l+1,m}\right],
\end{eqnarray}
and
\begin{eqnarray}
&& {\sin\theta } \frac{\partial Y_{lm}}{\partial \theta} \nonumber\\
&=& (-1)^m \sqrt{\frac{2l+1}{4\pi} \frac{(l-|m|)!}{(l+|m|)!}}e^{{\rm i} (m) \phi} \sin^2 \theta \left\{|m| (\sin^2 \theta)^{\frac{|m|}{2}-1}\cos \theta P_l^{(|m|)} -(\sin^2 \theta)^{\frac{|m|}{2}} P_l^{(|m|+1)}  \right\}\notag\\
&=&(-1)^m \sqrt{\frac{2l+1}{4\pi} \frac{(l-|m|)!}{(l+|m|)!}}e^{{\rm i} (m) \phi} \left\{|m| (\sin^2 \theta)^{\frac{|m|}{2}} \cos \theta P_l^{(|m|)} -(\sin^2 \theta)^{\frac{|m|}{2}+1}P_l^{(|m|+1)}  \right\}\notag\\
&=&(-1)^m \sqrt{\frac{2l+1}{4\pi} \frac{(l-|m|)!}{(l+|m|)!}}e^{{\rm i} (m) \phi}(\sin^2 \theta)^{\frac{|m|}{2}} \left\{|m| \cos \theta P_l^{(|m|)}
 +{(l+|m|)(l-|m|+1)P_l^{(|m|-1)} -2|m| \cos \theta P_l^{(|m|)}} \right\}\notag\\
&=&(-1)^m \sqrt{\frac{2l+1}{4\pi} \frac{(l-|m|)!}{(l+|m|)!}}e^{{\rm i} (m) \phi}(\sin^2 \theta)^{\frac{|m|}{2}} {\left\{(l+|m|)(l-|m|+1)P_l^{(|m|-1)} -|m| \cos \theta P_l^{(|m|)} \right\}}\notag\\
&=&(-1)^m \sqrt{\frac{2l+1}{4\pi} \frac{(l-|m|)!}{(l+|m|)!}}e^{{\rm i} (m) \phi}(\sin^2 \theta)^{\frac{|m|}{2}} {\left\{\frac{(l+|m|)(l-|m|+1)}{2l+1}(P_{l+1}^{(|m|)}-P_{l-1}^{(|m|)}) -|m| \cos \theta P_l^{(|m|)} \right\}}\notag\\
&=& {(l+|m|)(l-|m|+1)\left\{\sqrt{\frac{l+|m|+1}{(l-|m|+1)(2l+1)(2l+3)}}Y_{l+1,m} -\sqrt{\frac{l-|m|}{(l+|m|)(2l+1)(2l-1)}}Y_{l-1,m} \right\}-|m| \cos \theta Y_{lm}}\notag\\
&=& {l \sqrt{\frac{(l+|m|+1)(l-|m|+1)}{(2l+1)(2l+3)}}Y_{l+1,m}-(l+1)\sqrt{\frac{(l-|m|)(l+|m|)}{(2l+1)(2l-1)}}Y_{l-1,m}}.
\end{eqnarray}
Based on above, we can attain
\begin{eqnarray}
 \partial_z\,[F(r) Y_{lm}] &=&\sqrt{\frac{(l+m)(l-m)}{(2l+1)(2l-1)}} \left(\frac{\partial F(r)}{\partial r}+\frac{l+1}{r} F(r)\right) Y_{l-1,m} \notag\\
    &&+\sqrt{\frac{(l+m+1)(l-m+1)}{(2l+1)(2l+3)}} \left(\frac{\partial F(r)}{\partial r} -\frac{l}{r} F(r)\right) Y_{l+1,m}.
\end{eqnarray}
Thus \Eq{bas3} is proved.

(ii) Secondly, we consider the case of $m<0$.

For $ \,\partial_x -{\rm i} \partial_y $, based on \Eq{P4}, one has
\begin{eqnarray}
      &&\cos \theta e^{-{\rm i}  \phi} \dfrac{1}{r}\frac{\partial Y_{lm}}{\partial \theta} -e^{-\rm i \phi }
      \dfrac{{\rm i}}{r \sin \theta}\frac{ \partial Y_{lm} }{  \partial \phi }\notag\\
      &=&\sqrt{\frac{2l+1}{4\pi} \frac{(l-|m|)!}{(l+|m|)!}}e^{{\rm i} (m-1) \phi} \left\{|m| (\sin^2 \theta)^{\frac{|m|-1}{2}}(\cos^2 \theta -1 ) P_l^{(|m|)} -(\sin^2 \theta)^{\frac{|m|+1}{2}} \cos \theta P_l^{(|m|+1)}  \right\}\notag\\
      &=&\sqrt{\frac{2l+1}{4\pi} \frac{(l-|m|)!}{(l+|m|)!}}e^{{\rm i} (m-1) \phi} ((\sin^2 \theta)^{\frac{|m|+1}{2}}) \left\{-|m|  P_l^{(|m|)} -\cos \theta P_l^{(|m|+1)}  \right\}\notag\\
      &=&\sqrt{\frac{2l+1}{4\pi} \frac{(l-|m|)!}{(l+|m|)!}}e^{{\rm i} (m-1) \phi} ((\sin^2 \theta)^{\frac{|m-1|}{2}}) \left\{-|m|  P_l^{(|m|)} -\cos \theta P_l^{(|m|+1)}  \right\}.
\end{eqnarray}
Similar with \Eq{Equ1}, one obtains
\begin{eqnarray}
      &&{\cos \theta e^{{-\rm i}  \phi}\dfrac{1}{r} \frac{\partial Y_{lm}}{\partial \theta} -e^{-\rm i \phi }
      \dfrac{{\rm i} }{r \sin \theta}\frac{\partial Y_{lm} }{ \partial \phi }}\notag\\
      &=& -|m|\sqrt{\frac{(l+|m|+1)(l+|m|+2)}{(2l+3)(2l+1)}} Y_{l+1,m-1} +|m|\sqrt{\frac{(l-|m|-1)(l-|m|)}{(2l-1)(2l+1)}} Y_{l-1,m-1} \nonumber\\ &&
      {-\sqrt{(l-|m|)(l+|m|+1)} }\cos \theta Y_{l,m-1}\notag\\  \notag\\
      &=&-|m|\sqrt{\frac{(l+|m|+1)(l+|m|+2)}{(2l+3)(2l+1)}} Y_{l+1,m-1} + |m|\sqrt{\frac{(l-|m|-1)(l-|m|)}{(2l-1)(2l+1)}} Y_{l-1,m-1} \notag\\
      &&{-\sqrt{(l-|m|)(l+|m|+1)} }\left\{\sqrt{\frac{(l+|m|+2)(l-|m|)}{(2l+3)(2l+1)}} Y_{l+1,m-1} +\sqrt{\frac{(l-|m|-1)(l+|m|+1)}{(2l-1)(2l+1)}} Y_{l-1,m-1} \right\}\notag\\
      &=&-l\sqrt{\frac{(l+|m|+1)(l+|m|+2)}{(2l+3)(2l+1)}} Y_{l+1,m-1}- (l+1)\sqrt{\frac{(l-|m|-1)(l-|m|)}{(2l-1)(2l+1)}} Y_{l-1,m-1},
\end{eqnarray}
which is equivalent to
\begin{eqnarray}
      &&{\cos \theta e^{{-\rm i}  \phi}\dfrac{1}{r} \frac{\partial Y_{lm}}{\partial \theta} -e^{-\rm i \phi }
      \dfrac{{\rm i} }{r \sin \theta}\frac{\partial Y_{lm} }{ \partial \phi }}\notag\\
      &=&-l\sqrt{\frac{(l-m+1)(l-m+2)}{(2l+3)(2l+1)}} Y_{l+1,m-1}- (l+1)\sqrt{\frac{(l+m-1)(l+m)}{(2l-1)(2l+1)}} Y_{l-1,m-1}.
\end{eqnarray}
So we can still attain
\begin{eqnarray}
      (\partial_x -{\rm i} \partial_y )[F(r) Y_{lm}] &=&-\sqrt{\frac{(l+m)(l+m-1)}{(2l+1)(2l-1)}} \left(\frac{\partial F(r)}{\partial r}+\frac{l+1}{r} F(r)\right) Y_{l-1,m-1} \notag\\
      &&+\sqrt{\frac{(l-m+1)(l-m+2)}{(2l+1)(2l+3)}} \left(\frac{\partial F(r)}{\partial r} -\frac{l}{r} F(r)\right) Y_{l+1,m-1}.
  \end{eqnarray}

For $ \,\partial_x +{\rm i} \partial_y $, we can attain
\begin{eqnarray}
      && \cos \theta e^{{\rm i}  \phi} \dfrac{1}{r}\frac{\partial Y_{lm}}{\partial \theta} +e^{\rm i \phi }
      \dfrac{{\rm i}}{r \sin \theta}\frac{\partial Y_{lm} }{\partial \phi }\notag\\
      &=& \sqrt{\frac{2l+1}{4\pi} \frac{(l-|m|)!}{(l+|m|)!}}e^{{\rm i} (m+1) \phi} \left\{|m| (\sin^2 \theta)^{\frac{|m|-1}{2}}(\cos^2 \theta +1 ) P_l^{(|m|)} -(\sin^2 \theta)^{\frac{|m|+1}{2}} \cos \theta P_l^{(|m|+1)}  \right\}\notag\\
      &=&\sqrt{\frac{2l+1}{4\pi} \frac{(l-|m|)!}{(l+|m|)!}}e^{{\rm i} (m+1) \phi} ((\sin^2 \theta)^{\frac{|m|-1}{2}}) \left\{|m| (\cos^2 \theta +1 ) P_l^{(|m|)} -\cos \theta \sin^2 \theta P_l^{(|m|+1)}  \right\}\notag\\
      &=&\sqrt{\frac{2l+1}{4\pi} \frac{(l-|m|)!}{(l+|m|)!}}e^{{\rm i} (m+1) \phi} ((\sin^2 \theta)^{\frac{|m+1|}{2}}) \left\{|m| (\cos^2 \theta +1 ) P_l^{(|m|)} -\cos \theta \sin^2 \theta P_l^{(|m|+1)}  \right\}.
\end{eqnarray}
Similar with \Eq{Equ1'}, we have
\begin{eqnarray}
       &&\cos \theta e^{{\rm i}  \phi} \dfrac{1}{r}\frac{\partial Y_{lm}}{\partial \theta} +e^{\rm i \phi }
       \dfrac{{\rm i}}{r \sin \theta }\frac{ \partial Y_{lm} }{ \partial \phi }\notag\\
        &=&\sqrt{\frac{2l+1}{4\pi} \frac{(l-|m|)!}{(l+|m|)!}}e^{{\rm i} (m+1) \phi} ((\sin^2 \theta)^{\frac{|m|-1}{2}}) \left\{|m| (\cos^2 \theta +1 ) P_l^{(|m|)} -\cos \theta \sin^2 \theta  P_l^{(|m|+1)}  \right\}\notag\\
        &=&\sqrt{\frac{2l+1}{4\pi} \frac{(l-|m|)!}{(l+|m|)!}}e^{{\rm i} (m+1) \phi} ((\sin^2 \theta)^{\frac{|m|-1}{2}}) \left\{|m| (-\cos^2 \theta +1 ) P_l^{(|m|)} +\cos \theta [(l+1)l-(|m|-1)|m| ][ P_l^{(|m|-1)} ] \right\}\notag\\
        &=& |m| \sin \theta e^{{\rm i} \theta} Y_{lm} + \cos \theta [(l+1)l-(|m|-1)|m| ] \sqrt{\frac{1}{(l+|m|)(l-|m|+1)}} Y_{l,m+1}\notag\\
        &=& |m| \sin \theta e^{{\rm i} \theta} Y_{lm} + \cos \theta \sqrt{{(l+|m|)(l-|m|+1)}} Y_{l,m+1}\notag\\
        &=& \sqrt{\frac{(l-m)(l-m-1)}{(2l+1)(2l-1)}} ({l+1} ) Y_{l-1,m+1}
        +\sqrt{\frac{(l+m+1)(l+m+2)}{(2l+1)(2l+3)}} {l} Y_{l+1,m+1}.
    \end{eqnarray}
So we can still attain
\begin{eqnarray}
            (\partial_x +{\rm i} \partial_y) [F(r) Y_{lm}]& =&\sqrt{\frac{(l-m)(l-m-1)}{(2l+1)(2l-1)}} \left(\frac{\partial F(r)}{\partial r} +\frac{l+1}{r} F(r)\right) Y_{l-1,m+1}\notag\\
            &&-\sqrt{\frac{(l+m+2)(l+m+1)}{(2l+1)(2l+3)}} \left(\frac{\partial F(r)}{\partial r} -\frac{l}{r} F(r)\right) Y_{l+1,m+1}.
\end{eqnarray}

For $\partial_z$, when $m>0$, one has
\begin{eqnarray}
      {\sin\theta } \dfrac{\partial Y_{lm}}{\partial \theta } &=& (-1)^m \sqrt{\frac{2l+1}{4\pi} \frac{(l-|m|)!}{(l+|m|)!}}e^{{\rm i} (m) \phi} \sin^2 \theta \left\{|m| (\sin^2 \theta)^{\frac{|m|}{2}-1} \cos \theta P_l^{(|m|)} -(\sin^2 \theta)^{\frac{|m|}{2}} P_l^{(|m|+1)}  \right\}\notag\\
      &=& l \sqrt{\frac{(l+m+1)(l-m+1)}{(2l+1)(2l+3)}}Y_{l+1,m}-(l+1)\sqrt{\frac{(l-m)(l+m)}{(2l+1)(2l-1)}}Y_{l-1,m}.
\end{eqnarray}
When $m<0$ ,
\begin{eqnarray}
      {\sin\theta } \dfrac{\partial Y_{lm}}{\partial \theta } &=& \sqrt{\frac{2l+1}{4\pi} \frac{(l-|m|)!}{(l+|m|)!}}e^{{\rm i} (m) \phi} \sin^2 \theta \left\{|m| (\sin^2 \theta)^{\frac{|m|}{2}-1} \cos \theta P_l^{(|m|)} -(\sin^2 \theta)^{\frac{|m|}{2}} P_l^{(|m|+1)}  \right\}\notag\\
      &=& l \sqrt{\frac{(l+m+1)(l-m+1)}{(2l+1)(2l+3)}}Y_{l+1,m}-(l+1)\sqrt{\frac{(l-m)(l+m)}{(2l+1)(2l-1)}}Y_{l-1,m}.
\end{eqnarray}

So we can easily attain
\begin{eqnarray}
      \partial_z[F(r) Y_{lm}] &=&\sqrt{\frac{(l+m)(l-m)}{(2l+1)(2l-1)}} \left(\frac{\partial F(r)}{\partial r}+\frac{l+1}{r} F(r)\right) Y_{l-1,m} \notag\\
    &&+\sqrt{\frac{(l+m+1)(l-m+1)}{(2l+1)(2l+3)}} \left(\frac{\partial F(r)}{\partial r} -\frac{l}{r} F(r)\right) Y_{l+1,m}.
\end{eqnarray}
This ends the proof of \Eq{bas1}, \Eq{bas2}, and \Eq{bas3}.
\end{proof}

\subsection{Exact Solution of the Eigen-Problem}

The eigen-problem is given by
\begin{equation}
      \mathcal{H}\,\Psi(r,\theta,\phi)=E \, \Psi(r,\theta,\phi),
\end{equation}
where $\mathcal{H}$ is the Dirac's Hamiltonian of the hydrogen-like atom given in Eq. (\ref{eq:H-1a-2}), $E$ is the energy and the form of the wavefunction is as follows
\begin{eqnarray}
      \Psi(r,\theta,\phi) = \begin{pmatrix}
            F(r)\Phi^A_{ljm_j}(\theta, \phi) \\
            G(r)\Phi^B_{ljm_j}(\theta, \phi)
      \end{pmatrix}.
\end{eqnarray}
Due to Eq. (\ref{eq:alpha-2a}), we can easily attain
\begin{eqnarray}
      {\rm i }\hbar c\begin{pmatrix}
            \left(\frac{\partial G(r)}{\partial r}+\frac{l+2}{r} G(r)\right)\Phi^A_{ljm_j}(\theta, \phi) \\
            \left(\frac{\partial F(r)}{\partial r}-\frac{l}{r} F(r)\right)\Phi^B_{ljm_j}(\theta, \phi)
           \end{pmatrix}
           + \begin{pmatrix}
         \left( M c^2 +\frac{q \kappa }{r} \right)  F(r)\Phi^A_{ljm_j}(\theta, \phi) \\
         \left(-M c^2 +\frac{q \kappa }{r}\right)  G(r)\Phi^B_{ljm_j}(\theta, \phi)
      \end{pmatrix}=E \begin{pmatrix}
            F(r)\Phi^A_{ljm_j}(\theta, \phi) \\
            G(r)\Phi^B_{ljm_j}(\theta, \phi)
      \end{pmatrix},
\end{eqnarray}
i.e.,
\begin{subequations}
      \begin{eqnarray}
            {\rm i }\hbar c \left[\frac{\partial F(r)}{\partial r}-\frac{l}{r} F(r)\right] - \left(E+M c^2-\frac{q \kappa }{r}\right)G(r)=0,
      \end{eqnarray}
      \begin{eqnarray}
            {\rm i } \hbar c \left[\frac{\partial G(r)}{\partial r}+\frac{l+2}{r} G(r)\right] - \left(E- M c^2 -\frac{q \kappa }{r}\right) F(r)=0.
      \end{eqnarray}
\end{subequations}

When $r \rightarrow \infty $, we have
\begin{subequations}
      \begin{eqnarray}
            {\rm i }\hbar c  \frac{\partial F(r)}{\partial r} -(E+M c^2)G(r)=0,
      \end{eqnarray}
      \begin{eqnarray}
            {\rm i } \hbar c \frac{\partial G(r)}{\partial r}  -(E- M c^2 ) F(r)=0.
      \end{eqnarray}
\end{subequations}
If we designate
     \begin{eqnarray}\label{eq:FG-1}
      F = C_1 e^{-\varepsilon r }, \;\;\;\; G = C_2 e^{-\varepsilon r }
      \end{eqnarray}
then we have
\begin{subequations}
      \begin{eqnarray}
            -{\rm i }\hbar c  \,\varepsilon C_1 -(E+M c^2) C_2 =0,
      \end{eqnarray}
      \begin{eqnarray}
            -{\rm i }\hbar c  \, \varepsilon C_2 -(E-M c^2) C_1 =0,
      \end{eqnarray}
\end{subequations}
which lead to
\begin{eqnarray}\label{eq:C-1}
      && \varepsilon =  \frac{1}{\hbar c}\sqrt{M^2c^4 -E^2 }, \nonumber\\
      && C_1 = {\rm i } {\sqrt{\frac{M c^2+E}{M c^2-E} }}C_2={\rm i } \mu C_2.
\end{eqnarray}
Here for simplicity we have denoted
\begin{eqnarray}\label{eq:mu-1}
      \mu= \sqrt{\frac{M c^2+E}{M c^2-E} }.
\end{eqnarray}

When $r \rightarrow 0 $, we have
\begin{subequations}
      \begin{eqnarray}
            {\rm i }\hbar  c\left[\frac{\partial F(r)}{\partial r}-\frac{l}{r} F(r)\right] +\frac{q \kappa }{r}G(r)=0,
      \end{eqnarray}
      \begin{eqnarray}
            {\rm i }  \hbar  c \left[\frac{\partial G(r)}{\partial r}+\frac{l+2}{r} G(r)\right] +\frac{q \kappa }{r}F(r)=0.
      \end{eqnarray}
\end{subequations}
If we designate
      \begin{eqnarray}\label{eq:FG-2}
      F = C_1' r^{\nu-1}, \;\;\;\ G = C_2' r^{\nu-1}
      \end{eqnarray}
then we have
\begin{subequations}
      \begin{eqnarray}
            {\rm i } \hbar  c (\nu-1-l) C_1' +q\kappa C_2' =0,
      \end{eqnarray}
      \begin{eqnarray}
            {\rm i } \hbar  c (\nu+l+1) C_2' +q\kappa C_1' =0,
      \end{eqnarray}
\end{subequations}
which lead to
\begin{eqnarray}
     && \nu=\sqrt{(l+1)^2 -\frac{q^2 \kappa^2}{\hbar^2 c^2 }} = \sqrt{(l+1)^2 -\tau^2},\nonumber\\
     && C_1'={\rm i} \sqrt{\dfrac{l+1+\nu}{l+1-\nu}}\, C_2',
\end{eqnarray}
where  for simplicity we have denoted
\begin{eqnarray}
     && \tau = \frac{q \kappa}{\hbar c }.
\end{eqnarray}
Note that for the attractive potential, one has $\tau<0$.

Then we designate
\begin{eqnarray}
     && F(r) ={\rm i} \mu A(r) r^{\nu-1}e^{-\varepsilon r}, \;\;\;\;\;G(r) =B(r) r^{\nu-1}e^{-\varepsilon r},
\end{eqnarray}
then we have
\begin{subequations}
      \begin{eqnarray}
           \left[\frac{\partial A(r)}{\partial r}+\left(\frac{\nu-l-1}{r}-\varepsilon\right) A(r)\right] +\left(\varepsilon-\frac{\tau}{\mu r}\right)B(r)=0,
      \end{eqnarray}
      \begin{eqnarray}
            \left[\frac{\partial B(r)}{\partial r}+\left(\frac{\nu+l+1}{r}-\varepsilon\right) B(r)\right]  +\left(\varepsilon+\frac{\mu \tau  }{r}\right) A(r)=0.
      \end{eqnarray}
\end{subequations}
Now we set $h_1(r) = A(r)+B(r)$ and $h_2(r) =A(r)-B(r)$, then we obtain
\begin{subequations}
      \begin{eqnarray}
           \left[\frac{\partial h_1(r)}{\partial r}+\left(\frac{\nu+k_1  }{r}\right) h_1(r)\right] +\left(\frac{k_2  -(l+1) }{r}\right)h_2(r)=0,
      \end{eqnarray}
      \begin{eqnarray}
            \left[\frac{\partial h_2(r)}{\partial r}+\left(\frac{\nu-k_1 }{r}-2\varepsilon\right) h_2(r)\right] +\left(\frac{-k_2 -(l+1) }{ r}\right)h_1(r)=0,
      \end{eqnarray}
\end{subequations}
where
\begin{eqnarray}
      k_1 = \frac{\tau }{2} \left(\mu -\frac{1}{\mu}\right) \,\,,\,\, k_2 = \frac{\tau }{2} \left(\mu +\frac{1}{\mu}\right).
\end{eqnarray}
Now we set
\begin{eqnarray}
     && h_1 (r)=\sum_{K=0}^{\infty} a_K r^K= a_0 + a_1 r + \dots + a_K r^K +\dots, \nonumber\\
     && h_2 (r)=\sum_{K=0}^{\infty} b_K r^K= b_0 + b_1 r + \dots + b_K r^K +\dots,
\end{eqnarray}
then we have
\begin{subequations}
      \begin{eqnarray}
         \label{t11}  ( K+\nu+k_1) a_K +(k_2-l-1)b_K =0,
      \end{eqnarray}
      \begin{eqnarray}
            (K+\nu-k_1)b_K -2\varepsilon b_{K-1} -(k_2+l+1)a_K =0.
      \end{eqnarray}
\end{subequations}
Based on which, one can obtain
\begin{eqnarray}
      b_K &=& \frac{2\varepsilon (K+\nu+k_1)}{(K+\nu)^2 -k_1^2 +k_2^2 -(l+1)^2} b_{K-1}
      = \frac{2\varepsilon (K+\nu+k_1)}{(K+\nu)^2 +\tau^2 -(l+1)^2} b_{K-1} =\frac{2\varepsilon (K+\nu+k_1)}{(K+\nu)^2 -\nu^2} b_{K-1} \notag\\
&=& \frac{2\varepsilon (K+\nu+k_1)}{(K+2\nu)K} b_{K-1}.
\end{eqnarray}
Similarly, from
\begin{eqnarray}
       ( K+\nu+k_1)a_K &=& -(k_2-l-1)b_K= -(k_2-l-1) \frac{2\varepsilon (K+\nu+k_1)}{(K+2\nu)K} b_{K-1} = \frac{2\varepsilon (K+\nu+k_1)}{(K+2\nu)K}
       \left[-(k_2-l-1) b_{K-1} \right]\nonumber\\
       &=& \frac{2\varepsilon (K+\nu+k_1)}{(K+2\nu)K} \left[( K-1+\nu+k_1)a_{K-1}\right] = ( K+\nu+k_1) \frac{2\varepsilon (K+\nu+k_1-1)}{(K+2\nu)K}a_{K-1},
\end{eqnarray}
one has
\begin{eqnarray}
      a_K &=& \frac{2\varepsilon (K+\nu+k_1-1)}{(K+2\nu)K}a_{K-1}.
\end{eqnarray}
The above results implies that
\begin{eqnarray}
      h_1(r) =  {D_1}\,F(\nu+k_1;\, 2\nu+1;\,2\varepsilon {r}) \,\,,\,\,
      h_2(r) =  {D_2}\,F(\nu+k_1+1;\, 2\nu+1;\,2\varepsilon {r}).
\end{eqnarray}

Based on \Eq{t11}, we have
\begin{eqnarray}\label{eq:K-1}
      ( K+\nu+k_1) {D_1} \frac{(\nu+k_1)_K}{(2\nu+1)_K\, K!} = -(k_2-l-1) {D_2} \frac{(\nu+k_1+1)_K}{(2\nu+1)_K\, K!},
\end{eqnarray}
with $(\gamma)_K = \gamma(\gamma+1)\cdots(\gamma+K-1)$. From Eq. (\ref{eq:K-1}) we have
\begin{eqnarray}
      ( K+\nu+k_1) {D_1} {(\nu+k_1)_K} = -(k_2-l-1) {D_2} {(\nu+k_1+1)_K},
\end{eqnarray}
i.e.,
\begin{eqnarray}
      ( K+\nu+k_1) {D_1} {(\nu+k_1)} = -(k_2-l-1) {D_2} {(\nu+k_1+K)},
\end{eqnarray}
i.e.,
\begin{eqnarray}
      {D_1} = -\frac{k_2-l-1}{\nu+k_1} {D_2}.
\end{eqnarray}

To ensure the radial wavefunction is not divergent when $r\rightarrow \infty$, $h_1$ and $h_2$ must be truncated. Thus, we let
\begin{eqnarray}
      \nu+k_1=-N,  \,\,N = 0 ,1 ,2,\dots.
\end{eqnarray}
In this situation,
\begin{eqnarray}
      h_1(r) =  {D_1}\,F(-N;\, 2\nu+1;\,2\varepsilon {r}) \,\,,\,\,
      h_2 (r)=  {D_2}\,F(-N+1;\, 2\nu+1;\,2\varepsilon {r}).
\end{eqnarray}
Further, we can have
\begin{eqnarray}\label{eq:FG-3}
&& F(r) ={\rm i} \mu A(r) r^{\nu-1}e^{-\varepsilon r}={\rm i} \mu \dfrac{h_1(r)+h_2(r)}{2} r^{\nu-1}e^{-\varepsilon r},\nonumber\\
&&G(r) =B(r) r^{\nu-1}e^{-\varepsilon r}=\dfrac{h_1(r)-h_2(r)}{2} r^{\nu-1}e^{-\varepsilon r},
\end{eqnarray}
or
\begin{eqnarray}
   &&   F(r) =\frac{D_1}{2}{\rm i} \mu \,r^{\nu-1}e^{-\varepsilon r} \left\{F(-N;\, 2\nu+1;\,2\varepsilon r)
   -\frac{\nu+k_1}{k_2-l-1}F(-N+1;\, 2\nu+1;\,2\varepsilon r) \right\},\nonumber\\
   &&   G(r) =\frac{D_2}{2} r^{\nu-1}e^{-\varepsilon r} \left\{F(-N;\, 2\nu+1;\,2\varepsilon r)
   + \frac{\nu+k_1}{k_2-l-1}F(-N+1;\, 2\nu+1;\,2\varepsilon r) \right\}.
\end{eqnarray}

\begin{remark}Because
\begin{eqnarray}
    && N+ \nu+k_1=0,  \,\,(N = 0 ,1 ,2,\dots),\nonumber\\
    && a_K = \frac{2\varepsilon (K+\nu+k_1-1)}{(K+2\nu)K}a_{K-1}, \nonumber\\
    && b_K = \frac{2\varepsilon (K+\nu+k_1)}{(K+2\nu)K} b_{K-1},
\end{eqnarray}
i.e.,
\begin{eqnarray}
    && N+ \nu+k_1=0,  \,\,(N = 0 ,1 ,2,\dots),\nonumber\\
    && a_{N+1} = \frac{2\varepsilon (N+\nu+k_1)}{(N+1+2\nu)(N+1)}a_{N}, \nonumber\\
    && b_N = \frac{2\varepsilon (N+\nu+k_1)}{(N+2\nu)N} b_{N-1},
\end{eqnarray}
then for a fixed $N$, we have
\begin{eqnarray}
    && a_{i}=0, \;\; (i\geq N+1), \nonumber\\
    && b_j=0, \;\; (j \geq N),
\end{eqnarray}
thus
\begin{eqnarray}
    && h_1(r)=\sum_{i=0}^{N} a_i r^i, \;\;\;\; h_2(r)=\sum_{j=0}^{N-1} b_j r^j.
\end{eqnarray}

When $r\rightarrow \infty$, on one hand, from Eq. (\ref{eq:FG-1}) we have
\begin{eqnarray}\label{eq:FG-4a}
    && \dfrac{F}{G}=\dfrac{C_1 e^{-\varepsilon r }}{C_2 e^{-\varepsilon r }}=\dfrac{C_1}{C_2}={\rm i}\mu.
\end{eqnarray}
On one hand, from Eq. (\ref{eq:FG-3}) we have
\begin{eqnarray}\label{eq:FG-4b}
 \lim_{r\rightarrow \infty} \dfrac{F(r)}{G(r)} &=& \lim_{r\rightarrow \infty} \dfrac{{\rm i} \mu \dfrac{h_1(r)+h_2(r)}{2} r^{\nu-1}e^{-\varepsilon r}}
{\dfrac{h_1(r)-h_2(r)}{2} r^{\nu-1}e^{-\varepsilon r}}= {\rm i} \mu  \lim_{r\rightarrow \infty} \dfrac{h_1(r)+h_2(r)}{h_1(r)-h_2(r)}\nonumber\\
&=& {\rm i} \mu  \lim_{r\rightarrow \infty} \dfrac{\sum_{i=0}^{N} a_i r^i+\sum_{j=0}^{N-1} b_j r^j}{\sum_{i=0}^{N} a_i r^i-\sum_{j=0}^{N-1} b_j r^j}
= {\rm i} \mu  \lim_{r\rightarrow \infty}\dfrac{a_{N} r^N}{a_{N} r^N}={\rm i} \mu.
\end{eqnarray}
Namely, Eq. (\ref{eq:FG-4b}) coincides with Eq. (\ref{eq:FG-4a}).

When $r\rightarrow 0$, on one hand, from Eq. (\ref{eq:FG-2}) we have
\begin{eqnarray}\label{eq:FG-5a}
    && \dfrac{F}{G}=\dfrac{C_1' r^{\nu-1}}{C_2' r^{\nu-1}}=\dfrac{C_1'}{C_2'}={\rm i} \sqrt{\dfrac{l+1+\nu}{l+1-\nu}}.
\end{eqnarray}
On one hand, from Eq. (\ref{eq:FG-3}) we have
\begin{eqnarray}\label{eq:FG-5b}
 \lim_{r\rightarrow 0} \dfrac{F(r)}{G(r)} &=& \lim_{r\rightarrow 0} \dfrac{{\rm i} \mu \dfrac{h_1(r)+h_2(r)}{2} r^{\nu-1}e^{-\varepsilon r}}
{\dfrac{h_1(r)-h_2(r)}{2} r^{\nu-1}e^{-\varepsilon r}}= {\rm i} \mu  \lim_{r\rightarrow 0} \dfrac{h_1(r)+h_2(r)}{h_1(r)-h_2(r)}\nonumber\\
&=& {\rm i} \mu  \dfrac{D1+D_2}{D_1-D_2}={\rm i} \mu  \dfrac{\dfrac{D1}{D_2}+1}{\dfrac{D1}{D_2}-1}={\rm i} \mu
\dfrac{\left( -\dfrac{k_2-l-1}{\nu+k_1} \right)+1}{\left( -\dfrac{k_2-l-1}{\nu+k_1} \right)-1}\nonumber\\
&=& {\rm i} \mu \dfrac{-(k_2-l-1)+(\nu+k_1)}{-(k_2-l-1)-(\nu+k_1)}={\rm i} \mu \dfrac{(k_2-l-1)-(\nu+k_1)}{(k_2-l-1)+(\nu+k_1)}\nonumber\\
&=& {\rm i} \mu \dfrac{(k_2-k_1)-(l+1+\nu)}{(k_2+k_1)-(l+1-\nu)}= {\rm i} \mu \dfrac{\dfrac{\tau}{\mu}-(l+1+\nu)}{\tau\mu-(l+1-\nu)}
= {\rm i} \dfrac{\tau-(l+1+\nu)\mu}{\tau\mu-(l+1-\nu)}.
\end{eqnarray}
Namely, to make Eq. (\ref{eq:FG-5b}) coincide with Eq. (\ref{eq:FG-5a}), one must have
\begin{eqnarray}
  \dfrac{\tau-(l+1+\nu)\mu}{\tau\mu-(l+1-\nu)}=\sqrt{\dfrac{l+1+\nu}{l+1-\nu}},
\end{eqnarray}
i.e.,
\begin{eqnarray}
 \tau-(l+1+\nu)\mu=\sqrt{\dfrac{l+1+\nu}{l+1-\nu}} [\tau\mu-(l+1-\nu)],
\end{eqnarray}
i.e.,
\begin{eqnarray}
 \tau-(l+1+\nu)\mu=\sqrt{\dfrac{(l+1)^2-\nu^2}{(l+1-\nu)^2}} [\tau\mu-(l+1-\nu)],
\end{eqnarray}
i.e.,
\begin{eqnarray}
 \tau-(l+1+\nu)\mu=\dfrac{{|\tau|}}{l+1-\nu} [\tau\mu-(l+1-\nu)],
\end{eqnarray}
i.e.,
\begin{eqnarray}
 \tau-(l+1+\nu)\mu=\dfrac{-\tau}{l+1-\nu} [\tau\mu-(l+1-\nu)],
\end{eqnarray}
i.e.,
\begin{eqnarray}
(l+1-\nu) \tau-[(l+1)^2-\nu^2]\mu=-\tau [\tau\mu-(l+1-\nu)],
\end{eqnarray}
i.e.,
\begin{eqnarray}
(l+1-\nu) \tau-\tau^2\mu=-\tau [\tau\mu-(l+1-\nu)],
\end{eqnarray}
which is an identity. Thus Eq. (\ref{eq:FG-4b}) coincides with Eq. (\ref{eq:FG-4a}).  $\blacksquare$
\end{remark}

Next, let us come to determine the energy. Based on
\begin{eqnarray}
   &&   \nu+k_1=-N,  \,\,N = 0 ,1 ,2,\dots, \nonumber\\
   &&   k_1 = \frac{\tau }{2} \left(\mu -\frac{1}{\mu}\right), \nonumber\\
   &&    \mu= \sqrt{\frac{M c^2+E}{M c^2-E} },
\end{eqnarray}
we have
\begin{eqnarray}
   &&  \mu^2= \frac{M c^2+E}{M c^2-E},
\end{eqnarray}
i.e.,
\begin{eqnarray}
   &&  (\mu^2+1)E=(\mu^2-1)M c^2,
\end{eqnarray}
i.e.,
\begin{eqnarray}
   &&  E=\dfrac{\mu^2-1}{\mu^2+1} M c^2.
\end{eqnarray}
Here $\mu$ is determined by
\begin{eqnarray}
   &&  \frac{\tau }{2} \left(\mu -\frac{1}{\mu}\right)= k_1=-(N+\nu),
\end{eqnarray}
i.e.,
\begin{eqnarray}\label{eq:mu-2a}
   &&  \left(\mu -\frac{1}{\mu}\right)=-\dfrac{2}{\tau}(N+\nu),
\end{eqnarray}
i.e.,
\begin{eqnarray}\label{eq:mu-2b}
   &&  \mu^2 +\dfrac{2(N+\nu)}{\tau}\mu  -1=0.
\end{eqnarray}
By considering $\mu >0$, we then have
\begin{eqnarray}
     \mu &=& \dfrac{-\dfrac{2(N+\nu)}{\tau}+ \sqrt{\left(\dfrac{2(N+\nu)}{\tau}\right)^2+4}}{2}=-\dfrac{(N+\nu)}{\tau}+
   \sqrt{\left(\dfrac{(N+\nu)}{\tau}\right)^2+1}\nonumber\\
   &=& \dfrac{(N+\nu)}{|\tau|}+ \sqrt{\dfrac{(N+\nu)^2}{\tau^2}+1}.
   \end{eqnarray}
Due to Eq. (\ref{eq:mu-2a}) and Eq. (\ref{eq:mu-2b}), one obtains
\begin{eqnarray}\label{eq:mu-2c}
   &&  \mu^2 -1=-\dfrac{2(N+\nu)}{\tau}\mu, \;\;\;\; \mu^2 +1=-\dfrac{2(N+\nu)}{\tau}\mu+2, \;\;\; \frac{1}{\mu}=\mu +\dfrac{2}{\tau}(N+\nu),
\end{eqnarray}
then
\begin{eqnarray}
    \dfrac{\mu^2-1}{\mu^2+1}&=&\dfrac{{-}\dfrac{2(N+\nu)}{\tau}\mu }{{-}\dfrac{2(N+\nu)}{\tau}\mu +2}
    =\dfrac{-\dfrac{(N+\nu)}{\tau}\mu }{-\dfrac{(N+\nu)}{\tau}\mu +1}=\dfrac{1}{1-\dfrac{\tau}{(N+\nu)}\dfrac{1}{\mu}}\nonumber\\
   &=&\dfrac{1}{1-\dfrac{\tau}{(N+\nu)}\left[\mu +\dfrac{2}{\tau}(N+\nu)\right]}=\dfrac{1}{1+\dfrac{|\tau|}{(N+\nu)}
   \left[\mu -\dfrac{2}{|\tau|}(N+\nu)\right]}\nonumber\\
   &=& \dfrac{1}{\dfrac{|\tau|}{(N+\nu)}\mu-1}=\dfrac{1}{\dfrac{|\tau|}{(N+\nu)}\left[\dfrac{(N+\nu)}{|\tau|}
   + \sqrt{\dfrac{(N+\nu)^2}{\tau^2}+1}\right]-1}\nonumber\\
   &=&\dfrac{1}{\sqrt{1+\dfrac{\tau^2}{(N+\nu)^2}}}=\dfrac{1}{\sqrt{1+\dfrac{q^2 \kappa^2}{\hbar^2 c^2 }\dfrac{1}{(N+\nu)^2}}}.
\end{eqnarray}
Thus the energy is given by
\begin{eqnarray}\label{eq:hyd-3a}
   &&  E=\dfrac{\mu^2-1}{\mu^2+1} M c^2=\dfrac{M c^2}{\sqrt{1+\dfrac{q^2 \kappa^2}{\hbar^2 c^2 }\dfrac{1}{(N+\nu)^2}}},
\end{eqnarray}
with
\begin{eqnarray}
     && \nu=\sqrt{(l+1)^2 -\frac{q^2 \kappa^2}{\hbar^2 c^2 }}.
\end{eqnarray}

\begin{remark}\textcolor{blue}{The Non-Relativistic Limit.} Because particle 1 has a mass $M$, thus the internal energy of
the particle 1 is $E_{\rm int}=Mc^2$. In the non-relativistic limit, after subtracting the internal energy, the external
energy of particle 1 is given by
\begin{eqnarray}
     E_{\rm ext} &=& E-E_{\rm int}=\dfrac{Mc^2}{\sqrt{1+\dfrac{q^2 \kappa^2}{\hbar^2 c^2 }\dfrac{1}{(N+\nu)^2}}}-Mc^2\nonumber\\
     & \approx & Mc^2 \left(1-\dfrac{1}{2} \dfrac{q^2 \kappa^2}{\hbar^2 c^2 }\dfrac{1}{(N+\nu)^2} \right)-Mc^2\nonumber\\
     &=& - \dfrac{M q^2 \kappa^2}{2 \hbar^2 }\dfrac{1}{(N+\nu)^2}=-\dfrac{M q^2 \kappa^2}{2 \hbar^2 }\dfrac{1}
     {\left[N+\left(\sqrt{(l+1)^2 -\frac{q^2 \kappa^2}{\hbar^2 c^2 }}\right)\right]^2}\nonumber\\
     & \approx & - \dfrac{M q^2 \kappa^2}{2 \hbar^2 }\dfrac{1}{(N+l+1)^2}.
\end{eqnarray}
After denoting $E_{\rm ext} $ as $E_n$, and
    \begin{eqnarray}
    n=N+l+1,
    \end{eqnarray}
we then have
\begin{eqnarray}\label{eq:hyd-2a}
E_n=-\dfrac{M (q\kappa)^2}{2\hbar^2}\dfrac{1}{n^2},  \;\; (n=1, 2, 3,...),
\end{eqnarray}
which is just Eq. (\ref{eq:hyd-1a}), thus reducing to the energy of the non-relativistic hydrogen-like atom.
\end{remark}

\begin{remark}\textcolor{blue}{The Relation between the Fine Structure Constant $\alpha_{\rm e}$ and the Nuclear Charge Number $Z$.}
It is well-known that the fine structure constant $\alpha_{\rm e}$ is given by
\begin{eqnarray}\label{eq:fs-1}
\alpha_{\rm e} = \dfrac{e^2}{\hbar c} \approx \dfrac{1}{137.0359} \approx \dfrac{1}{137},
\end{eqnarray}
which is a mysterious dimensionless number.

If considering $\kappa=Q=Ze$, then Eq. (\ref{eq:hyd-2a}) can be rewritten as
\begin{eqnarray}\label{eq:hyd-2b}
E_n=-\dfrac{Z^2 M e^4}{2\hbar^2}\dfrac{1}{n^2}=-\dfrac{Z^2 M c^2}{2}\dfrac{e^4}{\hbar^2 c^2}\dfrac{1}{n^2}=-\dfrac{Z^2 M c^2}{2 n^2}
 (\alpha_{\rm e})^2,  \;\;\; (n=1, 2, 3,...).
\end{eqnarray}
From which one may find that in the non-relativistic quantum mechanics, the nuclear charge number $Z$ has nothing to do with the fine structure constant $\alpha_{\rm e}$. However, this situation is dramatically changed in the relativistic quantum mechanics. Let us focus on Eq. (\ref{eq:hyd-3a}), the relativistic energy of hydrogen-like atom is given by
\begin{eqnarray}\label{eq:hyd-3b}
   &&  E=\dfrac{Mc^2}{\sqrt{1+\dfrac{q^2 \kappa^2}{\hbar^2 c^2 }\dfrac{1}{(N+\nu)^2}}},
\end{eqnarray}
where $\nu$ is given by
\begin{eqnarray}
     && \nu=\sqrt{(l+1)^2 -\frac{q^2 \kappa^2}{\hbar^2 c^2 }} = \sqrt{(l+1)^2 -\frac{  (-e)^2 (Ze)^2}{\hbar^2 c^2 }}
     =\sqrt{(l+1)^2 -\frac{ Z^2 e^4}{\hbar^2 c^2 }}=\sqrt{(l+1)^2 - Z^2 (\alpha_{\rm e})^2 }.
\end{eqnarray}
Here $\nu$ contains a square-root, to make it be meaningful, one must have
\begin{eqnarray}
     && (l+1)^2 - Z^2 (\alpha_{\rm e})^2 \geq 0,
\end{eqnarray}
for any $l=0, 1, 2, \dots$. Because the minimum of $l$ is $l_{\rm min}=0$, thus one must have
\begin{eqnarray}
     && 1 - Z^2 (\alpha_{\rm e})^2 \geq 0,
\end{eqnarray}
i.e.,
\begin{eqnarray}
     &&  Z \leq  \dfrac{1}{\alpha_{\rm e}}\approx 137.0359.
\end{eqnarray}
\emph{Based on this finding, Richard Feynman asserted that the biggest number of $Z$ in the periodic table of  elements is $Z=137$.} $\blacksquare$
\end{remark}

\newpage

\section{Exactly Solving Dirac's Equation with Spin-Dependent Coulomb Interaction}

The Dirac-type Hamiltonian with spin-dependent Coulomb interaction (i.e., with the spin potentials $\mathcal{S}_{\rm YM}=\left\{\vec{\mathcal{A}}= k \dfrac{\vec{r} \times \vec{S}}{r^2},\; \varphi=\dfrac{\kappa}{r}\right\}$) is given by
\begin{eqnarray}\label{eq:H-1a-1a}
\mathcal{H}&=&\vec{\alpha_1}\cdot \vec{\Pi}+\beta_1 M c^2 + q \varphi
=\vec{\alpha_1}\cdot  \left(\vec{p}-\dfrac{q}{c}\vec{\mathcal{A}}\right)+\beta _1 M c^2+  \dfrac{q \kappa}{r},
\end{eqnarray}
where
\begin{eqnarray}
    &&  \vec{\mathcal{A}} = k \frac{\vec{r} \times \vec{S}_2}{r^2}.
\end{eqnarray}
and the constraint {$g\hbar k=1, 2$}. Here $\kappa=Q$, ($Q=Ze$, $Z$ is an integer, $Z=1$ for the hydrogen atom and $Z \geq 2$ for the hydrogen-like atom). For simplicity, we consider merely the spin-$1/2$ case, i.e.,
\begin{eqnarray}
    &&  \vec{S_2}=\frac{\hbar}{2}\vec{\sigma}_2,
\end{eqnarray}
which is written in the $2\times 2$ matrix representation.

\begin{remark}
Because the Dirac-type Hamilton is related to the spin operators of two particles, in order to distinguish them clearly, we now introduce the subscripts. Since the Dirac matrices $\{\vec{\alpha}, \beta\}$ correspond to particle 1, thus we rewrite them as
$\{\vec{\alpha}_1, \beta_1\}$. Explicitly, they are
\begin{equation}
            \vec{\alpha}_1=
            \left(\begin{array}{cc}
                    0 & \vec{\sigma}_1 \\
                  \vec{\sigma}_1 & 0 \\
                  \end{array}
                \right),\;\;\;\;\;
             \beta_1=
             \left(\begin{array}{cc}
                  \openone_1 & 0 \\
                  0 & -\openone_1 \\
                  \end{array}
                \right),
      \end{equation}
      where $\vec{\sigma}_1=(\sigma_{1x}, \sigma_{1y}, \sigma_{1z})$ is the Pauli-matrices vector for particle 1, and
      $\openone_1$ is the $2\times 2$ identity matrix of particle 1. Similarly, the original spin operator
      $\vec{S}=\dfrac{\hbar}{2}\vec{\sigma}$ corresponds to particle 2, thus we rewrite it as
      \begin{eqnarray}
      \vec{S}_2= \dfrac{\hbar}{2}\vec{\sigma}_2.
\end{eqnarray}
This explains why the Dirac-type Hamiltonian is written in the form of Eq. (\ref{eq:H-1a-1a}). $\blacksquare$
\end{remark}

\subsection{The Eigenfunctions of  ${\hat{r}} \cdot (\vec{\alpha }_1 \times \vec{\sigma}_2) $}

First we have known that
\begin{eqnarray}
    && \vec{\alpha}_1=\begin{bmatrix}
        0 & \vec{\sigma}_1 \\
        \vec{\sigma}_1 & 0
     \end{bmatrix}=\sigma_x\otimes\vec{\sigma}_1,\;\;\;\; \beta_1=\begin{bmatrix}
        \openone_1 & 0 \\
        0 & -\openone_1
     \end{bmatrix}=\sigma_z\otimes \openone_1,
\end{eqnarray}
We designate a new ``total'' angular momentum operator as
\begin{equation}
    \vec{J} =\vec{\ell}+\vec{S}_1+\vec{S}_2,
\end{equation}
where $\vec{\ell}$ is the orbital angular momentum operator for particle 1, and
\begin{equation}
    \vec{S}_1 =\frac{\hbar}{2}\begin{bmatrix}
        \vec{\sigma}_1 & 0\\
        0& \vec{\sigma}_1
     \end{bmatrix}=\frac{\hbar}{2} \left(\openone\otimes\vec{\sigma}_1\right)
\end{equation}
is the spin-$1/2$ angular momentum operator for particle 1, which is written in $4\times 4$ matrix representation. One can prove that
\begin{equation}\label{eq:alphar-1}
    [\vec{{r}} \cdot (\vec{\alpha }_1 \times \vec{\sigma}_2), \vec{J}] = 0,
\end{equation}
or
\begin{equation}\label{eq:alphar-2}
    [\hat{{r}} \cdot (\vec{\alpha }_1 \times \vec{\sigma}_2), \vec{J}] = 0,
\end{equation}
with $\hat{r}=\vec{r}/r$.
\begin{proof}
     According to \cite{1991Lange}, there are some useful properties about vectors and scalars.

\emph{Property (i).} Let $\vec{v}_1=(v_{1x}, v_{1y}, v_{1z})$ and $\vec{v}_2=(v_{2x}, v_{2y}, v_{2z})$ be two vectors with respect to the operator $\vec{\mathcal{L}}=(\mathcal{L}_{1x}, \mathcal{L}_{1y}, \mathcal{L}_{1z})$, then we have
\begin{eqnarray}
            &&  [\mathcal{L}_\alpha,\ v_{1 \beta}]={\rm i}\hbar\, \epsilon_{\alpha\beta\gamma} v_{1\gamma}, \nonumber\\
            &&  [\mathcal{L}_\alpha,\ v_{2 \beta}]={\rm i}\hbar\, \epsilon_{\alpha\beta\gamma} v_{2\gamma},
\end{eqnarray}
with $\epsilon_{\alpha\beta\gamma}$ is the Levi-Civita symbol ($\alpha, \beta, \gamma= x, y, z$). Then for
\begin{eqnarray}
            && \vec{v}= \vec{v}_1\times \vec{v}_2,
      \end{eqnarray}
      it is also a vector with respect to $\vec{\mathcal{L}}$ and it satisfies
\begin{eqnarray}
            &&  [\mathcal{L}_\alpha\,, (\vec{v}_1\times \vec{v}_2)_{\beta}]={\rm i}\hbar\, \epsilon_{\alpha\beta\gamma} (\vec{v}_1\times \vec{v}_2)_{\gamma}.
\end{eqnarray}

For examples, the position operator $\vec{r}$ and the momentum operator $\vec{p}$ are vectors with respect to the orbital angular momentum operator $\vec{\ell}=(\ell_x, \ell_y, \ell_z)$, i.e., they satisfied
\begin{eqnarray}
            &&  [\ell_\alpha,\ r_{\beta}]={\rm i}\hbar\, \epsilon_{\alpha\beta\gamma} r_{\gamma}, \nonumber\\
            &&  [\ell_\alpha,\ p_{ \beta}]={\rm i}\hbar\, \epsilon_{\alpha\beta\gamma} p_{\gamma},
\end{eqnarray}
then $\vec{\ell}=\vec{r}\times \vec{p}$ itself is also a vector with respect to $\vec{\ell}$, because
\begin{eqnarray}
            &&  [\ell_\alpha,\ \ell_{\beta}]={\rm i}\hbar\, \epsilon_{\alpha\beta\gamma} \ell_{\gamma}.
\end{eqnarray}
Similarly, the spin operator $\vec{S}$ is a vector with respective to itself because
\begin{eqnarray}
            &&  [S_\alpha,\ S_{\beta}]={\rm i}\hbar\, \epsilon_{\alpha\beta\gamma} S_{\gamma},
\end{eqnarray}
and the ``total'' angular momentum operator $\vec{J}$ is also a vector with respective to itself, because
\begin{eqnarray}
            &&  [J_\alpha,\ J_{\beta}]={\rm i}\hbar\, \epsilon_{\alpha\beta\gamma} J_{\gamma}.
\end{eqnarray}

\emph{Property (ii).} Let $\vec{v}_1$ and $\vec{v}_2$ be two vectors with respect to $\vec{\mathcal{L}}$, then one can easily prove that
\begin{eqnarray}
            && \mathcal{K}= \vec{v}_1\cdot \vec{v}_2
\end{eqnarray}
is  a scalar with respect to $\vec{\mathcal{L}}$, i.e.,
\begin{eqnarray}
      [\mathcal{L}_\alpha,\ \mathcal{K}]=0.
\end{eqnarray}
For examples, $\vec{r}\cdot\vec{r}$, $\vec{p}\cdot\vec{p}$, $\vec{\ell}\cdot\vec{\ell}$, $\vec{r}\cdot\vec{p}$,
$\vec{r}\cdot\vec{\ell}$, $\vec{p}\cdot\vec{\ell}$ are scalars with respect to $\vec{\ell}$.

Now we come to prove Eq. (\ref{eq:alphar-1}). It is obvious that
  \begin{eqnarray}
      [J_i, \alpha_{1j}] = {\rm i} \epsilon_{ijk} \alpha_k\,\,,\,\,
      [J_i, \sigma_{2j}] = {\rm i} \epsilon_{ijk} \sigma_{2k}\,\,,\,\, [J_i, r_i] = {\rm i} \epsilon_{ijk} r_k,
  \end{eqnarray}
namely, $\vec{\alpha}$, $\vec{\sigma}_2$, and $\vec{r}$ are vectors with respect to $\vec{J}$. Then, according to property (i), we have
      \begin{eqnarray}
           [J_i, (\vec{\alpha }_1 \times \vec{\sigma_2})_j] ={\rm i} \epsilon_{ijk} (\vec{\alpha }_1 \times \vec{\sigma_2})_k,
      \end{eqnarray}
i.e., $(\vec{\alpha }_1 \times \vec{\sigma_2})$ is a vector with respect to $\vec{J}$. Furthermore, according to property (ii), we have
      \begin{equation}
            [\vec{J},\, \vec{r} \cdot (\vec{\alpha }_1 \times \vec{\sigma}_2)] =0.
        \end{equation}
Since $r$ is a scalar with respect to $\vec{J}$, i.e., $[\vec{J}, \vec{r}\cdot\vec{r}]=[\vec{J}, r^2]=0$, $[\vec{J}, r]=0$, $[\vec{J}, 1/r]=0$,
then one easily has
      \begin{equation}
            [\vec{J},\, \hat{r} \cdot (\vec{\alpha }_1 \times \vec{\sigma}_2)] =0.
        \end{equation}
This ends the proof.
\end{proof}

One may verify that
\begin{equation}\label{eq:J-1'}
      [\vec{J},\mathcal{H}] =0, \;\;\; [\vec{J}^{\,2}, \mathcal{H}]=0,  \;\;\; [\vec{J}^{\,2}, \vec{J}]=0,
\end{equation}
which means that $\vec{J}$ is a conversed quantity for the Hamiltonian system of $\mathcal{H}$. Then one can choose
$\{\mathcal{H}, \vec{J}^{\,2}, J_z\}$ as a common set to study the eigen-problem.

For $\vec{J}$, it is coupled by $\vec{\ell }$, $\vec{S_1}$, and $\vec{S_2}$. For simplicity, we can designate
the ``total'' angular momentum operator for particle 1 as $\vec{J}_1$, i.e.,
\begin{equation}
      \vec{J}_1 = \vec{\ell} + \vec{S}_1,
\end{equation}
then
\begin{equation}
      \vec{J}=\vec{J_1}+\vec{S}_2.
\end{equation}
$\vec{J}$ can be regarded as the result of the coupling of $\vec{J_1}$ and $\vec{S_2}$, we then have the eigenstates of $\{\vec{J}^{\,2}, J_z\}$ as
\begin{eqnarray}
   && \vec{J}^{\,2} |j , m_j \rangle = j(j+1)\hbar^2 \, |j , m_j \rangle, \nonumber\\
   && J_z |j , m_j \rangle = m_j \hbar \, |j , m_j \rangle,
\end{eqnarray}
where
\begin{equation}
      \left |j , m_j \right \rangle =A \left | j_1 ,m_1, \frac{1}{2}  \right \rangle+B \left | j_1 ,m_1', -\frac{1}{2} \right \rangle +C \left | j_1' ,m_1, \frac{1}{2}  \right \rangle+D \left | j_1' ,m_1', -\frac{1}{2} \right \rangle  ,
\end{equation}
$j_1=j-\frac{1}{2}$ , $j_1'=j+\frac{1}{2}$, $m_1=m_j-\frac{1}{2}$ ,$m'_1=m_j+\frac{1}{2}$, and $A, B, C, D$ are some coefficients.

For convenience, let us denote $\hat{r}=\vec{r}/r=(\hat{x}, \hat{y}, \hat{z})$, we then have
\begin{eqnarray}
  && \left[ \hat{r}\cdot (\vec{\alpha }_1 \times \vec{\sigma}_2)\right]^2= \openone \otimes \left[ \hat{r}\cdot (\vec{\sigma }_1 \times \vec{\sigma}_2)\right]^2,
\end{eqnarray}
where
\begin{eqnarray}
  && \left[ \hat{r}\cdot (\vec{\sigma }_1 \times \vec{\sigma}_2)\right]^2= \left[\hat{x}\left({\sigma_{1y}} {\sigma_{2z}} -{\sigma_{1z}} {\sigma_{2y}}\right)+\hat{y}\left({\sigma_{1z}} {\sigma_{2x}} -{\sigma_{1x}} {\sigma_{2z}}\right)+\hat{z} \left({\sigma_{1x}} {\sigma_{2y}}  -{\sigma_y}_1 {\sigma_x}_2 \right)\right]^2.
\end{eqnarray}
To determine the eigenvalues of $\left[ \hat{r}\cdot (\vec{\sigma }_1 \times \vec{\sigma}_2)\right]^2$, it is sufficient to rotate $\hat{r}$ to $\hat{z}$. Namely,  $\left[ \hat{r}\cdot (\vec{\sigma }_1 \times \vec{\sigma}_2)\right]^2$ and $\left[ \hat{n}\cdot (\vec{\sigma }_1 \times \vec{\sigma}_2)\right]^2$ share the same eigenvalues with $\hat{n}=(0, 0, 1)$. One can have
\begin{eqnarray}
   \left[ \hat{n}\cdot (\vec{\sigma }_1 \times \vec{\sigma}_2)\right]^2 &=&  \left({\sigma_{1x}} {\sigma_{2y}}  -{\sigma_{1y}} {\sigma_{2x}} \right)^2=
  2 (\openone\times \openone) -\left[({\sigma_{1x}} {\sigma_{2y}})({\sigma_{1y}} {\sigma_{2x}})+({\sigma_{1y}} {\sigma_{2x}})({\sigma_{1x}} {\sigma_{2y}})\right]\nonumber\\
  &=& 2 (\openone\times \openone) -\left[({\rm i})(-{\rm i})(\sigma_{1z}\otimes \sigma_{2z})+(-{\rm i})({\rm i})(\sigma_{1z}\otimes \sigma_{2z})\right]=
  2\left(\openone\times \openone -\sigma_{1z}\otimes \sigma_{2z}\right),
\end{eqnarray}
thus the eigenvalues of $\left[ \hat{n}\cdot (\vec{\sigma }_1 \times \vec{\sigma}_2)\right]^2$ are 4 or 0.
This implies the eigenvalues of $\left[ \hat{n}\cdot (\vec{\sigma }_1 \times \vec{\sigma}_2)\right]$ are 2, $-2$, and 0.

Now we can designate the common eigenfunctions of $\{\vec{J}^{\,2}, J_z, \hat{r}\cdot (\vec{\alpha }_1 \times \vec{\sigma}_2)\}$ as
\begin{eqnarray}\label{eq:phi-1a}
    \Phi_{j,m_j}(\theta ,\phi) &= &\begin{pmatrix}
        a_1 \Phi^A_{l,j_1,m_1}+a_2 \Phi^A_{l+1,j'_1,m_1}+a_3 \Phi^B_{l,j_1,m_1}+a_4 \Phi^B_{l+1,j'_1,m_1} \\
        a_1' \Phi^A_{l,j_1,m_1}+a_2' \Phi^A_{l+1,j'_1,m_1}+a_3' \Phi^B_{l,j_1,m_1}+a_4' \Phi^B_{l+1,j'_1,m_1}
    \end{pmatrix}_1   \otimes
    \begin{pmatrix}
        1\\
        0
    \end{pmatrix}_2\notag\\
&&+
\begin{pmatrix}
    b_1 \Phi^A_{l,j_1,m_1'}+b_2 \Phi^A_{l+1,j_1',m_1'}+b_3 \Phi^B_{l,j_1,m_1'}+b_4 \Phi^B_{l+1,j_1',m_1'} \\
    b_1' \Phi^A_{l,j_1,m_1'}+b_2' \Phi^A_{l+1,j_1',m_1'}+b_3' \Phi^B_{l,j_1,m_1'}+b_4' \Phi^B_{l+1,j_1',m_1'}
\end{pmatrix}_1   \otimes
\begin{pmatrix}
    0\\
    1
\end{pmatrix}_2,
\end{eqnarray}
We have
\begin{eqnarray}
   && \vec{J}^{\,2} \, \Phi_{j,m_j}(\theta ,\phi) = j(j+1)\hbar^2 \, \, \Phi_{j,m_j}(\theta ,\phi), \nonumber\\
   && J_z \, \Phi_{j,m_j}(\theta ,\phi) = m_j \hbar \, \, \Phi_{j,m_j}(\theta ,\phi),
\end{eqnarray}
and
\begin{eqnarray}\label{eq:sigma-1a}
    \hat{r}\cdot (\vec{\alpha }_1 \times \vec{\sigma}_2)\, \Phi_{j,m_j}(\theta ,\phi) = \eta\, \Phi_{j,m_j}(\theta ,\phi),
\end{eqnarray}
with
\begin{eqnarray}
\eta=2, \;\; -2,  \;\;{\rm or} \;\;\; 0,
\end{eqnarray}
i.e., $\hat{r}\cdot (\vec{\alpha }_1 \times \vec{\sigma}_2)$ has eight eigenvalues, two of them are 2, two of them are $-2$, and the rest four are 0.

Let
\begin{eqnarray}
   x= r\sin \theta \cos \phi, \;\;\; y=r \sin \theta \sin \phi, \;\;\; z=r \cos\theta,
   \end{eqnarray}
we can attain
\begin{eqnarray}
   \hat{r}\cdot ( \vec{\sigma}_1\times \vec{\sigma}_2) &=& \left({\sigma_{1y}} {\sigma_{2z}} -{\sigma_{1z}} {\sigma_{2y}}\right) \sin \theta \cos \phi
   +\left({\sigma_{1z}} {\sigma_{2x}} -{\sigma_{1x}} {\sigma_{2z}}\right)\sin \theta \sin \phi
   +\left({\sigma_{1x}} {\sigma_{2y}}  -{\sigma_{1y}} {\sigma_{2x}} \right) \cos\theta.
   \end{eqnarray}
From Eq. (\ref{eq:sigma-1a}) we have
\begin{eqnarray}\label{eq:sig-1a}
&& \left[\hat{r}\cdot ( \vec{\sigma}_1\times \vec{\sigma}_2)\right]\biggr(  a_1' \Phi^A_{l,j_1,m_1}+a_2' \Phi^A_{l+1,j'_1,m_1}+a_3' \Phi^B_{l,j_1,m_1}+a_4' \Phi^B_{l+1,j'_1,m_1}\biggr)_1   \otimes
    \begin{pmatrix}
        1\\
        0
    \end{pmatrix}_2 \notag\\
&&+ \left[\hat{r}\cdot ( \vec{\sigma}_1\times \vec{\sigma}_2)\right] \biggr( b_1' \Phi^A_{l,j_1,m_1'}+b_2' \Phi^A_{l+1,j_1',m_1'}+b_3' \Phi^B_{l,j_1,m_1'}+b_4' \Phi^B_{l+1,j_1',m_1'}\biggr)_1   \otimes
\begin{pmatrix}
    0\\
    1
\end{pmatrix}_2 \nonumber\\
&=& \eta \, \Biggr[\biggr(  a_1 \Phi^A_{l,j_1,m_1}+a_2 \Phi^A_{l+1,j'_1,m_1}+a_3 \Phi^B_{l,j_1,m_1}+a_4 \Phi^B_{l+1,j'_1,m_1}\biggr)_1   \otimes
    \begin{pmatrix}
        1\\
        0
    \end{pmatrix}_2 \nonumber\\
&&+\biggr( b_1 \Phi^A_{l,j_1,m_1'}+b_2 \Phi^A_{l+1,j_1',m_1'}+b_3 \Phi^B_{l,j_1,m_1'}+b_4 \Phi^B_{l+1,j_1',m_1'}\biggr)_1   \otimes
\begin{pmatrix}
    0\\
    1
\end{pmatrix}_2\Biggr],
\end{eqnarray}
and
\begin{eqnarray}\label{eq:sig-1b}
&& \left[\hat{r}\cdot ( \vec{\sigma}_1\times \vec{\sigma}_2)\right]\biggr(  a_1 \Phi^A_{l,j_1,m_1}+a_2 \Phi^A_{l+1,j'_1,m_1}+a_3 \Phi^B_{l,j_1,m_1}+a_4 \Phi^B_{l+1,j'_1,m_1}\biggr)_1   \otimes
    \begin{pmatrix}
        1\\
        0
    \end{pmatrix}_2 \notag\\
&&+ \left[\hat{r}\cdot ( \vec{\sigma}_1\times \vec{\sigma}_2)\right] \biggr( b_1 \Phi^A_{l,j_1,m_1'}+b_2 \Phi^A_{l+1,j_1',m_1'}+b_3 \Phi^B_{l,j_1,m_1'}+b_4 \Phi^B_{l+1,j_1',m_1'}\biggr)_1   \otimes
\begin{pmatrix}
    0\\
    1
\end{pmatrix}_2 \nonumber\\
&=& \eta \, \Biggr[\biggr(  a'_1 \Phi^A_{l,j_1,m_1}+a'_2 \Phi^A_{l+1,j'_1,m_1}+a'_3 \Phi^B_{l,j_1,m_1}+a'_4 \Phi^B_{l+1,j'_1,m_1}\biggr)_1   \otimes
    \begin{pmatrix}
        1\\
        0
    \end{pmatrix}_2 \nonumber\\
&&+\biggr( b'_1 \Phi^A_{l,j_1,m_1'}+b'_2 \Phi^A_{l+1,j_1',m_1'}+b'_3 \Phi^B_{l,j_1,m_1'}+b'_4 \Phi^B_{l+1,j_1',m_1'}\biggr)_1   \otimes
\begin{pmatrix}
    0\\
    1
\end{pmatrix}_2\Biggr].
\end{eqnarray}
Due to the following relations
\begin{eqnarray}
 &&   \sigma_x \begin{pmatrix}
        1\\
        0
    \end{pmatrix}
    =\begin{pmatrix}
        0\\
        1
    \end{pmatrix}, \;\;\;
    \sigma_x \begin{pmatrix}
        0\\
        1
    \end{pmatrix}
    =\begin{pmatrix}
        1\\
        0
    \end{pmatrix}, \nonumber\\
  &&  \sigma_y\begin{pmatrix}
        1\\
        0
    \end{pmatrix}
    ={\rm i}\begin{pmatrix}
        0\\
        1
    \end{pmatrix}, \;\;\;
    \sigma_y \begin{pmatrix}
        0\\
        1
    \end{pmatrix}
    =-{\rm i}\begin{pmatrix}
        1\\
        0
    \end{pmatrix}, \nonumber\\
 &&   \sigma_z \begin{pmatrix}
        1\\
        0
    \end{pmatrix}
    =\begin{pmatrix}
        1\\
        0
    \end{pmatrix}, \;\;\;
    \sigma_z \begin{pmatrix}
        0\\
        1
    \end{pmatrix}
    =-\begin{pmatrix}
        0\\
        1
    \end{pmatrix},
\end{eqnarray}
from Eq. (\ref{eq:sig-1a}) and Eq. (\ref{eq:sig-1b}) we directly obtain
\begin{eqnarray}
    \label{aa1}&&\left(\sigma_{1y} {\sin} \theta {\cos} \phi - \sigma_{1x} {\sin}\theta {\sin} \phi \right)\left(  a_1' \Phi^A_{l,j_1,m_1}+a_2' \Phi^A_{l+1,j'_1,m_1}+a_3' \Phi^B_{l,j_1,m_1}+a_4' \Phi^B_{l+1,j'_1,m_1}\right) \notag\\
&&+ \left(\sigma_{1z} \sin \theta \sin \phi -\sigma_{1y} \cos \theta - {\rm i} \sigma_{1x} \cos \theta + {\rm i} \sigma_{1z} \sin \theta \cos  \phi\right)\left( b_1' \Phi^A_{l,j_1,m_1'}+b_2' \Phi^A_{l+1,j_1',m_1'}+b_3' \Phi^B_{l,j_1,m_1'}+b_4' \Phi^B_{l+1,j_1',m_1'}\right)\notag\\
&&=\eta \left(  a_1 \Phi^A_{l,j_1,m_1}+a_2 \Phi^A_{l+1,j'_1,m_1}+a_3 \Phi^B_{l,j_1,m_1}+a_4 \Phi^B_{l+1,j'_1,m_1}\right),
\end{eqnarray}

\begin{eqnarray}
    \label{aa2}   && \left(\sigma_{1y} {\sin} \theta {\cos} \phi - \sigma_{1x} {\sin} \theta {\sin} \phi \right)\left(  a_1 \Phi^A_{l,j_1,m_1}+a_2 \Phi^A_{l+1,j'_1,m_1}+a_3 \Phi^B_{l,j_1,m_1}+a_4 \Phi^B_{l+1,j'_1,m_1}\right) \notag\\
    &&+ \left(\sigma_{1z} \sin \theta \sin \phi -\sigma_{1y} \cos \theta - {\rm i} \sigma_{1x} \cos \theta + {\rm i} \sigma_{1z} \sin \theta \cos  \phi\right)\left(  b_1 \Phi^A_{l,j_1,m_1'}+b_2 \Phi^A_{l+1,j_1',m_1'}+b_3 \Phi^B_{l,j_1,m_1'}+b_4 \Phi^B_{l+1,j_1',m_1'}\right)\notag\\
    &&=\eta \left( a_1' \Phi^A_{l,j_1,m_1}+a_2' \Phi^A_{l+1,j'_1,m_1}+a_3' \Phi^B_{l,j_1,m_1}+a_4' \Phi^B_{l+1,j'_1,m_1}\right),
\end{eqnarray}

\begin{eqnarray}
    \label{aa3}      &&-\left(\sigma_{1y} {\sin} \theta {\cos} \phi - \sigma_{1x} {\sin} \theta {\sin} \phi \right)\left(  b_1' \Phi^A_{l,j_1,m_1'}+b_2' \Phi^A_{l+1,j_1',m_1'}+b_3' \Phi^B_{l,j_1,m_1'}+b_4' \Phi^B_{l+1,j_1',m_1'}\right) \notag\\
        &&+ \left(\sigma_{1z} {\sin} \theta {\sin} \phi -\sigma_{1y} {\cos} \theta +{\rm i} \sigma_{1x} {\cos} \theta - {\rm i} \sigma_{1z} {\sin} \theta {\cos}  \phi \right)\left(   a_1' \Phi^A_{l,j_1,m_1}+a_2' \Phi^A_{l+1,j'_1,m_1}+a_3' \Phi^B_{l,j_1,m_1}+a_4' \Phi^B_{l+1,j'_1,m_1} \right)\notag\\
        &&=\eta \left(  b_1 \Phi^A_{l,j_1,m_1'}+b_2 \Phi^A_{l+1,j_1',m_1'}+b_3 \Phi^B_{l,j_1,m_1'}+b_4 \Phi^B_{l+1,j_1',m_1'} \right),
\end{eqnarray}

\begin{eqnarray}
  \label{aa4}
    &&-\left(\sigma_{1y} {\sin}\theta {\cos} \phi - \sigma_{1x} {\sin} \theta {\sin} \phi \right)\left(    b_1 \Phi^A_{l,j_1,m_1'}+b_2 \Phi^A_{l+1,j_1',m_1'}+b_3 \Phi^B_{l,j_1,m_1'}+b_4 \Phi^B_{l+1,j_1',m_1'}\right) \notag\\
    &&+ \left(\sigma_{1z}{\sin} \theta {\sin} \phi -\sigma_{1y} {\cos} \theta +{\rm i} \sigma_{1x} {\cos} \theta -{ \rm i }\sigma_{1z} {\sin} \theta {\cos } \phi \right)\left(   a_1 \Phi^A_{l,j_1,m_1}+a_2 \Phi^A_{l+1,j'_1,m_1}+a_3 \Phi^B_{l,j_1,m_1}+a_4 \Phi^B_{l+1,j'_1,m_1} \right)\notag\\
    &&=\eta \left(  b_1' \Phi^A_{l,j_1,m_1'}+b_2' \Phi^A_{l+1,j_1',m_1'}+b_3' \Phi^B_{l,j_1,m_1'}+b_4' \Phi^B_{l+1,j_1',m_1'}\right).
\end{eqnarray}
In the behind calculation, we shall use the following relations
\begin{eqnarray}
    \sigma_{1y} \sin \theta \cos \phi - \sigma_{1x} \sin \theta \sin \phi = -{\rm i} \sin \theta  \begin{pmatrix}
        0 & e^{-{\rm i} \phi} \\
        -e^{{\rm i} \phi}&0
    \end{pmatrix},
\end{eqnarray}

\begin{eqnarray}
    \sigma_{1z} \sin \theta \sin \phi -\sigma_{1y} \cos \theta - {\rm i} \sigma_{1x} \cos \theta + {\rm i} \sigma_{1z} \sin \theta \cos  \phi = {\rm i} \, \sin \theta e^{- {\rm i} \phi } \begin{pmatrix}
        1& 0\\
        0&-1
    \end{pmatrix} -
\rm i \, \cos \theta \begin{pmatrix}
    0&0\\
    2&0
\end{pmatrix},
\end{eqnarray}

\begin{eqnarray}
    \sigma_{1z} \sin \theta \sin \phi -\sigma_{1y} \cos \theta +{\rm i} \sigma_{1x} \cos \theta - {\rm i} \sigma_{1z} \sin \theta \cos  \phi= -{\rm i} \, \sin \theta e^{{\rm i} \phi } \begin{pmatrix}
        1& 0\\
        0&-1
    \end{pmatrix} +
{\rm i} \, \cos \theta \begin{pmatrix}
    0&2\\
    0&0
\end{pmatrix}.
\end{eqnarray}

\subsubsection{The Case of $\eta =2$ or $-2$}

\begin{remark}
For \Eq{aa1},
it requires
\begin{eqnarray}\label{ab1}
    &&{\rm sin} \theta e^{-\rm i \phi} \left\{ \frac{a_1'\sqrt{l-m} -b_1'\sqrt{l+m+2}}{\sqrt{2l+1} } Y_{l,m+1} +\frac{a_2'\sqrt{l-m+1} -b_2'\sqrt{l+m+3}}{\sqrt{2l+3} } Y_{l+1,m+1}\right\}\notag\\
    && +{\rm sin}\theta e^{-\rm i \phi} \left\{\frac{a_3'\sqrt{l+m+2} +b_3'\sqrt{l-m}}{\sqrt{2l+3} } Y_{l+1,m+1}+\frac{a_4'\sqrt{l+m+3} +b_4'\sqrt{l-m+1}}{\sqrt{2l+5} } Y_{l+2,m+1}\right\}\notag\\
    &&={\rm i}\, \eta \left\{a_1 \sqrt{\frac{l+m+1}{2l+1}}Y_{lm}+a_2 \sqrt{\frac{l+m+2}{2l+3}}Y_{l+1,m}-a_3 \sqrt{\frac{l-m+1}{2l+3}}Y_{l+1,m} -a_4 \sqrt{\frac{l-m+2}{2l+5}}Y_{l+2,m}\right\} ,
\end{eqnarray}

and
\begin{eqnarray}\label{ab2}
    &&-{\rm sin}  \theta e^{\rm i \phi} \left\{ \frac{a_1'\sqrt{l+m+1} }{\sqrt{2l+1} } Y_{l,m} +\frac{a_2'\sqrt{l+m+2} }{\sqrt{2l+3} } Y_{l+1,m}-\frac{a_3'\sqrt{l-m+1} }{\sqrt{2l+3} } Y_{l+1,m}-\frac{a_4'\sqrt{l-m+2} }{\sqrt{2l+5} } Y_{l+2,m}\right\}\notag\\
    && +{\rm sin} \theta e^{-\rm i \phi} \left\{\frac{b_1'\sqrt{l-m-1}}{\sqrt{2l+1} } Y_{l,m+2} +\frac{ b_2'\sqrt{l-m}}{\sqrt{2l+3} } Y_{l+1,m+2}+\frac{b_3'\sqrt{l+m+3}}{\sqrt{2l+3} } Y_{l+1,m+2}+\frac{b_4'\sqrt{l+m+4}}{\sqrt{2l+5} } Y_{l+2,m+2}\right\}\notag\\
    &&+2 {\rm cos} \theta \left\{\frac{b_1'\sqrt{l+m+2}}{\sqrt{2l+1} } Y_{l,m+1} +\frac{ b_2'\sqrt{l+m+3}}{\sqrt{2l+3} } Y_{l+1,m+1}-\frac{b_3'\sqrt{l-m}}{\sqrt{2l+3} } Y_{l+1,m+1}-\frac{b_4'\sqrt{l-m+1}}{\sqrt{2l+5} } Y_{l+2,m+1}\right\}\notag\\
    &&={\rm i}\, \eta \left\{a_1 \sqrt{\frac{l-m}{2l+1}}Y_{l,m+1}+a_2 \sqrt{\frac{l-m+1}{2l+3}}Y_{l+1,m+1}+a_3 \sqrt{\frac{l+m+2}{2l+3}}Y_{l+1,m+1} +a_4 \sqrt{\frac{l+m+3}{2l+5}}Y_{l+2,m+1}\right\}.
\end{eqnarray}
Due to the following recursive formulas
\begin{subequations}
      \begin{align}
            & \cos\theta\,Y_{lm} =a_{l,m} Y_{l+1,m}
                  +a_{l-1,m} Y_{l-1,m}, \\
            & \sin\theta\,{\rm e}^{{\rm i}\,\phi} Y_{lm}
            =b_{l-1,-(m+1)} Y_{l-1,m+1} -b_{l,m} Y_{l+1,m+1}, \\
            & \sin\theta\,{\rm e}^{-{\rm i}\,\phi} Y_{lm}
            =-b_{l-1,m-1} Y_{l-1,m-1} +b_{l,-m} Y_{l+1,m-1},
      \end{align}
\end{subequations}
with
\begin{equation}
      a_{l,m}
            =\sqrt{\dfrac{(l+m+1)(l-m+1)}{
                  (2\,l+1)(2\,l+3)}},\qquad
      b_{l,m} =\sqrt{\dfrac{(l+m+1)(l+m+2)}{
            (2\,l+1)(2\,l+3)}},
\end{equation}
from \Eq{ab1} we must require that
\begin{eqnarray}
  &&  a_1'\sqrt{l-m} = b_1'\sqrt{l+m+2}, \nonumber\\
  &&  a_4'\sqrt{l+m+3} =-b_4'\sqrt{l-m+1}, \nonumber\\
  &&  a_2 \sqrt{{l+m+2}}=a_3 \sqrt{{l-m+1}},
\end{eqnarray}
thus \Eq{ab1} yields
\begin{eqnarray}\label{eq:1a}
    -\left\{\frac{a_2'\sqrt{l-m+1} -b_2'\sqrt{l+m+3}}{\sqrt{2l+3} } +\frac{a_3'\sqrt{l+m+2} +b_3'\sqrt{l-m}}{\sqrt{2l+3} }\right\} b_{lm} ={\rm i} \eta \, a_1 \sqrt{\frac{l+m+1}{2l+1}},
\end{eqnarray}
\begin{eqnarray}\label{eq:1b}
    -\left\{\frac{a_2'\sqrt{l-m+1} -b_2'\sqrt{l+m+3}}{\sqrt{2l+3} } +\frac{a_3'\sqrt{l+m+2} +b_3'\sqrt{l-m}}{\sqrt{2l+3} }\right\} b_{l+1,-m-1} ={\rm i} \eta \, a_4 \sqrt{\frac{l-m+2}{2l+5}},
\end{eqnarray}
which leads to
\begin{eqnarray}
    \sqrt{l-m+1} \,a_1 = \sqrt{l+m+2}\, a_4.
\end{eqnarray}

    \Eq{ab2} indicates
\begin{equation}
  - \frac{ a_1'\sqrt{l+m+1} }{\sqrt{2l+1} } b_{l-1, -m-1} -b_{l-1,m+1}\frac{b_1'\sqrt{l-m-1}}{\sqrt{2l+1} } +2a_{l-1,m+1}\frac{b_1'\sqrt{l+m+2}}{\sqrt{2l+1} } =0,
\end{equation}
\begin{equation}
    -b_{l+2,m}\frac{a_4'\sqrt{l-m+2} }{\sqrt{2l+5} }  + b_{l+2,-m-2}\frac{b_4'\sqrt{l+m+4}}{\sqrt{2l+5} } -2 a_{l+2,m+1}\frac{b_4'\sqrt{l-m+1}}{\sqrt{2l+5} } =0,
\end{equation}
which leads to
\begin{eqnarray}\label{eq:abab-1}
   && a_1'\sqrt{l-m} = b_1'\sqrt{l+m+2}, \nonumber\\
   &&a_4'\sqrt{l+m+3} =-b_4'\sqrt{l-m+1}.
\end{eqnarray}
We can attain from \Eq{ab2} that
\begin{eqnarray}
  \label{ac1} && \frac{ a_1'\sqrt{l+m+1} }{\sqrt{2l+1} } b_{l, m} +b_{l,-m-2}\frac{b_1'\sqrt{l-m-1}}{\sqrt{2l+1} } +2a_{l,m+1}\frac{b_1'\sqrt{l+m+2}}{\sqrt{2l+1} }  +b_{l+1,-m-1}\frac{a_4'\sqrt{l-m+2} }{\sqrt{2l+5} } \notag\\
    &&-b_{l+1,m+1}\frac{b_4'\sqrt{l+m+4}}{\sqrt{2l+5} }
   -2 a_{l+1,m+1}\frac{b_4'\sqrt{l-m+1}}{\sqrt{2l+5} } ={\rm i} \eta\, (a_2 \sqrt{\frac{l-m+1}{2l+3}}+a_3 \sqrt{\frac{l+m+2}{2l+3}}),
\end{eqnarray}
\begin{eqnarray}
    \label{ac2} &&-b_{l,-m-1} \left(\frac{a_2'\sqrt{l+m+2} }{\sqrt{2l+3} } -\frac{a_3'\sqrt{l-m+1} }{\sqrt{2l+3} }\right)-b_{l,m+1}\left(\frac{ b_2'\sqrt{l-m}}{\sqrt{2l+3} } +\frac{b_3'\sqrt{l+m+3}}{\sqrt{2l+3} }\right) \notag\\
   &&+2 a_{l,m+1}\left(\frac{ b_2'\sqrt{l+m+3}}{\sqrt{2l+3} } -\frac{b_3'\sqrt{l-m}}{\sqrt{2l+3} }\right)= {\rm i} \eta\, a_1 \sqrt{\frac{l-m}{2l+1}},
\end{eqnarray}
\begin{eqnarray}
    \label{ac3} &&b_{l+1,m} \left(\frac{a_2'\sqrt{l+m+2} }{\sqrt{2l+3} } -\frac{a_3'\sqrt{l-m+1} }{\sqrt{2l+3} }\right)+b_{l+1,-m-2}\left(\frac{ b_2'\sqrt{l-m}}{\sqrt{2l+3} } +\frac{b_3'\sqrt{l+m+3}}{\sqrt{2l+3} }\right) \notag\\
    &&+2 a_{l+1,m+1}\left(\frac{ b_2'\sqrt{l+m+3}}{\sqrt{2l+3} } -\frac{b_3'\sqrt{l-m}}{\sqrt{2l+3} }\right)= {\rm i} \eta\, a_4 \sqrt{\frac{l+m+3}{2l+5}}.
 \end{eqnarray}
 Based on \Eq{ac1}, we can easily attain
 \begin{eqnarray}
  \label{ac1'}  \frac{2(l+1)a_1'}{\sqrt{(l+m+2)(2l+3)} }  +  \frac{2(l+2)a_4'}{\sqrt{(l-m+1)(2l+3)} }  = {\rm i }\eta\, a_2 \sqrt{\frac{2l+3}{l-m+1}}.
 \end{eqnarray}
$\blacksquare$
\end{remark}

\begin{remark}\label{re1}
For \Eq{aa3}
it requires
\begin{eqnarray}\label{ad1}
    &&{\rm sin} \theta e^{\rm i \phi} \left\{ \frac{b_1'\sqrt{l+m+2} -a_1'\sqrt{l-m}}{\sqrt{2l+1} } Y_{l,m+1} +\frac{b_2'\sqrt{l+m+3} -a_2'\sqrt{l-m+1}}{\sqrt{2l+3} } Y_{l+1,m+1}\right\}\notag\\
    && {\rm sin}\theta e^{\rm i \phi} \left\{-\frac{b_3'\sqrt{l-m} +a_3'\sqrt{l+m+2}}{\sqrt{2l+3} } Y_{l+1,m+1}-\frac{b_4'\sqrt{l-m+1} +a_4'\sqrt{l+m+3}}{\sqrt{2l+5} } Y_{l+2,m+1}\right\}\notag\\
    &&={\rm i}\, \eta\,  \left\{b_1 \sqrt{\frac{l-m-1}{2l+1}}Y_{l,m+2}+b_2 \sqrt{\frac{l-m}{2l+3}}Y_{l+1,m+2}+b_3 \sqrt{\frac{l+m+3}{2l+3}}Y_{l+1,m+2} +b_4 \sqrt{\frac{l+m+4}{2l+5}}Y_{l+2,m+2}\right\} ,
\end{eqnarray}
and
\begin{eqnarray}\label{ad2}
    &&{\rm sin}  \theta e^{-\rm i \phi} \left\{ \frac{b_1'\sqrt{l-m-1} }{\sqrt{2l+1} } Y_{l,m+2} +\frac{b_2'\sqrt{l-m} }{\sqrt{2l+3} } Y_{l+1,m+2}+\frac{b_3'\sqrt{l+m+3} }{\sqrt{2l+3} } Y_{l+1,m+2}+\frac{b_4'\sqrt{l+m+4} }{\sqrt{2l+5} } Y_{l+2,m+2}\right\}\notag\\
    && -{\rm sin} \theta e^{\rm i \phi} \left\{\frac{a_1'\sqrt{l+m+1}}{\sqrt{2l+1} } Y_{l,m} +\frac{ a_2'\sqrt{l+m+2}}{\sqrt{2l+3} } Y_{l+1,m}-\frac{a_3'\sqrt{l-m+1}}{\sqrt{2l+3} } Y_{l+1,m}-\frac{a_4'\sqrt{l-m+2}}{\sqrt{2l+5} } Y_{l+2,m}\right\}\notag\\
    &&+2 {\rm cos} \theta \left\{\frac{a_1'\sqrt{l-m}}{\sqrt{2l+1} } Y_{l,m+1} +\frac{ a_2'\sqrt{l-m+1}}{\sqrt{2l+3} } Y_{l+1,m+1}+\frac{a_3'\sqrt{l+m+2}}{\sqrt{2l+3} } Y_{l+1,m+1}+\frac{a_4'\sqrt{l+m+3}}{\sqrt{2l+5} } Y_{l+2,m+1}\right\}\notag\\
    &&=-{\rm i}\, \eta\, \left\{b_1 \sqrt{\frac{l+m+2}{2l+1}}Y_{l,m+1}+b_2 \sqrt{\frac{l+m+3}{2l+3}}Y_{l+1,m+1}-b_3 \sqrt{\frac{l-m}{2l+3}}Y_{l+1,m+1} -b_4 \sqrt{\frac{l-m+1}{2l+5}}Y_{l+2,m+1}\right\}.
\end{eqnarray}
Based on \Eq{ad1}, we can attain
\begin{eqnarray}
    && {b_1'\sqrt{l+m+2} = a_1'\sqrt{l-m}}, \nonumber\\
    && b_4'\sqrt{l-m+1} =-a_4'\sqrt{l+m+3}, \nonumber\\
    && b_2 \sqrt{{l-m}}=-b_3 \sqrt{{l+m+3}},
\end{eqnarray}
and
\begin{equation}
  \left\{\frac{b_2'\sqrt{l+m+3} -a_2'\sqrt{l-m+1}}{\sqrt{2l+3} } -\frac{b_3'\sqrt{l-m} +a_3'\sqrt{l+m+2}}{\sqrt{2l+3} }\right\}  b_{l,-m-2} ={\rm i}\, \eta\, \left\{b_1 \sqrt{\frac{l-m-1}{2l+1}}\right\},
\end{equation}

\begin{equation}
    -\left\{\frac{b_2'\sqrt{l+m+3} -a_2'\sqrt{l-m+1}}{\sqrt{2l+3} } -\frac{b_3'\sqrt{l-m} +a_3'\sqrt{l+m+2}}{\sqrt{2l+3} }\right\}  b_{l+1,m+1} ={\rm i}\, \eta\,  \left\{b_4 \sqrt{\frac{l+m+4}{2l+5}}\right\},
  \end{equation}
which means
\begin{eqnarray}
  b_1 \sqrt{l+m+3} = -\sqrt{l-m}b_4.
\end{eqnarray}
Based on \Eq{ad2}, we can attain
\begin{equation}
    - \frac{ b_1'\sqrt{l-m-1} }{\sqrt{2l+1} } b_{l-1, m+1} -b_{l-1,-m-1}\frac{a_1'\sqrt{l+m+1}}{\sqrt{2l+1} } +2a_{l-1,m+1}\frac{a_1'\sqrt{l-m}}{\sqrt{2l+1} } =0,
  \end{equation}
  \begin{equation}
      b_{l+2,-m-2}\frac{b_4'\sqrt{l+m+4} }{\sqrt{2l+5} }  -b_{l+2,m}\frac{a_4'\sqrt{l-m+2}}{\sqrt{2l+5} } +2 a_{l+2,m+1}\frac{a_4'\sqrt{l+m+3}}{\sqrt{2l+5} } =0,
  \end{equation}
  which lead to
  \begin{eqnarray}
  &&  a_1'\sqrt{l-m} = b_1'\sqrt{l+m+2}, \nonumber\\
  &&  a_4'\sqrt{l+m+3} =-b_4'\sqrt{l-m+1}.
  \end{eqnarray}
The results are equivalent to those shown in Eq. (\ref{eq:abab-1}).

Based on \Eq{ad2}, we also can attain
  \begin{eqnarray}
    \label{ae1} && \frac{ b_1'\sqrt{l-m-1} }{\sqrt{2l+1} } b_{l,-m-2} +b_{l,m}\frac{a_1'\sqrt{l+m+1}}{\sqrt{2l+1} } +2a_{l,m+1}\frac{a_1'\sqrt{l-m}}{\sqrt{2l+1} }  -b_{l+1,m+1}\frac{b_4'\sqrt{l+m+4} }{\sqrt{2l+5} } \notag\\
      &&+b_{l+1,-m-1}\frac{a_4'\sqrt{l-m+2}}{\sqrt{2l+5} }
     +2 a_{l+1,m+1}\frac{a_4'\sqrt{l+m+3}}{\sqrt{2l+5} } =-{\rm i} \eta\, (b_2 \sqrt{\frac{l+m+3}{2l+3}}-b_3 \sqrt{\frac{l-m}{2l+3}}),
  \end{eqnarray}
  \begin{eqnarray}
      \label{ae2} &&-b_{l,m+1} \left(\frac{b_2'\sqrt{l-m} }{\sqrt{2l+3} } +\frac{b_3'\sqrt{l+m+3} }{\sqrt{2l+3} }\right)-b_{l,-m-1}\left(\frac{ a_2'\sqrt{l+m+2}}{\sqrt{2l+3} } -\frac{a_3'\sqrt{l-m+1}}{\sqrt{2l+3} }\right) \notag\\
     &&+2 a_{l,m+1}\left(\frac{ a_2'\sqrt{l-m+1}}{\sqrt{2l+3} } +\frac{a_3'\sqrt{l+m+2}}{\sqrt{2l+3} }\right)= -{\rm i} \eta\, b_1 \sqrt{\frac{l+m+2}{2l+1}},
  \end{eqnarray}
  \begin{eqnarray}
      \label{ae3}  &&b_{l+1,-m-2} \left(\frac{b_2'\sqrt{l-m} }{\sqrt{2l+3} } +\frac{b_3'\sqrt{l+m+3} }{\sqrt{2l+3} }\right)+b_{l+1,m}\left(\frac{ a_2'\sqrt{l+m+2}}{\sqrt{2l+3} } -\frac{a_3'\sqrt{l-m+1}}{\sqrt{2l+3} }\right) \notag\\
      &&+2 a_{l+1,m+1}\left(\frac{ a_2'\sqrt{l-m+1}}{\sqrt{2l+3} } +\frac{a_3'\sqrt{l+m+2}}{\sqrt{2l+3} }\right)= {\rm i} \eta\, b_4 \sqrt{\frac{l-m+1}{2l+1}}.
   \end{eqnarray}
According to \Eq{ae1}, we attain
\begin{eqnarray}
    \label{ae1'}\frac{2(l+1)a'_1}{\sqrt{(l+m+2)(2l+3)}}+ \frac{2(l+2)a'_4}{\sqrt{(l-m+1)(2l+3)}} = - {\rm i }\eta\,  b_2 \sqrt{\frac{2l+3 }{l+m+3}}.
\end{eqnarray}

For \Eq{aa2}, we can make transformation
 \begin{eqnarray}
 &&   a_1' \rightarrow a_1\,\,,\,a_2' \rightarrow a_2\,\,,\,a_3' \rightarrow a_3\,\,,\,a_4' \rightarrow a_4, \nonumber\\
 &&   b_1' \rightarrow b_1\,\,,\,b_2' \rightarrow b_2\,\,,\,b_3' \rightarrow b_3\,\,,\,b_4' \rightarrow b_4, \nonumber\\
 &&   a_1 \rightarrow a_1'\,\,,\,a_2 \rightarrow a_2'\,\,,\,a_3 \rightarrow a_3'\,\,,\,a_4 \rightarrow a_4',
 \end{eqnarray}
 then we can attain the similar conclusion as the one from \Eq{aa1}, i.e.,
    \begin{eqnarray}
 &&   a_1' \sqrt{l-m+1}  = a_4' \sqrt{l+m+2}, \nonumber\\
 &&    a_1\sqrt{l-m} = b_1\sqrt{l+m+2}, \nonumber\\
 &&    a_4\sqrt{l+m+3} =-b_4\sqrt{l-m+1},\nonumber\\
 &&    a_2' \sqrt{{l+m+2}}=a_3' \sqrt{{l-m+1}}.
 \end{eqnarray}
So \Eq{ac1'} becomes
 \begin{eqnarray}
    \frac{2(l+1)a_1'}{\sqrt{(l+m+2)(2l+3)} }  +  \frac{2(l+2)a_4'}{\sqrt{(l-m+1)(2l+3)} } =2 \sqrt{\frac{(2l+3)}{(l+m+2)}}a_1' = {\rm i }\eta\,  a_2 \sqrt{\frac{2l+3}{l-m+1}},
 \end{eqnarray}
 and \Eq{ae1'} becomes
 \begin{eqnarray}
    \frac{2(l+1)a'_1}{\sqrt{(l+m+2)(2l+3)}}+ \frac{2(l+2)a'_4}{\sqrt{(l-m+1)(2l+3)}} =2 \sqrt{\frac{(2l+3)}{(l+m+2)}}a_1'= - {\rm i }\eta\,  b_2 \sqrt{\frac{2l+3 }{l+m+3}},
\end{eqnarray}
which mean
\begin{eqnarray}
    a_2 \sqrt{l+m+3}= - b_2 \sqrt{l-m+1}.
\end{eqnarray}
$\blacksquare$
\end{remark}

Similarly, we can calculate \Eq{aa4} like \Eq{aa2}, and as a result, we can attain
\begin{subequations}
    \begin{eqnarray}
        \left\{\begin{matrix}
            a_1 \sqrt{l-m}=b_1 \sqrt{l+m+2}\\
           a_1\sqrt{l-m+1}=a_4 \sqrt{l+m+2}\\
           a_4\sqrt{l+m+3}=-b_4\sqrt{l-m+1}\\
           a_2 \sqrt{l+m+2}=a_3 \sqrt{l-m+1}
           \end{matrix}\right., \;\;\;\;\;\;\;
          \left\{\begin{matrix}
           a_2\sqrt{l+m+3}=-b_2\sqrt{l-m+1} \\
           b_2\sqrt{l-m} =-b_3 \sqrt{l+m+3}\\
           b_1 \sqrt{l+m+3}=-b_4 \sqrt{l-m}\\
           2 a_1 \sqrt{l+m+3} =-{\rm i} \eta\,   b_2'\sqrt{l+m+2}
           \end{matrix}\right.,
\end{eqnarray}
\begin{eqnarray}
    \left\{\begin{matrix}
        a_1' \sqrt{l-m}=b_1' \sqrt{l+m+2}\\
       a_1'\sqrt{l-m+1}=a_4' \sqrt{l+m+2}\\
       a_4'\sqrt{l+m+3}=-b_4'\sqrt{l-m+1}\\
       a_2' \sqrt{l+m+2}=a_3' \sqrt{l-m+1}
       \end{matrix}\right., \;\;\;\;\;\;\;
       \left\{\begin{matrix}
       a_2'\sqrt{l+m+3}=-b_2'\sqrt{l-m+1} \\
       b_2'\sqrt{l-m} =-b_3' \sqrt{l+m+3}\\
       b_1' \sqrt{l+m+3}=-b_4' \sqrt{l-m}\\
       2 a_1' \sqrt{l+m+3} =-{\rm i} \eta\,  b_2\sqrt{l+m+2}
       \end{matrix}\right..
\end{eqnarray}
\end{subequations}
Then for \Eq{ac2}, we have
\begin{eqnarray}
    2\sqrt{\frac{l+m+2}{l+m+3}} \, b_2'={\rm i} \eta\,  a_1
\end{eqnarray}
and based on \Eq{aa4}, we also have
\begin{eqnarray}
    2\sqrt{\frac{l+m+2}{l+m+3}}\, b_2={\rm i} \eta\, a_1'.
\end{eqnarray}

\emph{Analysis 1.---}Based on the above results, for $\eta=2$, we have the coefficients as
\begin{eqnarray}\label{eq:2}
&& b_1=  \dfrac{\sqrt{l-m}}{\sqrt{l+m+2}} \, a_1, \;\;\;   a_4= \dfrac{\sqrt{l-m+1}}{\sqrt{l+m+2}} \,a_1, \;\;\;
b_2'={\rm i} \frac{\sqrt{l+m+3}}{\sqrt{l+m+2}}\,  a_1, \nonumber\\
&&  b_4= - \dfrac{\sqrt{l+m+3}}{\sqrt{l-m+1}} \, a_4= - \dfrac{\sqrt{l+m+3}}{\sqrt{l+m+2}} \, a_1, \nonumber\\
&&  a_2'=-\dfrac{\sqrt{l-m+1}}{\sqrt{l+m+3}}\, b_2'= -{\rm i}\dfrac{\sqrt{l-m+1}}{\sqrt{l+m+2}}\, a_1, \nonumber\\
&&  a_3'  = \dfrac{\sqrt{l+m+2}}{\sqrt{l-m+1}} \, a_2'=-{\rm i}\, a_1, \nonumber\\
&& b_3' = -\dfrac{\sqrt{l-m}}{\sqrt{l+m+3}}\, b_2' =- {\rm i}\dfrac{\sqrt{l-m}}{\sqrt{l+m+2}}\, a_1, \nonumber\\
&& b'_1=  \dfrac{\sqrt{l-m}}{\sqrt{l+m+2}} \, a'_1, \;\;\;   a'_4= \dfrac{\sqrt{l-m+1}}{\sqrt{l+m+2}} \,a'_1, \;\;\;
b_2={\rm i} \frac{\sqrt{l+m+3}}{\sqrt{l+m+2}}\,  a'_1, \nonumber\\
&&  b'_4= - \dfrac{\sqrt{l+m+3}}{\sqrt{l-m+1}} \, a'_4= - \dfrac{\sqrt{l+m+3}}{\sqrt{l+m+2}} \, a'_1, \nonumber\\
&& a_2=-\dfrac{\sqrt{l-m+1}}{\sqrt{l+m+3}}\, b_2= -{\rm i}\dfrac{\sqrt{l-m+1}}{\sqrt{l+m+2}}\, a'_1\nonumber\\
&&  a_3  = \dfrac{\sqrt{l+m+2}}{\sqrt{l-m+1}} \, a_2=-{\rm i}\, a'_1, \nonumber\\
&& b_3 = -\dfrac{\sqrt{l-m}}{\sqrt{l+m+3}}\, b_2 =- {\rm i}\dfrac{\sqrt{l-m}}{\sqrt{l+m+2}}\, a'_1.
\end{eqnarray}
There are still two free coefficients (i.e., $a_1$ and $a'_1$) in the above equation (the reason is that the
eigenvalue $\eta=2$ is 2-fold degeneracy).
 Based on Eq. (\ref{eq:2}), one can verify that Eq. (\ref{eq:1a}), Eq. (\ref{eq:1b}), Eq. (\ref{ac2}),
 Eq. (\ref{ac3}), Eq. (\ref{ae2}), Eq. (\ref{ae3}) are valid.

From Eq. (\ref{eq:phi-1a}), if we require the wavefunction $\Phi_{j,m_j}(\theta ,\phi)$ is normalized, i.e.,
\begin{eqnarray}
   \sum_{j=1}^4 \left(|a_j|^2 + |a'_j|^2 + |b_j|^2 + |b'_j|^2\right)=1,
\end{eqnarray}
then we have
\begin{eqnarray}
 &&  \dfrac{1}{l+m+2} \biggr\{ \left[(1)+\left(l-m+1\right)+\left(l-m+1\right)+\left(1\right)+\left(l-m\right)+\left(l+m+3\right)+\left(l+m+3\right)+\left(l-m\right)\right]|a_1|^2\nonumber\\
  && +\left[\left(l-m+1\right)+\left(1\right)+\left(1\right)+\left(l-m+1\right)+\left(l+m+3\right)+\left(l-m\right)+\left(l-m\right)+\left(l+m+3\right)\right]|a'_1|^2\biggr\}=1,
\end{eqnarray}
i.e.,
\begin{eqnarray}
 &&  \dfrac{3l-m+5}{l+m+2} \left(|a_1|^2+|a'_1|^2 \right)=1,
\end{eqnarray}
i.e.,
\begin{eqnarray}
 &&  |a_1|^2+|a'_1|^2 = \dfrac{l+m+2}{3l-m+5}.
\end{eqnarray}

\emph{Analysis 2.---}Based on the above results, for $\eta=-2$, we have the coefficients as
\begin{eqnarray}\label{eq:3}
&& b_1=  \dfrac{\sqrt{l-m}}{\sqrt{l+m+2}} \, a_1, \;\;\;   a_4= \dfrac{\sqrt{l-m+1}}{\sqrt{l+m+2}} \,a_1, \;\;\;
b_2'={-}{\rm i} \frac{\sqrt{l+m+3}}{\sqrt{l+m+2}}\,  a_1, \nonumber\\
&&  b_4= - \dfrac{\sqrt{l+m+3}}{\sqrt{l-m+1}} \, a_4= - \dfrac{\sqrt{l+m+3}}{\sqrt{l+m+2}} \, a_1, \nonumber\\
&&  a_2'=-\dfrac{\sqrt{l-m+1}}{\sqrt{l+m+3}}\, b_2'= {+}{\rm i}\dfrac{\sqrt{l-m+1}}{\sqrt{l+m+2}}\, a_1, \nonumber\\
&&  a_3'  = \dfrac{\sqrt{l+m+2}}{\sqrt{l-m+1}} \, a_2'={+}{\rm i}\, a_1, \nonumber\\
&& b_3' = -\dfrac{\sqrt{l-m}}{\sqrt{l+m+3}}\, b_2' ={+}{\rm i}\dfrac{\sqrt{l-m}}{\sqrt{l+m+2}}\, a_1, \nonumber\\
&& b'_1=  \dfrac{\sqrt{l-m}}{\sqrt{l+m+2}} \, a'_1, \;\;\;   a'_4= \dfrac{\sqrt{l-m+1}}{\sqrt{l+m+2}} \,a'_1, \;\;\;
b_2= {-}{\rm i} \frac{\sqrt{l+m+3}}{\sqrt{l+m+2}}\,  a'_1, \nonumber\\
&&  b'_4= - \dfrac{\sqrt{l+m+3}}{\sqrt{l-m+1}} \, a'_4= - \dfrac{\sqrt{l+m+3}}{\sqrt{l+m+2}} \, a'_1, \nonumber\\
&& a_2=-\dfrac{\sqrt{l-m+1}}{\sqrt{l+m+3}}\, b_2= {+}{\rm i}\dfrac{\sqrt{l-m+1}}{\sqrt{l+m+2}}\, a'_1\nonumber\\
&&  a_3  = \dfrac{\sqrt{l+m+2}}{\sqrt{l-m+1}} \, a_2={+}{\rm i}\, a'_1, \nonumber\\
&& b_3 = -\dfrac{\sqrt{l-m}}{\sqrt{l+m+3}}\, b_2 ={+} {\rm i}\dfrac{\sqrt{l-m}}{\sqrt{l+m+2}}\, a'_1.
\end{eqnarray}
There are still two free coefficients (i.e., $a_1$ and $a'_1$) in the above equation (the reason is that the
eigenvalue $\eta=-2$ is 2-fold degeneracy). Under the normalized condition, similarly, we have
\begin{eqnarray}
 &&  |a_1|^2+|a'_1|^2 = \dfrac{l+m+2}{3l-m+5}.
\end{eqnarray}

\subsubsection{The Case of $\eta = 0$}

Similarly, we have the constrained conditions for the coefficients as
\begin{subequations}
    \label{con1}
     \begin{eqnarray}
        \left\{\begin{matrix}
            a_1 \sqrt{l-m}=b_1 \sqrt{l+m+2}\\
           a_1(l+1)\sqrt{l-m+1}=-a_4(l+2) \sqrt{l+m+2}\\
           a_4\sqrt{l+m+3}=-b_4\sqrt{l-m+1}
           \end{matrix}\right.,
           \;\;\;\;\;
           \left\{\begin{matrix}
           { b_2 (l+2)\sqrt{l-m}=b_3(l+1) \sqrt{l+m+3}}\\
           a_2(l+2)\sqrt{l+m+2}=-a_3(l+1) \sqrt{l-m+1}\\
           a_2\sqrt{l+m+3}=-b_2\sqrt{l-m+1}
           \end{matrix}\right.,
    \end{eqnarray}
    \begin{eqnarray}
        \left\{\begin{matrix}
            a_1' \sqrt{l-m}=b_1' \sqrt{l+m+2}\\
           a_1'(l+1)\sqrt{l-m+1}=-a_4'(l+2) \sqrt{l+m+2}\\
           a_4'\sqrt{l+m+3}=-b_4'\sqrt{l-m+1}
           \end{matrix}\right.,
           \;\;\;\;\;
           \left\{\begin{matrix}
           { b_2' (l+2)\sqrt{l-m}=b_3'(l+1) \sqrt{l+m+3}}\\
           a_2'(l+2)\sqrt{l+m+2}=-a_3'(l+1) \sqrt{l-m+1}\\
           a_2'\sqrt{l+m+3}=-b_2'\sqrt{l-m+1}
           \end{matrix}\right..
    \end{eqnarray}
\end{subequations}
From the above equation, we can have
\begin{eqnarray}\label{eq:ab-8a}
&& b_1=  \dfrac{\sqrt{l-m}}{\sqrt{l+m+2}} \, a_1, \;\;\;   a_4=-\frac{l+1}{l+2}\dfrac{\sqrt{l-m+1}}{\sqrt{l+m+2}} \,a_1, \;\;\;\nonumber\\
&& b_4= \frac{l+1}{l+2}\dfrac{\sqrt{l+m+3}}{\sqrt{l+m+2}} \,a_1, \nonumber\\
&&  b_3=  \frac{l+2}{l+1}\dfrac{\sqrt{l-m}}{\sqrt{l+m+3}} \, b_2 ,\; \;\; a_2= - \dfrac{\sqrt{l-m+1}}{\sqrt{l+m+3}} \, b_2, \nonumber\\
&&  a_3=-\frac{l+2}{l+1}\dfrac{\sqrt{l+m+2}}{\sqrt{l-m+1}}\, a_2= \frac{l+2}{l+1}\dfrac{\sqrt{l+m+2}}{\sqrt{l+m+3}}\, b_2, \nonumber\\
&& b'_1=  \dfrac{\sqrt{l-m}}{\sqrt{l+m+2}} \, a'_1, \;\;\;   a'_4=-\frac{l+1}{l+2}\dfrac{\sqrt{l-m+1}}{\sqrt{l+m+2}} \,a'_1, \;\;\;\nonumber\\
&& b'_4= \frac{l+1}{l+2}\dfrac{\sqrt{l+m+3}}{\sqrt{l+m+2}} \,a'_1, \nonumber\\
&&  b'_3=  \frac{l+2}{l+1}\dfrac{\sqrt{l-m}}{\sqrt{l+m+3}} \, b'_2 ,\; \;\; a'_2= - \dfrac{\sqrt{l-m+1}}{\sqrt{l+m+3}} \, b'_2, \nonumber\\
&&  a'_3=-\frac{l+2}{l+1}\dfrac{\sqrt{l+m+2}}{\sqrt{l-m+1}}\, a'_2= \frac{l+2}{l+1}\dfrac{\sqrt{l+m+2}}{\sqrt{l+m+3}}\, b'_2.
\end{eqnarray}
In the following, let us come to make a check. For example, for $\eta=0$, Eq. (\ref{ab1}) becomes
\begin{eqnarray}\label{ab1-1}
    &&{\rm sin} \theta e^{-\rm i \phi} \left\{ \frac{a_1'\sqrt{l-m} -b_1'\sqrt{l+m+2}}{\sqrt{2l+1} } Y_{l,m+1} +\frac{a_2'\sqrt{l-m+1} -b_2'\sqrt{l+m+3}}{\sqrt{2l+3} } Y_{l+1,m+1}\right\}\notag\\
    && +{\rm sin}\theta e^{-\rm i \phi} \left\{\frac{a_3'\sqrt{l+m+2} +b_3'\sqrt{l-m}}{\sqrt{2l+3} } Y_{l+1,m+1}+\frac{a_4'\sqrt{l+m+3} +b_4'\sqrt{l-m+1}}{\sqrt{2l+5} } Y_{l+2,m+1}\right\}=0,
\end{eqnarray}
which is valid based on Eq. (\ref{eq:ab-8a}). And for $\eta=0$, Eq. (\ref{ab2}) requires
\begin{eqnarray}
  &&  a_1'\sqrt{l-m} = b_1'\sqrt{l+m+2}, \nonumber\\
  &&  a_4'\sqrt{l+m+3} =-b_4'\sqrt{l-m+1}, \nonumber\\
  && {a_2'\sqrt{l-m+1} -b_2'\sqrt{l+m+3}} +{a_3'\sqrt{l+m+2} +b_3'\sqrt{l-m}}=0,
  \end{eqnarray}
which is valid based on Eq. (\ref{eq:ab-8a}). Similarly, one can check other related equations are also valid.

There are still four free coefficients (i.e., $a_1$, $a'_1$, $b_2$, and $b'_2$) in Eq. (\ref{eq:ab-8a}) (the reason is that the
eigenvalue $\eta=0$ is 4-fold degeneracy). If we require the wavefunction $\Phi_{j,m_j}(\theta ,\phi)$ is normalized, i.e.,
\begin{eqnarray}
   \sum_{j=1}^4 \left(|a_j|^2 + |a'_j|^2 + |b_j|^2 + |b'_j|^2\right)=1,
\end{eqnarray}
then we have
\begin{eqnarray}
 &&   \biggr\{ \left[\left(1\right)+\left(\frac{(l+1)^2}{(l+2)^2}\dfrac{{l-m+1}}{{l+m+2}}\right)+\left(\dfrac{{l-m}}{{l+m+2}}\right)
 +\left(\frac{(l+1)^2}{(l+2)^2}\dfrac{{l+m+3}}{{l+m+2}}\right)\right]|a_1|^2\nonumber\\
&& +\left[\left(1\right)+\left(\frac{(l+1)^2}{(l+2)^2}\dfrac{{l-m+1}}{{l+m+2}}\right)+\left(\dfrac{{l-m}}{{l+m+2}}\right)
+\left(\frac{(l+1)^2}{(l+2)^2}\dfrac{{l+m+3}}{{l+m+2}}\right)\right]|a'_1|^2\nonumber\\
  && +\left[\left(\dfrac{{l-m+1}}{{l+m+3}}\right)+\left(\frac{(l+2)^2}{(l+1)^2}\dfrac{{l+m+2}}{{l+m+3}}\right)+\left(1\right)
  +\left(\frac{(l+2)^2}{(l+1)^2}\dfrac{{l-m}}{{l+m+3}}\right)\right]|b_2|^2\nonumber\\
 && +\left[\left(\dfrac{{l-m+1}}{{l+m+3}}\right)+\left(\frac{(l+2)^2}{(l+1)^2}\dfrac{{l+m+2}}{{l+m+3}}\right)+\left(1\right)
 +\left(\frac{(l+2)^2}{(l+1)^2}\dfrac{{l-m}}{{l+m+3}}\right)\right]|b'_2|^2 \biggr\}=1,
 \end{eqnarray}
i.e.,
\begin{eqnarray}
 &&   \left[\frac{2(l+1)(2l+3)}{(l+2)(l+m+2)}\right]\left(|a_1|^2+|a'_1|^2\right)+\left[\frac{2(l+2)(2l+3)}{(l+1)(l+m+3)}\right]\left(|b_2|^2+|b'_2|^2\right)
 =1.
 \end{eqnarray}

\subsection{Exact Solution of Dirac's Equation with Spin-Dependent Coulomb Interaction}

Based on Eq. (\ref{eq:H-1a-1a}), we can have Dirac's Hamiltonian as
\begin{eqnarray}
    \mathcal{H} = c \vec{\alpha_1 } \cdot \left[\vec{p} - \frac{q k\hbar(\hat{r} \times \vec{\sigma}_2) }{2cr} \right]
    + \beta_1 M c^2 + \frac{q\kappa}{r},
\end{eqnarray}
and the eigen-problem is given by
\begin{eqnarray}
    \mathcal{H} \Psi_{j,m_j}(r,\theta ,\phi) = E\, \Psi_{j,m_j}(r,\theta ,\phi),
\end{eqnarray}
where $E$ is the eigenvalue and $\Psi_{j,m_j}(r,\theta ,\phi)$ is the eigenfunction. For convenience, we designate
\begin{eqnarray}
    \frac{E}{\hbar c } =\overline{E}, \;\;\;\; \frac{ M c^2}{\hbar c } = \overline{M}, \;\;\;\;
    \frac{q k}{2c}=\overline{k},\;\;\;\; \frac{q \kappa }{\hbar c }= \overline{\kappa}.
\end{eqnarray}
We now assume that the form of the eigenfunction $\Psi_{j,mj}(r,\theta ,\phi)$  is as follows
\begin{eqnarray}
    \Psi_{j,mj}(r, \theta ,\phi) &= &\left[f_1(r) \Phi_{1,l,m}(\theta,\phi)+f_2(r) \Phi_{2,l,m}(\theta,\phi)+f_3(r) \Phi_{3,l,m}(\theta,\phi)+f_4(r) \Phi_{4,l,m}(\theta,\phi)\right] \notag\\
    &&+\left[g_1(r) \Phi'_{1,l,m}(\theta,\phi)+g_2(r) \Phi'_{2,l,m}(\theta,\phi)+g_3(r) \Phi'_{3,l,m}(\theta,\phi)+g_4(r) \Phi'_{4,l,m}(\theta,\phi)\right],
\end{eqnarray}
where
\begin{subequations}
    \begin{eqnarray}
        \Phi_{1,l,m}(\theta,\phi)&=&\begin{pmatrix}
            \Phi^A_{l,j_1,m_1}+ \sqrt{\frac{l-m+1}{l+m+2}}\Phi^B_{l+1,j_1',m_1} \\
            -{\rm i}\left\{\Phi^B_{l,j_1,m_1} + \sqrt{\frac{l-m+1}{l+m+2}}\Phi^A_{l+1,j_1',m_1}\right\}
        \end{pmatrix}
        \otimes \begin{pmatrix}
            1\\
            0
        \end{pmatrix}\notag\\
        &&+\sqrt{\frac{l-m}{l+m+2}}\begin{pmatrix}
            \Phi^A_{l,j_1,m_1'}- \sqrt{\frac{l+m+3}{l-m}}\Phi^B_{l+1,j_1',m_1'} \\
            -{\rm i}\left\{\Phi^B_{l,j_1,m_1'} - \sqrt{\frac{l+m+3}{l-m}}\Phi^A_{l+1,j_1',m_1'}\right\}
        \end{pmatrix}
        \otimes \begin{pmatrix}
            0\\
            1
        \end{pmatrix},
    \end{eqnarray}
    \begin{eqnarray}
      \Phi'_{1,l,m}(\theta,\phi)&=&\begin{pmatrix}
            \Phi^A_{l,j_1,m_1}+ \sqrt{\frac{l-m+1}{l+m+2}}\Phi^B_{l+1,j_1',m_1} \\
            {\rm i}\left\{\Phi^B_{l,j_1,m_1} + \sqrt{\frac{l-m+1}{l+m+2}}\Phi^A_{l+1,j_1',m_1}\right\}
        \end{pmatrix}
        \otimes \begin{pmatrix}
            1\\
            0
        \end{pmatrix}\notag\\
        &&+\sqrt{\frac{l-m}{l+m+2}}\begin{pmatrix}
            \Phi^A_{l,j_1,m_1'}- \sqrt{\frac{l+m+3}{l-m}}\Phi^B_{l+1,j_1',m_1'} \\
            {\rm i}\left\{\Phi^B_{l,j_1,m_1'} - \sqrt{\frac{l+m+3}{l-m}}\Phi^A_{l+1,j_1',m_1'}\right\}
        \end{pmatrix}
        \otimes \begin{pmatrix}
            0\\
            1
        \end{pmatrix},
    \end{eqnarray}
    \begin{eqnarray}
        \Phi_{2,l,m}(\theta,\phi)&=&\begin{pmatrix}
            \Phi^A_{l+1,j_1',m_1}+ \sqrt{\frac{l+m+2}{l-m+1}}\Phi^B_{l,j_1,m_1} \\
            {\rm i}\left\{\Phi^B_{l+1,j_1',m_1} + \sqrt{\frac{l+m+2}{l-m+1}}\Phi^A_{l,j_1,m_1}\right\}
        \end{pmatrix}
        \otimes \begin{pmatrix}
            1\\
            0
        \end{pmatrix}\notag\\
        &&-\sqrt{\frac{l+m+3}{l-m+1}}\begin{pmatrix}
            \Phi^A_{l+1,j_1',m_1'}-\sqrt{\frac{l-m}{l+m+3}}\Phi^B_{l,j_1,m_1'} \\
            {\rm i}\left\{\Phi^B_{l+1,j_1',m_1'} - \sqrt{\frac{l-m}{l+m+3}}\Phi^A_{l,j_1,m_1'}\right\}
        \end{pmatrix}
        \otimes \begin{pmatrix}
            0\\
            1
        \end{pmatrix},
    \end{eqnarray}
    \begin{eqnarray}
      \Phi'_{2,l,m}(\theta,\phi)&=&\begin{pmatrix}
            \Phi^A_{l+1,j_1',m_1}+ \sqrt{\frac{l+m+2}{l-m+1}}\Phi^B_{l,j_1,m_1} \\
            -{\rm i}\left\{\Phi^B_{l+1,j_1',m_1} + \sqrt{\frac{l+m+2}{l-m+1}}\Phi^A_{l,j_1,m_1}\right\}
        \end{pmatrix}
        \otimes \begin{pmatrix}
            1\\
            0
        \end{pmatrix}\notag\\
        &&-\sqrt{\frac{l+m+3}{l-m+1}}\begin{pmatrix}
            \Phi^A_{l+1,j_1',m_1'}-\sqrt{\frac{l-m}{l+m+3}}\Phi^B_{l,j_1,m_1'} \\
            -{\rm i}\left\{\Phi^B_{l+1,j_1',m_1'} - \sqrt{\frac{l-m}{l+m+3}}\Phi^A_{l,j_1,m_1'}\right\}
        \end{pmatrix}
        \otimes \begin{pmatrix}
            0\\
            1
        \end{pmatrix},
    \end{eqnarray}
    \begin{eqnarray}
        \Phi_{3,l,m}(\theta,\phi)&=&\begin{pmatrix}
            \Phi^A_{l,j_1,m_1}-\frac{l+1}{l+2}\sqrt{\frac{l-m+1}{l+m+2}}\Phi^B_{l+1,j_1',m_1} \\
          0
        \end{pmatrix}
        \otimes \begin{pmatrix}
            1\\
            0
        \end{pmatrix}\notag\\
        &&+\sqrt{\frac{l-m}{l+m+2}}\begin{pmatrix}
            \Phi^A_{l,j_1,m_1'}+\frac{l+1}{l+2} \sqrt{\frac{l+m+3}{l-m}}\Phi^B_{l+1,j_1',m_1'} \\
           0
        \end{pmatrix}
        \otimes \begin{pmatrix}
            0\\
            1
        \end{pmatrix},
    \end{eqnarray}
    \begin{eqnarray}
        \Phi'_{3,l,m}(\theta,\phi)&=&\begin{pmatrix}
            0\\
            -{\rm i}\left\{\Phi^B_{l,j_1,m_1} -\frac{l+1}{l+2}\sqrt{\frac{l-m+1}{l+m+2}}\Phi^A_{l+1,j_1',m_1}\right\}
        \end{pmatrix}
        \otimes \begin{pmatrix}
            1\\
            0
        \end{pmatrix}\notag\\
        &&+\sqrt{\frac{l-m}{l+m+2}}\begin{pmatrix}
            0 \\
            -{\rm i}\left\{\Phi^B_{l,j_1,m_1'} +\frac{l+1}{l+2}\sqrt{\frac{l+m+3}{l-m}}\Phi^A_{l+1,j_1',m_1'}\right\}
        \end{pmatrix}
        \otimes \begin{pmatrix}
            0\\
            1
        \end{pmatrix},
    \end{eqnarray}
    \begin{eqnarray}
        \Phi_{4,l,m}(\theta,\phi)&=&\begin{pmatrix}
            \Phi^A_{l+1,j_1',m_1}-\frac{l+2}{l+1} \sqrt{\frac{l+m+2}{l-m+1}}\Phi^B_{l,j_1,m_1} \\
           0
        \end{pmatrix}
        \otimes \begin{pmatrix}
            1\\
            0
        \end{pmatrix}\notag\\
        &&-\sqrt{\frac{l+m+3}{l-m+1}}\begin{pmatrix}
            \Phi^A_{l+1,j_1',m_1'}+\frac{l+2}{l+1}\sqrt{\frac{l-m}{l+m+3}}\Phi^B_{l,j_1,m_1'} \\
           0
        \end{pmatrix}
        \otimes \begin{pmatrix}
            0\\
            1
        \end{pmatrix},
    \end{eqnarray}
    \begin{eqnarray}
        \Phi'_{4,l,m}(\theta,\phi)&=&\begin{pmatrix}
            0 \\
            {\rm i}\left\{{\Phi^B_{l+1,j_1',m_1}} -\frac{l+2}{l+1}\sqrt{\frac{l+m+2}{l-m+1}}\Phi^A_{l,j_1,m_1}\right\}
        \end{pmatrix}
        \otimes \begin{pmatrix}
            1\\
            0
        \end{pmatrix}\notag\\
        &&-\sqrt{\frac{l+m+3}{l-m+1}}\begin{pmatrix}
            0\\
            {\rm i}\left\{\Phi^B_{l+1,j_1',m_1} +\frac{l+2}{l+1}\sqrt{\frac{l-m}{l+m+3}}\Phi^A_{l,j_1,m_1'}\right\}
        \end{pmatrix}
        \otimes \begin{pmatrix}
            0\\
            1
        \end{pmatrix},
    \end{eqnarray}
\end{subequations}
and
\begin{subequations}
    \begin{eqnarray}
       \left[\hat{r} \cdot \left( \vec{\alpha}_1 \times \vec{\sigma}_2\right)\right]\,\Phi_{1,l,m}(\theta,\phi)=2\Phi_{1,l,m}(\theta,\phi),\;\;\;\;\;  \left[\hat{r} \cdot \left( \vec{\alpha}_1 \times \vec{\sigma}_2\right)\right]\Phi'_{1,l,m}(\theta,\phi)=-2\Phi'_{1,l,m}(\theta,\phi),
    \end{eqnarray}
    \begin{eqnarray}
      \left[\hat{r} \cdot \left( \vec{\alpha}_1 \times \vec{\sigma}_2\right)\right]\Phi_{2,l,m}(\theta,\phi)=2\Phi_{2,l,m}(\theta,\phi),\;\;\;\;\;
       \left[\hat{r} \cdot \left( \vec{\alpha}_1 \times \vec{\sigma}_2\right)\right]\Phi'_{2,l,m}(\theta,\phi)=-2\Phi'_{2,l,m}(\theta,\phi),
   \end{eqnarray}
    \begin{eqnarray}
     &&   \left[\hat{r} \cdot \left( \vec{\alpha}_1 \times \vec{\sigma}_2\right)\right]\Phi_{3,l,m}(\theta,\phi)=0, \;\;\;\;\;  \
     \left[\hat{r} \cdot \left( \vec{\alpha}_1 \times \vec{\sigma}_2\right)\right]\Phi'_{3,l,m}(\theta,\phi)=0, \;\;\;\;\;\nonumber\\
     && \left[\hat{r} \cdot \left( \vec{\alpha}_1 \times \vec{\sigma}_2\right)\right]\Phi_{4,l,m}(\theta,\phi)=0,\;\;\;\;\;
     \left[\hat{r} \cdot \left( \vec{\alpha}_1 \times \vec{\sigma}_2\right)\right]\Phi'_{4,l,m}(\theta,\phi)=0.
    \end{eqnarray}
\end{subequations}
Then we can attain
\begin{subequations}\label{De11}
    \begin{eqnarray}
    \label{de1}    \frac{\partial (f_1 -g_1+g_3) }{\partial r}+ \frac{l+2}{r}(f_1 -g_1+g_3) + \frac{2 \overline{k}}{r}(f_1 -g_1)
    + \left(\frac{\overline{\kappa}}{r}-\overline{E}+\overline{M}\right)(f_1 +g_1+f_3)=0,
    \end{eqnarray}
    \begin{eqnarray}
        \label{de2}     \frac{\partial (f_1 +g_1+f_3) }{\partial r}- \frac{l}{r}(f_1 +g_1+f_3) - \frac{2\overline{k}}{r}(f_1 +g_1)
        - \left(\frac{\overline{\kappa}}{r}-\overline{E}-\overline{M}\right)(f_1 -g_1+g_3)=0,
    \end{eqnarray}
    \begin{eqnarray}
        \label{de3}   \frac{\partial (f_1 -g_1-\frac{l+1}{l+2}g_3) }{\partial r}- \frac{l+1}{r}\left(f_1 -g_1-\frac{l+1}{l+2}g_3\right)
        + \frac{2\overline{k}}{r}(f_1 -g_1)+ \left(\frac{\overline{\kappa}}{r}-\overline{E}+\overline{M}\right)\left(f_1 +g_1-\frac{l+1}{l+2}f_3\right)=0,
    \end{eqnarray}
    \begin{eqnarray}
        \label{de4}   \frac{\partial (f_1 +g_1-\frac{l+1}{l+2}f_3) }{\partial r}+ \frac{l+3}{r}\left(f_1 +g_1-\frac{l+1}{l+2}f_3\right)
         - \frac{2\overline{k}}{r}(f_1 +g_1)- \left(\frac{\overline{\kappa}}{r}-\overline{E}-\overline{M}\right)\left(f_1 -g_1-\frac{l+1}{l+2}g_3\right)=0,
    \end{eqnarray}
\end{subequations}
and
\begin{subequations}\label{De12}
      \begin{eqnarray}
      \label{de1'}    \frac{\partial (f_2 -g_2+g_4) }{\partial r}+ \frac{l+3}{r}(f_2 -g_2+g_4) - \frac{2\overline{k}}{r}(f_2 -g_2)
      - \left(\frac{\overline{\kappa}}{r}-\overline{E}+\overline{M}\right)(f_2 +g_2+f_4)=0,
      \end{eqnarray}
      \begin{eqnarray}
          \label{de2'}     \frac{\partial (f_2 +g_2+f_4) }{\partial r}- \frac{l+1}{r}(f_2 +g_2+f_4) + \frac{2\overline{k}}{r}(f_2 +g_2)
          +\left(\frac{\overline{\kappa}}{r}-\overline{E}-\overline{M}\right)(f_2 -g_2+g_4)=0,
      \end{eqnarray}
      \begin{eqnarray}
          \label{de3'}   \frac{\partial (f_2 -g_2-\frac{l+2}{l+1}g_4) }{\partial r}- \frac{l}{r}\left(f_2 -g_2-\frac{l+2}{l+1}g_4\right)
          - \frac{2\overline{k}}{r}(f_2 -g_2)- \left(\frac{\overline{\kappa}}{r}-\overline{E}+\overline{M}\right)\left(f_2 +g_2-\frac{l+2}{l+1}f_4\right)=0,
      \end{eqnarray}
      \begin{eqnarray}
          \label{de4'}   \frac{\partial (f_2 +g_2-\frac{l+2}{l+1}f_4) }{\partial r}+ \frac{l+2}{r}\left(f_2 +g_2-\frac{l+2}{l+1}f_4\right)
          + \frac{2\overline{k}}{r}(f_2 +g_2)+\left(\frac{\overline{\kappa}}{r}-\overline{E}-\overline{M}\right)\left(f_2 -g_2-\frac{l+2}{l+1}g_4\right)=0.
      \end{eqnarray}
  \end{subequations}
For \Eq{De11}, we designate
\begin{eqnarray}
    F_1 = f_1 -g_1\,\,,\,\, F_2 = f_1 +g_1.
\end{eqnarray}
Based on \Eq{de1} and \Eq{de3}, we can attain
\begin{subequations}
    \begin{eqnarray}
     \label{dd1}   \frac{\partial F_1}{\partial r}+ \frac{2 \overline{k}}{r}F_1 +\frac{l+1}{r} g_3
      +\left(\frac{\overline{\kappa}}{r}-\overline{E} + \overline{M} \right)F_2 =0,
    \end{eqnarray}
    \begin{eqnarray}
        \label{dd2}  \frac{\partial g_3}{\partial r}+ \frac{1}{r}g_3 +\frac{l+2}{r} F_1
        +\left(\frac{\overline{\kappa}}{r}-\overline{E} + \overline{M} \right)f_3 =0.
    \end{eqnarray}
\end{subequations}
Similarly, based on \Eq{de2} and \Eq{de4}, we can attain
\begin{subequations}
    \begin{eqnarray}
        \label{dd3}   \frac{\partial F_2}{\partial r}- \frac{2(\overline{k}-1)}{r}F_2 -\frac{l+1}{r} f_3
        -\left(\frac{\overline{\kappa}}{r}-\overline{E} - \overline{M} \right)F_1 =0,
    \end{eqnarray}
    \begin{eqnarray}
        \label{dd4} \frac{\partial f_3}{\partial r}+ \frac{1}{r}f_3 -\frac{l+2}{r} F_2
        -\left(\frac{\overline{\kappa}}{r}-\overline{E} - \overline{M} \right)g_3 =0.
    \end{eqnarray}
\end{subequations}

Firstly, we consider the behavior of $r \rightarrow \infty$. For \Eq{dd2} and \Eq{dd4}, we acquire
\begin{eqnarray}
    \left\{\begin{matrix}
        \dfrac{\partial g_3}{\partial r} +(\overline{M}-\overline{E}) f_3 =0,\\
        \dfrac{\partial f_3}{\partial r} +(\overline{M}+\overline{E}) g_3 =0.
       \end{matrix}\right.
\end{eqnarray}
If we designate $g_3 = C_1 e^{-\varepsilon r }$ and $f_3 = C_2 e^{-\varepsilon r }$ ($\varepsilon$ is real number), and require the solution is a bound state,
we then  get $\overline{E} < \overline{M}$, i.e. $E<M c^2$, and
\begin{eqnarray}
   && \varepsilon= \sqrt{\overline{M}^2 -\overline{E}^2}=\frac{1}{\hbar c}\sqrt{M^2c^4 -E^2 }, \nonumber\\
   && f_3 =\sqrt{\frac{\overline{M}+\overline{E}}{\overline{M}-\overline{E}}} g_3 =\mu\, g_3,
\end{eqnarray}
with
\begin{eqnarray}
     \mu= \sqrt{\dfrac{\overline{M}+\overline{E}}{\overline{M}-\overline{E}}}=\sqrt{\frac{Mc^2+E}{Mc^2-E} }.
\end{eqnarray}
For \Eq{dd1} and \Eq{dd3}, we acquire
\begin{eqnarray}
    \left\{\begin{matrix}
        \dfrac{\partial F_1}{\partial r} +(\overline{M}-\overline{E}) F_2 =0,\\
        \dfrac{\partial F_2}{\partial r} +(\overline{M}+\overline{E}) F_1 =0.
       \end{matrix}\right.
\end{eqnarray}
If we designate $F_1 = C_1 e^{-\varepsilon r }$ and $F_2 = C_2 e^{-\varepsilon r }$ and require the solution is a bound state,
we can also get $\overline{E} < \overline{M}$ (i.e. $E<M c^2$), and
\begin{eqnarray}
 && \varepsilon= \sqrt{\overline{M}^2 -\overline{E}^2}, \;\;\;\;\; F_2 = \mu\, F_1.
\end{eqnarray}

Secondly, we consider the behavior of $r \rightarrow 0$. We can have
\begin{subequations}
    \begin{eqnarray}
         \frac{\partial F_1}{\partial r}+ \frac{2 \overline{k}}{r}F_1 +\frac{l+1}{r} g_3 +\frac{\overline{\kappa}}{r}F_2 =0,
       \end{eqnarray}
       \begin{eqnarray}
           \frac{\partial F_2}{\partial r}- \frac{2(\overline{k}-1)}{r}F_2 -\frac{l+1}{r} f_3 -\frac{\overline{\kappa}}{r}F_1 =0,
    \end{eqnarray}
    \begin{eqnarray}
        \frac{\partial f_3}{\partial r}+ \frac{1}{r}f_3 -\frac{l+2}{r} F_2 -\frac{\overline{\kappa}}{r}g_3 =0,
    \end{eqnarray}
       \begin{eqnarray}
            \frac{\partial g_3}{\partial r}+ \frac{1}{r}g_3 +\frac{l+2}{r} F_1 +\frac{\overline{\kappa}}{r}f_3 =0.
       \end{eqnarray}
\end{subequations}
If we designate $F_1= d_1 r^{\nu-1}$, $F_2= d_2 r^{\nu-1}$, $f_3= d_3 r^{\nu-1}$, $g_3= d_4 r^{\nu-1}$,
then
\begin{eqnarray}
  &&      (\nu+2 \overline{k}-1)d_1+ (l+1)d_4+\overline{\kappa} d_2 =0, \nonumber\\
    &&   -\overline{\kappa} d_1 -(l+1)d_3 +(\nu-2 \overline{g}+1) d_2 =0, \nonumber\\
      && \nu d_3 -\overline{\kappa}d_4 -(l+2)d_2=0, \nonumber\\
       && (l+2)d_1 +\overline{\kappa} d_3 +\nu d_4 =0.
\end{eqnarray}
We can get the solution of $\nu$ as follows
\begin{eqnarray}
  && \nu =\pm \sqrt{\left(\frac{2 \overline{k}-1 +\sqrt{(2 \overline{k}-1)^2 +4(l+1)(l+2)}}{2}\right)^2 -\overline{\kappa}^2}, \nonumber\\
   && {\rm or }\;\;\;\; \nu =\pm \sqrt{\left(\frac{2 \overline{k}-1 -\sqrt{(2 \overline{k}-1)^2 +4(l+1)(l+2)}}{2}\right)^2 -\overline{\kappa}^2}.
\end{eqnarray}
Because $\nu$ reduces to $\nu = \sqrt{ (l+1)^2 - \overline{\kappa}^2}$ when $\overline{k}=0$ (or $k$ tends to 0),  and $\nu$ should satisfy $\nu>1$, we then choose
\begin{eqnarray}
      \nu =\sqrt{\left(\frac{2 \overline{k}-1 +\sqrt{(2\overline{k} -1)^2 +4(l+1)(l+2)}}{2}\right)^2 -\overline{\kappa}^2} .
\end{eqnarray}

Finally, we designate
\begin{eqnarray}
  &&  F_1 =\frac{1}{\mu} h_1(r)\, e^{-\varepsilon r } r^{\nu-1}, \;\;\;\;F_2 = h_2(r)\, e^{-\varepsilon r } r^{\nu-1}, \nonumber\\
  &&  f_3 =h_3(r)\, e^{-\varepsilon r } r^{\nu-1}, \;\;\;\; g_3 =\frac{1}{\mu} h_4(r)\, e^{-\varepsilon r } r^{\nu-1},
\end{eqnarray}
where $h_j(r)$ ($j=1, 2, 3, 4$) are functions of $r$. Then we obtain
\begin{subequations}
     \begin{eqnarray}
        \frac{\partial h_1}{\partial r}+ \left(\frac{2 \overline{k}+\nu-1}{r}-\varepsilon\right)h_1 +\frac{l+1}{r} h_4
        +\left(\frac{\overline{\kappa}\mu}{r}+\varepsilon\right)h_2 =0,
      \end{eqnarray}
      \begin{eqnarray}
        \frac{\partial h_2}{\partial r}- \left(\frac{2 \overline{k}-\nu-1}{r}+\varepsilon\right)h_2 -\frac{l+1}{r} h_3 -\left(\frac{\overline{\kappa}}{\mu r}-\varepsilon\right)h_1 =0,
   \end{eqnarray}
   \begin{eqnarray}
    \frac{\partial h_3}{\partial r}+ \left(\frac{\nu}{r}-\varepsilon\right)h_3 -\frac{l+2}{r} h_2
    -\left(\frac{\overline{\kappa}}{\mu r}-\varepsilon\right)h_4 =0,
   \end{eqnarray}
      \begin{eqnarray}
        \frac{\partial h_4}{\partial r}+ \left(\frac{\nu}{r}-\varepsilon\right)h_4 +\frac{l+2}{r} h_1
        +\left(\frac{\overline{\kappa}\mu}{ r}+\varepsilon\right)h_3 =0.
      \end{eqnarray}
\end{subequations}
We continue to designate
\begin{eqnarray}
    h_1 +h_2 =A,\,\,\,\,h_1 -h_2 =B,\,\,\,\,h_3+h_4 =C,\,\,\,\,h_3-h_4=D,\,\,\,\, \overline{p}=\frac{\overline{\kappa}}{2}\left(\mu +\frac{1}{\mu}\right),\,\,\,\,\overline{q}=\frac{\overline{\kappa}}{2}\left(\mu -\frac{1}{\mu}\right),
\end{eqnarray}
and then
\begin{subequations}
    \begin{eqnarray}
     \label{f1''}   \frac{\partial A }{\partial r}+ \frac{\nu+\overline{q}}{r}A +\frac{2 \overline{k}-1-\overline{p}}{r} B -\frac{l+1}{r}D =0,
      \end{eqnarray}
      \begin{eqnarray}
        \label{f2''}    \frac{\partial B}{\partial r}+ \left(\frac{\nu-\overline{q}}{r}-2 \varepsilon\right)B +\frac{l+1}{r} C +\frac{2 \overline{k}-1+\overline{p}}{r}A =0,
   \end{eqnarray}
   \begin{eqnarray}
      \label{f3''}  \frac{\partial C}{\partial r}+ \frac{\nu+\overline{q}}{r}C +\frac{l+2}{r} B +\frac{\overline{p}}{r}D =0,
   \end{eqnarray}
      \begin{eqnarray}
            \label{f4''}      \frac{\partial D}{\partial r}+ \left(\frac{\nu-\overline{q}}{r}-2 \varepsilon\right)D -\frac{l+2}{r} A -\frac{\overline{p}}{r}C =0.
      \end{eqnarray}
\end{subequations}
If we assume $A= C_1 C$ and $B= C_2 D$, it should satisfy
\begin{subequations}
    \begin{eqnarray}
    (l+2)C_1C_2 + \overline{p} \,C_1-(2 \overline{k}-1- \overline{p})C_2 +l+1=0,
    \end{eqnarray}
    \begin{eqnarray}
        (l+2)C_1C_2+(2 \overline{k}-1+ \overline{p})C_1 +\overline{p}\, C_2+l+1=0.
    \end{eqnarray}
\end{subequations}
Based on which, we can attain
\begin{eqnarray}
    C_1 =-C_2.
\end{eqnarray}
Then we have
\begin{eqnarray}
     (l+2)C_1^2 -(2\overline{k}-1)C_1 -(l+1)=0,
\end{eqnarray}
which leads to
\begin{eqnarray}
      C_1 = \frac{(2\overline{k}-1) \pm \sqrt{ (2\overline{k}-1)^2 +4(l+1)(l+2)}}{2(l+2)}.
\end{eqnarray}
So if we designate
\begin{eqnarray}
&& C= c_0 +c_1 r +\dots + c_n r^n +\dots, \nonumber\\
&& D= d_0 +d_1 r +\dots + d_n r^n +\dots,
\end{eqnarray}
based on \Eq{f3''} and \Eq{f4''}, we have
\begin{eqnarray}
 (n+1+\nu+\overline{q}) c_{n+1} +\left\{-(l+2)C_1 + \overline{p}  \right\}d_{n+1}=0,
\end{eqnarray}
and
\begin{eqnarray}
    (n+1+\nu-\overline{q})d_{n+1}-2\varepsilon\,d_n-[(l+2)C_1 + \overline{p}]c_{n+1}=0.
\end{eqnarray}
As a result,
\begin{eqnarray}
    d_{n+1} = 2\varepsilon\,\frac{n+1+\nu+\overline{q}}{(n+1+\nu)^2+\overline{\kappa}^2-(l+2)^2 {C_1}^2} d_n.
\end{eqnarray}
Because when $\overline{k}=0$,
\begin{eqnarray}
  {  (n+1+\nu)^2+\overline{\kappa}^2-(l+2)^2 {C_1}^2 = (n+1)(n+2\nu+1),}
\end{eqnarray}
thus $C_1 $ should be
\begin{eqnarray}
      C_1 = \frac{l+1}{l+2}.
\end{eqnarray}
So we choose
\begin{eqnarray}
      C_1 = \frac{(2\overline{k}-1) +\sqrt{ (2\overline{k}-1)^2 +4(l+1)(l+2)}}{2(l+2)},
\end{eqnarray}
which causes
\begin{eqnarray}
      d_{n+1} = 2\varepsilon\,\frac{n+1+\nu+\overline{q}}{(n+1+\nu)^2+\overline{\kappa}^2-(l+2)^2 {C_1}^2} d_n
      = 2\varepsilon\, \frac{n+1+\nu+\overline{q}}{(n+1+2\nu)(n+1)} d_n.
\end{eqnarray}
In addition, from
\begin{eqnarray}
      (n+1+\nu+\overline{q}) c_{n+1} &=& 2 \varepsilon\,\frac{n+1+\nu+\overline{q}}{(n+1+2\nu)(n+1)} (n+\nu+\overline{q}) c_n,
\end{eqnarray}
one has
\begin{eqnarray}
      c_{n+1} &=& 2\varepsilon\,\frac{n+\nu+\overline{q}}{(n+1+2\nu)(n+1)} c_n.
\end{eqnarray}
As a result, the coefficients are
\begin{eqnarray}
      &&A= C_1T_1 F(\nu+\overline{q}; 2\nu+1;2 \varepsilon r),\nonumber\\
      &&B= -C_1 T_2 F(\nu+\overline{q}+1; 2\nu+1;2 \varepsilon r),\nonumber\\
      &&C= T_1 F(\nu+\overline{q}; 2\nu+1;2 \varepsilon r),\nonumber\\
      &&D= T_2 F(\nu+\overline{q}+1; 2\nu+1;2 \varepsilon r),
\end{eqnarray}
and
\begin{subequations}
      \begin{eqnarray}
            f_1 &=& \frac{1}{4} \left\{\left(\frac{1}{\mu } +1\right)A +\left(\frac{1}{\mu } -1\right)B\right\} e^{-\varepsilon r } r^{\nu-1}\notag\\
&=& \frac{1}{4} e^{-\varepsilon r } r^{\nu-1} \left\{\left(\frac{1}{\mu }
+1\right)C_1T_1 F(\nu+\overline{q}; 2\nu+1;2\varepsilon r)
-\left(\frac{1}{\mu } -1\right)C_1 T_2 F(\nu+\overline{q}+1; 2\nu+1;2\varepsilon r) \right\},
      \end{eqnarray}
      \begin{eqnarray}
            g_1 &=& \frac{1}{4} \left\{\left(-\frac{1}{\mu } +1\right)A -\left(\frac{1}{\mu } +1\right)B\right\} e^{-\varepsilon r } r^{\nu-1}\notag\\
&=& \frac{1}{4} e^{-\varepsilon r } r^{\nu-1} \left\{\left(-\frac{1}{\mu }
+1\right)C_1T_1 F(\nu+\overline{q}; 2\nu+1;2 \varepsilon r) +\left(\frac{1}{\mu }
+1\right)C_1 T_2 F(\nu+\overline{q}+1; 2\nu+1;2\varepsilon r) \right\},
      \end{eqnarray}
      \begin{eqnarray}
            f_3 =\frac{1}{2} \left\{C+D\right\} e^{-\varepsilon r } r^{\nu-1}
=&\dfrac{1}{2} e^{-\varepsilon r } r^{\nu-1} \left\{T_1 F(\nu+\overline{q}; 2\nu+1;2\varepsilon r)
+ T_2 F(\nu+\overline{q}+1; 2\nu+1;2\varepsilon r) \right\},
      \end{eqnarray}
      \begin{eqnarray}
            g_3 =\frac{1}{2\mu } \left\{C-D\right\} e^{-\varepsilon r } r^{\nu-1}
=&\dfrac{1}{2\mu } e^{-\varepsilon r } r^{\nu-1} \left\{T_1 F(\nu+\overline{q}; 2\nu+1;2\varepsilon r)
- T_2 F(\nu+\overline{q}+1; 2\nu+1;2\varepsilon r) \right\}.
      \end{eqnarray}
\end{subequations}
Here $T_1 $ and $T_2$ is the pending constant.

For \Eq{De12}, we designate
\begin{eqnarray}
    F_1' = f_2 -g_2,\;\;\;\;\; F_2' = f_2 +g_2.
\end{eqnarray}
Based on \Eq{de1'} and \Eq{de3'}, we can attain
\begin{subequations}
    \begin{eqnarray}
     \label{dd1'}   \frac{\partial F_1'}{\partial r}-\frac{2(\overline{k}-1)}{r}F_1' +\frac{l+2}{r} g_4
     -\left(\frac{\overline{\kappa}}{r}-\overline{E} + \overline{M} \right)F_2' =0
    \end{eqnarray}
    \begin{eqnarray}
        \label{dd2'}  \frac{\partial g_4}{\partial r}+ \frac{1}{r}g_4 +\frac{l+1}{r} F_1'
        -\left(\frac{\overline{\kappa}}{r}-\overline{E} + \overline{M} \right)f_4 =0.
    \end{eqnarray}
\end{subequations}
Based on \Eq{de2'} and \Eq{de4'}, we can attain
\begin{subequations}
    \begin{eqnarray}
        \label{dd3'}   \frac{\partial F_2'}{\partial r}+\frac{2\overline{k}}{r}F_2'
        -\frac{l+2}{r} f_4 +\left(\frac{\overline{\kappa}}{r}-\overline{E} - \overline{M} \right)F_1' =0,
    \end{eqnarray}
    \begin{eqnarray}
        \label{dd4'} \frac{\partial f_4}{\partial r}+ \frac{1}{r}f_4 -\frac{l+1}{r} F_2'
        +\left(\frac{\overline{\kappa}}{r}-\overline{E} - \overline{M} \right)g_4 =0.
    \end{eqnarray}
\end{subequations}

We consider the behavior of $r \rightarrow \infty$. For \Eq{dd2'} and \Eq{dd4'}, we acquire
\begin{eqnarray}
    \left\{\begin{matrix}
        \dfrac{\partial g_4}{\partial r} -(\overline{M}-\overline{E}) f_4 =0,\\
        \dfrac{\partial f_4}{\partial r} -(\overline{M}+\overline{E}) g_4 =0.
       \end{matrix}\right.,
\end{eqnarray}
If we designate $g_4 = D_1 e^{-\varepsilon r }$ and $f_4 = D_2 e^{-\varepsilon r }$ and require the solution is a bound state,
we can  get $\overline{E} < \overline{M}$, i.e. $E<Mc^2$, and similarly
\begin{eqnarray}
    \varepsilon= \sqrt{\overline{M}^2 -\overline{E}^2}, \;\;\;\;\;  f_4 =-\mu g_4.
\end{eqnarray}
For \Eq{dd1'} and \Eq{dd3'}, we acquire
\begin{eqnarray}
    \left\{\begin{matrix}
        \dfrac{\partial F_1'}{\partial r} -(\overline{M}-\overline{E}) F_2' =0,\\
        \dfrac{\partial F_2'}{\partial r} -(\overline{M}+\overline{E}) F_1' =0.
       \end{matrix}\right.,
\end{eqnarray}
If we designate $F_1' = D_1 e^{-\varepsilon r }$ and $F_2' = D_2 e^{-\varepsilon r }$ and require the solution is a bound state,
we can also get $\overline{E} < \overline{M}$, i.e. $E<Mc^2$, and
\begin{eqnarray}
    \varepsilon = \sqrt{\overline{M}^2 -\overline{E}^2},\;\;\;\;\; F_2' =-\mu F_1'.
\end{eqnarray}

We continue to consider the behavior of $r \rightarrow 0$. We have
\begin{subequations}
    \begin{eqnarray}
         \frac{\partial F_1'}{\partial r}- \frac{2(\overline{k}-1)}{r}F_1' +\frac{l+2}{r} g_4 -\frac{\overline{\kappa}}{r}F_2' =0,
       \end{eqnarray}
       \begin{eqnarray}
           \frac{\partial F_2'}{\partial r}+ \frac{2 \overline{k}}{r}F_2' -\frac{l+2}{r} f_4 +\frac{\overline{\kappa}}{r}F_1' =0,
    \end{eqnarray}
    \begin{eqnarray}
        \frac{\partial f_4}{\partial r}+ \frac{1}{r}f_4 -\frac{l+1}{r} F_2' +\frac{\overline{\kappa}}{r}g_4 =0,
    \end{eqnarray}
       \begin{eqnarray}
            \frac{\partial g_4}{\partial r}+ \frac{1}{r}g_4 +\frac{l+1}{r} F_1' -\frac{\overline{\kappa}}{r}f_4 =0.
       \end{eqnarray}
\end{subequations}
If we designate $F_1'= d_1' r^{\nu'-1}$, $F_2'= d_2' r^{\nu'-1}$, $f_4= d_3' r^{\nu'-1}$, $g_4= d_4' r^{\nu'-1}$,
then
\begin{eqnarray}
    \left\{\begin{matrix}
        (\nu'-2\overline{k}+1)d_1'+ (l+2)d_4'-\overline{\kappa} \,d_2' =0, \\
       \overline{\kappa} \, d_1' -(l+2)d_3' +(\nu'+2\overline{k}-1) d_2' =0, \\
       \nu'd_3' + \overline{\kappa} d_4' -(l+1)d_2'=0, \\
       (l+1)d_1' -\overline{\kappa} d_3' +\nu' d_4' =0.
       \end{matrix}\right.
\end{eqnarray}
We can get the solution of $\nu'$,
\begin{eqnarray}
&&  \nu' =\pm \sqrt{\left(\frac{2 \overline{k}-1 +\sqrt{(2 \overline{k}-1)^2 +4(l+1)(l+2)}}{2}\right)^2 -\overline{\kappa}^2}, \nonumber\\
&& {\rm or }\;\; \nu' =\pm \sqrt{\left(\frac{2\overline{k}-1 -\sqrt{(2\overline{k}-1)^2 +4(l+1)(l+2)}}{2}\right)^2 -\overline{\kappa}^2}.
\end{eqnarray}
Similarly, since $\nu'$ reduces to $\nu' = \sqrt{ (l+1)^2 - \overline{\kappa}^2}$ when $\overline{k}=0$ (or $k$ tends to 0),
and $\nu'$ should satisfy $\nu'>1$,
we then choose
\begin{eqnarray}
      \nu' =\nu=\sqrt{\left(\frac{2\overline{k}-1 +\sqrt{(2\overline{k}-1)^2 +4(l+1)(l+2)}}{2}\right)^2 -\overline{\kappa}^2}.
\end{eqnarray}

Finally, we designate
\begin{eqnarray}
 &&   F_1' =-\frac{1}{\mu} h_1' e^{-\varepsilon r } r^{\nu-1},\;\;\;\; F_2' = h_2' e^{-\varepsilon r } r^{\nu-1}, \nonumber\\
 &&   f_3' =h_3' e^{-\varepsilon r } r^{\nu-1},\;\;\;\; g_3' =-\frac{1}{\mu} h_4' e^{-\varepsilon r } r^{\nu-1},
\end{eqnarray}
where $h'_j(r)$ ($j=1, 2, 3, 4$) are functions of $r$. Then we obtain
\begin{subequations}
     \begin{eqnarray}
        \frac{\partial h_1'}{\partial r}+ \left(\frac{-2\overline{k}+\nu+1}{r}-\varepsilon\right)h_1'
        +\frac{l+2}{r} h_4' +\left(\frac{\overline{\kappa}\mu}{r}+\varepsilon\right)h_2' =0,
      \end{eqnarray}
      \begin{eqnarray}
        \frac{\partial h_2'}{\partial r}+ \left(\frac{2\overline{k}+\nu-1}{r}-\varepsilon\right)h_2'
        -\frac{l+2}{r} h_3' -\left(\frac{\overline{\kappa}}{\mu r}-\varepsilon\right)h_1' =0,
   \end{eqnarray}
   \begin{eqnarray}
    \frac{\partial h_3'}{\partial r}+ \left(\frac{\nu}{r}-\varepsilon\right)h_3' -\frac{l+1}{r} h_2'
    -\left(\frac{\overline{\kappa}}{\mu r}-\varepsilon\right)h_4' =0,
   \end{eqnarray}
      \begin{eqnarray}
        \frac{\partial h_4'}{\partial r}+ \left(\frac{\nu}{r}-\varepsilon\right)h_4'
        +\frac{l+1}{r} h_1 +\left(\frac{\overline{\kappa}\mu}{ r}+\varepsilon\right)h_3' =0,
      \end{eqnarray}
\end{subequations}

We continue to designate
\begin{eqnarray}
    h'_1 +h'_2 =A', \;\;\; h'_1 -h'_2 =B', \;\;\; h'_3+h'_4 =C', \;\;\; h'_3-h'_4=D', \;\;\;
     \overline{p}=\frac{\overline{\kappa}}{2}\left(\mu +\frac{1}{\mu}\right), \;\;\;  \overline{q}=\frac{\overline{\kappa}}{2}\left(\mu -\frac{1}{\mu}\right),
\end{eqnarray}
and then
\begin{subequations}
    \begin{eqnarray}
     \label{f1'''}   \frac{\partial A' }{\partial r}+ \frac{\nu+\overline{q}}{r}A' -\frac{2 \overline{k}-1+\overline{p}}{r} B' -\frac{l+2}{r}D' =0,
      \end{eqnarray}
      \begin{eqnarray}
        \label{f2'''}    \frac{\partial B'}{\partial r}+ \left(\frac{\nu-\overline{q}}{r}-2\varepsilon\right)B' +\frac{l+2}{r} C' -\frac{2 \overline{k}-1-\overline{p}}{r}A' =0,
   \end{eqnarray}
   \begin{eqnarray}
      \label{f3'''}  \frac{\partial C'}{\partial r}+ \frac{\nu+\overline{q}}{r}C' +\frac{l+1}{r} B' +\frac{\overline{p}}{r}D' =0,
   \end{eqnarray}
      \begin{eqnarray}
            \label{f4'''}     \frac{\partial D'}{\partial r}+ \left(\frac{\nu-\overline{q}}{r}-2\varepsilon \right)D' -\frac{l+1}{r} A' -\frac{\overline{p}}{r}C' =0.
      \end{eqnarray}
\end{subequations}
If we assume $A'= C_1' C'$ and $B'= C_2' D'$, it should satisfy
\begin{subequations}
    \begin{eqnarray}
    (l+1)C_1'C_2' +\overline{p}\, C_1'+(2\overline{k}-1+\overline{p})C_2' +l+2=0,
    \end{eqnarray}
    \begin{eqnarray}
        (l+1)C_1'C_2'-(2\overline{k}-1-\overline{p})C_1' + \overline{p}\, C_2'+l+2=0.
    \end{eqnarray}
\end{subequations}
We can attain
\begin{eqnarray}
    C_1' =-C_2',
\end{eqnarray}
and
\begin{eqnarray}
     (l+1)C_1'^2 -(2\overline{k}-1)C_1' -(l+2)=0.
\end{eqnarray}
We can attain
\begin{eqnarray}
      C_1' = \frac{(2\overline{k}-1) \pm \sqrt{ (2\overline{k}-1)^2 +4(l+1)(l+2)}}{2(l+1)}.
\end{eqnarray}
So if we designate
\begin{eqnarray}
&& C'= c'_0 +c'_1 r +\dots + c'_n r^n +\dots, \nonumber\\
&& D'= d'_0 +d'_1 r +\dots + d'_n r^n +\dots,
\end{eqnarray}
based on \Eq{f3'''} and \Eq{f4'''}, we have
\begin{eqnarray}
 (n+1+\nu+\overline{q}) c'_{n+1} +\left\{-(l+2)C'_1 + \overline{p}  \right\}d'_{n+1}=0
\end{eqnarray}
and
\begin{eqnarray}
    (n+1+\nu-\overline{q})d'_{n+1}-2ad'_n-[(l+1)C'_1 + \overline{p}]c'_{n+1}=0.
\end{eqnarray}
As a result,
\begin{eqnarray}
    d'_{n+1} =2\varepsilon\, \frac{n+1+\nu+\overline{q}}{(n+1+\nu)^2+ \overline{\kappa}^2-(l+1)^2 {C_1'}^2} d'_n.
\end{eqnarray}
Because when $\overline{k}=0$,
\begin{eqnarray}
     { (n+1+\nu)^2+\overline{\kappa}^2-(l+1)^2 {C_1'}^2 = (n+1)(n+2\nu+1),}
\end{eqnarray}
$C_1 $ should be
\begin{eqnarray}
      C_1' = 1.
\end{eqnarray}
So we choose
\begin{eqnarray}
      C_1' = \frac{(2\overline{k}-1) +\sqrt{ (2\overline{k}-1)^2 +4(l+1)(l+2)}}{2(l+1)}.
\end{eqnarray}
It causes
\begin{eqnarray}
      d'_{n+1} =2\varepsilon\, \frac{n+1+\nu+\overline{q}}{(n+1+\nu)^2+k'^2-(l+2)^2 {C'_1}^2} d'_n
      = 2\varepsilon\, \frac{n+1+\nu+\overline{q}}{(n+1+2\nu)(n+1)} d'_n.
\end{eqnarray}
In addition, we have
\begin{eqnarray}
     (n+1+\nu+\overline{q}) c'_{n+1} &=& 2\varepsilon\, \frac{n+1+\nu+\overline{q}}{(n+1+2\nu)(n+1)} (n+\nu+\overline{q}) c'_n,
\end{eqnarray}
i.e.,
\begin{eqnarray}
    c'_{n+1} &=& 2\varepsilon\, \frac{n+\nu+\overline{q}}{(n+1+2\nu)(n+1)} c'_n.
\end{eqnarray}
As a result,
\begin{eqnarray}
      &&A'= T_1' C_1' F(\nu+\overline{q}; 2\nu+1;2\varepsilon r),\nonumber\\
      &&B= -C_1' T_2' F(\nu+\overline{q}+1; 2\nu+1;2\varepsilon r), \nonumber\\
      &&C'= T_1' F(\nu+\overline{q}; 2\nu+1;2\varepsilon r), \nonumber\\
      && D'= T_2' F(\nu+\overline{q}+1; 2\nu+1;2\varepsilon r),
\end{eqnarray}
and
\begin{subequations}
      \begin{eqnarray}
            f_2 &=& \frac{1}{4} \left\{\left(-\frac{1}{\mu } +1\right)A'
            +\left(-\frac{1}{\mu } -1\right)B'\right\} e^{-\varepsilon r } r^{\nu-1}\notag\\
&=& -\frac{1}{4} e^{-\varepsilon r } r^{\nu-1} \left\{\left(\frac{1}{\mu } -1\right)T_1'C_1' F(\nu+\overline{q}; 2\nu+1;2\varepsilon r)
+\left(\frac{1}{\mu } +1\right)C_1' T_2' F(\nu+\overline{q}+1; 2\nu+1;2\varepsilon r) \right\},
      \end{eqnarray}
      \begin{eqnarray}
            g_2 &=& \frac{1}{4} \left\{\left(\frac{1}{\mu } +1\right)A' -\left(-\frac{1}{\mu } +1\right)B'\right\} e^{- \varepsilon r } r^{\nu-1}\notag\\
&=& \frac{1}{4} e^{-\varepsilon r } r^{\nu-1} \left\{\left(\frac{1}{\mu } +1\right)T_1'C_1' F(\nu+\overline{q}; 2\nu+1;2\varepsilon r)
-\left(\frac{1}{\mu } -1\right)T_2' C_1' F(\nu+\overline{q}+1; 2\nu+1;2\varepsilon r) \right\},
      \end{eqnarray}
      \begin{eqnarray}
            f_4 =\dfrac{1}{2} \left\{C'+D'\right\} e^{-\varepsilon r } r^{\nu-1}
=&\dfrac{1}{2} e^{-\varepsilon r } r^{\nu-1} \left\{T_1' F(\nu+\overline{q}; 2\nu+1;2\varepsilon r)
+ T_2' F(\nu+\overline{q}+1; 2\nu+1;2\varepsilon r) \right\},
      \end{eqnarray}
      \begin{eqnarray}
            g_4 =-\dfrac{1}{2\mu } \left\{C'-D'\right\} e^{-\varepsilon r } r^{\nu-1}
=&-\dfrac{1}{2\mu } e^{-\varepsilon r } r^{\nu-1} \left\{T_1' F(\nu+\overline{q}; 2\nu+1;2 \varepsilon r)
-T_2' F(\nu+\overline{q}+1; 2\nu+1;2\varepsilon r) \right\},
      \end{eqnarray}
\end{subequations}
where $T_1' $ and $T_2'$ is the pending constant.

When
\begin{eqnarray}
&& \nu+\overline{q}=-N, \;\;\; (N=0, 1, 2, \dots),
\end{eqnarray}
$A\,,\,B\,,\,C\,,\,D$ and $A'\,,\,B'\,,\,C'\,,\,D'$ become  polynomials with respect to $r$.
Based on
\begin{eqnarray}
&& N+\nu+\overline{q}=0, \nonumber\\
&& \overline{q}=\frac{\overline{\kappa}}{2}\left(\mu -\frac{1}{\mu}\right), \nonumber\\
&&  \mu= \sqrt{\frac{M c^2+E}{M c^2-E}},
\end{eqnarray}
one can obtain the energy spectrum of the bound state as
\begin{eqnarray}
    E=\frac{M c^2}{\sqrt{1 + \dfrac{\overline{\kappa}^2}{(N+{\nu})^2}}}, \;\;\;\;\; (N=0, 1, 2, \dots),
\end{eqnarray}
i.e.,
\begin{eqnarray}\label{eq:ernery-2}
    E=\frac{M c^2}{\sqrt{1 + \dfrac{q^2 \kappa^2}{\hbar^2 c^2 }\dfrac{1}{(N+{\nu})^2}}}, \;\;\;\;\; (N=0, 1, 2, \dots),
\end{eqnarray}
with
\begin{eqnarray}\label{eq:s-1}
\nu=\sqrt{\left(\frac{2\overline{k}-1 +\sqrt{(2\overline{k}-1)^2 +4(l+1)(l+2)}}{2}\right)^2 -\overline{\kappa}^2}.
\end{eqnarray}
When $k$ tends to 0 (i.e., there is no spin vector potential $\vec{\mathcal{A}}$), the parameter $\nu$ reduces to $\nu = \sqrt{ (l+1)^2 - \overline{\kappa}^2}$. In this case, Eq. (\ref{eq:ernery-2}) naturally reduces to Eq. (\ref{eq:hyd-3a}), i.e., the energy of the usual relativistic hydrogen-like atom.

\begin{remark}\textcolor{blue}{The Relation between the Fine Structure Constant $\alpha_{\rm e}$ and the Nuclear Charge Number
$Z$ Modified by the Spin Vector Potential.}
Due to
\begin{eqnarray}
    \frac{q k}{2c}=\overline{k},\;\;\;\; \frac{q \kappa }{\hbar c }= \overline{\kappa},
\end{eqnarray}
From Eq. (\ref{eq:s-1}) we can have
\begin{eqnarray}\label{eq:s-2}
\nu &=& \sqrt{\left(\frac{\dfrac{q k}{c}-1 +\sqrt{\left(\dfrac{q k}{c}-1\right)^2 +4(l+1)(l+2)}}{2}\right)^2 -\left(\frac{q \kappa }{\hbar c }\right)^2}.
\end{eqnarray}
As a square-root, one should have
\begin{eqnarray}
\left(\frac{\dfrac{q k}{c}-1 +\sqrt{\left(\dfrac{q k}{c}-1\right)^2 +4(l+1)(l+2)}}{2}\right)^2 -\left(\frac{q \kappa }{\hbar c }\right)^2 \geq 0,
\end{eqnarray}
for any quantum numbers $l=0, 1, 2, \dots$. By considering the minimum $l_{\rm min}=0$, one has
\begin{eqnarray}
\left(\frac{\dfrac{q k}{c}-1 +\sqrt{\left(\dfrac{q k}{c}-1\right)^2 +8}}{2}\right)^2 -\left(\frac{q \kappa }{\hbar c }\right)^2 \geq 0.
\end{eqnarray}
By considering $q=-e$ and $\kappa=Q=Ze$, one has
\begin{eqnarray}
\left(\frac{-\dfrac{e k}{c}-1 +\sqrt{\left(\dfrac{e k}{c}+1\right)^2 +8}}{2}\right)^2 -\left(\frac{Z e^2 }{\hbar c }\right)^2 \geq 0.
\end{eqnarray}
Due to \begin{eqnarray}
\alpha_{\rm e} = \dfrac{e^2}{\hbar c} \approx \dfrac{1}{137.0359} \approx \dfrac{1}{137},
\end{eqnarray}
one has
\begin{eqnarray}\label{eq:Zk-1}
\left(\frac{-\dfrac{e k}{c}-1 +\sqrt{\left(\dfrac{e k}{c}+1\right)^2 +8}}{2}\right)^2 -\left(Z \alpha_{\rm e}\right)^2 \geq 0,
\end{eqnarray}
i.e.,
\begin{eqnarray}\label{eq:z-1}
Z \leq  \frac{\sqrt{\left(\dfrac{e k}{c}+1\right)^2 +8}-\left(\dfrac{e k}{c}+1\right)}{2}\, \dfrac{1}{\alpha_{\rm e}}.
\end{eqnarray}
This is the modified relation between the fine structure constant $\alpha_{\rm e}$ and $Z$ by considering the spin vector potential.

\emph{Discussion.---} For $k=0$ (or $k$ tends to 0), we recover the form result, i.e.,
\begin{eqnarray}
     &&  Z \leq  \dfrac{1}{\alpha_{\rm e}}\approx 137.0359.
\end{eqnarray}
As we have mentioned in the previous section, based on this finding, Richard Feynman asserted that the biggest number of
$Z$ in the periodic table of  elements is $Z=137$.

For $k> 0$. When $k$ increases, the right-hand-side of Eq. (\ref{eq:z-1}) decreases, i.e., $Z$ cannot reach 137 any more. This means
that the spin-dependent Coulomb interaction can further suppress the the biggest number of $Z$.
At present, the biggest number $Z$ listed in the periodic table of elements is $Z=118$ (i.e., the element ${\rm Oq}$). Why scientists cannot create
the element with $Z>118$ at present? A possible reason is due to the existence of the spin-dependent Coulomb interaction (with $k>0$).

Suppose $k>0$, let us focus on Eq. (\ref{eq:Zk-1}). For convenience, we denote $(e/c) k$ by $k_{\rm cr}$. By solving
\begin{eqnarray}\label{eq:Zk-2}
\left(\frac{-(k_{\rm cr}+1) +\sqrt{\left(k_{\rm cr}+1\right)^2 +8}}{2}\right)^2 -\left(Z \alpha_{\rm e}\right)^2 =0,
\end{eqnarray}
we obtain
\begin{eqnarray}\label{eq:Zk-3}
&&  k_{\rm cr}= \dfrac{1-Z \alpha_{\rm e} -Z^2\, \alpha_{\rm e}^2}{Z \alpha_{\rm e}}.
\end{eqnarray}
In Table \ref{tab:Zk1}, we have listed the relation between $Z$ and $k_{\rm cr}$. For a fixed $Z$ (e.g., $Z=100$), if $(e/c) k$ exceeds $k_{\rm cr}$, (e.g., $k_{\rm cr}=1.0110$), then the atom system will become unstable.

        \begin{table}[h]
	\centering
\caption{The relation between $Z$ and $k_{\rm cr}$.}
\begin{tabular}{llllllllllllll}
\hline\hline
Z & 1 &  2 & 3 & 4 & 5 & 6& 7& 8 & 9 & 10 &20 &40 & 60 \\
  \hline
$k_{\rm cr}$ & 273.0645 &  136.0213 & 90.3354 & 67.4888 & 53.7779 & 44.6348 & 38.1020 & 33.2006 & 29.3867 & 26.3342 & 12.5576 & 5.5599 & 3.1300\\
 \hline
Z & 100 &  102 & 104 & 106 & 108 & 110& 115 & 118 & 120 & 125 &130 & 135 & 137 \\
  \hline
$k_{\rm cr}$ & 1.0110 &  0.9426 & 0.8764 & 0.8121 & 0.7496 & 0.6889& 0.5440& 0.4616 & 0.4082 & 0.2804 &0.1596 &0.0450 & 0.0008 \\
 \hline\hline
\end{tabular}\label{tab:Zk1}
\end{table}

$\blacksquare$
\end{remark}

\newpage

\section{Discussion}

\subsection{From Dirac's Equation to Pauli's Equation and Schr{\" o}dinger's Equation}

Dirac's equation is a fundamental equation in relativistic quantum mechanics. In the non-relativistic limit, it can reduce firstly to Pauli's equation and then to Schr{\" o}dinger's equation. In the following, we make a brief review for the procedure of such reductions.

Dirac's Hamiltonian with vector potential $\vec{\mathcal{A}}$ and scalar potential $\varphi$ is given by
\begin{equation}
            \mathcal{H}_{\rm Dirac} =c\,\vec{\alpha} \cdot \vec{\Pi} +\beta  M c^2 + q \varphi,
\end{equation}
where $M$ is the mass of particle, $q$ is the charge of the particle, and
\begin{eqnarray}
\vec{\Pi}=\vec{p}-\frac{q}{c}\vec{\mathcal{A}}
\end{eqnarray}
is the canonical momentum. Then Dirac's equation reads
      \begin{equation}\label{eq:DiracEqE}
            {\rm i}\hbar\p{}{t} \Psi=\mathcal{H}_{\rm Dirac} \Psi,
      \end{equation}
where $\Psi$ is the time-dependent wavefunction. Next we study the Dirac equation in the representation of
      \begin{equation}
            \Psi=\begin{bmatrix}
                  \tilde{\eta} \\ \tilde{\chi}
            \end{bmatrix},
      \end{equation}
      then \Eq{eq:DiracEqE} becomes
      \begin{equation}
            {\rm i}\hbar\p{}{t} \begin{bmatrix}
                        \tilde{\eta} \\ \tilde{\chi}
                  \end{bmatrix}
            =c\,\begin{bmatrix}
                              0 & \vec{\sigma} \\
                              \vec{\sigma} & 0
                        \end{bmatrix}\cdot\vec{\Pi}\begin{bmatrix}
                                    \tilde{\eta} \\ \tilde{\chi}
                              \end{bmatrix}
                  +\begin{bmatrix}
                              \openone & 0 \\
                              0 & -\openone
                        \end{bmatrix}\,M c^2 \begin{bmatrix}
                              \tilde{\eta} \\ \tilde{\chi}
                        \end{bmatrix}
                  +q\,\varphi\begin{bmatrix}
                              \tilde{\eta} \\ \tilde{\chi}
                        \end{bmatrix},
      \end{equation}
      i.e.,
      \begin{equation}\label{eq:DiracEExpand-0}
            {\rm i}\hbar\p{}{t} \begin{bmatrix}
                        \tilde{\eta} \\ \tilde{\chi}
                  \end{bmatrix}
            =c(\vec{\sigma}\cdot\vec{\Pi})\begin{bmatrix}
                              \tilde{\chi} \\ \tilde{\eta}
                        \end{bmatrix}
                  +Mc^2 \,\openone\begin{bmatrix}
                              \tilde{\eta} \\ -\tilde{\chi}
                        \end{bmatrix}
                  +q\,\varphi\begin{bmatrix}
                              \tilde{\eta} \\ \tilde{\chi}
                        \end{bmatrix}.
      \end{equation}

      Now, we consider the non-relativistic approximation of Dirac's equation. In the case of low energy,
      the term involving $M c^2$ is far larger than other ones. Then a part of time factor can be separated as follows.
      \begin{equation}
            \begin{bmatrix}
                  \tilde{\eta} \\ \tilde{\chi}
            \end{bmatrix}=\begin{bmatrix}
                  \eta \\ \chi
            \end{bmatrix}{\rm e}^{-{\rm i}\frac{Mc^2}{\hbar} t},
      \end{equation}
      thus \Eq{eq:DiracEExpand-0} is transformed as
      \begin{equation}
            {\rm i}\hbar\Biggl\{\p{}{t} \begin{bmatrix}
                        \eta \\ \chi
                  \end{bmatrix}\Biggr\}{\rm e}^{-{\rm i}\frac{Mc^2}{\hbar} t}
            +{\rm i}\begin{bmatrix}
                        \eta \\ \chi
                  \end{bmatrix}\hbar\Bigl(\p{}{t} {\rm e}^{-{\rm i}\frac{Mc^2}{
                        \hbar} t}\Bigr)
            =c(\vec{\sigma}\cdot\vec{\Pi})\begin{bmatrix}
                              \chi \\ \eta
                        \end{bmatrix}{\rm e}^{-{\rm i}\frac{Mc^2}{\hbar} t}
                  +M c^2 \openone\begin{bmatrix}
                              \eta \\ -\chi
                        \end{bmatrix}{\rm e}^{-{\rm i}\frac{M c^2}{\hbar} t}
                  +q\,\varphi\begin{bmatrix}
                              \eta \\ \chi
                        \end{bmatrix}{\rm e}^{-{\rm i}\frac{Mc^2}{\hbar} t},
      \end{equation}
      i.e.,
      \begin{equation}
            {\rm i}\hbar\p{}{t} \begin{bmatrix}
                        \eta \\ \chi
                  \end{bmatrix}
            +Mc^2 \openone\begin{bmatrix}
                        \eta \\ \chi
                  \end{bmatrix}
            =c(\vec{\sigma}\cdot\vec{\Pi})\begin{bmatrix}
                              \chi \\ \eta
                        \end{bmatrix}
                  +Mc^2 \openone\begin{bmatrix}
                              \eta \\ -\chi
                        \end{bmatrix}
                  +q\,\varphi\begin{bmatrix}
                              \eta \\ \chi
                        \end{bmatrix},
      \end{equation}
      i.e.,
      \begin{equation}\label{eq:DiracELowE-0}
            {\rm i}\hbar\p{}{t} \begin{bmatrix}
                        \eta \\ \chi
                  \end{bmatrix}
            =c(\vec{\sigma}\cdot\vec{\Pi})\begin{bmatrix}
                              \chi \\ \eta
                        \end{bmatrix}
                  +Mc^2 \openone\begin{bmatrix}
                              0 \\ -2\,\chi
                        \end{bmatrix}
                  +q\,\varphi\begin{bmatrix}
                              \eta \\ \chi
                        \end{bmatrix}.
      \end{equation}
      For the second row of \Eq{eq:DiracELowE-0}, we have
      \begin{equation}\label{eq:chi-1}
            {\rm i}\hbar\p{\chi}{t} =c(\vec{\sigma}\cdot\vec{\Pi})\eta
                  -2\,Mc^2 \openone\chi+q\,\varphi\,\chi.
      \end{equation}
      Since
      \begin{eqnarray}
        && Mc^2 \,\chi\gg   {\rm i}\hbar\p{\chi}{t}, \nonumber\\
        && Mc^2 \,\chi\gg q\,\phi\,\chi,
      \end{eqnarray}
      then from Eq. (\ref{eq:chi-1}) we arrive at
      \begin{equation}
            c(\vec{\sigma}\cdot\vec{\Pi})\eta=2\,Mc^2 \openone\chi,
      \end{equation}
      which means
      \begin{equation}\label{eq:chieta}
            \chi=\dfrac{c(\vec{\sigma}\cdot\vec{\Pi})}{2\,Mc^2} \eta,
      \end{equation}
      i.e., the wavefunction $\eta$ can be connected to the wavefunction $\chi$ through the relation (\ref{eq:chieta}).

      After that, for the first row, we obtain
      \begin{equation}
            {\rm i}\hbar\p{\eta}{t} =c(\vec{\sigma}\cdot\vec{\Pi})\chi
                  +q\,\varphi\,\eta,
      \end{equation}
      i.e.,
      \begin{equation}\label{eq:DS-1}
            {\rm i}\hbar\p{\eta}{t} =\dfrac{1}{2M} (\vec{\sigma}\cdot\vec{\Pi})(
                        \vec{\sigma}\cdot\vec{\Pi})\eta
                  +q\,\varphi\,\eta.
      \end{equation}
     Because
\begin{equation}
    (\vec{\sigma} \cdot \vec{\Pi})(\vec{\sigma} \cdot \vec{\Pi}) = \vec{\Pi}^2 + {\rm i} \vec{\sigma} \cdot (\vec{\Pi} \times \vec{\Pi}),
\end{equation}
\begin{eqnarray}
    \vec{\Pi} \times \vec{\Pi} &= &\left[\left(\vec{p}-\frac{q}{c} \vec{\mathcal{A}}\right)\times \left(\vec{p}-\frac{q}{c} \vec{\mathcal{A}}\right)\right] =\frac{q}{c}\left[{-}\vec{p} \times \vec{\mathcal{A}} - \vec{\mathcal{A}}\times\vec{p} \right] +\frac{q^2}{c^2} \vec{\mathcal{A}} \times \vec{\mathcal{A}}\notag\\
    &=&{+}{\rm i}\hbar \frac{q}{c} \vec{\nabla} \times \vec{\mathcal{A}} +\frac{q^2}{c^2} \vec{\mathcal{A}}\times \vec{\mathcal{A}},
\end{eqnarray}
then for the spin vector potential given in Eq. (\ref{eq:A1}), i.e.,
\begin{equation}
      \vec{\mathcal{A}} = k \frac{\vec{r} \times \vec{S}}{r^2},
\end{equation}
one can have
\begin{eqnarray}\label{eq:M-5-02}
&& \nabla\times\vec{\mathcal{A}} =  -2 k\,\dfrac{\vec{r}\cdot\vec{S} }{r^4} \vec{r}, \;\;\;\;\;\;
 \vec{\mathcal{A}} \times \vec{\mathcal{A}}= {\rm i} \hbar \; k^2 \; \frac{\vec{r}\cdot\vec{S}}{r^4} \; \vec{r},\nonumber\\
&&  {\rm i}\left(\vec{\Pi} \times \vec{\Pi}\right)=
{\rm i}\left[{+}{\rm i}\hbar \frac{q}{c} \vec{\nabla} \times \vec{\mathcal{A}} +\frac{q^2}{c^2} \vec{\mathcal{A}}\times \vec{\mathcal{A}}\right]
= {\rm i}\left\{{+}{\rm i}\hbar \frac{q}{c} \left[ -2 k\,\dfrac{\vec{r}\cdot\vec{S} }{r^4} \vec{r}\right]
+\frac{q^2}{c^2} \left[{\rm i} \hbar \; k^2 \; \frac{\vec{r}\cdot\vec{S}}{r^4} \; \vec{r}\right]\right\}\nonumber\\
&&= {\rm i} \left\{{\rm i}\hbar \frac{qk}{c} \left[{-}2+\frac{qk}{c}\right]\frac{\vec{r}\cdot\vec{S}}{r^4} \; \vec{r}\right\}
=-\hbar \frac{qk}{c} \left[{-}2+\frac{qk}{c}\right]\frac{\vec{r}\cdot\vec{S}}{r^4} \; \vec{r}
=\hbar \frac{qk}{c} \left[2-\frac{qk}{c}\right]\frac{\vec{r}\cdot\vec{S}}{r^4} \; \vec{r},
\end{eqnarray}
and the ``magnetic'' filed
\begin{eqnarray}
      \vec{\mathcal{B}} &=& \vec{\nabla}\times\vec{\mathcal{A}}
                        -{\rm i}\,g\left(\vec{\mathcal{A}}\times\vec{\mathcal{A}}\right)= \left[ -2 k\,\dfrac{\left(\vec{r}\cdot\vec{S} \right)}{r^4} \vec{r}\right]
      -{\rm i}\,g \left[{\rm i} \hbar \; k^2 \; \frac{\vec{r}\cdot\vec{S}}{r^4} \; \vec{r}\right]= k (g \hbar k -2)\frac{\vec{r}\cdot\vec{S}}{r^4} \; \vec{r}.
\end{eqnarray}

In this work, we restrict to the case of $k\neq 0$. The constraint is given by {$g\hbar k=1, 2$}.  When $g\hbar k= 2$, we have the  ``magnetic'' filed $\vec{\mathcal{B}}=0$; while {$g\hbar k= 1$}, we have the  ``magnetic'' filed as
\begin{eqnarray}
      \vec{\mathcal{B}} &=& {- k \frac{\vec{r}\cdot\vec{S}}{r^4} \; \vec{r}}.
\end{eqnarray}
In this case, one has
\begin{eqnarray}
      \vec{\mathcal{B}} &=&  \dfrac{ (g \hbar k -2)}{\hbar \dfrac{q}{c} \left[2-\dfrac{qk}{c}\right]} \left[{\rm i}\left(\vec{\Pi} \times \vec{\Pi}\right)\right]=\dfrac{1}{\gamma}\, \left[{\rm i}\left(\vec{\Pi} \times \vec{\Pi}\right)\right],
\end{eqnarray}
with
\begin{eqnarray}
    \gamma= \dfrac{\hbar \dfrac{q}{c} \left[2-\dfrac{qk}{c}\right]}{ (g \hbar k -2)} .
\end{eqnarray}
Then Eq. (\ref{eq:DS-1}) becomes
\begin{equation}\label{eq:pauli-1a}
    {\rm i}\hbar\frac{\partial \eta}{\partial t}  =\left[\frac{\vec{\Pi}^2 + \gamma\, \vec{\sigma} \cdot \vec{\mathcal{B}}}{2M} +q \varphi\right] \eta,
\end{equation}
which is just Pauli's equation in the non-relativistic quantum mechanics. Based on Eq. (\ref{eq:pauli-1a}) one may extract the Pauli Hamiltonian as
      \begin{eqnarray}\label{eq:pauli-1b}
            \mathcal{H}_{\rm Pauli} &=&\left[\frac{\vec{\Pi}^2 + \gamma\, \vec{\sigma} \cdot \vec{\mathcal{B}}}{2M} +q \varphi\right]
            =\dfrac{1}{2M} \Bigl(\vec{p}
                        -\dfrac{q}{c} \vec{\mathcal{A}}\Bigr)^2
                  +\dfrac{\gamma}{2M} \vec{\sigma}\cdot\vec{\mathcal{B}}
                  +q\,\varphi.
      \end{eqnarray}
When $\gamma$ is very small and can be neglected, then Pauli's equation reduces further to Schr{\" o}dinger's equation in non-relativistic quantum mechanics, i.e., the corresponding  Schr{\" o}dinger Hamiltonian with spin potential is given by
\begin{eqnarray}
 \mathcal{H}_{\rm Schr} &=& \frac{\vec{\Pi}^2 }{2M} +q \varphi
            =\dfrac{1}{2M} \Bigl(\vec{p}
                        -\dfrac{q}{c} \vec{\mathcal{A}}\Bigr)^2  +q\,\varphi.
\end{eqnarray}

\begin{remark}\textcolor{blue}{The Tensor Interaction.} The relation between Pauli's Hamiltonian and Schr{\" o}dinger's Hamiltonian is
 \begin{eqnarray}\label{eq:pauli-1c}
            \mathcal{H}_{\rm Pauli} &=& \mathcal{H}_{\rm Schr}
                  +\boxed{\dfrac{\gamma}{2M} \vec{\sigma}\cdot\vec{\mathcal{B}}}.
      \end{eqnarray}
Namely, in comparison to Schr{\" o}dinger's Hamiltonian, there is an extra term in Pauli's Hamiltonian, which is
\begin{eqnarray}
 \mathcal{H}_{\rm extra} &=& \dfrac{\gamma}{2\,M} \vec{\sigma}\cdot\vec{\mathcal{B}},
\end{eqnarray}
or
\begin{eqnarray}
 \mathcal{H}_{\rm extra} &=& \dfrac{\gamma}{2\,M} \vec{\sigma}\cdot\vec{\mathcal{B}}=\dfrac{\dfrac{\hbar \dfrac{q}{c} \left[2-\dfrac{qk}{c}\right]}{ (g \hbar k -2)}}{2\,M} \vec{\sigma}\cdot \left[ k (g \hbar k -2)\frac{\vec{r}\cdot\vec{S}}{r^4} \; \vec{r}\right]\nonumber\\
 &=& \dfrac{-\hbar \dfrac{qk}{c} \left[{-}2+\dfrac{qk}{c}\right]}{2\,M} \vec{\sigma}\cdot \left[\frac{\vec{r}\cdot\vec{S}}{r^4} \; \vec{r}\right]
 = \dfrac{\hbar \tilde{k} \left(2-\tilde{k}\right)}{2\,M} \times \frac{(\vec{r}\cdot \vec{\sigma})(\vec{r}\cdot\vec{S})}{r^4}.
\end{eqnarray}
This extra term can be rewritten as
\begin{eqnarray}
 \mathcal{H}_{\rm extra} &=& V(r) \mathcal{T}_{12},
\end{eqnarray}
where
\begin{eqnarray}
  V(r)\propto \dfrac{\gamma}{2M}\dfrac{1}{r^2},
\end{eqnarray}
and
\begin{eqnarray}
\mathcal{T}_{12}=\dfrac{\left(\vec{r}\cdot\vec{\sigma}_1\right)\left(\vec{r}\cdot\vec{\sigma}_2\right)}{r^2}
\end{eqnarray}
represents a tensor interaction between two spins. Here the ``$\vec{\sigma}\;$'' in $\vec{\sigma}\cdot\vec{\mathcal{B}}$ is $\vec{\sigma}_1$
(representing the spin of the first particle,i.e., $\vec{S}_1=\dfrac{\hbar}{2} \vec{\sigma}$), and
\begin{eqnarray}
      \vec{\mathcal{B}} &=&  k (g \hbar k -2)\frac{\vec{r}\cdot\vec{S}_2}{r^4} \; \vec{r}
\end{eqnarray}
is the ``magnetic'' field induced by the spin of the second particle. The physical meaning of the tensor interaction $\mathcal{T}_{12}$ is
\emph{the dipole-dipole spin interaction} between two particles, which is very common in nuclear physics \cite{1999Flugge}. Based on the above derivation,
 one may find that the dipole-dipole spin interaction is a natural result of Yang-Mills equations. More explicitly, one can predict the
 spin-dependent Coulomb interaction based on Yang-Mills equations, by considering the non-relativistic approximation of Dirac's equation,
 from the spin-dependent Coulomb interaction one can naturally obtain the dipole-dipole spin interaction. $\blacksquare$
\end{remark}

\subsection{From Heisenberg's Equation to Spin Forces}

\subsubsection{Heisenberg's Equation}

Heisenberg's equation of motion is a fundamental equation in quantum mechanics, which is given by
\begin{eqnarray}\label{eq:HeisenbergEq-1}
\dfrac{{\rm d}\,\hat{\mathcal{O}}_H (t)}{{\rm d}\,t}
&=& \dfrac{1}{{\rm i}\hbar} \left(\Bigl[\hat{\mathcal{O}}(t),\ \mathcal{H}\Bigr]\right)_H +\left(\dfrac{\partial\,\hat{\mathcal{O}}}{\partial\,t}\right)_H,
\end{eqnarray}
where ``$\mathcal{H}$'' denotes the Hamiltonian of physical system, and ``$H$'' represents the Heisenberg picture.

\begin{remark}\textcolor{blue}{\emph{Derivation of Heisenberg's Equation.}} Let $\hat{\mathcal{O}}$ be a physical observable, which is a time-independent hermitian operator. The time-dependent operator $\hat{\mathcal{O}}_H (t)$ is defined through
\begin{eqnarray}
&&\hat{\mathcal{O}}_H (t)= U^\dag(t) \hat{\mathcal{O}} U (t),
\end{eqnarray}
where
\begin{eqnarray}\label{eq:U1}
&&U(t)={\rm e}^{\frac{-{\rm i}}{\hbar}\mathcal{H}t }
\end{eqnarray}
is a unitary evolution operator, and
\begin{eqnarray}
&&U^\dag(t)={\rm e}^{\frac{{\rm i}}{\hbar}\mathcal{H}t }, \;\;\;\; U(t)U^\dag(t)=U^\dag(t)   U(t)=\mathbb{I},
\end{eqnarray}
with $\mathbb{I}$ being the identity operator.

From Eq. (\ref{eq:U1}) one can have
\begin{eqnarray}\label{eq:U1a}
&& \frac{\partial U(t)}{\partial t}=\frac{1}{{\rm i}\hbar} \mathcal{H}U(t)=\frac{1}{{\rm i}\hbar} U(t)\mathcal{H}, \nonumber\\
&& \frac{\partial U^\dag(t)}{\partial t}=-\frac{1}{{\rm i}\hbar} \mathcal{H}U^\dag(t)=-\frac{1}{{\rm i}\hbar} U^\dag(t)\mathcal{H},
\end{eqnarray}
which yields


\begin{eqnarray}
            \dfrac{{\rm d}\,\hat{\mathcal{O}}_H (t)}{{\rm d}\,t}
            &=& \dfrac{{\rm d}\,\left[U^\dag(t) \hat{\mathcal{O}} U (t)\right]}{{\rm d}\,t} = \dfrac{\partial\,\hat{U}^\dagger}{\partial\,t} \hat{\mathcal{O}} U
                  +\hat{U}^\dagger \dfrac{\partial\,\hat{\mathcal{O}}}{\partial\,t} U
                  +\hat{U}^\dagger \hat{\mathcal{O}} \dfrac{\partial\,U}{\partial\,t} \nonumber\\
            &=&\ -\dfrac{1}{{\rm i}\hbar} \hat{U}^\dagger \mathcal{H}U\hat{U}^\dagger \hat     {\mathcal{O}}U
                  +\left(\dfrac{\partial\,\hat{\mathcal{O}}}{\partial\,t}\right)_H
                  +\dfrac{1}{{\rm i}\hbar} \hat{U}^\dagger \hat{\mathcal{O}}U\hat{U}^\dagger \mathcal{H}U \nonumber\\
            &=&\ -\dfrac{1}{{\rm i}\hbar}  \mathcal{H} \hat{U}^\dagger \hat     {\mathcal{O}}U
                  +\left(\dfrac{\partial\,\hat{\mathcal{O}}}{\partial\,t}\right)_H
                  +\dfrac{1}{{\rm i}\hbar} \hat{U}^\dagger \hat{\mathcal{O}}U \nonumber\\
            &=&\ -\dfrac{1}{{\rm i}\hbar} \left[ \mathcal{H} \hat{\mathcal{O}}(t)- \hat{\mathcal{O}}(t) \mathcal{H}\right]
                  +\left(\dfrac{\partial\,\hat{\mathcal{O}}}{\partial\,t}\right)_H
                   \nonumber\\
            &=&\ \dfrac{1}{{\rm i}\hbar} \left(\Bigl[\hat{\mathcal{O}}(t),\ H\Bigr]\right)_H +\left(\dfrac{\partial\,\hat{\mathcal{O}}}{\partial\,t}\right)_H.
\end{eqnarray}
Moreover, when $\hat{\mathcal{O}}$ does not apparently contain time $t$, one has
\begin{equation}
      \left(\dfrac{\partial\,\hat{\mathcal{O}}}{\partial\,t}\right)_H =0.
\end{equation}
This ends the derivation. $\blacksquare$
\end{remark}

\subsubsection{Spin Force: The Case of Non-Relativistic Quantum Mechanics}

Firstly, for simplicity let us consider the case of non-relativistic quantum mechanics. In such a case, the Hamiltonian $\mathcal{H}$ in Heisenberg' equation (\ref{eq:HeisenbergEq-1}) is chosen as the Schr{\" o}dinger-type Hamiltonian, which is given by
   \begin{equation}
     \label{Ham1}   \mathcal{H}=\dfrac{1}{2M} \vec{\Pi}^2 +q \varphi,
   \end{equation}
   with
   \begin{equation}
   \vec{\Pi}=\vec{p}-\frac{q}{c}\vec{\mathcal{A}}.
   \end{equation}
Here $M$ is the mass of the particle, $\vec{p}$ is the usual linear momentum operator, $\vec{\Pi}$ is the canonical momentum operator, $q$ is the ``charge'', $c$ is the speed of light in vacuum, $\vec{\mathcal{A}}$ is the vector potential, and $\varphi$ is the scalar potential.

We now select the physical observable $\hat{\mathcal{O}}$ as the position operator $\vec{r}$.  Based on Heisenberg's equation, we obtain
the ``velocity'' operator as
\begin{eqnarray}
  \vec{v}:&=&\frac{{\rm d}\,\vec{r}}{{\rm d}\,t} = \frac{1}{{\rm i}\,\hbar}\left[\vec{r}, \mathcal{H}\right] +\dfrac{\partial\,\vec{r}}{\partial\,t}= \frac{1}{{\rm i}\,\hbar}\left[\vec{r}, \mathcal{H}\right] \nonumber\\
  &=& \dfrac{1}{M} \left(\vec{p}-\frac{q}{c}\vec{\mathcal{A}}\right)
  \equiv \dfrac{1}{M} \;\vec{\Pi}.
\end{eqnarray}
\textcolor{black}{Here we have assumed that the vector potential $\vec{\mathcal{A}}$ does not depend on $\vec{p}$}, thus $\vec{r}$ and $\vec{\mathcal{A}}$ are mutually commutative.

Furthermore, we can have the ``acceleration'' operator as
\begin{eqnarray}
    \vec{a}:&=&\dfrac{{\rm d}\,\vec{v}}{{\rm d}\,t} =
                        \dfrac{1}{{\rm i}\,\hbar} \left[
                              \vec{v},\mathcal{H}\right]
                        +\dfrac{\partial\,\vec{v}}{\partial\,t}  \nonumber\\
                        &=&\dfrac{1}{2{\rm i}\,\hbar\,M^2} \left[
                            \vec{\Pi},\vec{\Pi}^2\right]
                      +\dfrac{1}{{\rm i}\,\hbar} \left[\vec{v},\ q\varphi\right]
                      +\dfrac{1}{M} \dfrac{\partial\,\vec{\Pi}}{\partial\,t}\nonumber\\
      &=&\dfrac{1}{2{\rm i}\,\hbar\,M^2} \left[
                            \vec{\Pi},\vec{\Pi}^2\right]
                      +\dfrac{q}{{\rm i}\,\hbar\, M} \left[\vec{\Pi},\ \varphi\right]
                      -\dfrac{q}{Mc} \dfrac{\partial\,\vec{\mathcal{A}}}{\partial\,t}\nonumber\\
       &=&\dfrac{1}{2{\rm i}\,\hbar\,M^2} \left[
                            \vec{\Pi},\vec{\Pi}^2\right]
                      -\dfrac{q}{M} (\vec{\nabla} \varphi) -\dfrac{q^2}{{\rm i}\,\hbar\, Mc} \left[\vec{\mathcal{A}},\ \varphi\right]
                      -\dfrac{q}{Mc} \dfrac{\partial\,\vec{\mathcal{A}}}{\partial\,t}.
                     \end{eqnarray}
Because
\begin{eqnarray}
    \left[{\Pi}_{z},\vec{\Pi}^2\right] &=&\left[\Pi_z,\ \Pi_x\right]\Pi_x +\Pi_x \left[\Pi_z,\ \Pi_x\right]+ \Pi_y \left[\Pi_z,\ \Pi_y\right]+\left[\Pi_z,\ \Pi_y\right] \Pi_y.
\end{eqnarray}
If we designate
\begin{eqnarray}\label{eq:B-1a}
   && [\Pi_{a},\Pi_{b}]={\rm i} \hbar \, \epsilon_{abc} G_c,
\end{eqnarray}
with $\epsilon_{abc}$ is the Levi-Civita symbol, or
\begin{eqnarray}
   && \vec{\Pi} \times \vec{\Pi} ={\rm i} \hbar \, \vec{G},
\end{eqnarray}
then we can get
\begin{equation}
    \left[\vec{\Pi}_{z},\vec{\Pi}^2\right] ={\rm i}\hbar \left[\left(\vec{\Pi}\times \vec{G}\right)_z-\left(\vec{G}\times \vec{\Pi}\right)_z\right],
\end{equation}
which means
\begin{eqnarray}
    \left[\vec{\Pi},\vec{\Pi}^2\right] &=&{\rm i}\hbar \left[\left(\vec{\Pi}\times \vec{G}\right)-\left(\vec{G}\times \vec{\Pi}\right)\right]={\rm i}\hbar\, {M} \left[\left(\vec{v}\times \vec{G}\right)-\left(\vec{G}\times \vec{v}\right)\right].
\end{eqnarray}
Therefore we have
\begin{eqnarray}
    \vec{a} &=& \dfrac{1}{M} \dfrac{1}{2}\left(\vec{v}\times \vec{G}-\vec{G}\times \vec{v}\right)+\dfrac{q}{M}\biggl[-\frac{1}{c}\dfrac{\partial\,\vec{\mathcal{A}}}{\partial\,t}
          -\vec{\nabla}\,\varphi
          -\dfrac{q}{c}\dfrac{{\rm i}}{\hbar} \left[\varphi, \vec{\mathcal{A}}\right]\biggr].
\end{eqnarray}

Because
\begin{eqnarray}\label{eq:Coul-1-1f}
    && \vec{\mathcal{A}} = k \frac{\vec{r} \times \vec{S}}{r^2},\;\;\;\;\; \varphi =  \dfrac{\kappa}{r}.
\end{eqnarray}
\begin{eqnarray}
      \vec{\mathcal{B}} &=&\dfrac{1}{\gamma}\, \left[{\rm i}\left(\vec{\Pi} \times \vec{\Pi}\right)\right]
      =\dfrac{{\rm i}}{\gamma}\, {\rm i}\hbar \vec{G}=-\dfrac{\hbar}{\gamma}\, \vec{G}, \;\;\;\;\;
      \gamma= \dfrac{\hbar \dfrac{q}{c} \left[2-\dfrac{qk}{c}\right]}{ (g \hbar k -2)}.
\end{eqnarray}
Then we have
\begin{eqnarray}
    \vec{a} &=& \dfrac{1}{M} \dfrac{1}{2}\left(\vec{v}\times \vec{G}-\vec{G}\times \vec{v}\right)+\dfrac{q}{M}\biggl[-\frac{1}{c}\dfrac{\partial\,\vec{\mathcal{A}}}{\partial\,t}
          -\vec{\nabla}\,\varphi
          -\dfrac{q}{c}\dfrac{{ i}}{\hbar} \left[\varphi, \vec{\mathcal{A}}\right]\biggr]\nonumber\\
    &=& -\dfrac{1}{M} \dfrac{1}{2} \dfrac{\gamma}{\hbar}\left(\vec{v}\times \vec{\mathcal{B}}-\vec{\mathcal{B}}\times \vec{v}\right)+\dfrac{q}{M}\biggl[-\frac{1}{c}\dfrac{\partial\,\vec{\mathcal{A}}}{\partial\,t}
          -\vec{\nabla}\,\varphi
          -\dfrac{q}{c}\dfrac{{ i}}{\hbar} \left[\varphi, \vec{\mathcal{A}}\right]\biggr]
\end{eqnarray}
and the spin force as
\begin{eqnarray}
   \vec{F}=M\vec{a}=- \dfrac{1}{2} \dfrac{\gamma}{\hbar}\left(\vec{v}\times \vec{\mathcal{B}}-\vec{\mathcal{B}}\times \vec{v}\right)+q\biggl[-\frac{1}{c}\dfrac{\partial\,\vec{\mathcal{A}}}{\partial\,t}
          -\vec{\nabla}\,\varphi
          -\dfrac{q}{c}\dfrac{{ i}}{\hbar} \left[\varphi, \vec{\mathcal{A}}\right]\biggr].
\end{eqnarray}
Due to the ``electric'' field
\begin{eqnarray}
\vec{\mathcal{E}} &=& -\dfrac{1}{c}
\dfrac{\partial\,\vec{\mathcal{A}}}{\partial\,t}
-\vec{\nabla}\varphi-{\rm i}\,g\left[\varphi,\ \vec{\mathcal{A}}\right],
\end{eqnarray}
one has
\begin{eqnarray}
   \vec{F}&=&- \dfrac{1}{2} \dfrac{\gamma}{\hbar} \left(\vec{v}\times \vec{\mathcal{B}}-\vec{\mathcal{B}}\times \vec{v}\right)+q \biggl[-\frac{1}{c}\dfrac{\partial\,\vec{\mathcal{A}}}{\partial\,t}
          -\vec{\nabla}\,\varphi -{\rm i}\,g\left[\varphi,\ \vec{\mathcal{A}}\right]\biggr]+
        q \biggl[{\rm i}\,g\left[\varphi,\ \vec{\mathcal{A}}\right] -\dfrac{q}{c}\dfrac{{ i}}{\hbar} \left[\varphi, \vec{\mathcal{A}}\right]\biggr]\nonumber\\
   &=&- \dfrac{1}{2} \dfrac{\gamma}{\hbar}\left(\vec{v}\times \vec{\mathcal{B}}-\vec{\mathcal{B}}\times \vec{v}\right)+q \vec{\mathcal{E}}
   + q \biggl[{\rm i}\,g\left[\varphi,\ \vec{\mathcal{A}}\right] -\dfrac{q}{c}\dfrac{{ i}}{\hbar} \left[\varphi, \vec{\mathcal{A}}\right]\biggr]\nonumber\\
      &=&- \dfrac{1}{2} \dfrac{\gamma}{\hbar} \left(\vec{v}\times \vec{\mathcal{B}}-\vec{\mathcal{B}}\times \vec{v}\right)+q \vec{\mathcal{E}}
   + q\left(g-\dfrac{q}{\hbar c}\right) {\rm i}\,\left[\varphi,\ \vec{\mathcal{A}}\right].
\end{eqnarray}
Notice the following commutator
\begin{eqnarray}
{\rm i}\,\left[\varphi,\ \vec{\mathcal{A}}\right] &=& {\rm i}\,\left[  \dfrac{\kappa}{r},\ k \frac{\vec{r} \times \vec{S}}{r^2}\right]=0.
\end{eqnarray}
Then one has the spin force as
\begin{eqnarray}
   \vec{F} &=& - \dfrac{1}{2} \dfrac{\gamma}{\hbar} \left(\vec{v}\times \vec{\mathcal{B}}-\vec{\mathcal{B}}\times \vec{v}\right)+q \vec{\mathcal{E}}.
\end{eqnarray}
One may note that the spin force consists two parts: the first part represents the magnetic-like force (i.e., the Lorentz-like force) and the second part represents the electric-like force, i.e.,
\begin{eqnarray}
   \vec{F}&=& \vec{F}_{\rm magnetic}+\vec{F}_{\rm electric},
\end{eqnarray}
with
\begin{subequations}
\begin{eqnarray}
&&\vec{F}_{\rm magnetic}=- \dfrac{1}{2} \dfrac{\gamma}{\hbar} \left(\vec{v}\times \vec{\mathcal{B}}-\vec{\mathcal{B}}\times \vec{v}\right), \nonumber\\
&&\vec{F}_{\rm electric}=q \vec{\mathcal{E}}.
\end{eqnarray}
\end{subequations}

\begin{remark}
Note that for
\begin{eqnarray}
    && \vec{\mathcal{A}} = k \frac{\vec{r} \times \vec{S}}{r^2},\;\;\;\;\; \varphi =  \dfrac{\kappa}{r},
\end{eqnarray}
one can have the ``electric'' field as
\begin{eqnarray}
\vec{\mathcal{E}} &=& -\dfrac{1}{c}
\dfrac{\partial\,\vec{\mathcal{A}}}{\partial\,t}
-\vec{\nabla}\varphi-{\rm i}\,g\left[\varphi,\ \vec{\mathcal{A}}\right]=-\dfrac{1}{c}
\dfrac{\partial\,\vec{\mathcal{A}}}{\partial\,t}-\vec{\nabla}\varphi\nonumber\\
&=& -\vec{\nabla}\varphi=\dfrac{\kappa}{r^3} \vec{r}=\dfrac{\kappa}{r^2} \hat{e}_r,
\end{eqnarray}
which is just the usual electric force. $\blacksquare$
\end{remark}

\subsubsection{Spin Force: The Case of Relativistic Quantum Mechanics}

Secondly, let us consider the case of relativistic quantum mechanics. In such a case, the Hamiltonian $\mathcal{H}$ in Heisenberg' equation (\ref{eq:HeisenbergEq-1}) is chosen as the Dirac-type Hamiltonian, which is given by
\begin{equation}\label{ham2}
\mathcal{H} =c\,\vec{\alpha} \cdot \vec{\Pi} +\beta M c^2 +q\varphi,
\end{equation}
where
             \begin{eqnarray}
              && \vec{\alpha}=\begin{bmatrix}
                  0 & \vec{\sigma} \\
                  \vec{\sigma} & 0
               \end{bmatrix}, \;\;\;\;\; \beta=\begin{bmatrix}
                  \openone & 0 \\
                  0 & -\openone
               \end{bmatrix},
            \end{eqnarray}
are the Dirac matrices, $\openone$ is the $2\times 2$ identity matrix, $\vec{\sigma}=(\sigma_x, \sigma_y, \sigma_z)$ is the vector of Pauli matrices. Because the vector potential $\vec{\mathcal{A}}$ and the scalar potential $\varphi$ commute with Dirac's matrices $\vec{\alpha}$ and $\beta$, i.e.,
\begin{eqnarray}
&&\left[\mathcal{A}_j, \vec{\alpha}\right] =0, \;\;\;  \left[\mathcal{A}_j, \beta\right] =0, \;\;\; (j=x, y, z), \nonumber\\
&& \left[\varphi, \vec{\alpha}\right] =0, \;\;\;  \left[\varphi, \beta\right] =0,
\end{eqnarray}
then one easily derives the ``velocity'' operator as
\begin{eqnarray}
    &&  \vec{v}:=\frac{d \vec{r}}{d t} =
      \frac{1}{{\rm i}\,\hbar}\left[\vec{r}, \mathcal{H}\right] +\dfrac{\partial\,\vec{r}}{\partial\,t} =\frac{1}{{\rm i}\,\hbar}\,c\, \vec{\alpha} \cdot \left[\vec{r}, \vec{p}\right]= c\, \vec{\alpha}.
\end{eqnarray}
However, in special relativity, the mass of the particle is not a constant (i.e., it is velocity-dependent). Due to this reason, we shall attain the ``force'' operator directly through the momentum operator $\vec{\Pi}$ as follows
\begin{eqnarray}\label{eq:F-1}
      \vec{F}&:=& \frac{d \vec{\Pi}}{d t}=\frac{1}{{\rm i}\hbar}\left[\vec{\Pi}, \mathcal{H}\right] +\dfrac{\partial\,\vec{\Pi}}{\partial\,t} \notag\\
   &=& \frac{1}{{\rm i}\hbar} \left\{c\,  \left[\vec{\Pi},\vec{\alpha} \cdot \vec{\Pi} \right]+q\left[\vec{\Pi} ,\varphi\right]\right\}  -\frac{q}{c}\dfrac{\partial \vec{\mathcal{A}}}{\partial t}        \notag \\
   &=&\frac{1}{{\rm i}\hbar} \left\{ -c\, \left[\vec{\alpha} \cdot \vec{\Pi}, \vec{\Pi} \right]+q\left[\vec{p} ,\varphi\right]-\frac{q^2}{c}\left[\vec{\mathcal{A}} ,\varphi\right]\right\}  -\frac{q}{c}\dfrac{\partial \vec{\mathcal{A}}}{\partial t} \notag\\
      &=&{\frac{1}{{\rm i}\hbar}}\left\{-c\, \left[\vec{\alpha} \cdot \vec{\Pi}, \vec{\Pi} \right] {-{\rm i}\hbar}\, q\vec{\nabla} \varphi{+}\frac{q^2}{c}[\varphi ,\vec{\mathcal{A}}]\right\} -\frac{q}{c}\frac{\partial \vec{\mathcal{A}}}{\partial t}\nonumber\\
  &=&{\frac{-1}{{\rm i}\hbar}}\left\{c\, \left[\vec{\alpha} \cdot \vec{\Pi}, \vec{\Pi} \right] {+{\rm i}\hbar}\, q\vec{\nabla} \varphi{-}\frac{q^2}{c}[\varphi ,\vec{\mathcal{A}}]\right\} -\frac{q}{c}\frac{\partial \vec{\mathcal{A}}}{\partial t}.
\end{eqnarray}
If we still designate
\begin{eqnarray}
      && [\Pi_{a},\Pi_{b}]={\rm i} \hbar \, \epsilon_{abc} G_c,
\end{eqnarray}
or
\begin{eqnarray}
   && \vec{\Pi} \times \vec{\Pi} ={\rm i} \hbar \, \vec{G},
\end{eqnarray}
we then have
\begin{eqnarray}
       [\vec{\alpha} \cdot \vec{\Pi}, {\Pi}_z]&=&[\alpha_x \Pi_x+\alpha_y \Pi_y+\alpha_z \Pi_z, {\Pi}_z]\nonumber\\
       &=&\alpha_x [\Pi_x, {\Pi}_z] +\alpha_y [\Pi_y, {\Pi}_z]\nonumber\\
       &=& -{\rm i} \hbar \, \alpha_x G_y + {\rm i} \hbar \, \alpha_y G_x\nonumber\\
       &=& -{\rm i} \hbar \, \left(\vec{\alpha}\times \vec{G}\right)_z,
\end{eqnarray}
which means
\begin{eqnarray}\label{eq:alpha-1}
       [\vec{\alpha} \cdot \vec{\Pi}, \vec{\Pi}]&=& -{\rm i} \hbar \, \left(\vec{\alpha}\times \vec{G}\right).
\end{eqnarray}
After substituting Eq. (\ref{eq:alpha-1}) into Eq. (\ref{eq:F-1}), we have
\begin{eqnarray}
      \vec{F} &=&{\frac{-1}{{\rm i}\hbar}}\left\{  -{\rm i} \hbar \,c\, \left(\vec{\alpha}\times \vec{G}\right) {+{\rm i}\hbar}\, q\vec{\nabla} \varphi{-}\frac{q^2}{c}[\varphi ,\vec{\mathcal{A}}]\right\} -\frac{q}{c}\frac{\partial \vec{\mathcal{A}}}{\partial t}\nonumber\\
      &=& c\, \left(\vec{\alpha}\times \vec{G}\right) {-}\, q\vec{\nabla} \varphi {-\frac{{\rm i}}{\hbar}}\frac{q^2}{c}[\varphi ,\vec{\mathcal{A}}] -\frac{q}{c}\frac{\partial \vec{\mathcal{A}}}{\partial t}\nonumber\\
       &=& \left(\vec{v}\times \vec{G}\right) +q \left\{-\frac{1}{c}\dfrac{\partial\,\vec{\mathcal{A}}}{\partial\,t}
          -\vec{\nabla}\,\varphi
          -\dfrac{q}{c}\dfrac{{ i}}{\hbar} \left[\varphi, \vec{\mathcal{A}}\right]\right\}.
\end{eqnarray}
In this case, similarly one has
\begin{eqnarray}
   \vec{F}   &=& - \dfrac{1}{2} \dfrac{\gamma}{\hbar} \left(\vec{v}\times \vec{\mathcal{B}}-\vec{\mathcal{B}}\times \vec{v}\right)+q \vec{\mathcal{E}},
\end{eqnarray}
which shares the same form with the non-relativistic case. The only difference is that $\vec{v}$ commutes with $\vec{\mathcal{B}}$ in the relativistic case (i.e., $\left[\vec{v}, \vec{\mathcal{B}}\right]=0$), while they do not commute in the non-relativistic case (i.e., $\left[\vec{v}, \vec{\mathcal{B}}\right] \neq 0$).

\onecolumngrid